\documentclass[smallextended]{svjour3}
\usepackage{graphicx}
\usepackage{amsmath}
\usepackage{amssymb}
\usepackage{tikz}
\usepackage{multicol}
\usepackage{hyperref}
\reversemarginpar%
\usepackage[ruled, vlined]{algorithm2e}
\usepackage[giveninits=true,doi=true, url=false, isbn=false, eprint=false,
backend=biber,
hyperref=true, maxnames=15,maxbibnames=9,maxcitenames=2]{biblatex}

\DeclareFieldFormat{doi}{%
    \small%
  \mkbibacro{DOI}\addcolon\space
  \ifhyperref
    {\href{https://doi.org/#1}{\nolinkurl{http://doi.org/#1}}}
    {\nolinkurl{http://doi.org/#1}}}
    
\usetikzlibrary{fit, arrows, positioning, shapes}
\usepackage{pgfplotstable}
\usepackage{booktabs}
\usepackage{array}
\usepackage{colortbl}

\pgfplotsset{compat=1.18}

\pgfkeys{/pgf/number format/.cd,sci,std={-1:1},sci generic={mantissa sep={\times},exponent={10^{#1}}}}
\def\num#1{\pgfkeys{/pgf/number format/.cd,std, precision=4}\pgfmathprintnumber{#1}}
\def\rnum#1{\pgfkeys{/pgf/number format/.cd,std, precision=2, zerofill}\pgfmathprintnumber{#1}}

\title{Determination of Navier's slip parameter and the inflow velocity using variational data assimilation}

\titlerunning{Determination of Navier's slip parameter using data assimilation}

\author{Alena Jarolímová         \and
        Jaroslav Hron
}

\institute{A. Jarolímová (ORCID: 0000-0003-2704-2211) \at
              Charles University, Faculty of Mathematics and Physics, Mathematical Institute, Sokolovská 83, 186 75 Prague, Czech Republic\\
              \email{jarolimova@karlin.mff.cuni.cz}
           \and
           J. Hron (ORCID: 0000-0001-5862-2353)\at
              Charles University, Faculty of Mathematics and Physics, Mathematical Institute, Sokolovská 83, 186 75 Prague, Czech Republic \\
              \email{hron@karlin.mff.cuni.cz}
}

\date{Received: date / Accepted: date}

\newcommand{\vin}{\mathbf{v}_{\text{in}}}
\newcommand{\bndry}[1]{\Gamma_\text{#1}}
\newcommand{\dx}[1]{\,\text{d}{#1}}
\newcommand{\thet}[1]{\theta_{\text{#1}}}

\begin{document}

\maketitle

\begin{abstract}
One of the crucial aspects of patient-specific blood flow simulations is to specify material parameters and boundary conditions. 
The choice of boundary conditions can have a substantial impact on the character of the flow. While no-slip is the most popular wall boundary condition, some amount of slip, which determines how much fluid is allowed to flow along the wall, might be beneficial for better agreement with flow patterns in medical images.   
However, even if one assumes the simple Navier's boundary conditions on the wall, in which the relationship between tangential components of the normal traction and the velocity is linear, the determination of the specific value of the slip parameter is often difficult. 

In this work, we present and test an optimal control method to estimate Navier's slip parameter on the wall and the velocity profile at the inlet using artificially generated flow domain and flow data. 
The results show that it is possible to recover the flow patterns and Navier's slip parameter by using sufficiently accurate discretization even from data containing a substantial amount of noise.

\begin{acknowledgements}
    The work was supported by the Czech Science Foundation, project 23-05207S. In addition, Alena Jarolímová was supported by the Grant Agency of Charles University, grant nr. 282222 and project SVV-2023-260711.

    We thank Josef Málek and Patrick Farrell for their help and insightful discussions.
\end{acknowledgements}
\end{abstract}

\section{Introduction}
Performing patient-specific simulations can be a helpful tool for diagnostic purposes in clinical practice. 
However, the choice of an appropriate model can have crucial consequences on the results of the simulation. 
More specifically, it is necessary to choose all the parameters in the model precisely enough to reflect the complex biological processes of a particular patient. 
Since some of the model parameters are patient-specific and hard or even impossible to measure directly, one can use data from medical imaging to compensate for the lack of such information.

The phase-contrast magnetic resonance imaging (PC MRI) technique can provide 4D data of the velocity field (3D space + time). 
However, its spatial and temporal resolution is limited compared to the resolution of computational fluid simulations, and it also contains a substantial amount of noise. 
Therefore, using such data to determine the unknown quantities directly may be unreliable. 
Instead, it is possible to use the data to recover the model parameters by solving an inverse problem. 

Several data assimilation methods have been used to solve inverse problems in cardiovascular settings, some of which are discussed, for example, in \cite{Bertagna2014}.
Kalman filter based approach, which is usually used to estimate lumped parameters of the model as done in \cite{Nolte2019, Pant2014}, represents one of the popular approaches.
Another option, which we follow in this work, is to reformulate the problem in an optimal control setting, minimizing an error functional constrained to the initial- and boundary-value problem governed by a system of partial differential equations (PDE), describing steady or unsteady fluid flows.
For example, various options for solving optimal control problems were discussed in \cite{Gunzburger2000} and \cite{Herzog2010}. 
A PDE-constrained optimization can be reformulated as one large system expressing the Karush--Kuhn--Tucker optimality conditions (KKT). 
The complete KKT system based on a steady-state flow problem is solved using Picard's and Newton's method in \cite{DElia2012}. 
In recent years, sequential quadratic programming algorithms gained popularity for solving the KKT systems in cardiovascular settings \cite{Guerra2014, Guerra2018, Tiago2017}.
The KKT system for the steady state can also be solved iteratively as done in \cite{Koltukluoglu2018}.  
For time-dependent problems, the KKT system is usually transformed into an iterative process, see for example \cite{Funke2019} or \cite{juniper2022}.

A crucial part of any data assimilation method is the choice of a fluid flow model. 
While the choice of the Navier--Stokes fluid model seems reasonable for the description of flow in aortic vessels (as non-Newtonian variants where the viscosity depends on the shear rate have rather minor effects), it is known, see for example \cite{Hron2008, Chabiniok2022}, that the choice of boundary conditions can substantially impact the blood flow patterns. 
Even though the no-slip boundary condition is the most popular wall boundary condition in the cardiovascular modelling community, some partial slip along the wall can be observed in certain experiments \cite{nubar}. Such choice of boundary condition can also partially compensate the errors caused by the inaccuracy of image segmentation, which has been addressed in \cite{Nolte2019} using a special choice of boundary condition and in \cite{juniper2022} by treating the shape of the geometry as an unknown determined using the optimal control. 

We are now in the position to formulate our basic question motivating this study: Assuming that the Navier--Stokes fluid slips on the wall according to Navier's slip boundary condition, can we determine the slip parameter using 4D MRI data? 
In addition, can we also use the same set of 4D MRI data to determine the velocity of the inlet?
In this context, it is worth mentioning that, in their recent study \cite{MR1}, Málek and Rajagopal found criteria (expressed in terms of macroscopic quantities such as the pressure gradient, the volumetric flow rate, the viscosity, the density and geometric dimensions) stating/determining when the Navier--Stokes fluid in a five simple unidirectional flows exhibits no-slip. Even more, in their following study \cite{MR2}, assuming that the flow rate is bigger than the "critical" flow rate corresponding to the no-slip boundary condition and assuming Navier's slip boundary condition, they developed a methodology to determine the value of the slip parameter. 

In this work, we test the performance of an adjoint-based variational assimilation method for 4D PC-MRI data. 
We aim to assess whether Navier's slip boundary condition is more suitable for modelling blood in large arteries. 
Moreover, we present numerical experiments to verify the reliability of the approach. In particular, the developed code is able to determine the slip parameter for simple flows in accordance with the methodology developed in \cite{MR2}.
For the sake of simplicity, we restrict ourselves to steady flows. 
Our main objective is to develop an efficient method supported by its careful testing that could have the potential to be used on actual patient-specific data and geometry. 

The structure of this work is the following. In Section \ref{sec:methodology}, we present our approach based on a general optimal control methodology as described in  \cite{Bertagna2014}. We also formulate the underlying boundary-value problem governed by the Navier-Stokes equations. The setup of the numerical experiments is presented in Section \ref{sec:setup}, and the results are discussed in Section \ref{sec:experiments}. Section \ref{sec:conclusion} consists of a conclusion and discussion of future work.
 
\section{Methodology}\label{sec:methodology}

\subsection{Fluid flow model}\label{sec:fluid flow model}

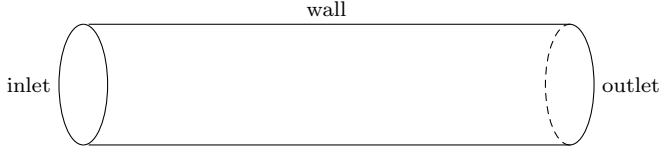
\begin{figure}
    \centering
    \begin{tikzpicture}[>=latex,shorten >=2pt,shorten <=2pt,shape aspect=1, scale=0.8]
\begin{scope}
    \coordinate (A) at (0mm,10mm);
    \coordinate (B) at (80mm,10mm);
    \coordinate (C) at (80mm,-10mm);
    \coordinate (D) at (0mm,-10mm);
    \draw [] (A) -- node[midway, above] {wall} (B) -- ++(1mm,0mm);
    \draw [] (D) -- (C) -- ++(1mm,0mm);
    \draw (B) arc [x radius = 4mm, y radius = 10mm, start angle=90, end angle=-90] node[midway, right] {outlet} (C);
    \draw (D) arc [x radius =-4mm, y radius = 10mm, start angle=-90, end angle=90] node[midway, left] {inlet} (A);
    \draw (A) arc [x radius =4mm, y radius = 10mm, start angle=90, end angle=-90] (D);
  \draw [densely dashed] (C) arc [x radius = -4mm, y radius = 10mm, start angle=-90, end angle=90] (B);
\end{scope}
\end{tikzpicture}
    \caption{A scheme of the domain with the boundary types.}
    \label{fig:geometry}
\end{figure}

Since we plan to apply the method to blood flow in large arteries, we consider a tubular domain $\Omega\subset\mathbb{R}^3$ representing an artery containing no bifurcations.
The boundary of the domain can be split into three parts: inlet $\bndry{in}$, outlet $\bndry{out}$ and wall $\bndry{wall}$, see Figure \ref{fig:geometry}. 

We consider the steady-state incompressible Navier--Stokes equations for the velocity $\mathbf{v}: \Omega\to\mathbb{R}^3$ and the pressure $p: \Omega\to\mathbb{R}$.
A Dirichlet boundary condition is prescribed at the inlet $\bndry{in}$ using a given function $\vin:\bndry{in}\to\mathbb{R}^3$ and a directional do-nothing boundary condition is prescribed at the outlet $\bndry{out}$. 
At the wall $\bndry{wall}$, Navier's slip boundary condition is used to describe the behaviour of the fluid in the tangential direction along the wall, accompanied by the impermeability condition in the normal direction.
Together, the governing equations describing the underlying boundary-value problem read

\begin{align}\label{strong form NS}
    \begin{split}
        \rho(\nabla\mathbf{v})\mathbf{v} - \text{div}\,\mathbb{T}(\mathbf{v}, p) = \mathbf{0} \quad &\text{in }\Omega,\\
        \mathbb{T}(\mathbf{v}, p) = -p\,\mathbb{I} + 2\mu\mathbb{D}(\mathbf{v}) \quad &\text{in }\Omega,\\
        \text{div}\,\mathbf{v}=0 \quad &\text{in }\Omega,\\
        \mathbf{v} = \vin \quad &\text{on }\bndry{in},\\
        (\mathbb{T}(\mathbf{v}, p)\,\mathbf{n})\cdot\mathbf{n} = \frac{1}{2}\,\rho\,(\mathbf{v}\cdot\mathbf{n})_-^2 \quad \text{and} \quad \mathbf{v}_\text{t} = \mathbf{0} \quad &\text{on }\bndry{out},\\
        \theta\,\mathbf{v}_{\text{t}} + \gamma_*(1-\theta)(\mathbb{T}(\mathbf{v}, p)\,\mathbf{n})_\text{t} = \mathbf{0}  \quad \text{and} \quad \mathbf{v}\cdot\mathbf{n} = 0 \quad &\text{on }\bndry{wall},
    \end{split}
\end{align}
where $\mathbb{T}$ is the Cauchy stress tensor, $\mathbb{D}(\mathbf{v})=\frac{1}{2}(\nabla\mathbf{v}+(\nabla\mathbf{v})^T)$ is the symmetric part of the velocity gradient, $\rho>0$ and $\mu>0$ are constant density and viscosity prescribed to match the properties of human blood and $\mathbf{n}$ is the vector normal to a given part of the boundary. 
Moreover, for $x\in\mathbb{R}$, we denote the positive and negative part by $x_+=\text{max}\{x, 0\}$ and $x_-=\text{min}\{x, 0\}$ respectively. 
Similarly, for $\mathbf{u}\in\mathbb{R}^3$ we denote the normal component $\mathbf{u}_\text{n}=(\mathbf{u}\cdot\mathbf{n})\,\mathbf{n}$ and the tangential component $\mathbf{u}_\text{t}=\mathbf{u}-\mathbf{u}_\text{n}$. 
The parameter $\theta\in [0,1]$ defines how much slip can be permitted along the wall, where $\theta=0$ and $\theta=1$ correspond to free slip and no-slip, respectively. 
The parameter $\gamma_*\in (0,\infty)$ is set experimentally, and its optimal choice has been discussed in \cite{Chabiniok2021}.

The value of $\theta$ and the velocity profile $\vin$ defining the flow at the inlet of the computational geometry are unknown patient-specific parameters which we want to estimate.
Let us denote them as $\mathbf{m}=(\theta, \vin)$. 
In the context of data assimilation, the unknown parameters $\mathbf{m}$ are often called control variables.
Using this simplified notation, we can denote the equations defined in \eqref{strong form NS} by $F(\mathbf{w}, \mathbf{m}) = 0$, where $\mathbf{w}=(\mathbf{v}, p)$ is the solution for given control variables $\mathbf{m}=(\theta, \vin)$.

\subsection{Variational data assimilation}

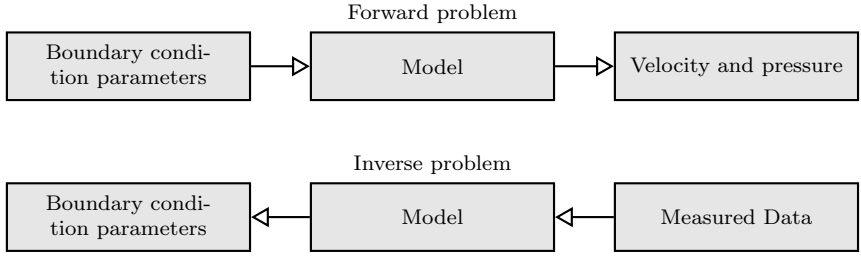
\begin{figure}
    \centering

\tikzstyle{box} = [rectangle, draw, fill=black!10, node distance=1cm, text width=3cm, text centered, minimum height=3em, thick]
\tikzstyle{line} = [draw, -open triangle 60, thick]

    \begin{tikzpicture}[auto]
  \node[box] (f1) {Boundary condition parameters};
  \node[box, right of=f1, xshift=3cm, label={Forward problem}] (f2) {Model};
  \node[box, right of=f2, xshift=3cm] (f3) {Velocity and pressure};
    
  \path[line] (f1) --  (f2);
  \path[line] (f2) --  (f3);

  \node[box, below of=f3, yshift=-1cm] (i1) {Measured Data};
  \node[box, below of=f2, yshift=-1cm, label={Inverse problem}] (i2) {Model};
  \node[box, below of=f1, yshift=-1cm] (i3) {Boundary condition parameters};
    
  \path[line] (i1) --  (i2);
  \path[line] (i2) --  (i3);

  \end{tikzpicture}
    
    \caption{A diagram of the difference between forward and inverse problem}
    \label{fig:inverse_problem_diagram}
\end{figure}

The model contains several parameters that need to be provided to solve the system \eqref{strong form NS} for a specific patient. 
It is possible to use tabulated values for density $\rho$ and viscosity $\mu$ of human blood or even perform experiments to obtain more accurate values for the given patient. 
On the other hand, the inlet velocity $\vin$ and the slip parameter $\theta$ have to be determined by solving an inverse problem instead, as illustrated in Figure \ref{fig:inverse_problem_diagram}.

The variational assimilation is a technique to determine unknown parameters of a model from physical measurements. 
For real patient-specific cases, we want to use 4D phase-contrast magnetic resonance images (4D PC-MRI), which contain information about the velocity field in 3D space and time.
The data received using this technique have low resolution (around 2 mm) and contain substantial noise. 
Since we are solving a steady problem, we would use only a single temporal slice of the 4D PC-MRI data.
Therefore, we obtain a vector-valued function defined over the whole domain $\Omega$,
which we denote by $\mathbf{d}_\text{MRI}\in (L^{\infty}(\Omega))^3$. 
However, in this work, we will use only artificially generated data to be able to compare the results with the ground truth. These artificially generated data are obtained by solving \eqref{strong form NS} with apriori chosen data $\theta$ and $\vin$ and then extrapolated to a coarser mesh and convoluted with random noise.

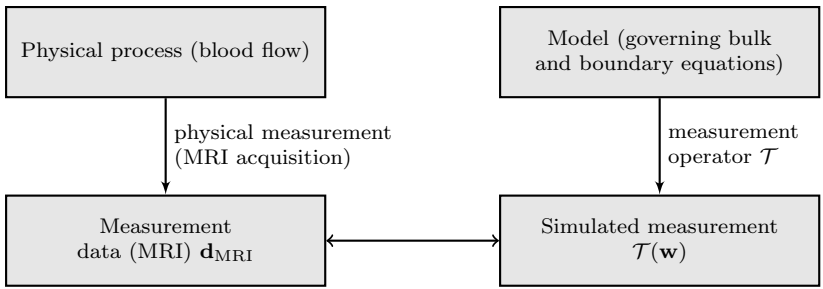
\begin{figure}
    \centering

\tikzstyle{box} = [rectangle, draw, fill=black!10, node distance=25mm, text width=40mm, text centered, minimum height=4em, thick]
\tikzstyle{line} = [draw, -latex', thick]

    \begin{tikzpicture}[auto]
  \node[box] (proc) {Physical process (blood flow)};
  \node[box, below of=proc] (measure) {Measurement data (MRI) $\mathbf{d}_{\text{MRI}}$};
  \node[box, right of=measure, xshift=4cm] (simul) {Simulated measurement\\$\mathcal{T}(\mathbf{w})$};
  \node[box, above of=simul] (model) {Model (governing bulk and boundary equations)};

  \path[line] (proc) --  node [text width=30mm,midway,right] {physical measurement (MRI acquisition)} (measure);
  \path[line, <->] (measure) -- (simul);
  \path[line] (model) --  node [text width=20mm,midway,right] {measurement operator $\mathcal{T}$} (simul);
  \end{tikzpicture}
    \caption{A scheme of variational data assimilation}
    \label{fig:diagram_T}
\end{figure}

The governing boundary-value problem $F(\mathbf{w}, \mathbf{m})=0$, see equation \eqref{strong form NS}, is used to describe the physical process of the flow.
Since we want to compare the solution $\mathbf{w}$ to the measured data $\mathbf{d}_\text{MRI}$, we define an operator $\mathcal{T}$, which describes the measurement process. Then, we can apply the operator to $\mathbf{w}$ to obtain simulated measurements $\mathcal{T}(\mathbf{w})$ as shown in Figure \ref{fig:diagram_T}.

The main idea of variational assimilation is to minimize a certain distance of the simulated measurements $\mathcal{T}(\mathbf{w})$ from the actual measurements $\mathbf{d}_{\text{MRI}}$ with respect to the unknown model parameters $\mathbf{m}$. 
The governing equations defined by $F(\mathbf{w}, \mathbf{m})=0$ act as a constraint to the optimization establishing the relation between $\mathbf{m}$ and $\mathbf{w}$. 
Let us denote the error functional as 
\begin{equation}\label{eq: J}
    \mathcal{J}(\mathbf{w}) = \frac{1}{2\ell^3}||\mathcal{T}(\mathbf{w})-\mathbf{d}_{\text{MRI}}||_{L^2(\Omega)}^2,
\end{equation}
where $\ell$ denotes a characteristic length of the problem geometry.\footnote{Since our computations are done in SI units, a characteristic length $\ell=0.01\text{ m}$ has been introduced to normalize the value of the error functional and unify the units in $\mathcal{J}_R$ to $\text{m}^2\cdot\text{s}^{-2}$.}
The optimization problem then can be stated as

\begin{align}
         \min_{(\mathbf{w},\mathbf{m})\in \mathcal{A}} \mathcal{J}(\mathbf{w})\quad\textrm{ where } \mathcal{A}:= \{(\mathbf{w}, \mathbf{m})\in W\times M; F(\mathbf{w}, \mathbf{m}) = 0\},
\end{align}
for $W = [H^1(\Omega)]^3\times L^2(\Omega)$ and $M = [0, 1]\times H^1(\bndry{in})$.
In other words, we need to solve a PDE-constrained optimization problem.

Since we are working with a steady case, we chose the measurement operator simply as
\begin{equation}\label{measurement operator}
    \mathcal{T}(\mathbf{w}) = \mathbf{v}.
\end{equation}

The error functional $\mathcal{J}(\mathbf{w})$ is often combined with a Tikhonov regularization $\mathcal{R}(\mathbf{m})$:
\begin{equation}\label{regularized J}
    \mathcal{J}_R(\mathbf{w}, \mathbf{m})=\mathcal{J}(\mathbf{w})+\mathcal{R}(\mathbf{m}).
\end{equation}
The regularization of a given parameter is usually a norm of the parameter or its distance from some prescribed value. 
A correct choice of regularization can help the parameters to be in the desired function space and make the problem well-posed.
It also helps to resolve the situation where multiple solutions would be viable otherwise.

In our approach, the regularization $\mathcal{R}(\mathbf{m})$ consists of two terms:
\begin{equation}
    \mathcal{R}(\mathbf{m})=\mathcal{R}_{\alpha}(\mathbf{m})+\mathcal{R}_{\beta}(\mathbf{m}).
\end{equation}
The first term is designed to penalize the gradient of inlet velocity $\vin$ in the tangential direction of the inlet, and its purpose is to encourage the algorithm to select a smooth inlet velocity profile, i.e.,
\begin{equation}\label{regularization alpha}
    \mathcal{R}_{\alpha}(\mathbf{m})=\frac{\alpha}{2}||\nabla\vin||^2_{L^2(\bndry{in})}.
\end{equation}
The second term also corresponds to $\vin$, and it is defined as a $L^2$ norm of the distance of $\vin$ from a given velocity profile $\mathbf{v}_\text{analytic}$ over the inlet boundary $\bndry{in}$. 
 \begin{equation}\label{regularization beta}
     \mathcal{R}_{\beta}(\mathbf{m})=\frac{\beta}{2\ell^2}||\vin-\mathbf{v}_{\text{analytic}}(V,\theta)||^2_{L^2(\bndry{in})}
 \end{equation}
The velocity profile $\mathbf{v}_\text{analytic}$ is the analytical solution to the steady flow in a straight tube with a given average inflow velocity $V$ and Navier's slip on the walls given by parameter $\theta$, see \cite{Hron2008}, and it represents our estimate of how the inlet velocity profile might look like:
\begin{equation}\label{analytic profile}
\mathbf{v}_\text{analytic}(V,\theta) = V\frac{4\mu\gamma_*(1-\theta)R+2\theta(R^2-r^2)}{4\mu\gamma_*(1-\theta)R+\theta R^2}\,\mathbf{n}.
\end{equation}
Here, $R$ is the radius of the inlet, $r$ is the distance from the centre axis, and $V$ is computed from $\mathbf{d}_{\text{MRI}}$ as the average inflow through the inlet. 
The motivation behind $\mathcal{R}_{\beta}$ is to increase the link between $\theta$ and $\vin$ in the optimization process. 
It also acts as a counterweight to $\mathcal{R}_{\alpha}$, which enforces $\vin$ to be a constant function.

Substituting the relation \eqref{eq: J} for $\mathcal{J}$ and \eqref{regularization alpha}-\eqref{regularization beta} for $\mathcal{R}$ into the definition of the regularized error functional \eqref{regularized J} then gives us
\begin{multline}\label{error functional}
\mathcal{J}_R\left(
(\mathbf{v},p), (\theta,\vin)
\right) = \frac{1}{2\,\ell^3}||\mathbf{v}-\mathbf{d}_\text{MRI}||^2_{L^2(\Omega)}
    + \frac{\alpha}{2}||\nabla\vin||^2_{L^2(\bndry{in})}\\
    + \frac{\beta}{2\,\ell^2} ||\vin-\mathbf{v}_{\text{analytic}}(V, \theta)||^2_{L^2(\bndry{in})}
\end{multline}
The weights $\alpha$ and $\beta$ have to be selected carefully so that the regularization terms $\mathcal{R}(\mathbf{m})$ in \eqref{error functional} are in good balance with the first term $\mathcal{J}(\mathbf{w})$.
The choice of the weights $\alpha$ and $\beta$ is discussed in more detail in section \ref{subsec: reg}.

\subsection{Numerical solution}

In order to solve the PDE-constrained optimization problem
\begin{equation}\label{pde constrained}
\min_{(\mathbf{w},\mathbf{m})\in \mathcal{A}} \mathcal{J_R}(\mathbf{w}, \mathbf{m})\quad\textrm{ where } \mathcal{A}:= \{(\mathbf{w}, \mathbf{m})\in W\times M; F(\mathbf{w}, \mathbf{m}) = 0\},   
\end{equation}
the constraint can be removed by enforcing it implicitly in the functional.
Since $\mathbf{w}$ is the solution of model $F(\mathbf{w},\mathbf{m}) = 0$,  $\mathbf{w}$ can be considered an implicit function of the control variables as $\mathbf{w}(\mathbf{m})$, assuming the implicit function theorem applies. 
Therefore, we can define a reduced functional as
\begin{equation}\label{reduced functional def}
    \hat{\mathcal{J}_R}(\mathbf{m}) := \mathcal{J}_R(\mathbf{w}(\mathbf{m}),\mathbf{m}) = \mathcal{J}(\mathbf{w}(\mathbf{m}))+\mathcal{R}(\mathbf{m}). 
\end{equation}
The reduced functional $\hat{\mathcal{J}_R}(\mathbf{m})$ already contains the PDE constraint implicitly, which means that the original problem \eqref{pde constrained} is equivalent to an unconstrained optimization problem
\begin{equation}
    \min_{\mathbf{m}\in M} \hat{\mathcal{J}_R}(\mathbf{m}).
\end{equation}

There are many iterative algorithms to solve an unconstrained optimization problem. 
Since the model $F(\mathbf{w}, \mathbf{m}) = 0$ has to be solved in each iteration to evaluate the reduced functional $\mathcal{J}_R(\mathbf{w}(\mathbf{m}^i),\mathbf{m}^i)$, it is reasonable to choose an algorithm that can minimize the error functional in an efficient way. 
Therefore, a gradient-based optimization algorithm is used since it is known to converge in fewer iterations than gradient-free algorithms. 
The iterative process can then be summarized as follows:

\begin{algorithm}[H]
    \KwData{$\mathbf{m}^0$}
    \For{$i = 0, 1, 2 \dots$}{
        \eIf{stopping criterion is fulfilled}{
        stop the optimization
    }{
        find $\mathbf{w}(\mathbf{m}^i)$ by solving the PDE equations with $\mathbf{m}^i$\;
        evaluate $\hat{\mathcal{J}_R}(\mathbf{m}^i) = \mathcal{J}_R(\mathbf{w}(\mathbf{m}^i), \mathbf{m}^i)$\;
        compute $\frac{\partial\hat{\mathcal{J}_R}}{\partial\mathbf{m}}(\mathbf{m}^i)$\;
        determine $\mathbf{m}^{i+1}$ using the chosen optimization algorithm\;
      }
    }
    \caption{Optimization iterative process}
\end{algorithm}

The optimization algorithm has to be able to fulfil the constraint on the control variable $\theta\in[0,1]$. Therefore, we chose the L-BFGS-B algorithm \cite{Byrd1996}, which can support box constraints. 
The goal of the algorithm is to find a minimum of a nonlinear function $f(\mathbf{x}):\mathbb{R}^n\to\mathbb{R}$ subject to $\mathbf{l}\leq\mathbf{x}\leq\mathbf{u}$, where the vectors $\mathbf{l}$ and $\mathbf{u}$ are the lower and upper bound for the variable $\mathbf{x}$, respectively. The algorithm is an extension of a quasi-Newtonian unconstrained optimization algorithm BFGS, an iterative method based on approximation of the Hessian matrix. 

Each optimization iteration requires solving not only the PDE represented by $F(\mathbf{m},\mathbf{w})=0$ but also computing the gradient $\frac{\partial\hat{\mathcal{J}_R}}{\partial\mathbf{m}}(\mathbf{m}^i)$, which can be done using the adjoint equations.
The first step in deriving the adjoint equations is to differentiate the definition of the reduced functional \eqref{reduced functional def} with respect to the control variables $\mathbf{m}$.
\begin{equation}\label{chain rule J}
    \frac{\partial\hat{\mathcal{J}_R}}{\partial \mathbf{m}} = \frac{\partial\mathcal{J}_R}{\partial \mathbf{w}}\left(\frac{\partial\mathbf{w}}{\partial \mathbf{m}}\right) + \frac{\partial\mathcal{J}_R}{\partial\mathbf{m}}.
\end{equation}
Since the operator $\mathbf{w}(\mathbf{m})$ has been defined implicitly using the governing equations $F(\mathbf{m},\mathbf{w})=0$, we cannot explicitly evaluate $\frac{\partial\mathbf{w}}{\partial \mathbf{m}}$. 
However, we can differentiate the relation $F(\mathbf{w}(\mathbf{m}),\mathbf{m})=0$ to obtain a relation for $\frac{\partial\mathbf{w}}{\partial \mathbf{m}}$:
\begin{equation}\label{chain rule F}
    \frac{\partial F}{\partial\mathbf{w}}\left(\frac{\partial\mathbf{w}}{\partial\mathbf{m}}\right)+\frac{\partial F}{\partial\mathbf{m}} = 0.
\end{equation}
Let us define an adjoint operator $\frac{\partial F}{\partial \mathbf{w}}^*$ as
\begin{equation}\label{adjoint operator}
    \left\langle\left(\frac{\partial F}{\partial \mathbf{w}}\right)^*(\lambda), \mathbf{y}\right\rangle = \left\langle\lambda, \frac{\partial F}{\partial \mathbf{w}}(\mathbf{y})\right\rangle\quad\forall\lambda,\mathbf{y}\in Y,
\end{equation}
for the appropriate function space $Y$.
Then, we consider an adjoint problem 
\begin{equation}\label{adjoint problem}
    \left\langle\left(\frac{\partial F}{\partial \mathbf{w}}\right)^*(\lambda), \mathbf{y}\right\rangle = \frac{\partial\mathcal{J}_R}{\partial \mathbf{w}}(\mathbf{y})\quad\forall \mathbf{y}\in Y.
\end{equation}
for an adjoint variable $\lambda\in Y$.
Combining relations \eqref{chain rule J}, \eqref{adjoint problem}, the definition of the adjoint operator \eqref{adjoint operator} and the relation \eqref{chain rule F}, we arrive at
\begin{align}\label{gradient formula derivation}
    \begin{split}
    \frac{\partial\hat{\mathcal{J}_R}}{\partial \mathbf{m}} &= \frac{\partial\mathcal{J}_R}{\partial \mathbf{w}}\left(\frac{\partial\mathbf{w}}{\partial \mathbf{m}}\right) + \frac{\partial\mathcal{J}_R}{\partial\mathbf{m}}\\
    &= \left\langle\left(\frac{\partial F}{\partial \mathbf{w}}\right)^*(\lambda),\frac{\partial\mathbf{w}}{\partial \mathbf{m}}\right\rangle + \frac{\partial\mathcal{J}_R}{\partial\mathbf{m}}\\
    &= \left\langle\lambda,\frac{\partial F}{\partial \mathbf{w}}\left(\frac{\partial\mathbf{w}}{\partial \mathbf{m}}\right)\right\rangle + \frac{\partial\mathcal{J}_R}{\partial\mathbf{m}}\\
    &= -\left\langle\lambda,\frac{\partial F}{\partial \mathbf{m}}\right\rangle + \frac{\partial\mathcal{J}_R}{\partial\mathbf{m}}.
    \end{split}
\end{align}
To conclude, the computation of the gradient requires one solve of the adjoint problem \eqref{adjoint problem}. Then, the adjoint variable $\lambda$ can be used to evaluate the gradient using the relation 
\begin{equation}\label{gradient relation}
    \frac{\partial\hat{\mathcal{J}_R}}{\partial \mathbf{m}} = -\left\langle\lambda,\frac{\partial F}{\partial \mathbf{m}}\right\rangle + \frac{\partial\mathcal{J}_R}{\partial\mathbf{m}}
\end{equation}
derived in \eqref{gradient formula derivation}. 

\subsection{Weak formulation and finite element discretization}

The problem \eqref{strong form NS} is solved using the finite element method. 
The domain $\Omega$ is approximated by a polyhedral domain $\Omega_h$. The impermeability condition on the wall and zero tangential part of the velocity condition at the outflow are implemented using the non-symmetric Nitsche method with penalization \cite{Burman2012, Nitsche1971}. The choice of the particular variant of the Nitsche method was discussed in \cite{Chabiniok2021}. Denoting $\beta_*>0$ as the penalization parameter and $h>0$ as the cell diameter of the mesh, the weak formulation reads
\begin{align}\label{weak_form}
    \begin{split}
    &\text{Find }(\mathbf{v}-\vin^*, p)\in V_h\times P_h \text{ such that}\\
        &\int_\Omega\rho(\nabla\mathbf{v})\mathbf{v}\cdot\mathbf{\phi}\dx{x}
        +\int_\Omega\mathbb{T}(\mathbf{v}, p):\nabla\mathbf{\phi}\dx{x}
        +\int_\Omega q\,\text{div}\,\mathbf{v}\dx{x}\\
        &+\int_{\bndry{wall}}\frac{\theta}{\gamma_*(1-\theta)}\mathbf{v}_\text{t}\cdot\mathbf{\phi}_\text{t}\dx{s} -\int_{\bndry{wall}}(\mathbb{T}(\mathbf{v}, p)\,\mathbf{n})_\text{n}\cdot\mathbf{\phi}_\text{n}\dx{s}\\
        &+\int_{\bndry{wall}}\mathbf{v}_{\text{n}}\cdot(\mathbb{T}(\phi, q)\,\mathbf{n})_\text{n}\dx{s}+\frac{\beta_*\mu}{h}\int_{\bndry{wall}}\mathbf{v}_{\text{n}}\cdot\mathbf{\phi}_{\text{n}}\dx{s}\\
        &-\int_{\bndry{out}}\frac{1}{2}\,\rho\,(\mathbf{v}\cdot\mathbf{n})_-^2\cdot(\mathbf{\phi}\cdot\mathbf{n})\dx{s} -\int_{\bndry{out}}(\mathbb{T}(\mathbf{v}, p)\,\mathbf{n})_\text{t}\cdot\mathbf{\phi}_\text{t}\dx{s}\\
        &+\int_{\bndry{out}}\mathbf{v}_{\text{t}}\cdot(\mathbb{T}(\phi, q)\,\mathbf{n})_\text{t}\dx{s}+\frac{\beta_*\mu}{h}\int_{\bndry{out}}\mathbf{v}_{\text{t}}\cdot\mathbf{\phi}_{\text{t}}\dx{s}  = 0 \\
        &\text{for all }(\phi, q) \in V_h\times P_h,
    \end{split}
\end{align}
where $V_h\subset\{ \mathbf{v}\in H^1(\Omega_h), \mathbf{v} = 0 \text{ on }\bndry{in} \}$, $P_h\subset L^2(\Omega_h)$ and $\vin^*$ is the finite element extension of $\vin$ to the whole $\overline{\Omega_h}$.
The unit boundary normal $\mathbf{n}$ is computed as the continuous vertex normal as defined in \cite{Chabiniok2021}. 

The discrete $V_h, P_h$ spaces are defined using either the MINI or $P_1/P_1$ finite elements. 
The advantage of these elements is that there are significantly fewer degrees of freedom than, for example, the Taylor--Hood element used in \cite{Chabiniok2021,Chabiniok2022}. 
On the other hand, the $P_1/P_1$ element is not inf-sup stable by itself, so we have to add appropriate stabilization. 
In this case, we use the interior penalty stabilization \cite{Burman2007}, which is achieved by adding the following terms to the weak formulation:
\begin{align}\label{ip stabilization}
\alpha_v\sum_{K\in\mathcal{F}}\int_K \rho h^2 [\nabla\mathbf{v}]\cdot[\nabla\mathbf{\phi}]\dx{s} +        \alpha_p\sum_{K\in\mathcal{F}}\int_K \frac{h^2}{\rho} [\nabla p]\cdot[\nabla q]\dx{s},
\end{align}
where $\mathcal{F}$ denotes the set of all interior faces in $\Omega_h$ and $[.]$ denotes the jump of any quantity across given face.
The interior penalty stabilization is used because of the simplicity of the implementation compared to, for example, GLS or SUPG stabilization.
The effect of stabilization in optimal control was previously studied in \cite{Abraham2004} for GLS stabilization and in \cite{Collis2002} for SUPG stabilization.
The adjoint equations for the weak formulation \eqref{weak_form} with stabilization \eqref{ip stabilization} are included in the appendix.

\section{Setup of the numerical experiments}\label{sec:setup}
The numerical experiments are designed to test whether the method can robustly reconstruct the control variables $\vin$ and $\theta$ from given subsampled, noisy data in various geometries.  

\subsection{Geometries}
\begin{figure}
    \centering
    \begin{minipage}{0.63\textwidth}
        \centering
        \includegraphics[width=0.85\textwidth]{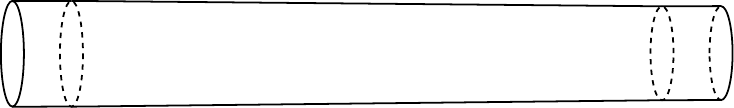}\\
        Straight tube
        \includegraphics[width=0.98\textwidth]{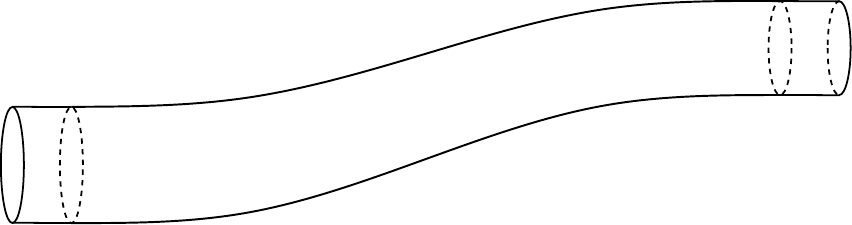}\\
        Bent tube
    \end{minipage}
    \begin{minipage}{0.3\textwidth}
        \centering
        \includegraphics[width=0.98\textwidth]{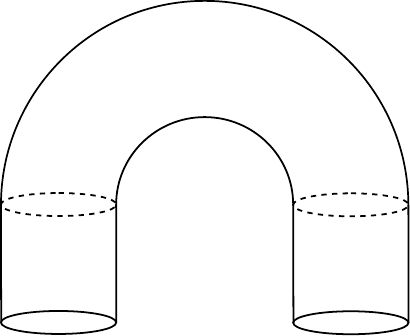}\\
        Arch
    \end{minipage}
    \caption{
        Schemes of the three geometries created to test the method's performance are presented in section \ref{sec:methodology}.
        The length of the straight tube is 0.12 m, and its inlet and outlet radii are 9 mm and 8 mm, respectively. The bent tube measures 0.14 mm in length, with inlet and outlet radii set to 10 mm and 8 mm, respectively. Both tube geometries include a 10 mm long region at each end with a constant radius. The arch geometry has a constant radius of 10 mm and straight 20 mm long sections at each end of the 180-degree bend.
    }
    \label{fig:geometries}
\end{figure}

Three artificial 3D geometries, shown in Figure \ref{fig:geometries}, are created using GMSH \cite{gmsh} and VMTK \cite{vmtk} libraries to test the performance of the method discussed in section \ref{sec:methodology}. The first geometry is a straight narrowing tube with a circular cross-section. The width and length of the geometry were selected to approximately agree with the dimensions of a real descending aorta. The second geometry is a slightly bent, extended version of the first geometry. The third geometry consists of a tube with a constant radius with a 180-degree bend mimicking the shape of the aortic arch. 

\subsection{Reference velocity fields}\label{sec:reference velocity fields}
The ground truth data are obtained by solving for velocity and pressure fields for given (prescribed)  $\theta$ and $\vin$. To ensure good resolution of the results, the edge length $h$ of the meshes is set uniformly to 0.75 mm.

%

    

Using the generated meshes, the reference fields are computed for each of the following values of $\theta\in[0.2,0.5,0.8,1.0]$. 
The inlet velocity $\vin$ is prescribed to be the analytical profile defined by \eqref{analytic profile} for each corresponding $\theta$, and the magnitude of the inlet velocity $V$ is set to $0.1\,\text{m}\cdot\text{s}^{-1}$, which approximately corresponds to the diastolic velocity of blood in descending aorta. 
The MINI element without stabilization \eqref{ip stabilization} is used to discretize the weak formulation \eqref{weak_form} to compute the ground truth solution. 
The density $\rho$ and viscosity $\mu$ are set to $1050\,\text{kg}\cdot\text{m}^{-3}$ and $3.8955\cdot10^{-3}\,\text{kg}\cdot\text{m}^{-1}\cdot\text{s}^{-1}$ respectively to agree with values used in literature for human blood, these parameters are summarized in Table \ref{tab:parameters}. 
The parameter $\gamma_*$ is set to $0.25\,\text{m}^2\cdot\text{s}\cdot\text{kg}^{-1}$ to maximize the boundary dissipation 
\begin{equation}
        \frac{\theta}{\gamma_*(1-\theta)}\int_{\bndry{wall}}|\mathbf{v}|^2 \dx{s}       
\end{equation}
for $\theta=0.5$ in the case of Poiseuille flow. The effect of $\gamma_*$ has been discussed in more detail in our previous work \cite{Chabiniok2021}. 

\begin{table}
    \centering
    \begin{tabular}{c c c c}
        \hline\noalign{\smallskip}
        Symbol & Name & Value & Unit \\
        \noalign{\smallskip}\hline\noalign{\smallskip}
        $\mu$ & dynamic viscosity  & $\num{3.8955e-3}$ & $\text{kg}\cdot\text{m}^{-1}\text{s}^{-1}$ \\ 
        $\rho$ & density & $\num{1050}$ & $\text{kg}\cdot\text{m}^{-1}$ \\
        $V$ & mean inlet velocity & $\num{0.1}$ & $\text{m}\cdot\text{s}^{-1}$ \\
        $\gamma_*$ & - & $\num{0.25}$ & $\text{m}^2\cdot\text{s}\cdot\text{kg}^{-1}$ \\
        $\beta_*$ & Nitsche penalty parameter & $\num{1000}$ & - \\
        $\theta$ & slip parameter & $0.0-1.0$ & - \\
        \noalign{\smallskip}\hline
    \end{tabular}
    \caption{Parameters used to generate reference velocity and pressure fields.}
    \label{tab:parameters}
\end{table}

\subsection{Preparing data for assimilation}

\begin{figure}
    \centering
     \begin{minipage}{0.6\textwidth}
     \centering
        \includegraphics[width=0.87\textwidth]{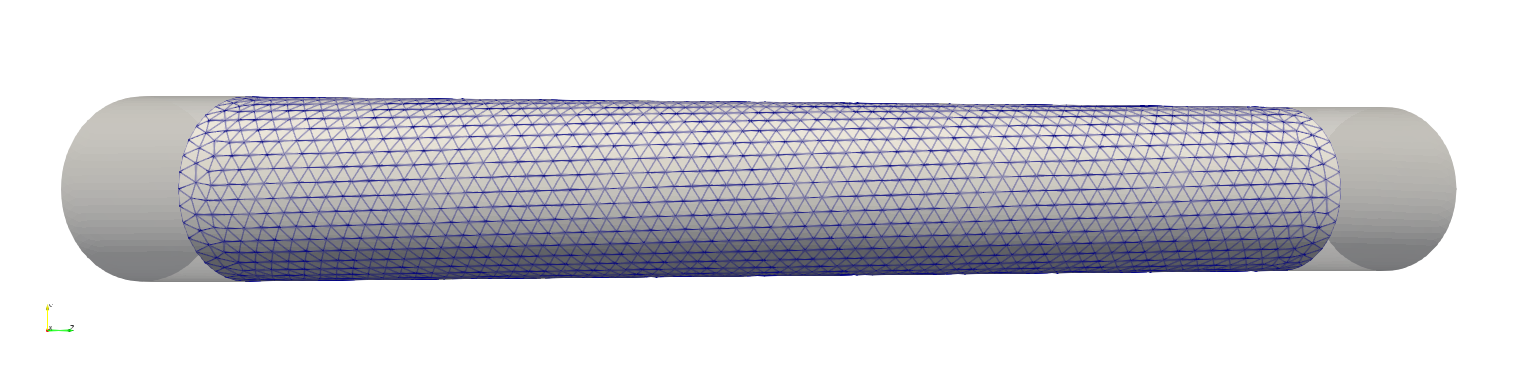}
        \includegraphics[width=0.98\textwidth]{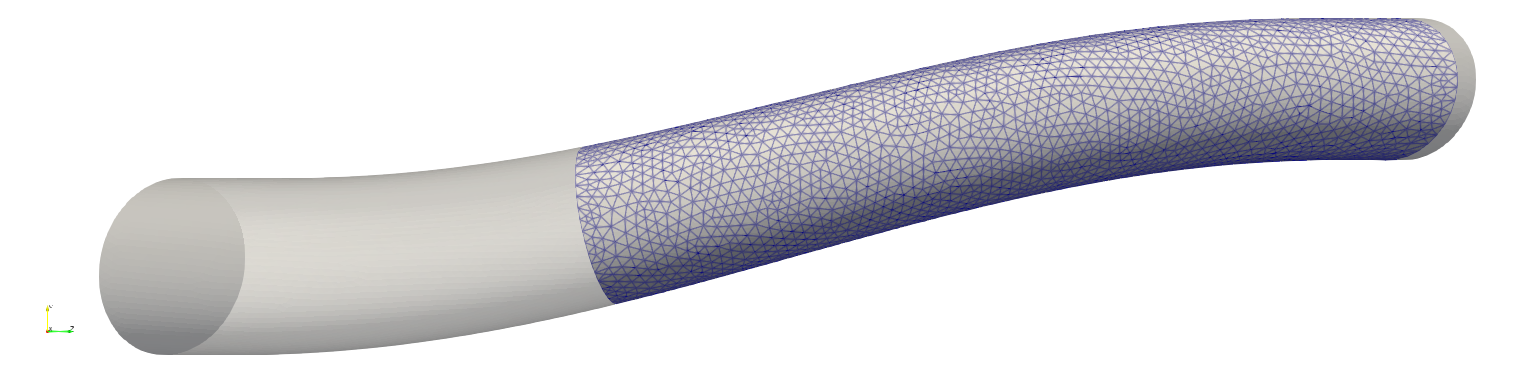}
    \end{minipage}
    \begin{minipage}{0.3\textwidth}
    \centering
        \includegraphics[width=0.98\textwidth]{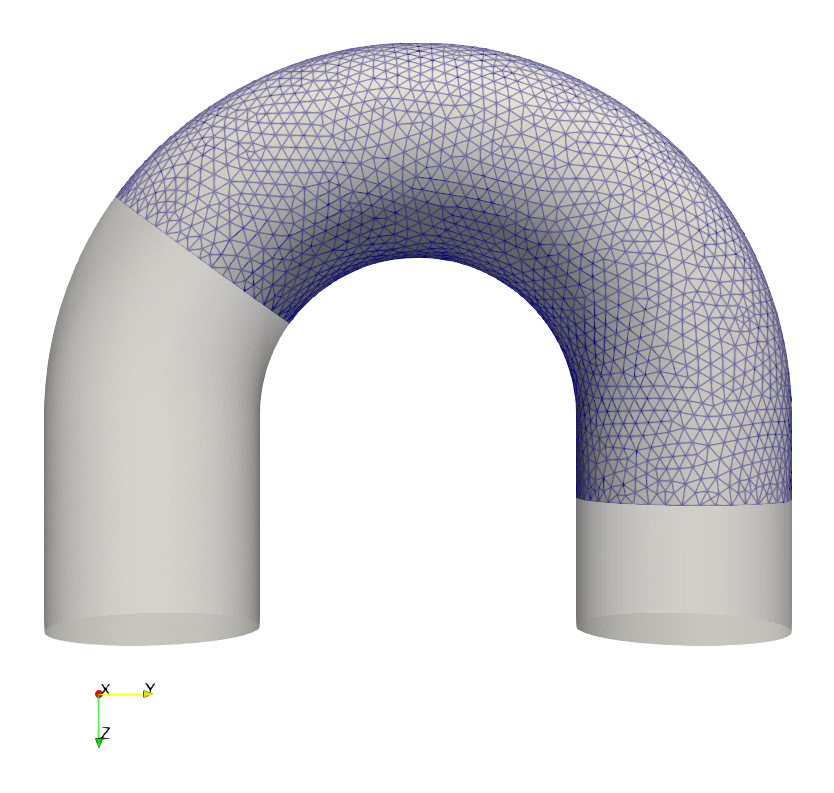}
    \end{minipage}
    \caption{Shorter segments of the geometries used for the assimilation with edge length set to 1.5 mm.}
    \label{fig:clipped_geometries}
\end{figure}

A shorter segment of each geometry was created, as shown in Figure \ref{fig:clipped_geometries}, to eliminate the effect of inlet and outlet from the reference velocity fields. 
These segments are meshed with an edge length set uniformly to 0.8 mm, 1 mm and 1.5 mm to acquire meshes with different densities, in order to test the sensitivity of the assimilation process. 
Using coarser meshes also helps to mimic reality, where the computational domain and real domain are not identical.
Moreover, it also corresponds better to the real-patient cases where the typical voxel size of the 4D PC-MRI data is usually between 1.5 and 2.5 mm.
The numbers of degrees of freedom of the generated meshes are provided in Tables \ref{tab:ref_dofs015}, \ref{tab:ref_dofs01} and \ref{tab:ref_dofs008}.
\begin{table}
    \centering

\pgfplotstableread[col sep=comma]{
Name,          Vertices, Faces,   Cells
Straight tube, 14790,   169526,  83227
Bent tube,     24482,   287316,  141664
Arch,          26513,   315267,  155762
}\loadedtable

    \pgfplotstabletypeset[ 
every even row/.style={
before row={\rowcolor[gray]{0.9}}},
every head row/.style={
before row=\toprule,after row=\midrule},
every last row/.style={
after row=\bottomrule},
create on use/mini/.style={create col/expr={3*\thisrow{Vertices}+3*\thisrow{Cells}+\thisrow{Vertices}}},
create on use/p1p1/.style={create col/expr={3*\thisrow{Vertices}+\thisrow{Vertices}}},
columns={Name, Vertices, Faces, Cells, mini, p1p1},
columns/Name/.style={column name={Geometry}, string type},
columns/Vertices/.style={int detect,  dec sep align},
columns/Faces/.style={int detect,  dec sep align},
columns/Cells/.style={int detect,  dec sep align},
columns/mini/.style={column name={MINI dofs}, fixed, precision=0,  dec sep align},
columns/p1p1/.style={column name={$P_1/P_1$ dofs}, fixed, precision=0,  dec sep align},
    ]\loadedtable
    
    \caption{Table of degrees of freedom of the shorter meshes with $h=1.5\,\text{mm}$.}
    \label{tab:ref_dofs015}
\end{table}

\begin{table}
    \centering

\pgfplotstableread[col sep=comma]{
Name,          Vertices, Faces,   Cells
Bent tube,     80716, 974956, 483018
Arch,          89554, 1088000, 539561
}\loadedtable

    \pgfplotstabletypeset[ 
every even row/.style={
before row={\rowcolor[gray]{0.9}}},
every head row/.style={
before row=\toprule,after row=\midrule},
every last row/.style={
after row=\bottomrule},
create on use/mini/.style={create col/expr={3*\thisrow{Vertices}+3*\thisrow{Cells}+\thisrow{Vertices}}},
create on use/p1p1/.style={create col/expr={3*\thisrow{Vertices}+\thisrow{Vertices}}},
columns={Name, Vertices, Faces, Cells, mini, p1p1},
columns/Name/.style={column name={Geometry}, string type},
columns/Vertices/.style={int detect,  dec sep align},
columns/Faces/.style={int detect,  dec sep align},
columns/Cells/.style={int detect,  dec sep align},
columns/mini/.style={column name={MINI dofs}, fixed, precision=0,  dec sep align},
columns/p1p1/.style={column name={$P_1/P_1$ dofs}, fixed, precision=0,  dec sep align},
    ]\loadedtable
    
    \caption{Table of degrees of freedom of the shorter meshes with $h=1\,\text{mm}$.}
    \label{tab:ref_dofs01}
\end{table}

\begin{table}
    \centering

\pgfplotstableread[col sep=comma]{
Name,          Vertices, Faces,   Cells
Bent tube,     155772, 1901386, 943686
Arch,          164745, 2028419, 1008139
}\loadedtable

    \pgfplotstabletypeset[ 
every even row/.style={
before row={\rowcolor[gray]{0.9}}},
every head row/.style={
before row=\toprule,after row=\midrule},
every last row/.style={
after row=\bottomrule},
create on use/mini/.style={create col/expr={3*\thisrow{Vertices}+3*\thisrow{Cells}+\thisrow{Vertices}}},
create on use/p1p1/.style={create col/expr={3*\thisrow{Vertices}+\thisrow{Vertices}}},
columns={Name, Vertices, Faces, Cells, mini, p1p1},
columns/Name/.style={column name={Geometry}, string type},
columns/Vertices/.style={int detect,  dec sep align},
columns/Faces/.style={int detect,  dec sep align},
columns/Cells/.style={int detect,  dec sep align},
columns/mini/.style={column name={MINI dofs}, fixed, precision=0,  dec sep align},
columns/p1p1/.style={column name={$P_1/P_1$ dofs}, fixed, precision=0,  dec sep align},
    ]\loadedtable
    
    \caption{Table of degrees of freedom of the shorter meshes with $h=0.8\,\text{mm}$.}
    \label{tab:ref_dofs008}
\end{table}

The corresponding reference velocity fields for each $\theta$ are then interpolated to the coarser meshes, and an additional Gaussian noise is added based on the given signal-to-noise ratio (SNR). 
The noise is scaled by $v_{\text{max}}$/SNR and its standard deviation is set to 
\begin{equation}\label{eq:standard deviation}
    \sigma=0.45\,\frac{v_\text{max}}{\text{SNR}},
\end{equation}
where $v_{\text{max}}$ is a maximal characteristic velocity (inspired by velocity encoding parameter for MRI), which represents an upper bound for the velocity components of $\mathbf{d}_{\text{MRI}}$ for each of the geometries.
The relation \eqref{eq:standard deviation} is used for the same purpose in \cite{Koltukluoglu2018}. 

\subsection{Solver setup}
The finite element implementation of the problem is done using the FEniCS framework \cite{fenics}.
The forward discrete problem is solved by the Newton method with line search as implemented in the PETSc library \cite{petsc,petsc4py}. 
The linear sub-problems are solved by sparse direct LU factorization applied to the full system using the MUMPS library \cite{mumps}. 

The optimization step is done using SciPy \cite{scipy} implementation of the L-BFGS-B algorithm and the dolfin-adjoint library \cite{Mitusch2019}, which generates the adjoint equation 
automatically from the weak formulation of the forward problem \eqref{weak_form}-\eqref{ip stabilization}. 
In each optimization iteration, the solution from the previous iteration is used as the starting point for the nonlinear solver to reduce computational time.
The stopping criteria $\texttt{ftol}$ and $\texttt{gtol}$ for the L-BFGS-B algorithm are set to $10^{-7}$ and $10^{-5}$, respectively, where $\texttt{ftol}$ measures the relative decrease of the functional and $\texttt{gtol}$ controls the norm of the gradient of the functional.

\section{Results of numerical experiments}\label{sec:experiments}
\begin{figure}
\centering
\begin{tabular}{c c c c c}
     \rotatebox{90}{data} & 
     \includegraphics[width=.19\textwidth]{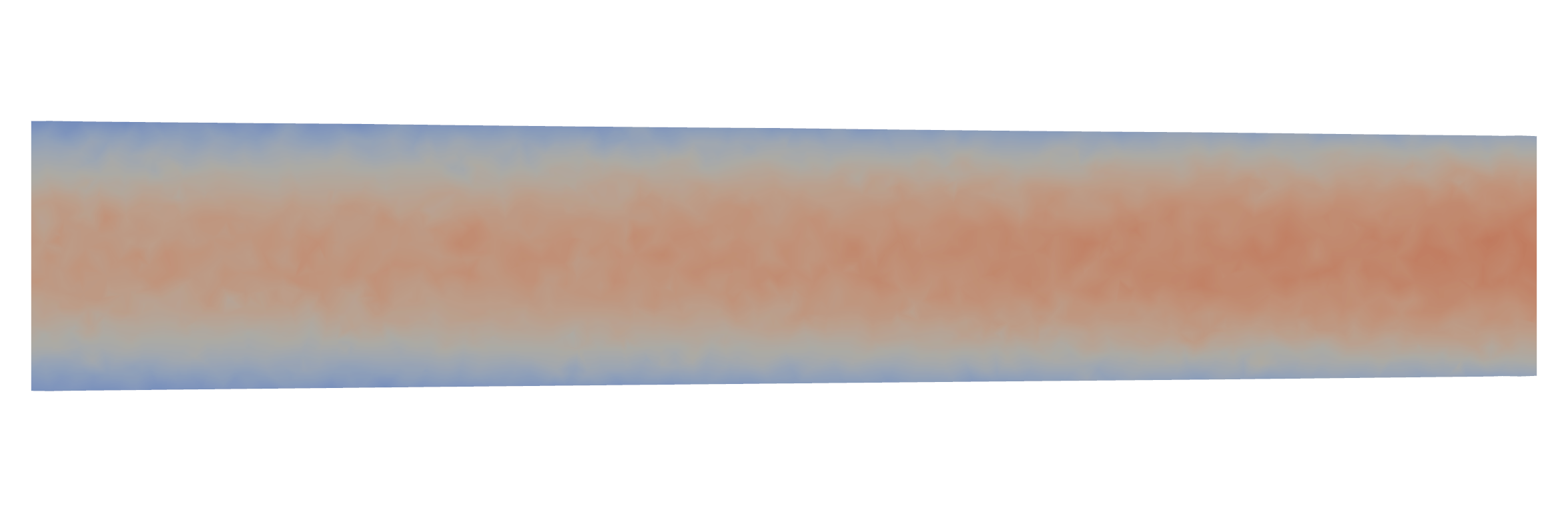} &
     \includegraphics[width=.19\textwidth]{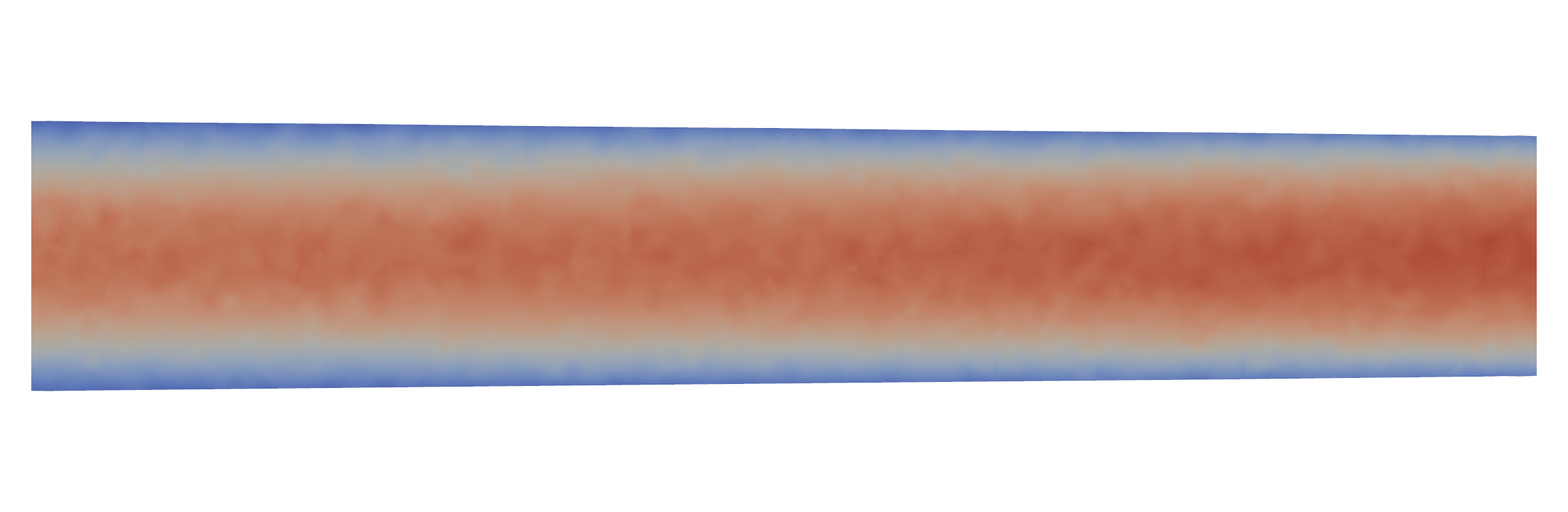} &
     \includegraphics[width=.19\textwidth]{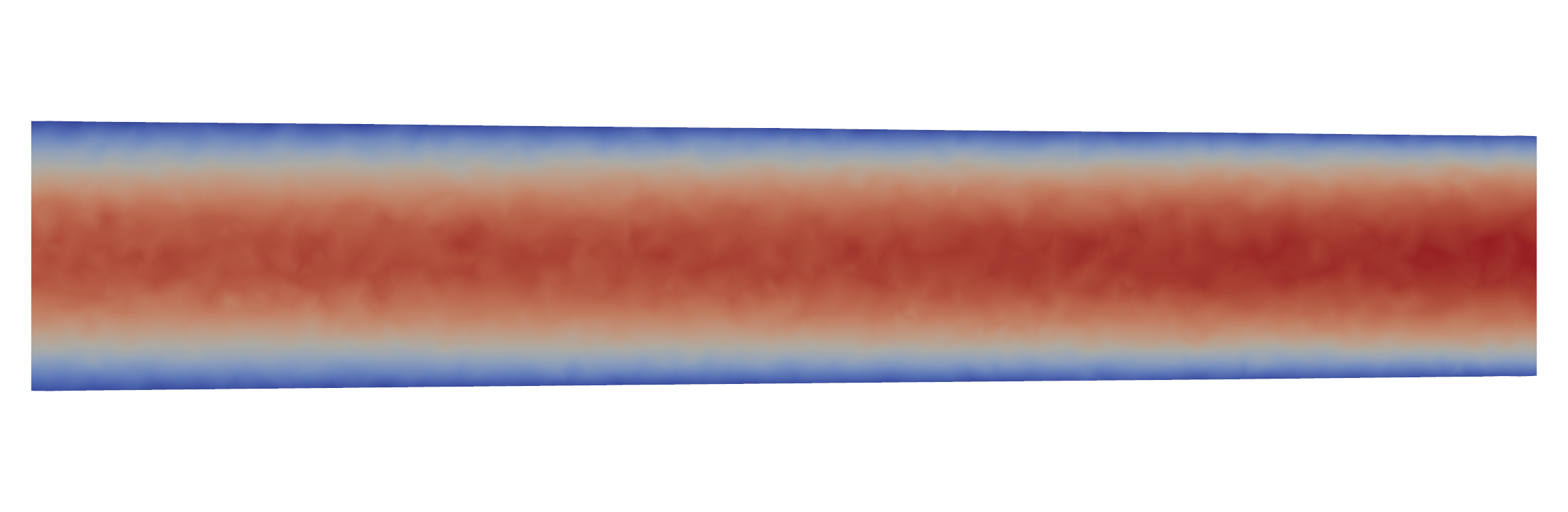} &
     \includegraphics[width=.19\textwidth]{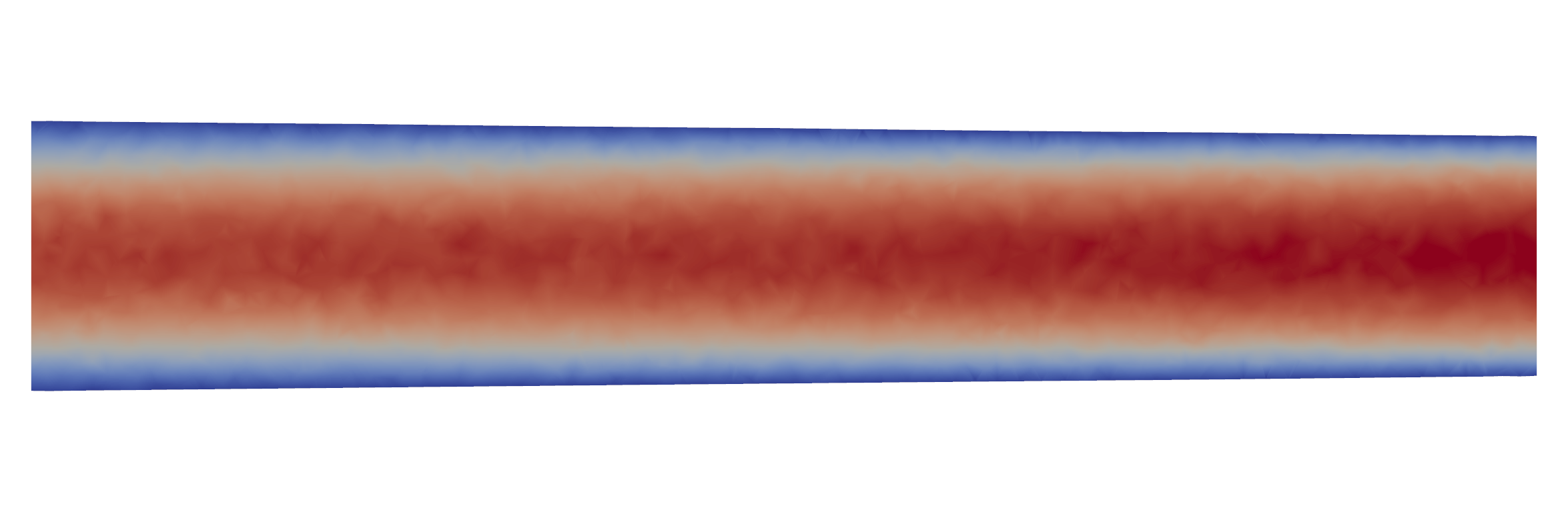} \\
     & $\theta=0.2$ & $\theta=0.5$ & $\theta=0.8$ & $\theta=1.0$ \\ 
     \rotatebox{90}{MINI} &
     \includegraphics[width=.19\textwidth]{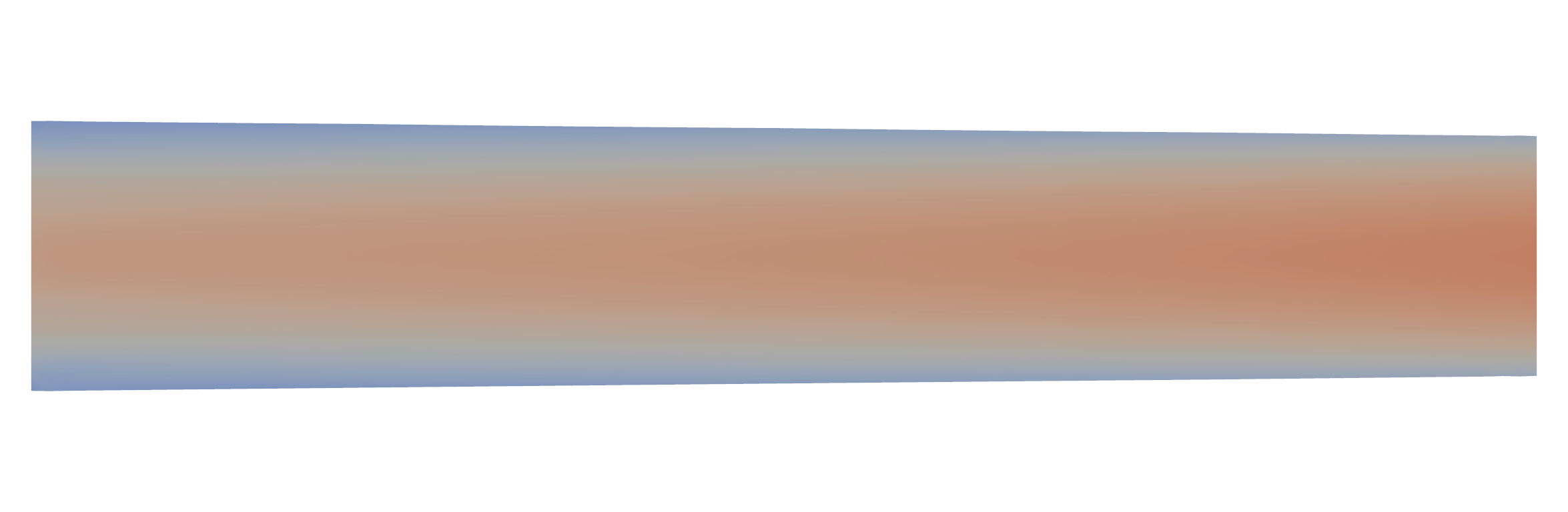} & 
     \includegraphics[width=.19\textwidth]{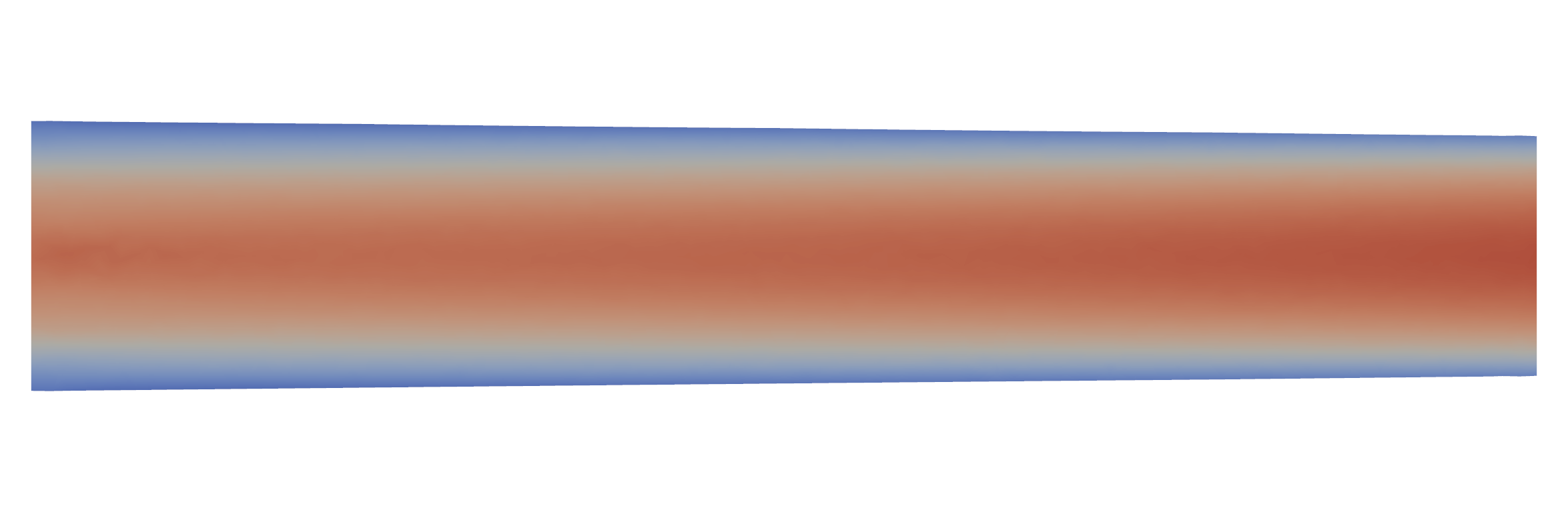} &
     \includegraphics[width=.19\textwidth]{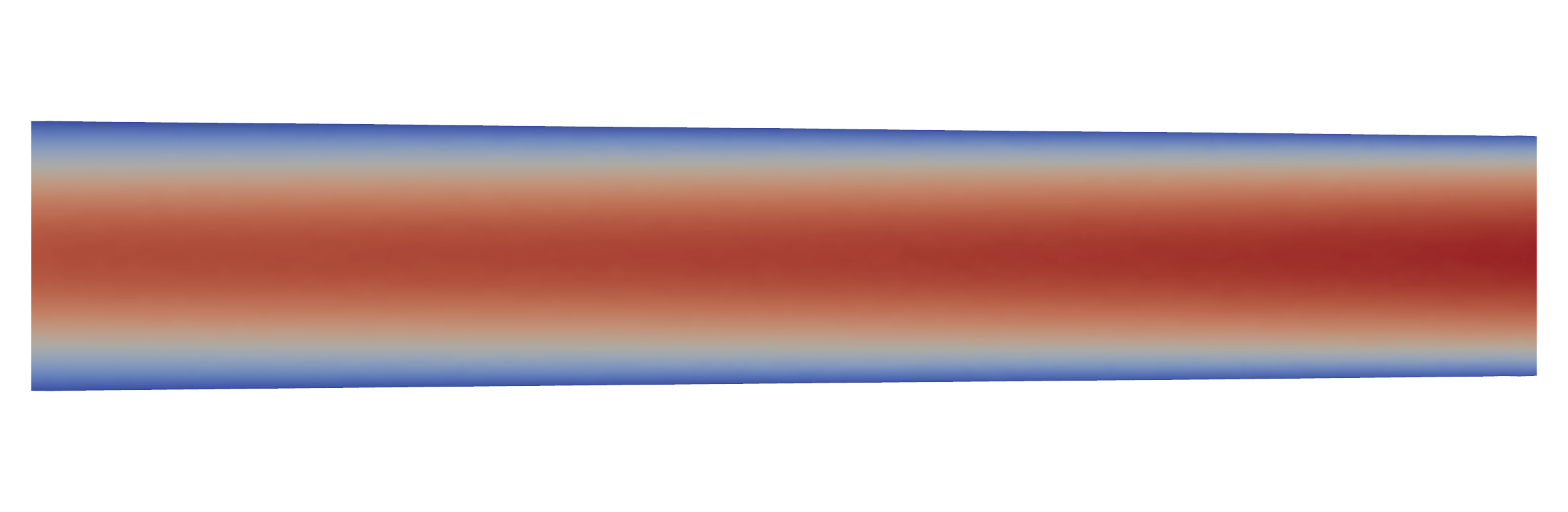} &
     \includegraphics[width=.19\textwidth]{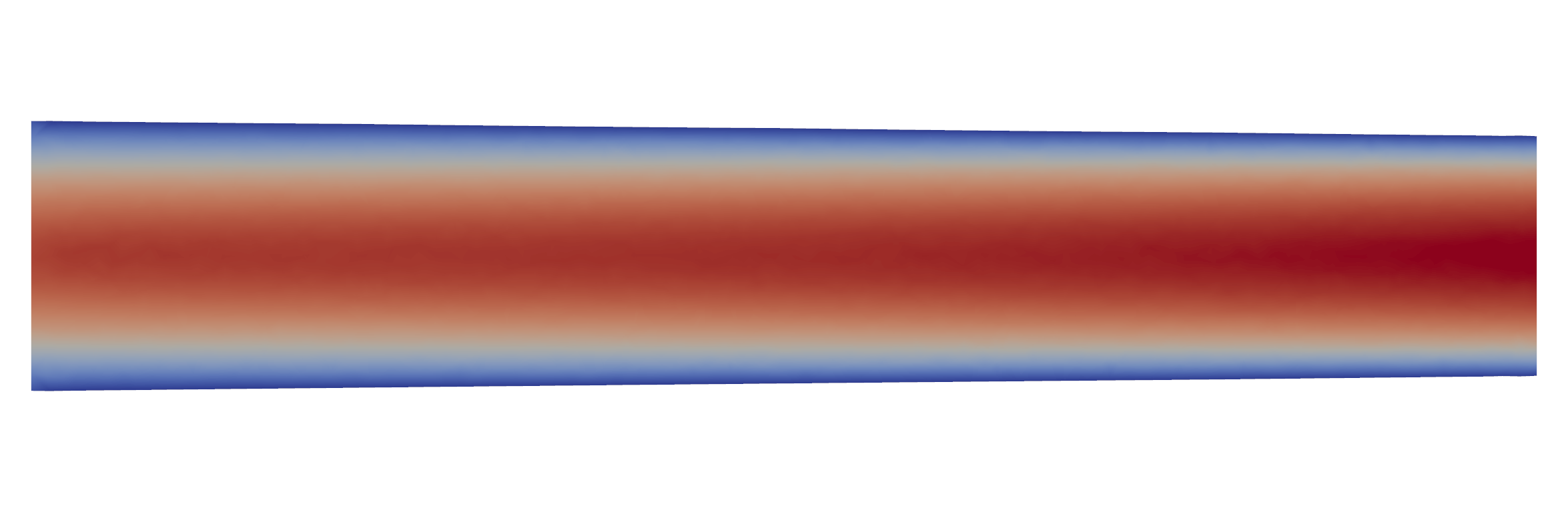} \\
     & $\thet{opt}=0.199$ & $\thet{opt}=0.494$ & $\thet{opt}=0.762$ & $\thet{opt}=0.985$ \\
     & $\mathcal{J}=\rnum{0.000223}$ & $\mathcal{J}=\rnum{0.0002262}$ & $\mathcal{J}=\rnum{0.0002331}$ & $\mathcal{J}=\rnum{0.00023}$ \\
     & $\mathcal{R}=\rnum{1.738e-05}$ & $\mathcal{R}=\rnum{5.033e-05}$ & $\mathcal{R}=\rnum{9.402e-05}$ & $\mathcal{R}=\rnum{9.606e-05}$ \\
     & iterations: 25 & iterations: 23 & iterations: 5 & iterations: 17 \\
        \rotatebox{90}{\begin{tabular}{c} 
            $P_1/P_1$\\
            stab.
    \end{tabular}}&
     \includegraphics[width=.19\textwidth]{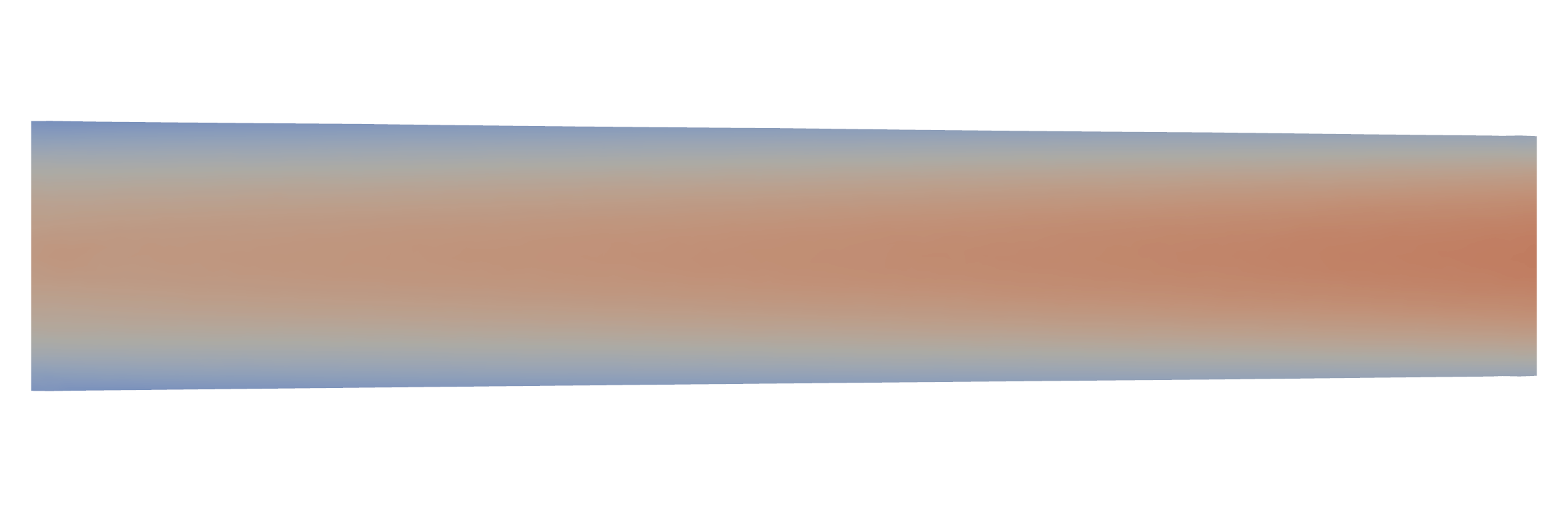} &
     \includegraphics[width=.19\textwidth]{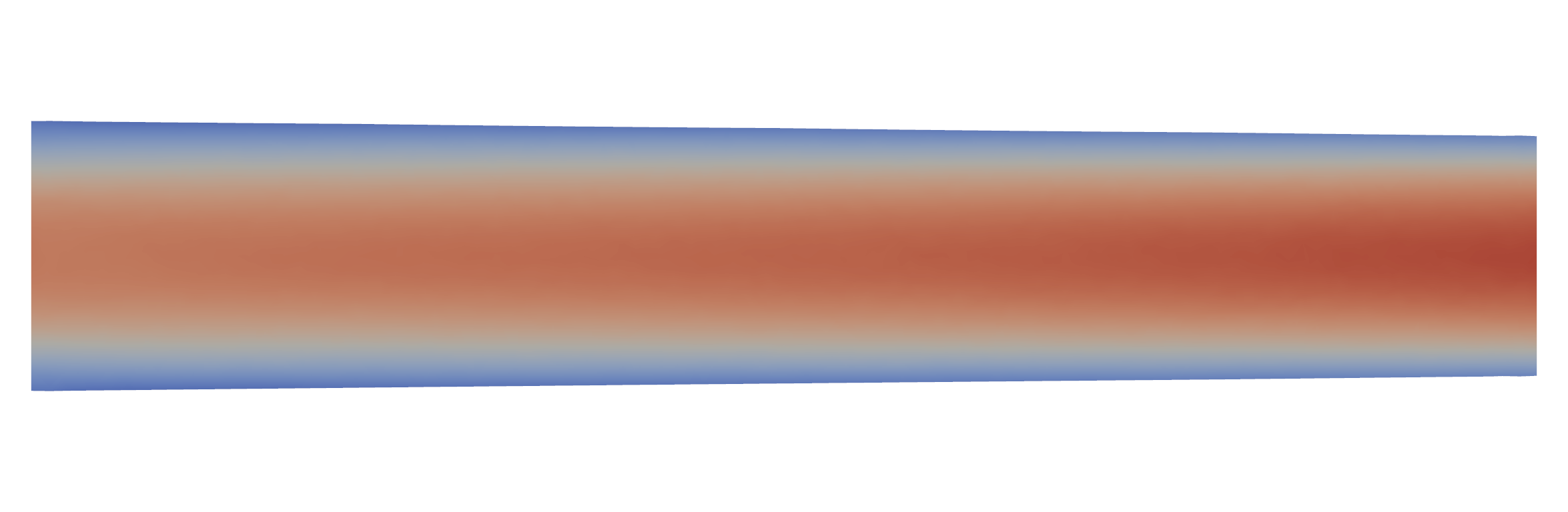} &
     \includegraphics[width=.19\textwidth]{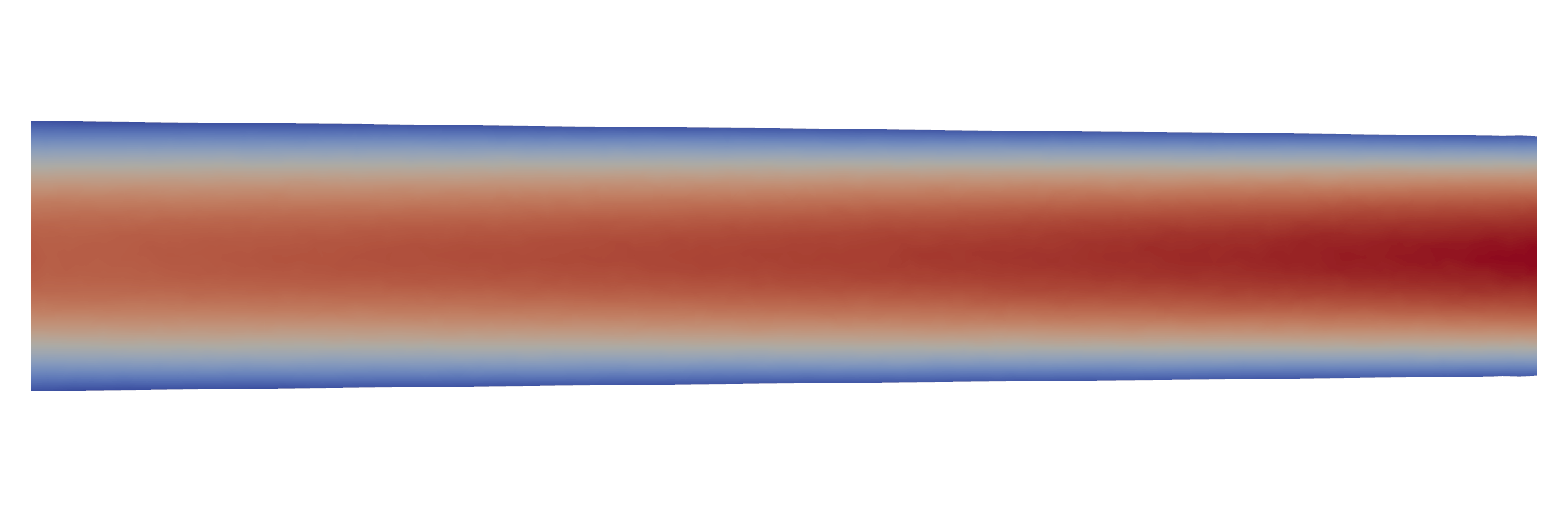} &
     \includegraphics[width=.19\textwidth]{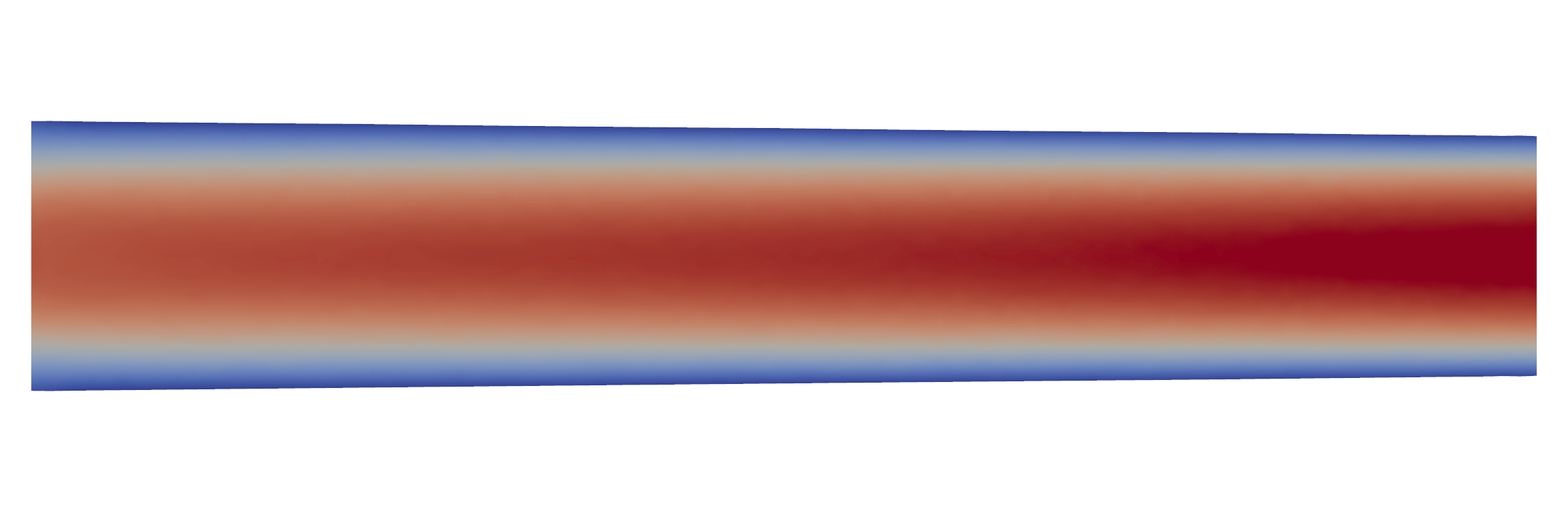} \\
     & $\thet{opt}=0.217$ & $\thet{opt}=0.515$ & $\thet{opt}=0.778$ & $\thet{opt}=0.925$ \\
     & $\mathcal{J}=\rnum{0.0002258}$ & $\mathcal{J}=\rnum{0.000251}$ & $\mathcal{J}=\rnum{0.0002843}$ & $\mathcal{J}=\rnum{0.0003189}$ \\
     & $\mathcal{R}=\rnum{1.602e-05}$ & $\mathcal{R}=\rnum{4.901e-05}$ & $\mathcal{R}=\rnum{8.844e-05}$ & $\mathcal{R}=\rnum{9.56e-05}$ \\
     & iterations: 22 & iterations: 16 & iterations: 9 & iterations: 10 \\
     & \multicolumn{4}{c}{\includegraphics[width=0.5\textwidth]{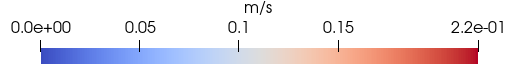}}
\end{tabular}
\caption{Comparison of noisy data on straight tube geometry ($h=1.5\text{ mm}$) with $\text{SNR}=2$ (first row) with the assimilation velocity results using MINI element (second row) and stabilized $P_1/P_1$ element (third row) for multiple values of $\theta$.}
\label{fig:straight}
\medskip
\centering
\begin{tabular}{c c c c c}
    & $\theta=0.2$ & $\theta=0.5$ & $\theta=0.8$ & $\theta=1.0$ \\ 
     \rotatebox{90}{MINI} &
     \includegraphics[width=.19\textwidth]{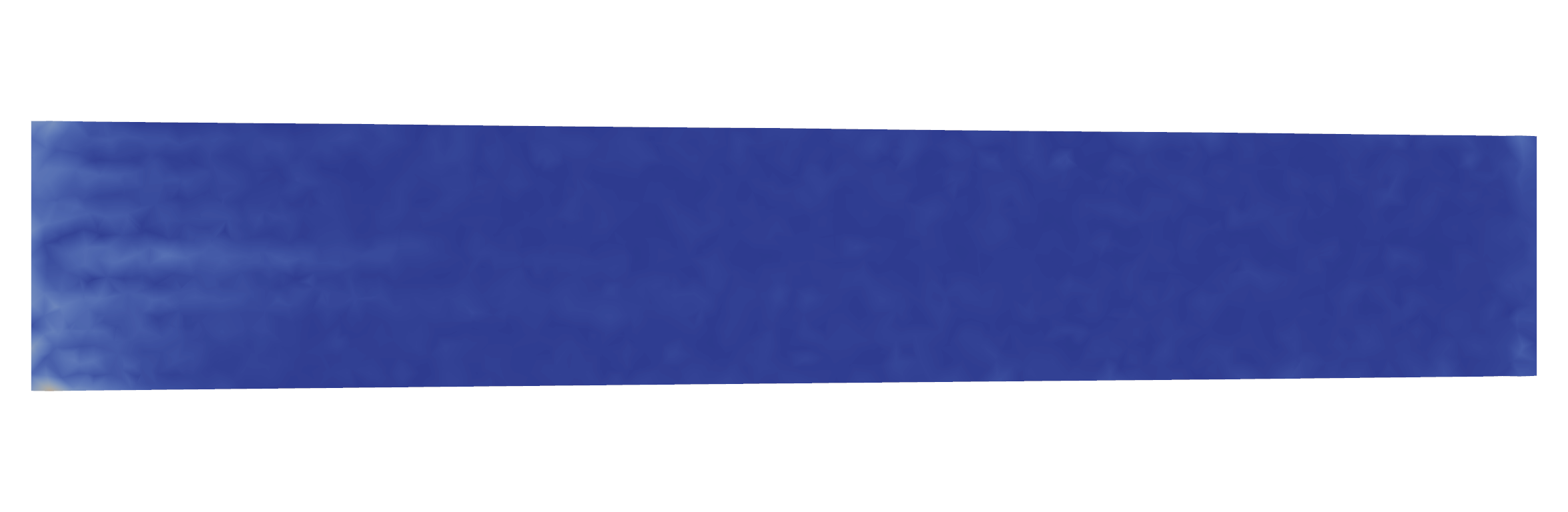} & 
     \includegraphics[width=.19\textwidth]{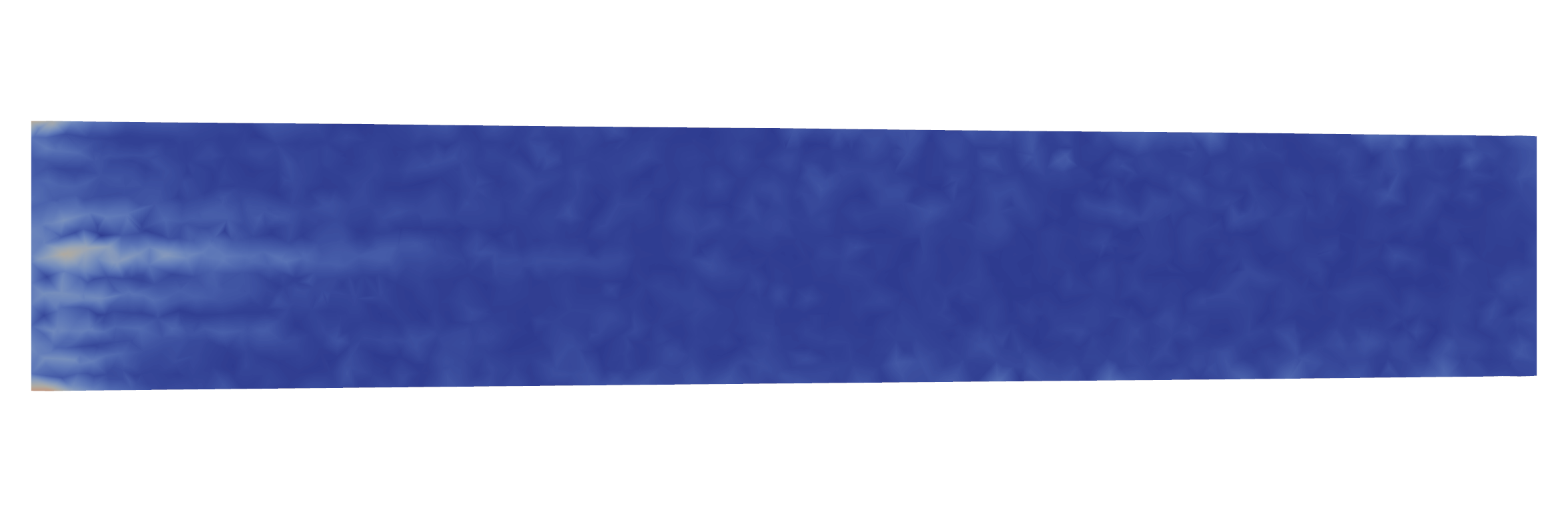} &
     \includegraphics[width=.19\textwidth]{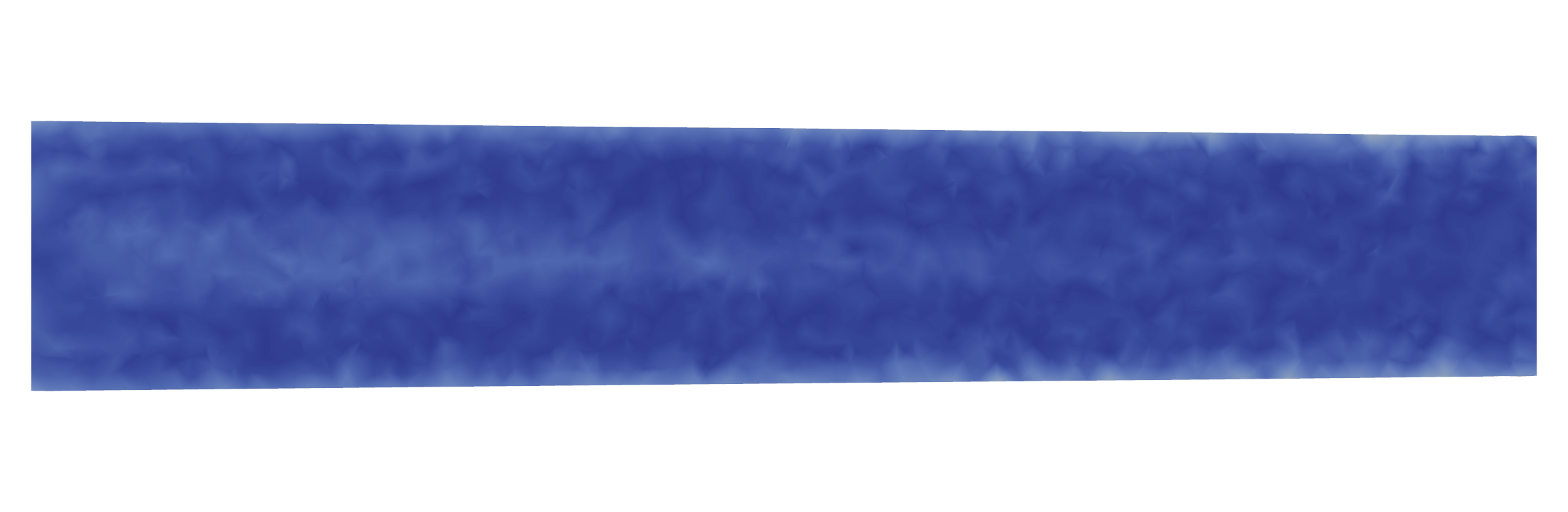} &
     \includegraphics[width=.19\textwidth]{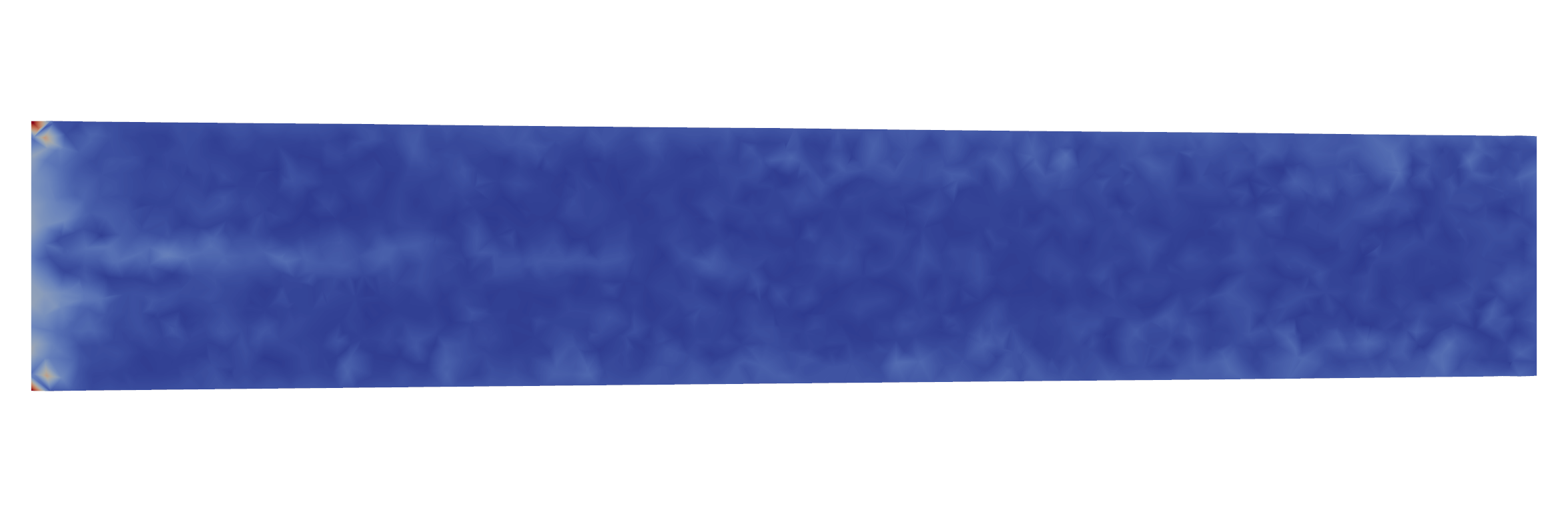} \\
     \rotatebox{90}{\begin{tabular}{c} 
        MINI\\
        stab.
    \end{tabular}}&
     \includegraphics[width=.19\textwidth]{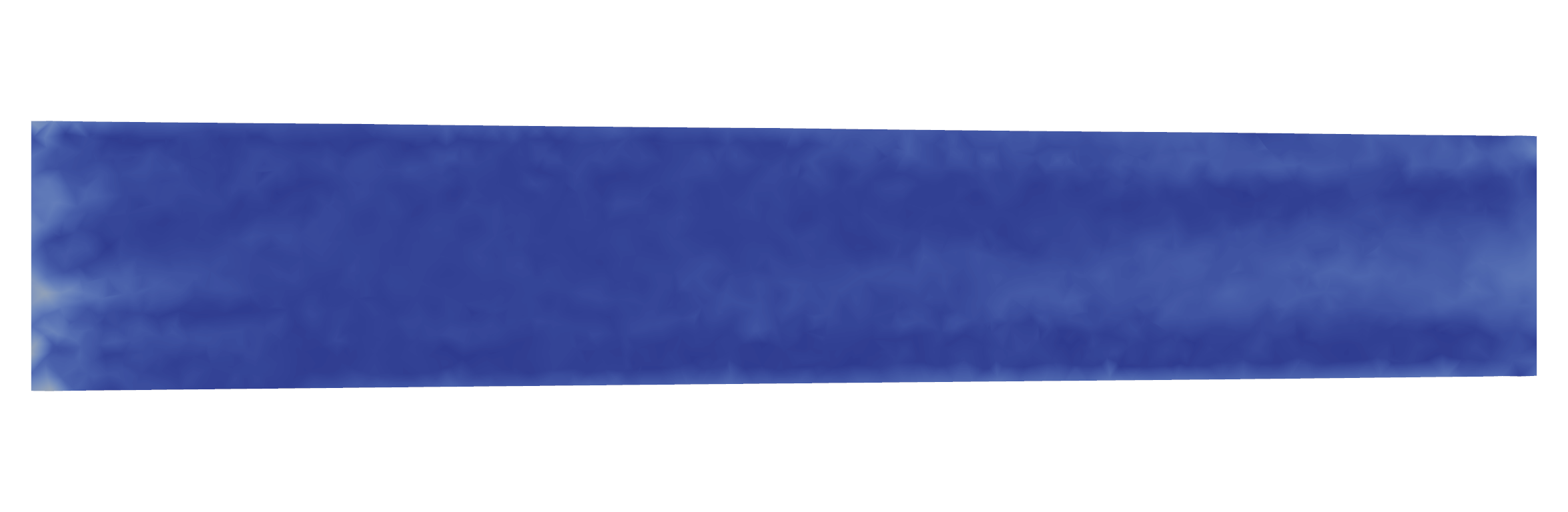} &
     \includegraphics[width=.19\textwidth]{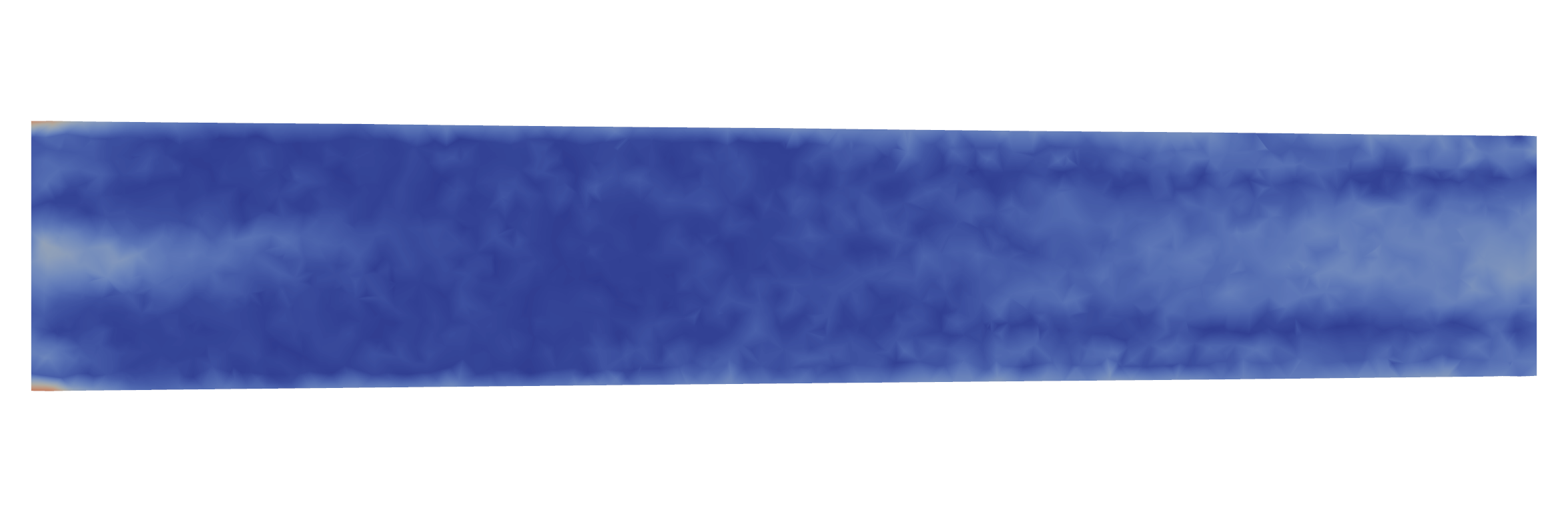} &
     \includegraphics[width=.19\textwidth]{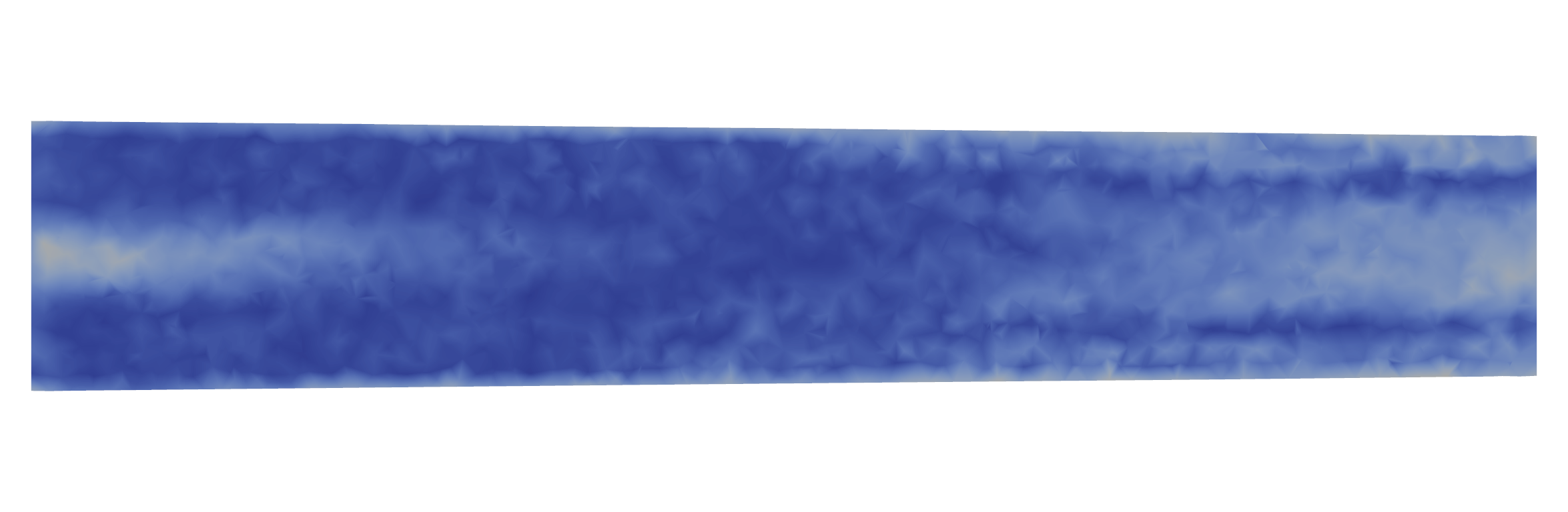} &
     \includegraphics[width=.19\textwidth]{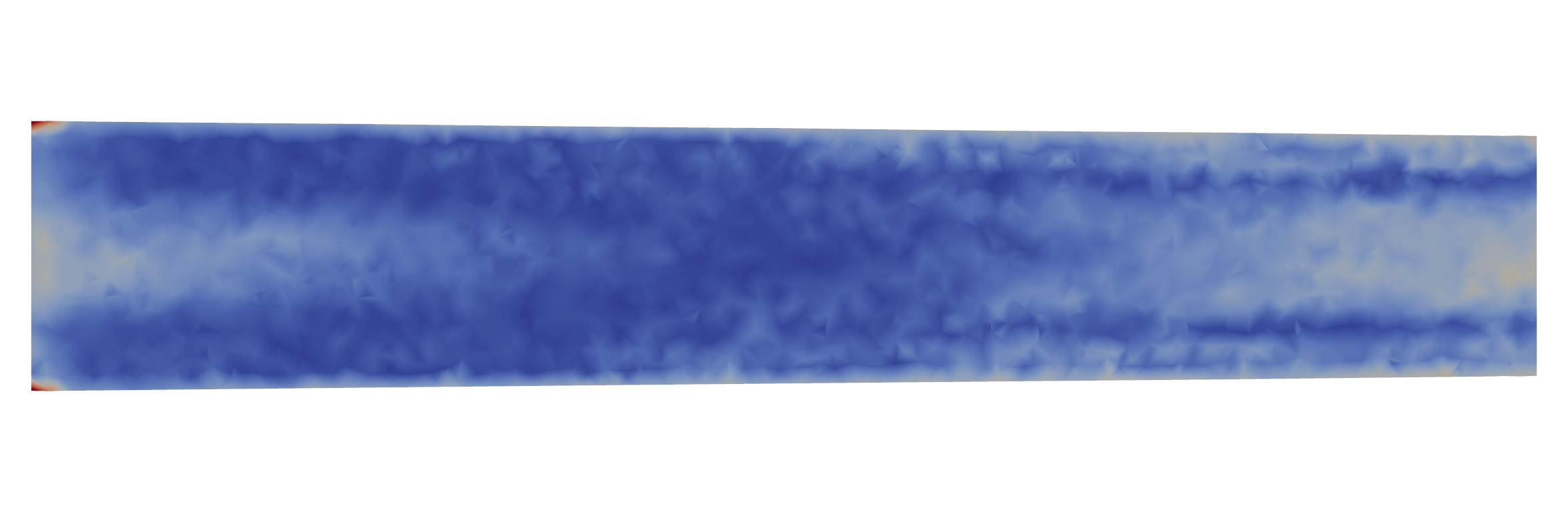} \\
        \rotatebox{90}{\begin{tabular}{c} 
            $P_1/P_1$\\
            stab.
    \end{tabular}}&
     \includegraphics[width=.19\textwidth]{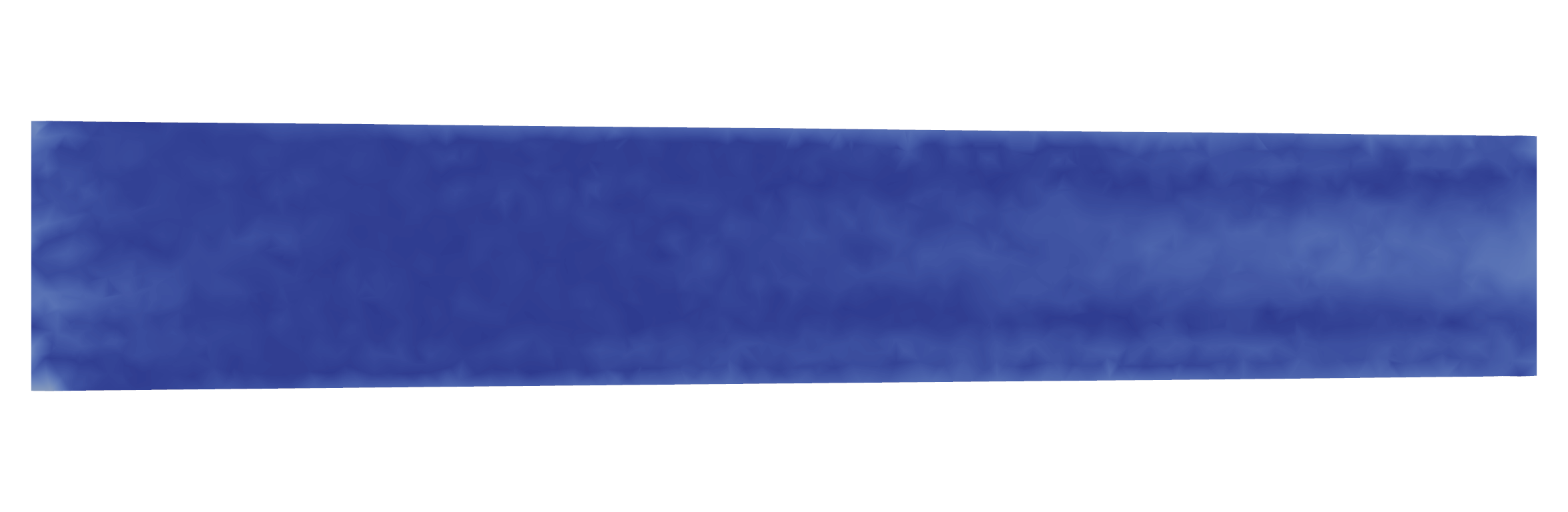} &
     \includegraphics[width=.19\textwidth]{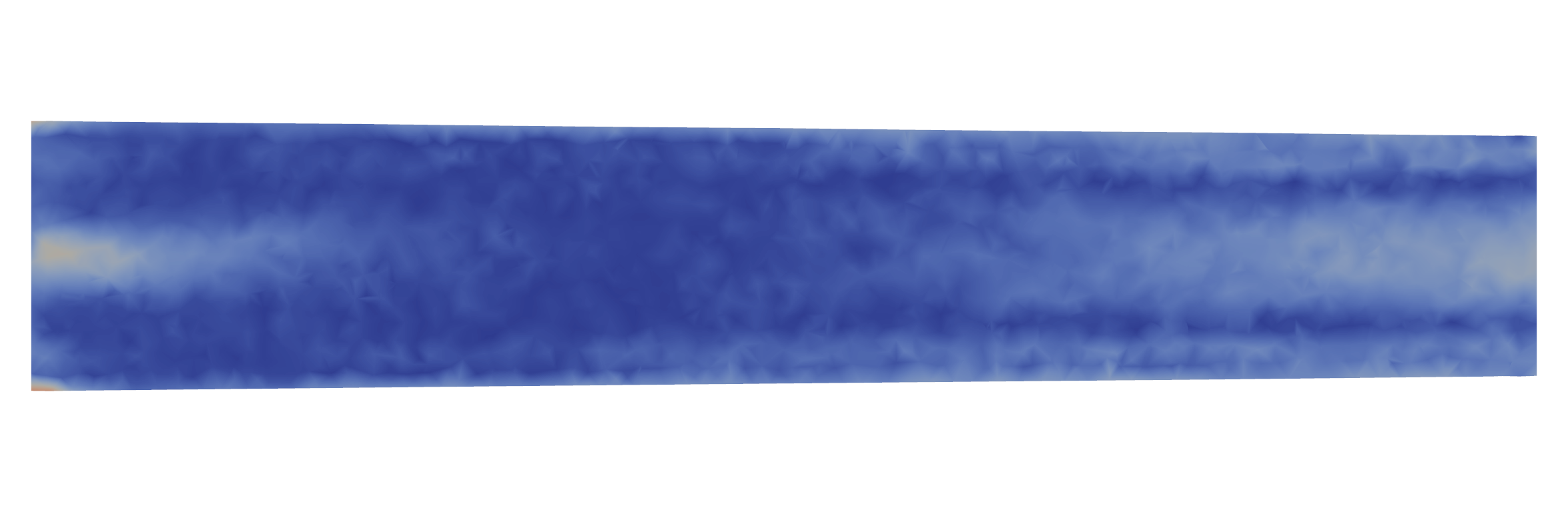} &
     \includegraphics[width=.19\textwidth]{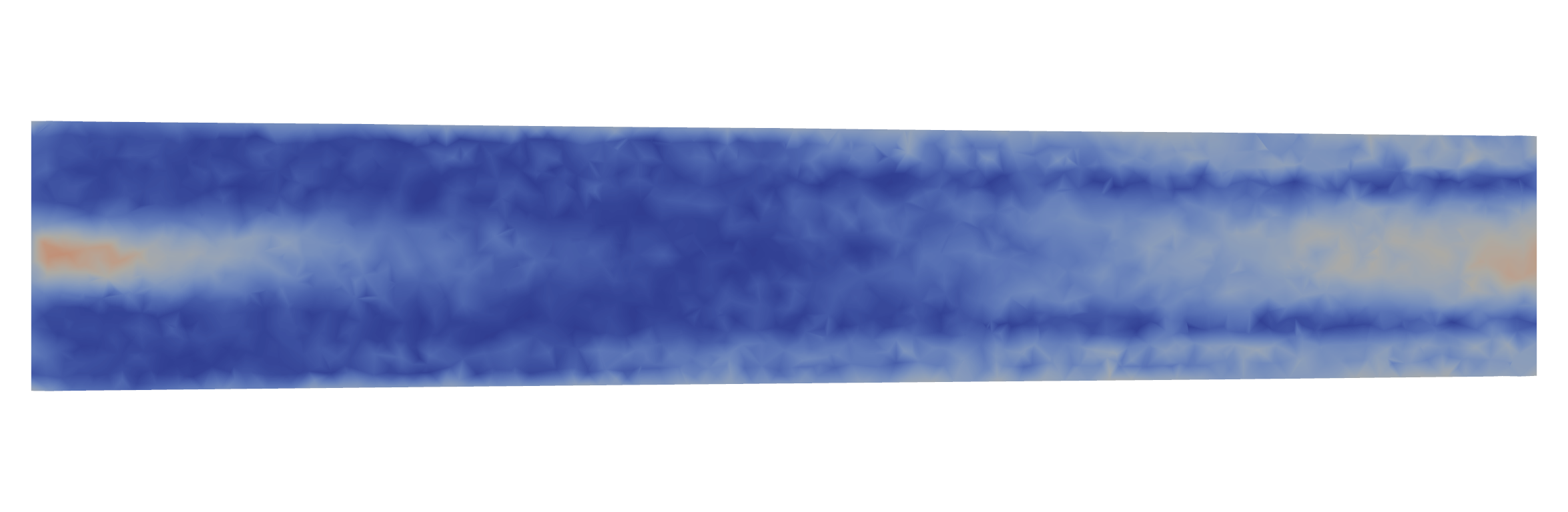} &
     \includegraphics[width=.19\textwidth]{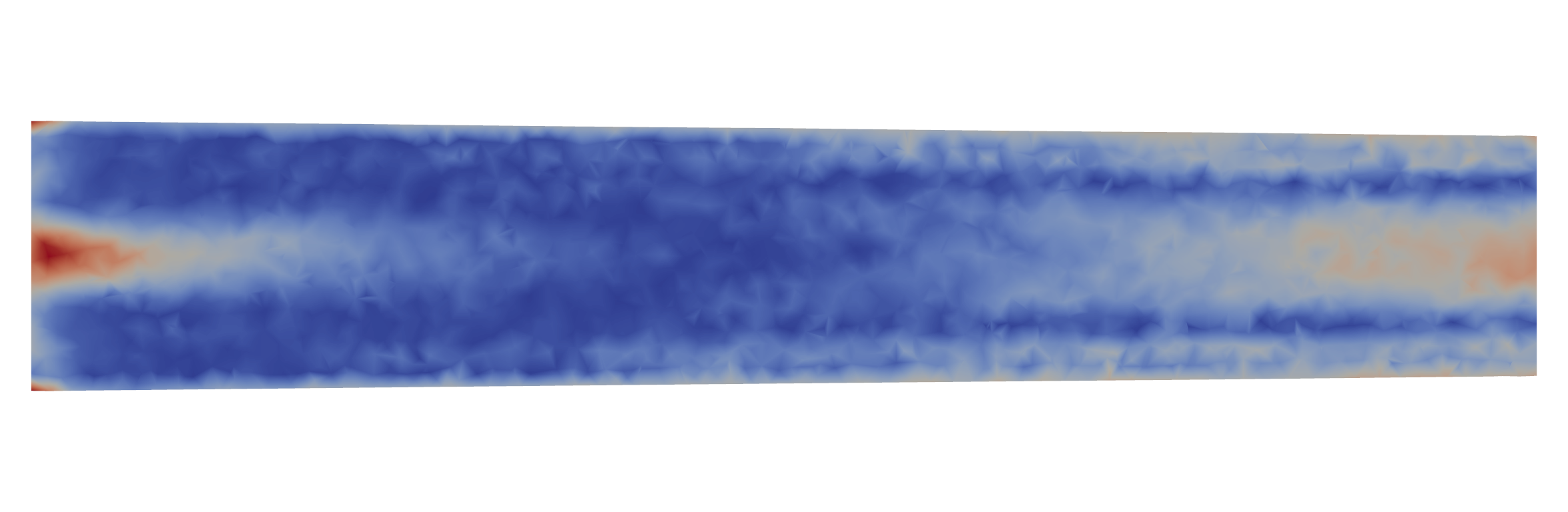} \\
     & \multicolumn{4}{c}{\includegraphics[width=0.5\textwidth]{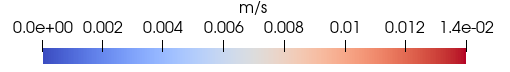}}
\end{tabular}
\caption{Visulatization of the difference between reference data (without noise) and computed velocity fields on straight tube geometry ($h=1.5\text{ mm}$) using MINI element (first row), stabilized MINI element (second row) and stabilized $P_1/P_1$ element (third row) for multiple values of $\theta$.}
\label{fig:errors}
\end{figure}

The first test is performed on the straight tube mesh with edge length $h=1.5\text{ mm}$. Data with $\text{SNR}=2$ were used for the assimilation with the regularization weights $\alpha$ and $\beta$ set to $0.001$ and $0.1$, respectively. 
A few different initial guesses were tested, and their choice did not significantly influence the optimal $\theta$ and $\vin$. 
However, a good initial guess helped reduce the number of optimization iterations and, in some cases, reduce the risk of divergence of the nonlinear solver.
The comparison of the data and the assimilation results for the MINI element and stabilized $P_1/P_1$ element with stabilization weights $\alpha_v=0.01$ and $\alpha_p=0.01$ are shown in Figure \ref{fig:straight} and the differences between the assimilation results and ground truth velocity are shown in Figure \ref{fig:errors}.
We observe that the MINI element performed better in terms of recovering the value of $\theta$ and the ground truth velocity field compared to the stabilized $P_1/P_1$ element.
This is expected since the stabilized $P_1/P_1$ element is less accurate and has significantly fewer degrees of freedom than the MINI element on the same mesh, as shown in Table \ref{tab:ref_dofs015}.
It is also important to point out that the MINI element has been used to generate the ground truth data and, therefore, has an advantage in the reconstruction process.
Figure \ref{fig:errors} also shows that the error for stabilized $P_1/P_1$ element has a similar character to the error for stabilized MINI element, suggesting that the stabilization might be the main cause of the inaccuracies.

In the following subsection, we present and comment on various aspects of the assimilation results for the bent and arch geometries.

\subsection{Finite elements and stabilization}
\begin{figure}
\centering
\begin{tabular}{c c c c c}
    \rotatebox{90}{\begin{tabular}{c}
         data:
    \end{tabular}} &  
    \includegraphics[width=.2\textwidth]{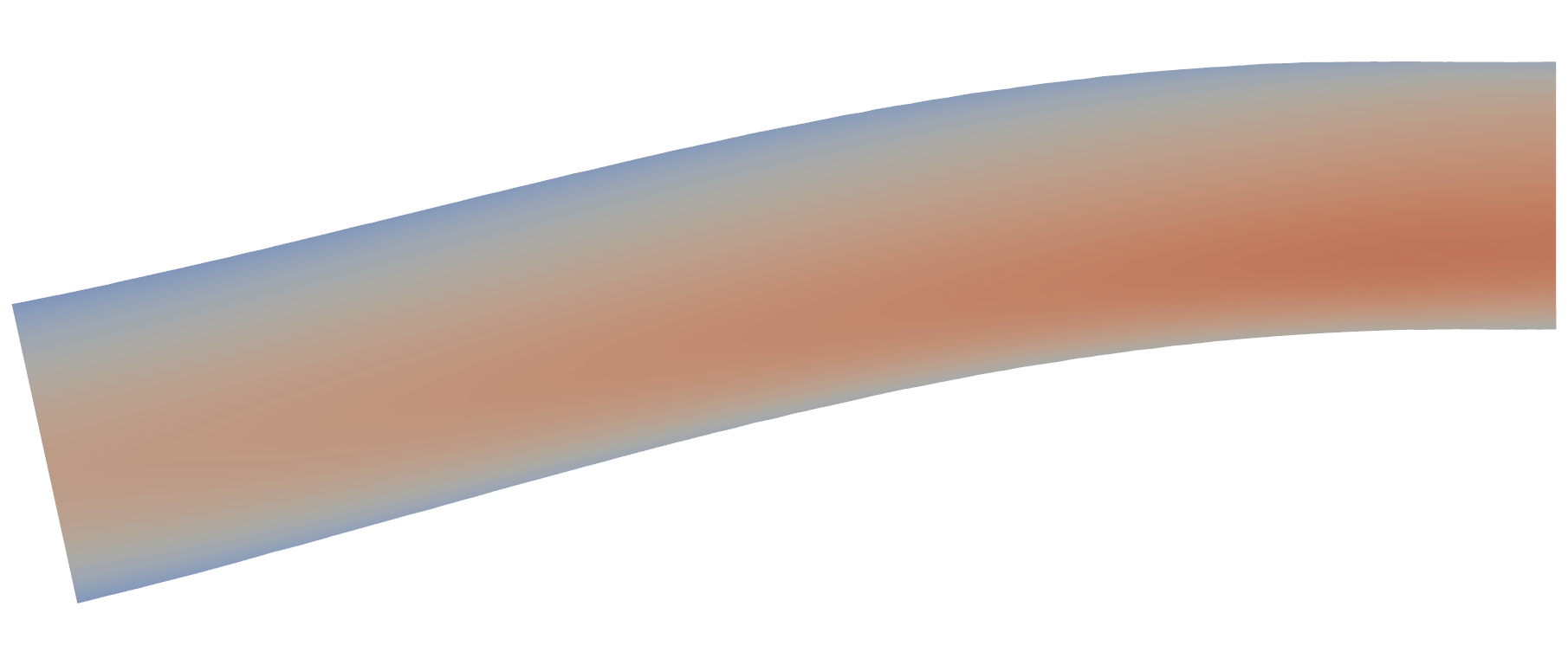} &
    \includegraphics[width=.2\textwidth]{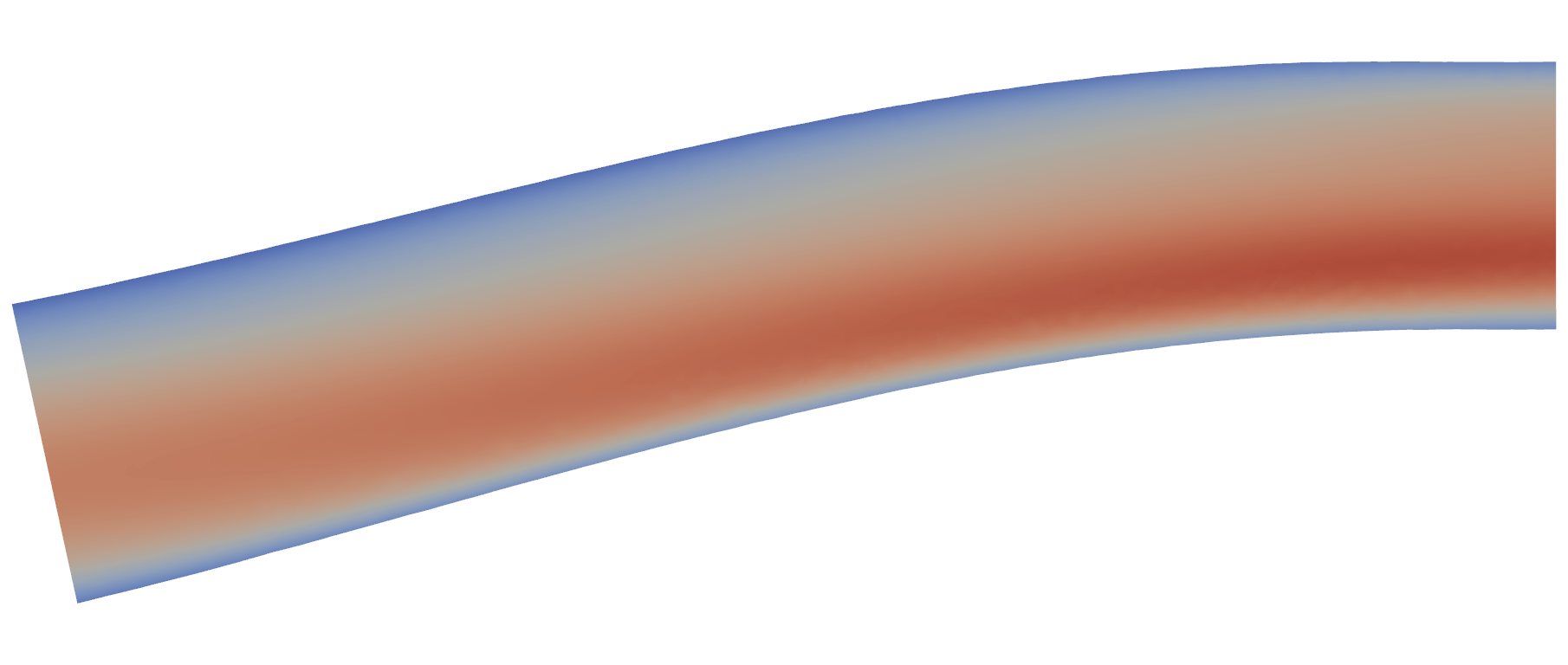} &
    \includegraphics[width=.2\textwidth]{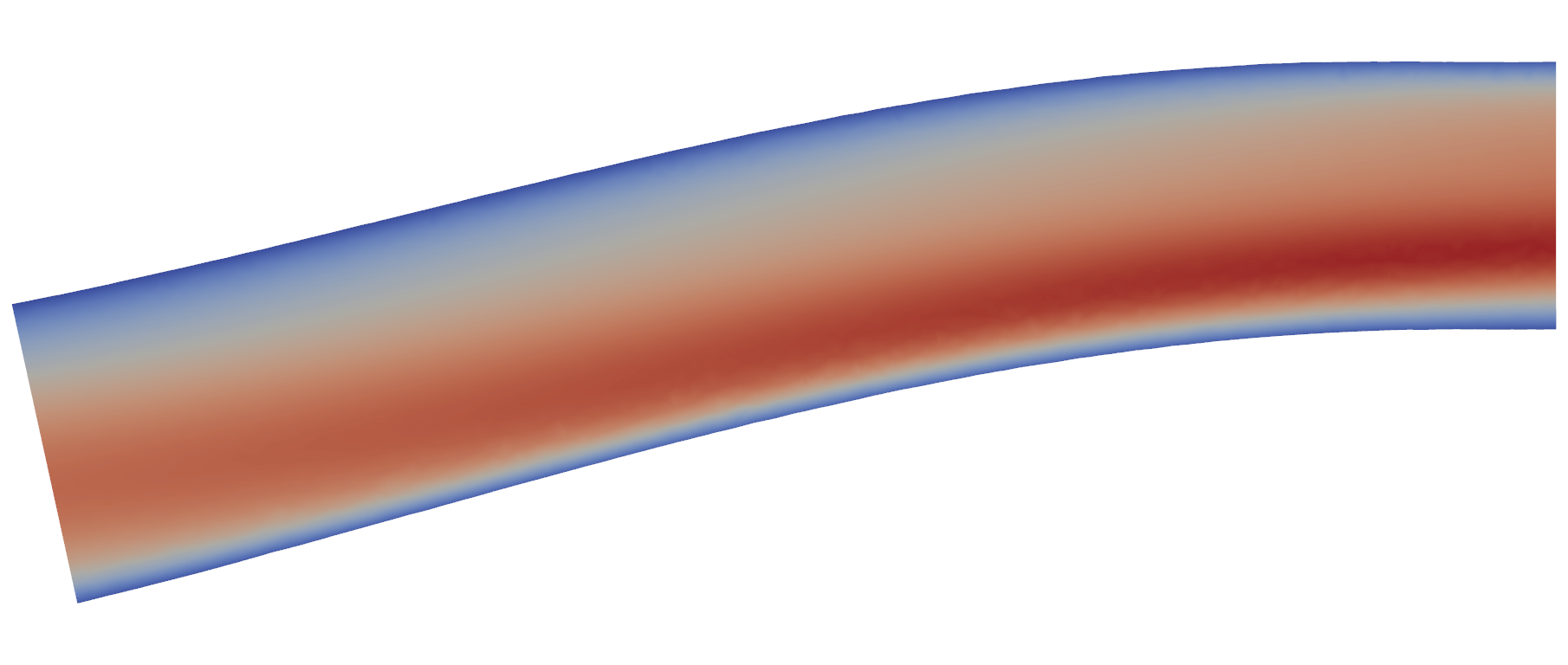} &
    \includegraphics[width=.2\textwidth]{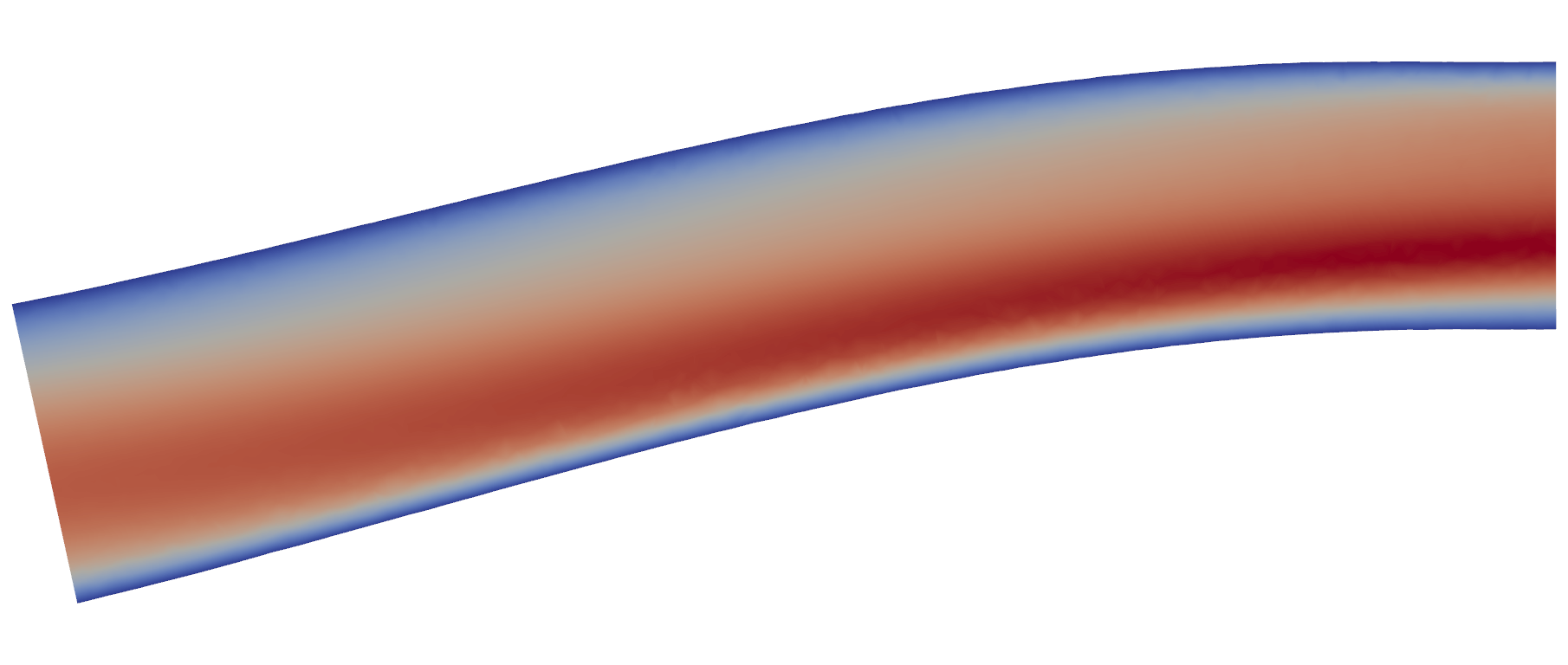} \\
         & $\theta=0.2$ & $\theta=0.5$ & $\theta=0.8$ & $\theta=1$ \\
    \rotatebox{90}{\begin{tabular}{c}
         MINI
    \end{tabular}} & 
    \includegraphics[width=.2\textwidth]{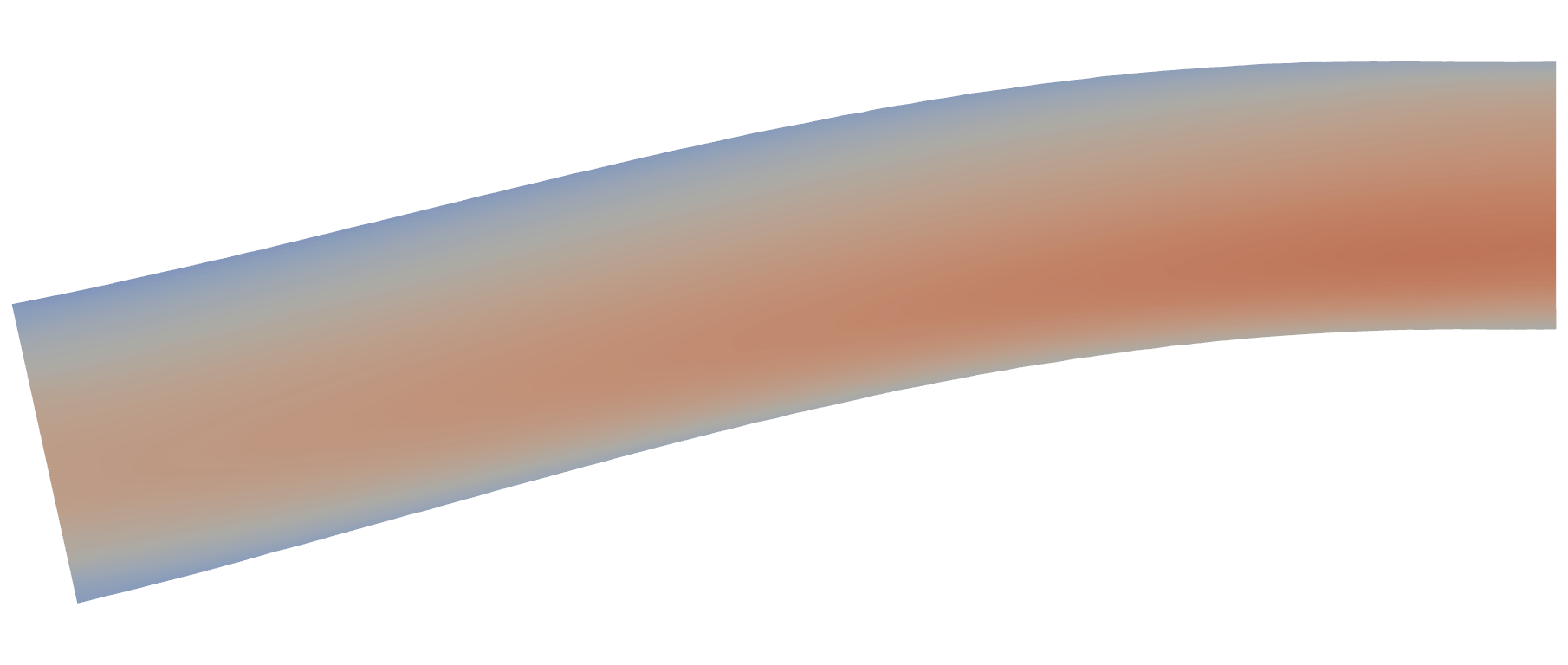} &
    \includegraphics[width=.2\textwidth]{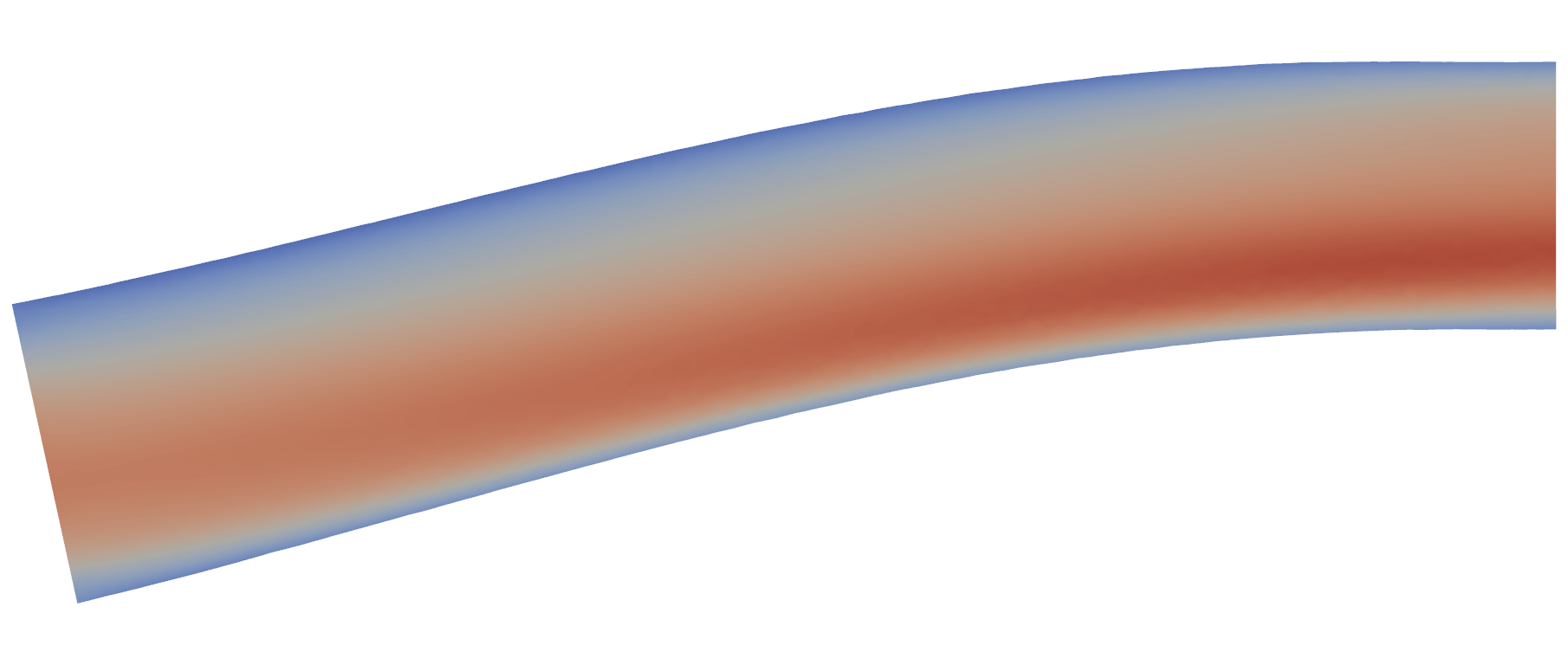} &
    \includegraphics[width=.2\textwidth]{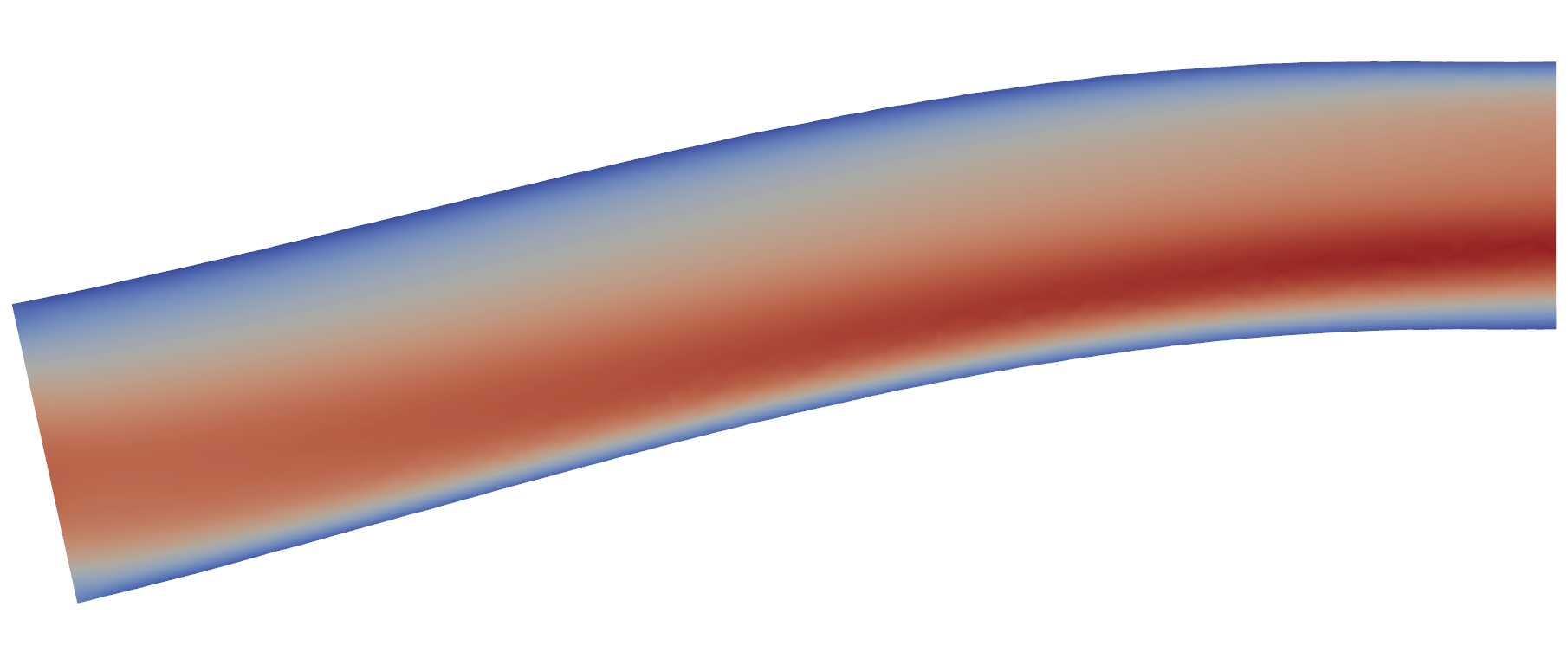} &
    \includegraphics[width=.2\textwidth]{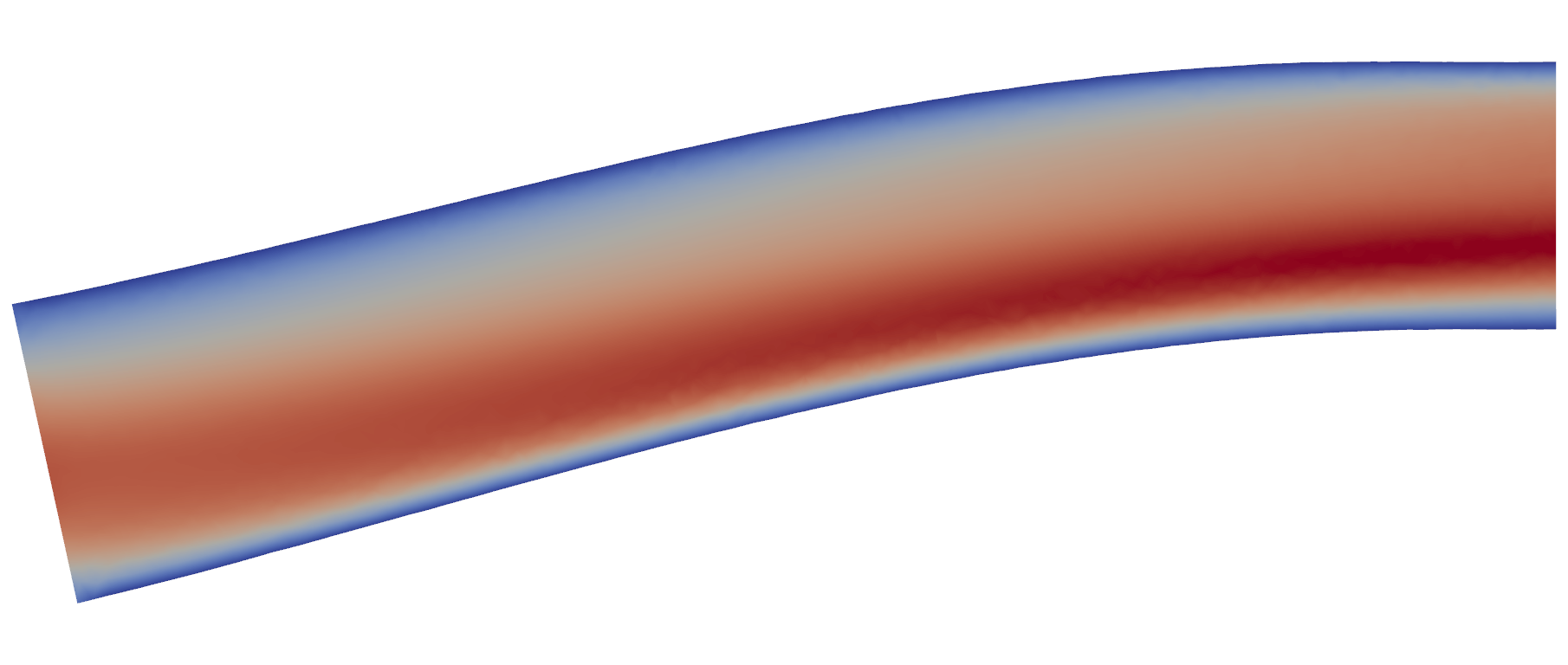} \\
     & $\thet{opt}=0.194$ & $\thet{opt}=0.486$ & $\thet{opt}=0.779$ & $\thet{opt}=0.968$ \\
     & $\mathcal{J}=\rnum{3.032e-06}$ & $\mathcal{J}=\rnum{1.188e-05}$ & $\mathcal{J}=\rnum{2.407e-05}$ & $\mathcal{J}=\rnum{3.533e-05}$ \\
     & $\mathcal{R}=\rnum{2.557e-05}$ & $\mathcal{R}=\rnum{9.724e-05}$ & $\mathcal{R}=\rnum{0.0001622}$ & $\mathcal{R}=\rnum{0.0001882}$ \\
     & iterations: 28 & iterations: 29 & iterations: 25 & iterations: 22 \\
        \rotatebox{90}{\begin{tabular}{c}
         MINI\\
         stab.
    \end{tabular}} & 
    \includegraphics[width=.2\textwidth]{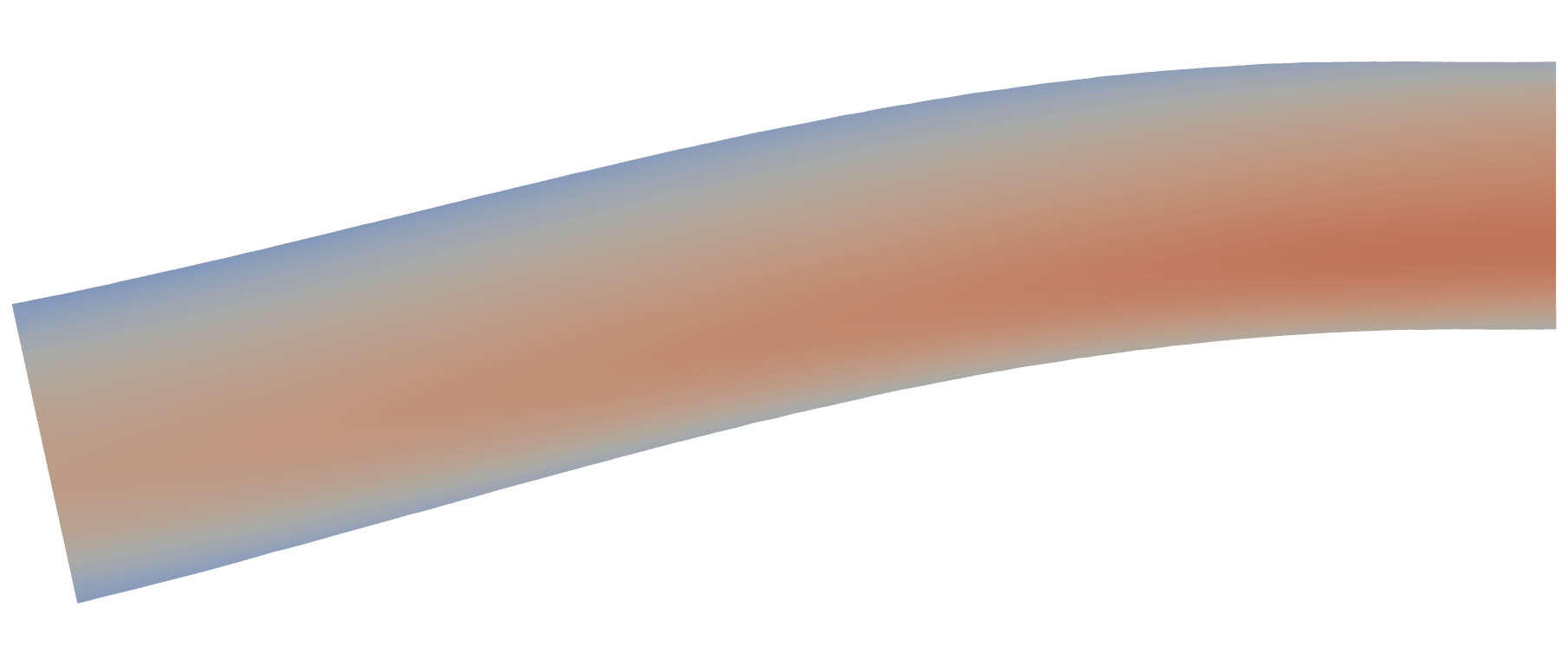} &
    \includegraphics[width=.2\textwidth]{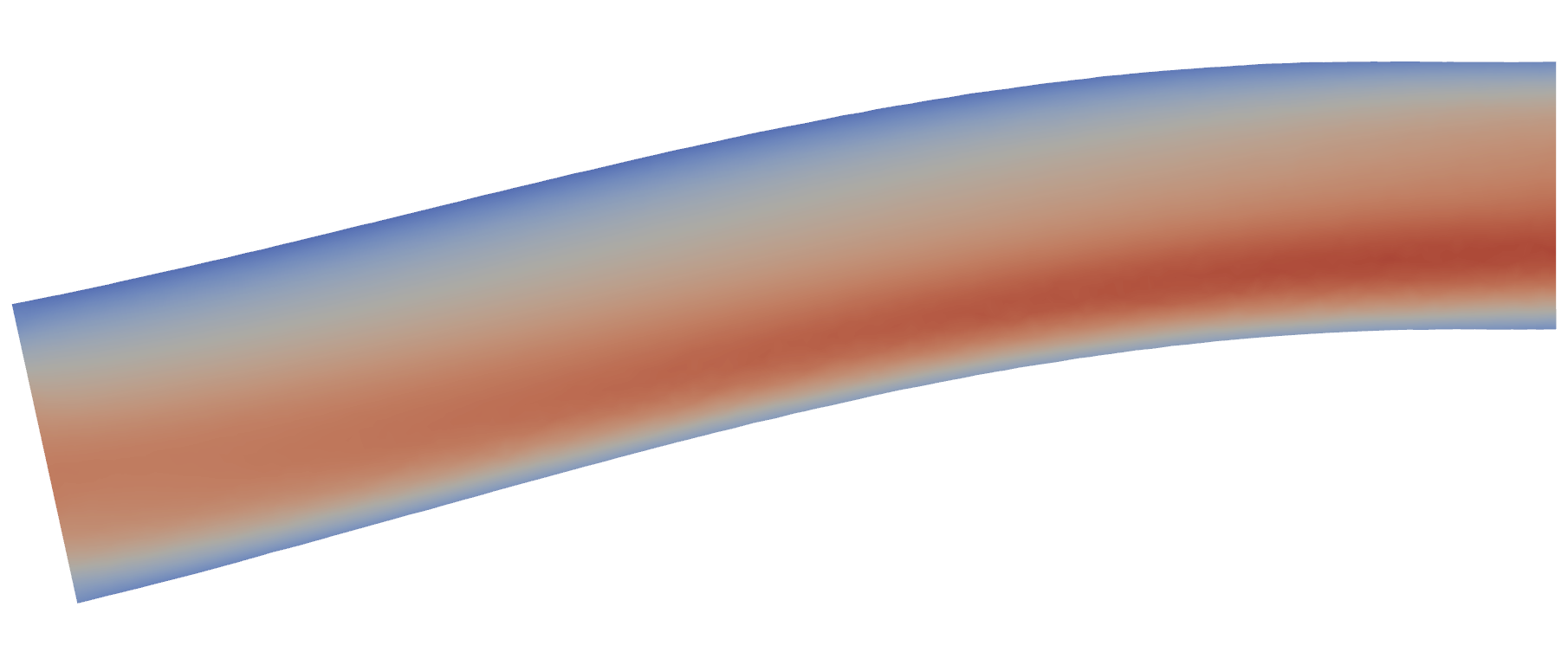} &
    \includegraphics[width=.2\textwidth]{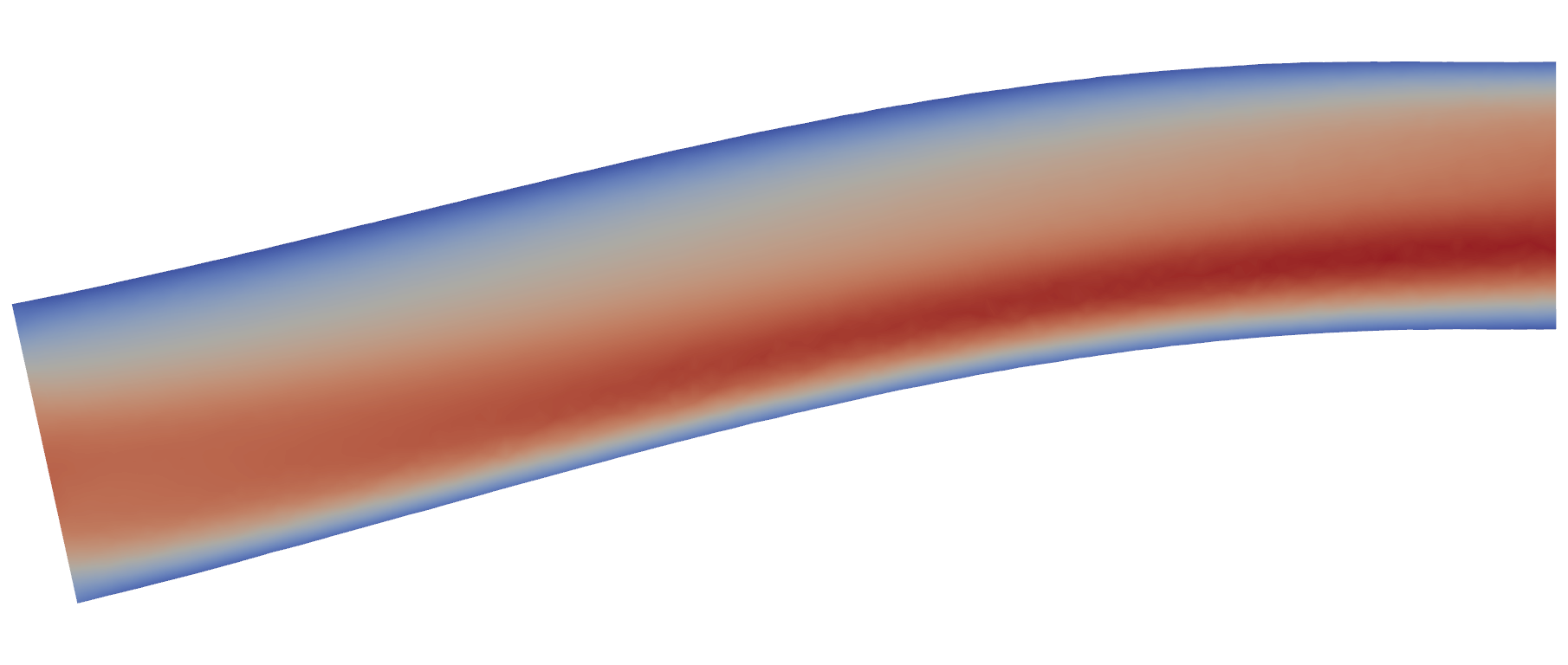} &
    \includegraphics[width=.2\textwidth]{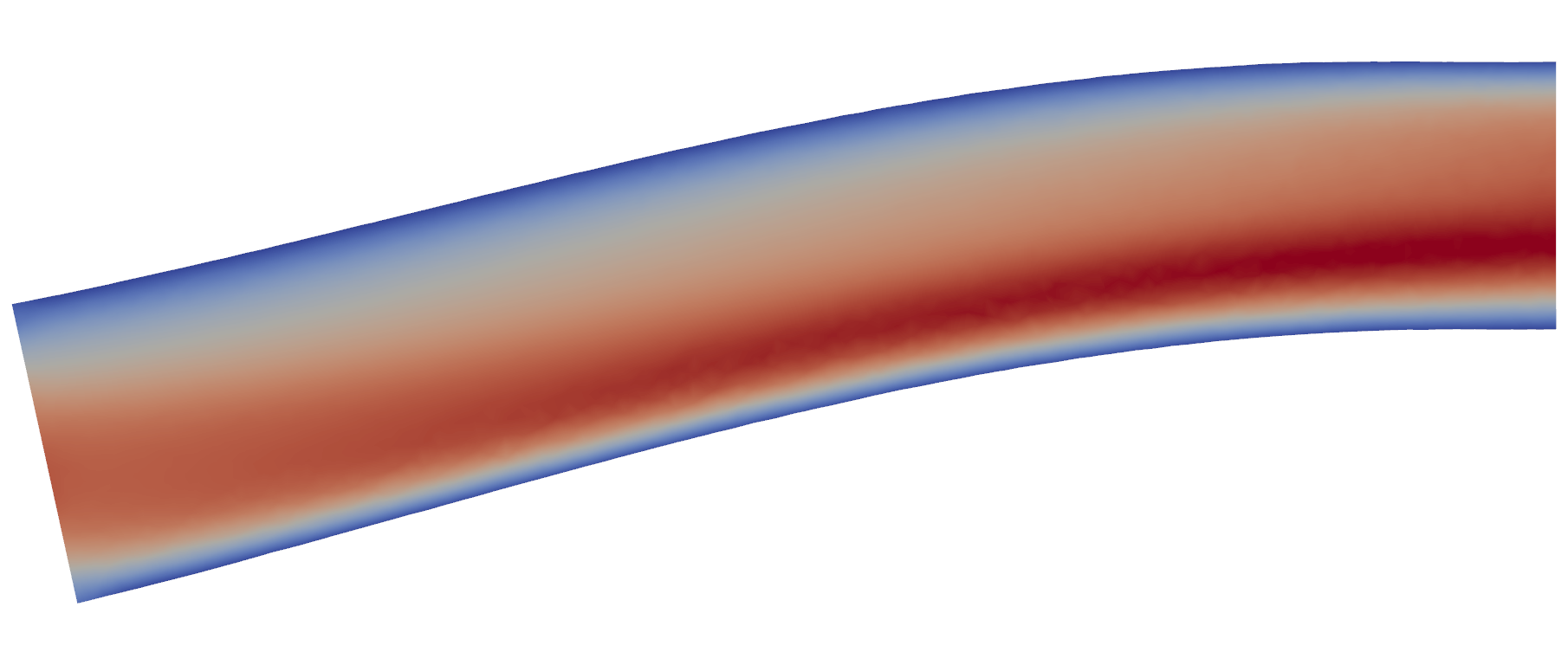} \\
     & $\thet{opt}=0.194$ & $\thet{opt}=0.485$ & $\thet{opt}=0.746$ & $\thet{opt}=0.908$ \\
     & $\mathcal{J}=\rnum{1.498e-05}$ & $\mathcal{J}=\rnum{8.122e-05}$ & $\mathcal{J}=\rnum{0.0001858}$ & $\mathcal{J}=\rnum{0.0002541}$ \\
     & $\mathcal{R}=\rnum{2.708e-05}$ & $\mathcal{R}=\rnum{0.0001004}$ & $\mathcal{R}=\rnum{0.0001657}$ & $\mathcal{R}=\rnum{0.0001922}$ \\
     & iterations: 23 & iterations: 29 & iterations: 21 & iterations: 20 \\
    \rotatebox{90}{\begin{tabular}{c}
         $P_1/P_1$\\
         stab.
    \end{tabular}} &  
    \includegraphics[width=.2\textwidth]{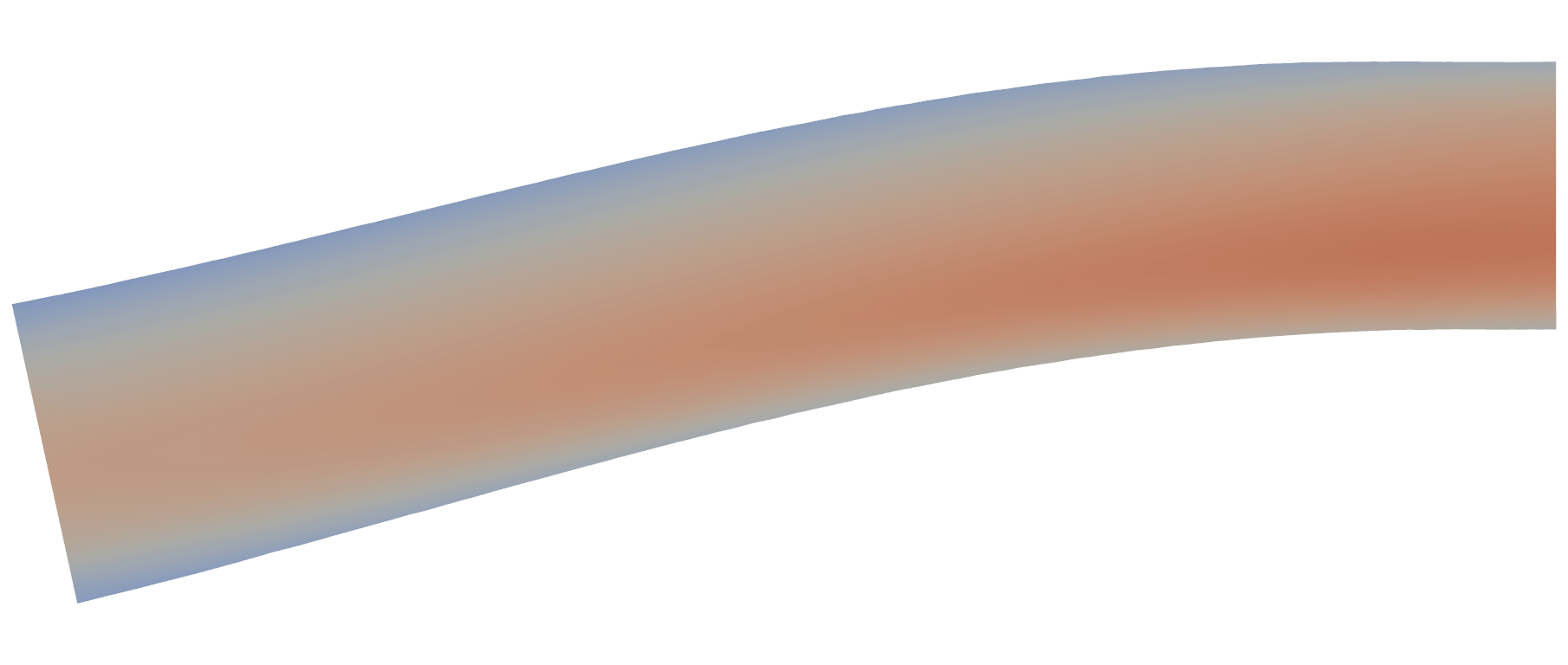} &
    \includegraphics[width=.2\textwidth]{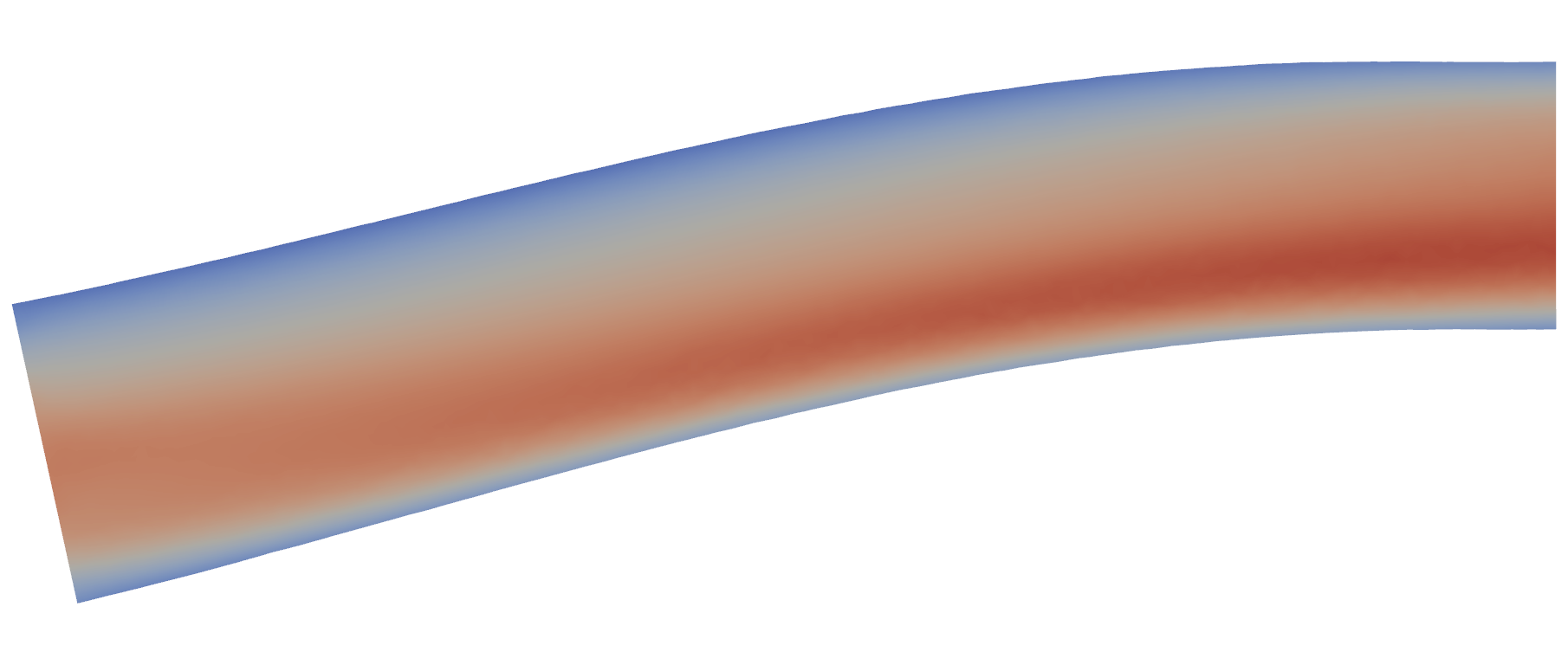} &
    \includegraphics[width=.2\textwidth]{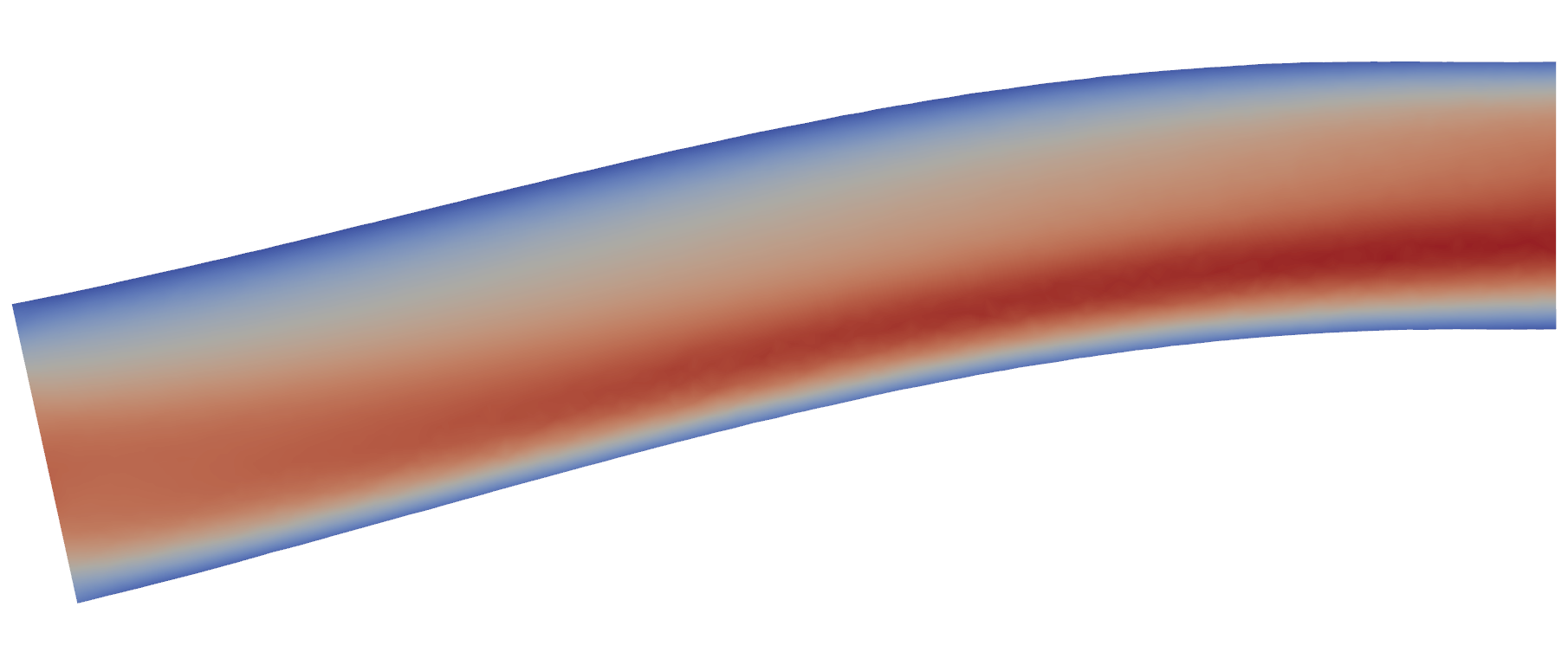} &
    \includegraphics[width=.2\textwidth]{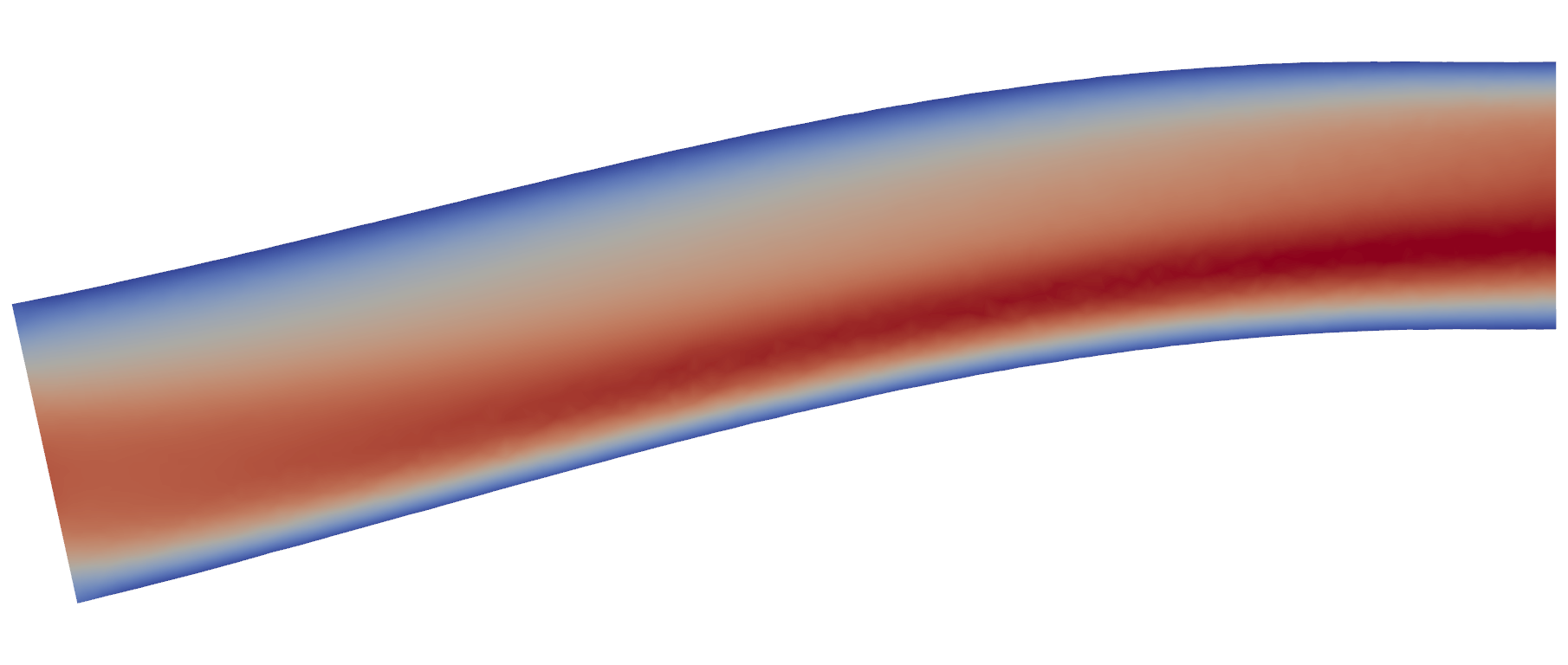} \\
     & $\thet{opt}=0.199$ & $\thet{opt}=0.488$ & $\thet{opt}=0.745$ & $\thet{opt}=0.9$ \\
     & $\mathcal{J}=\rnum{1.606e-05}$ & $\mathcal{J}=\rnum{0.0001248}$ & $\mathcal{J}=\rnum{0.0002822}$ & $\mathcal{J}=\rnum{0.0003827}$ \\
     & $\mathcal{R}=\rnum{2.697e-05}$ & $\mathcal{R}=\rnum{0.0001044}$ & $\mathcal{R}=\rnum{0.000168}$ & $\mathcal{R}=\rnum{0.0001949}$ \\
     & iterations: 25 & iterations: 23 & iterations: 21 & iterations: 21 \\
     & \multicolumn{4}{c}{\includegraphics[width=0.5\textwidth]{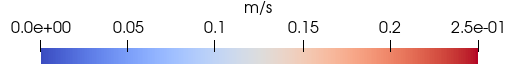}}
\end{tabular}
\caption{Comparison of data without noise with assimilation velocity results in bent tube geometry with edge length $h=1.5\text{ mm}$ using MINI element, stabilized MINI element ($\alpha_v=0.01, \alpha_p=0$) and stabilized $P_1/P_1$ element ($\alpha_v=\alpha_p=0.01$) for multiple values of $\theta$.}
\label{fig:elem_bent}
\medskip
\centering
\begin{tabular}{c c c c c}
    & $\theta=0.2$ & $\theta=0.5$ & $\theta=0.8$ & $\theta=1.0$ \\ 
     \rotatebox{90}{MINI} &
     \includegraphics[width=.2\textwidth]{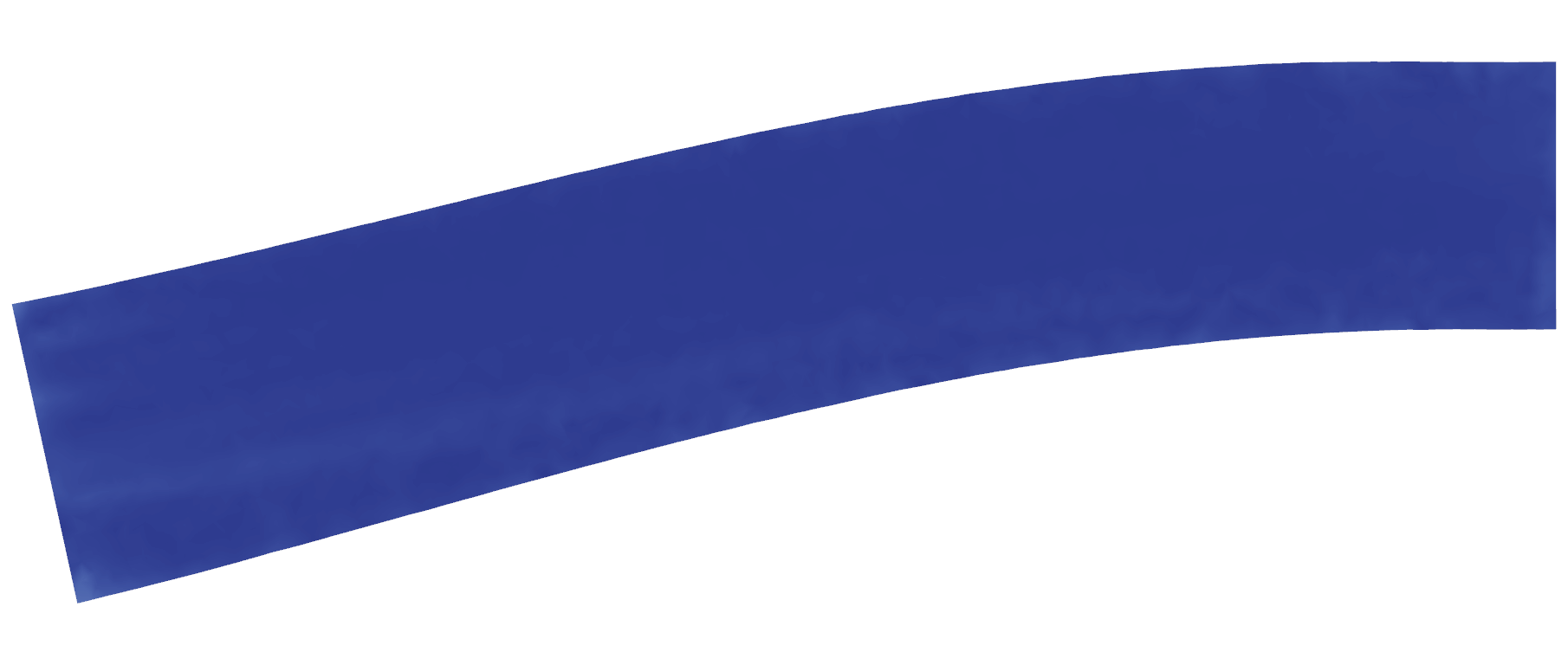} & 
     \includegraphics[width=.2\textwidth]{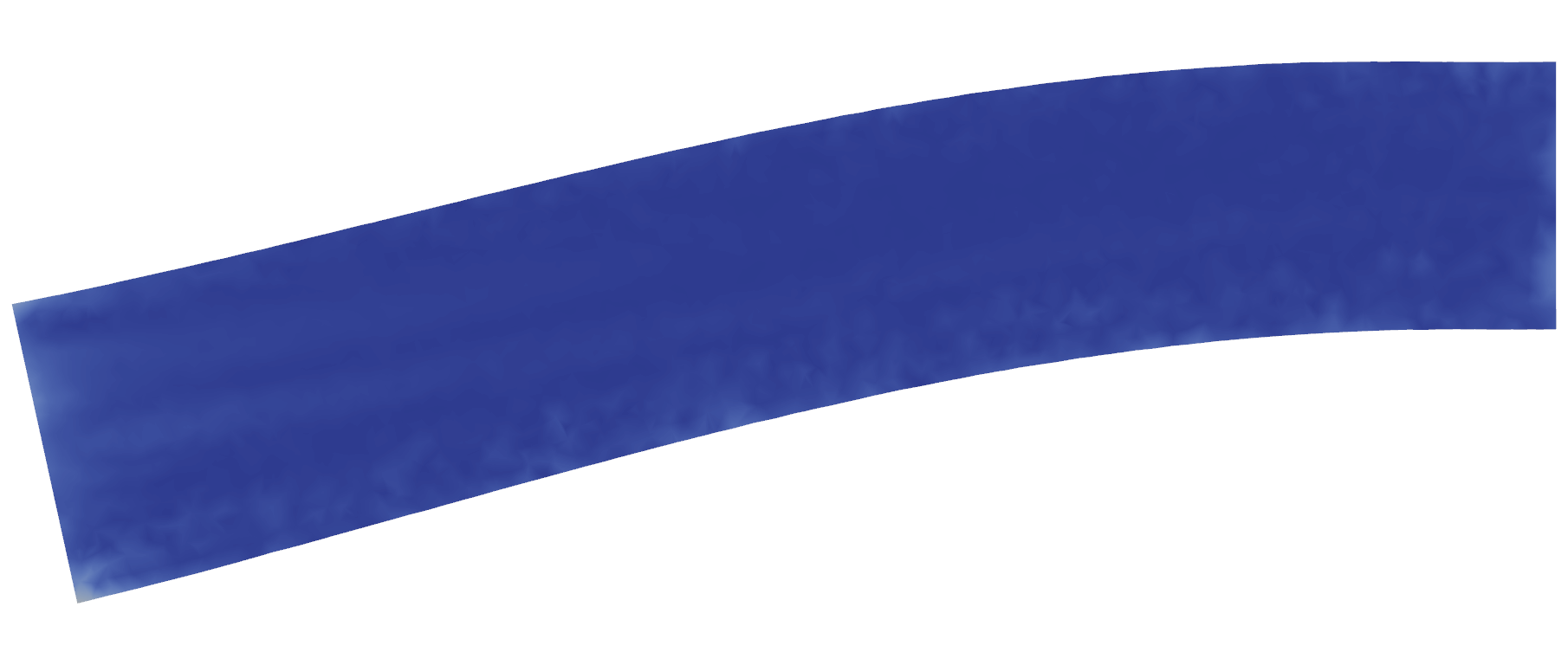} &
     \includegraphics[width=.2\textwidth]{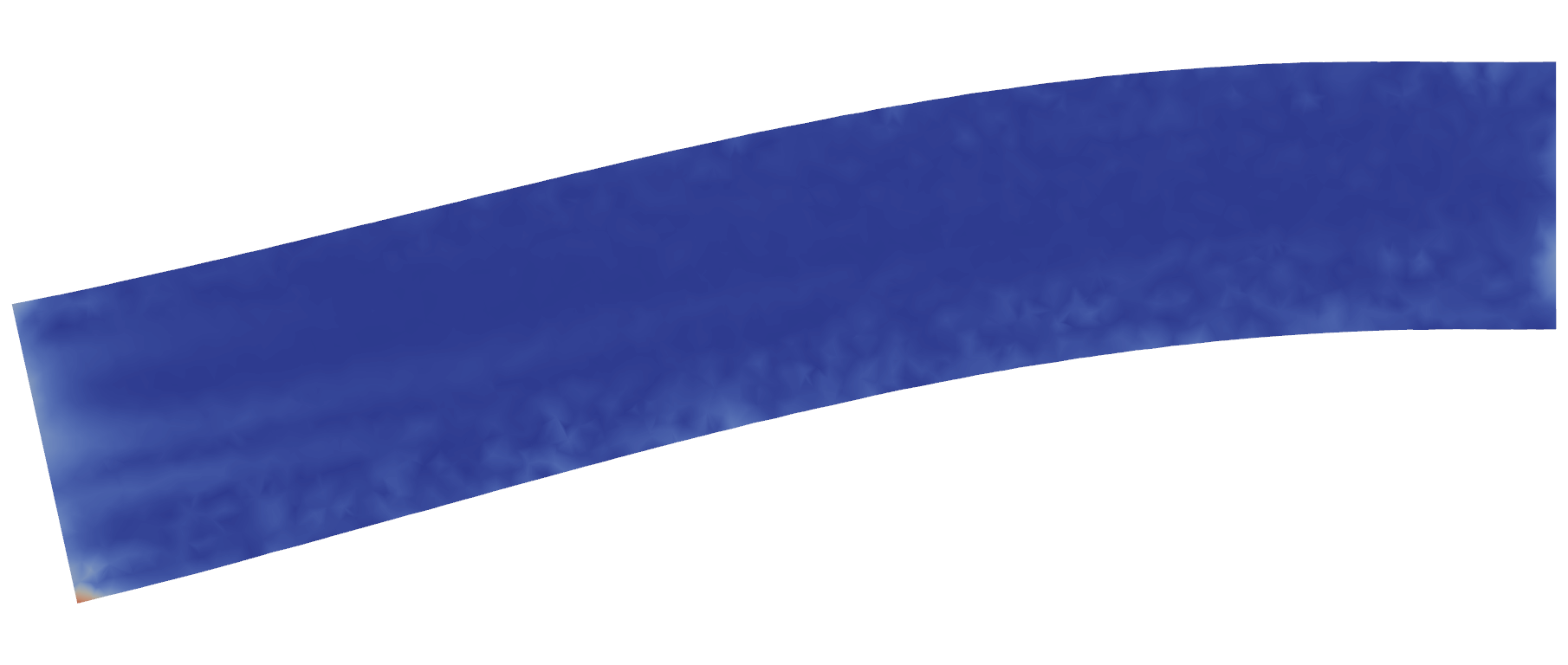} &
     \includegraphics[width=.2\textwidth]{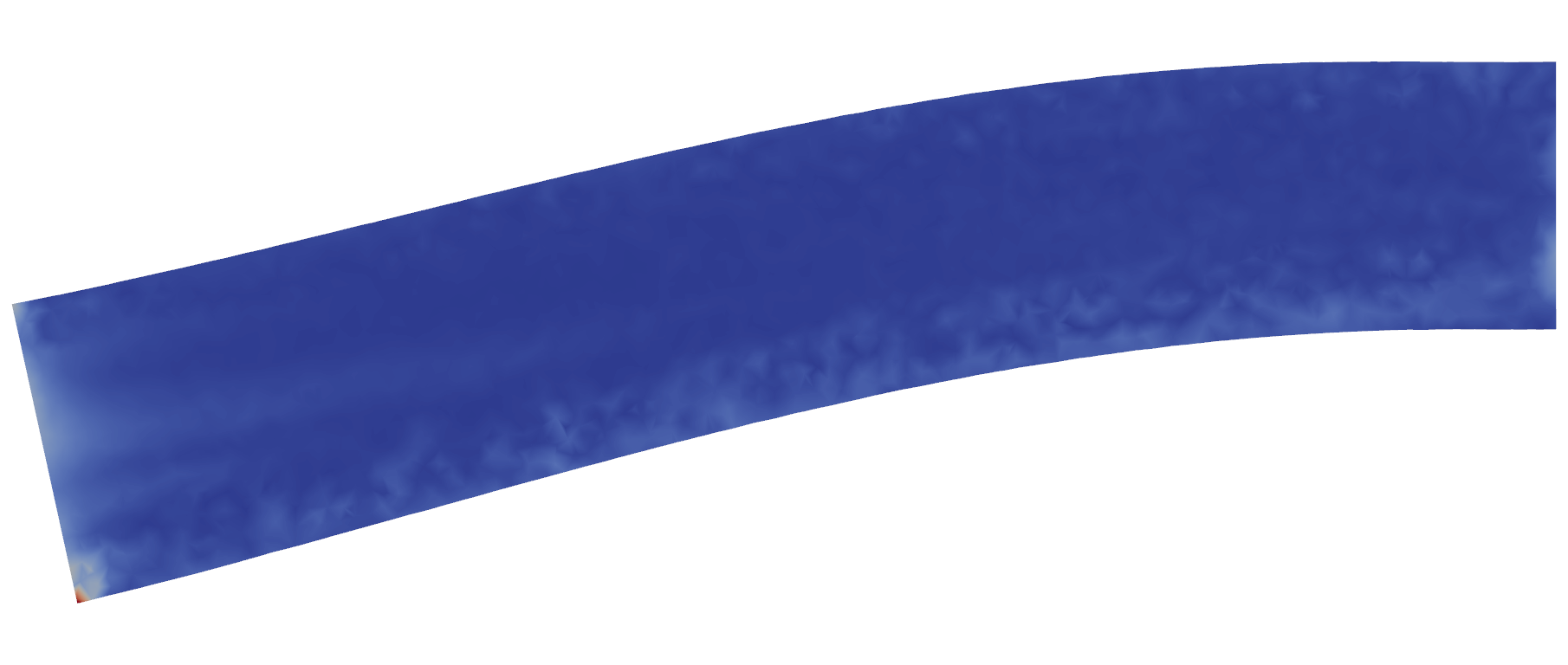} \\
     \rotatebox{90}{\begin{tabular}{c} 
        MINI\\
        stab.
    \end{tabular}}&
     \includegraphics[width=.19\textwidth]{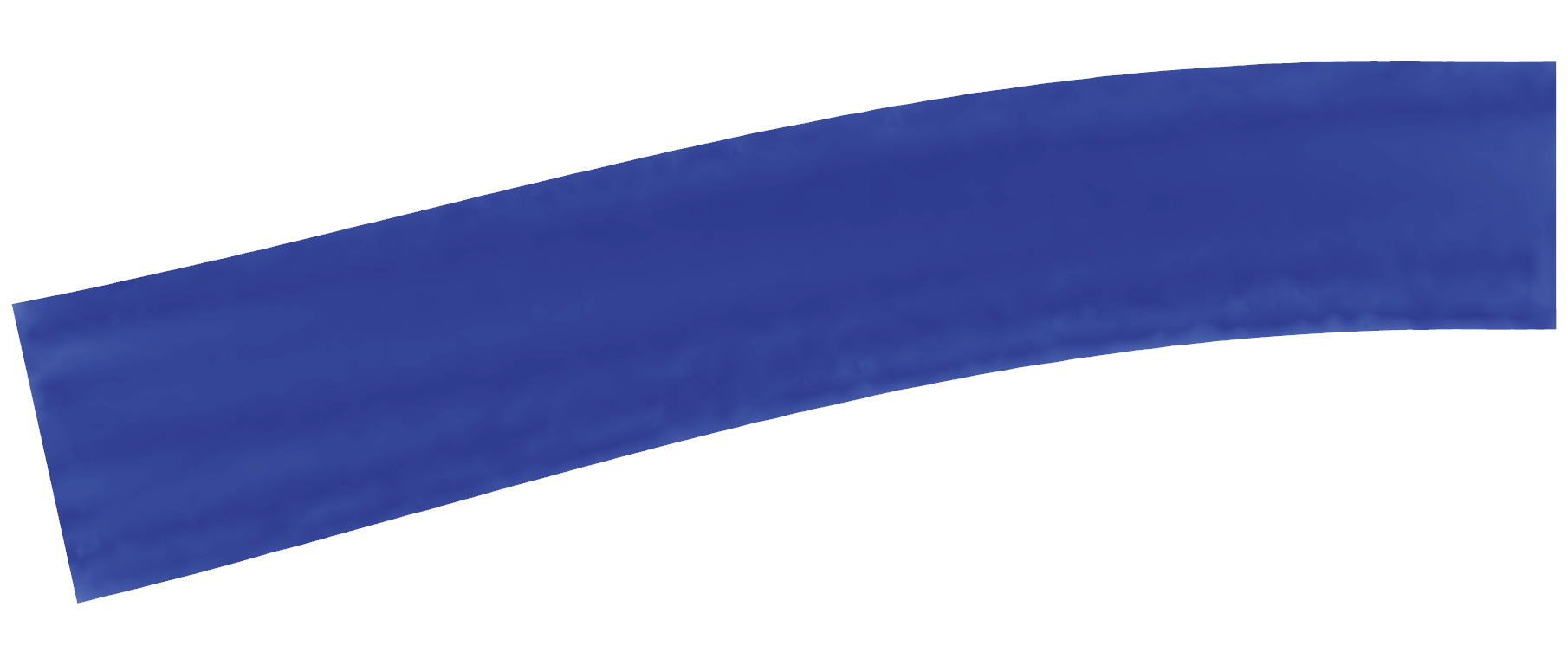} &
     \includegraphics[width=.19\textwidth]{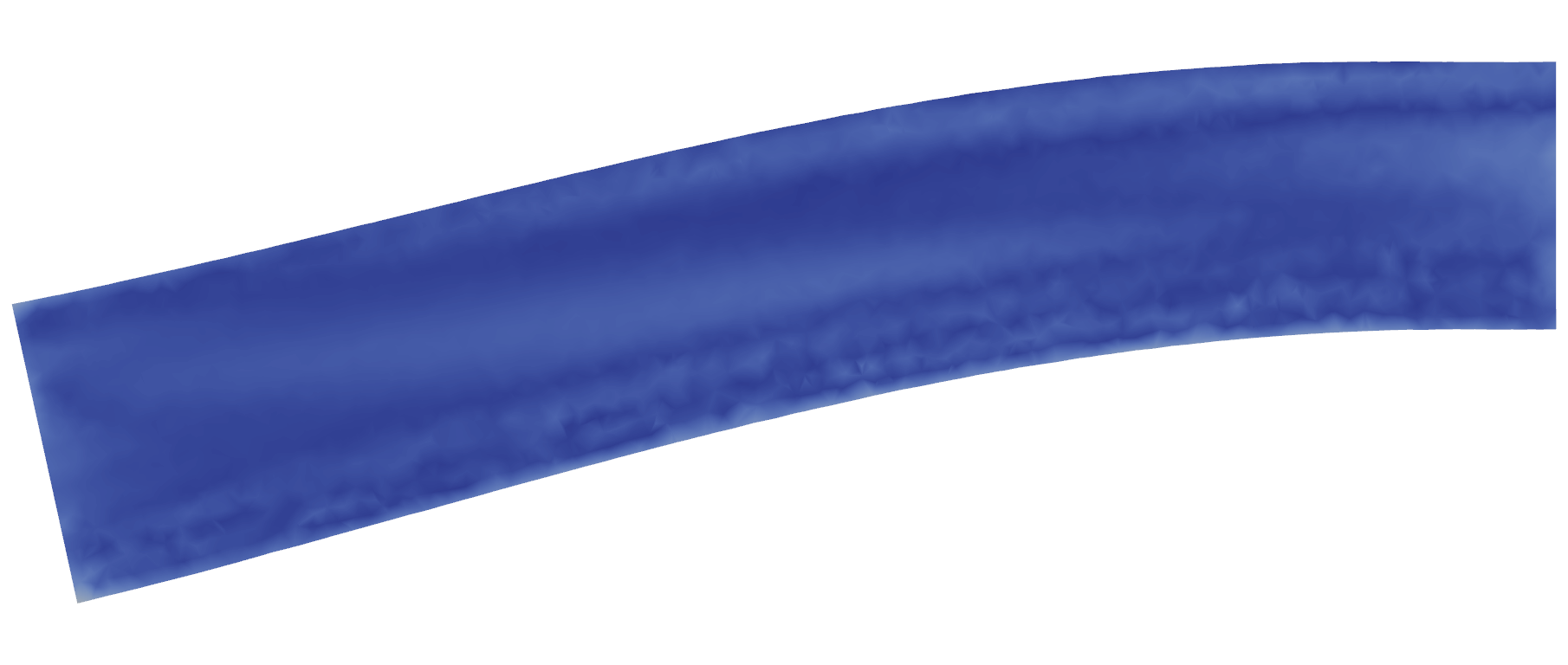} &
     \includegraphics[width=.19\textwidth]{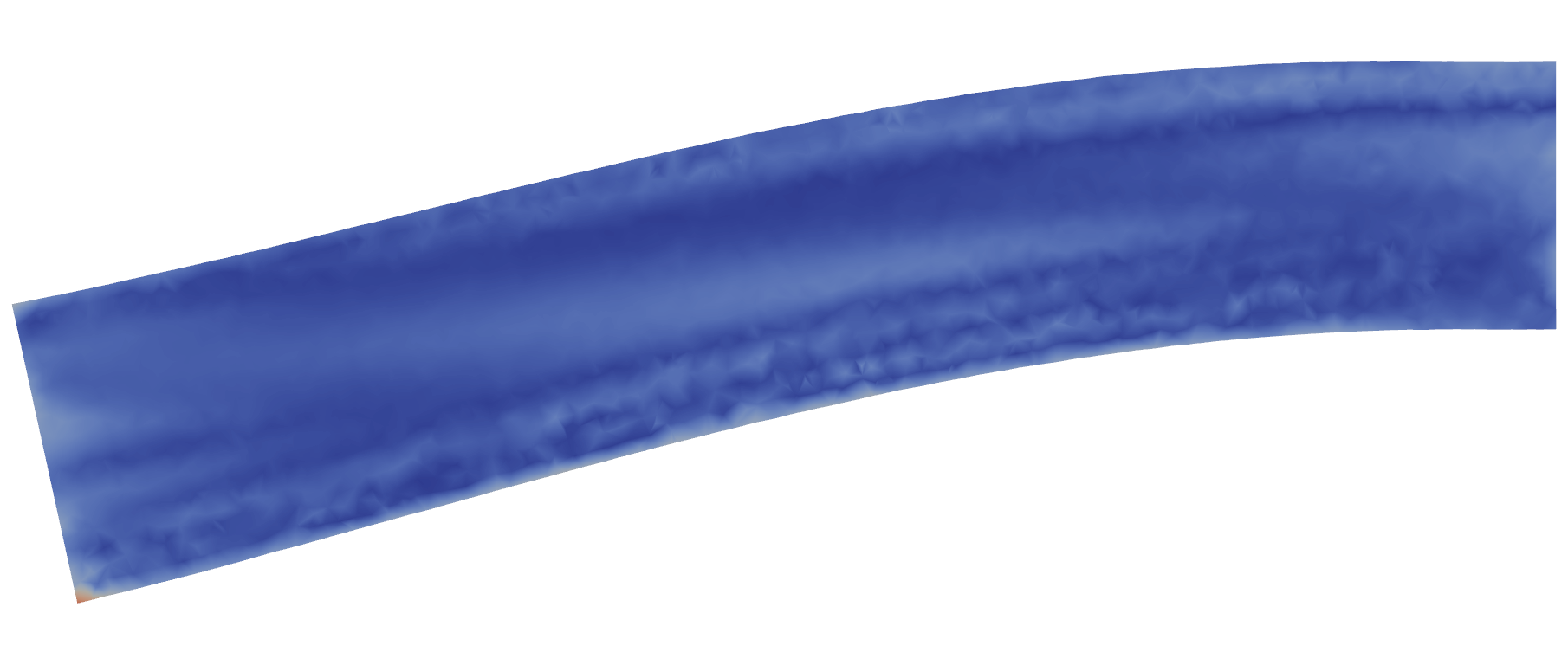} &
     \includegraphics[width=.19\textwidth]{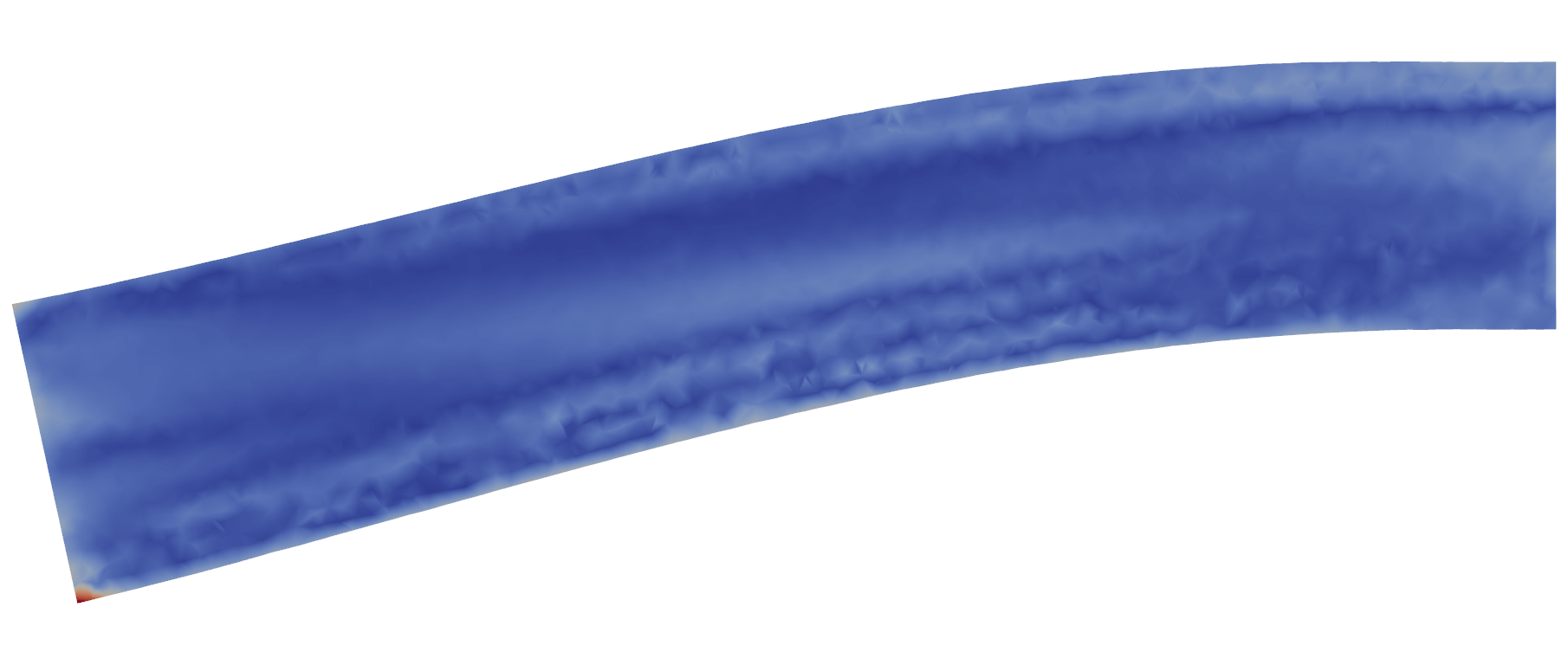} \\
        \rotatebox{90}{\begin{tabular}{c} 
            $P_1/P_1$\\
            stab.
    \end{tabular}}&
     \includegraphics[width=.19\textwidth]{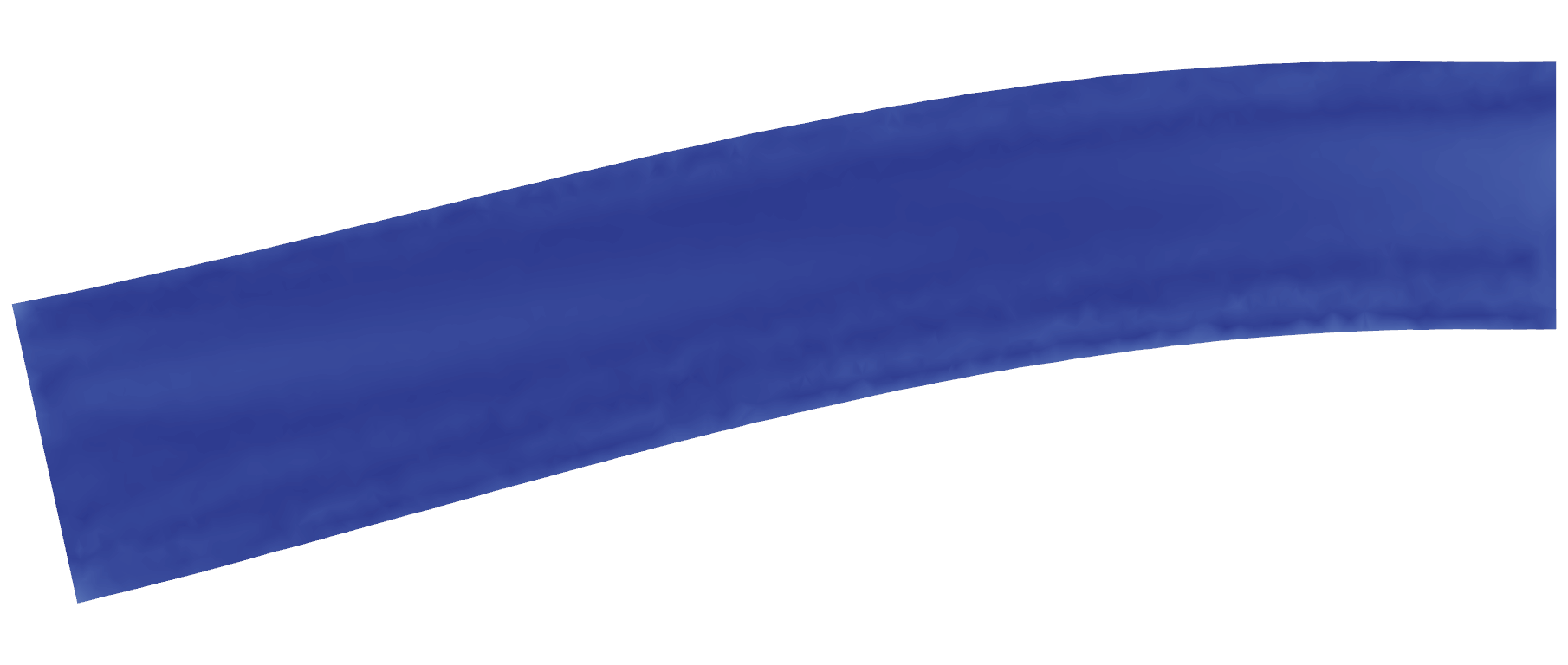} &
     \includegraphics[width=.19\textwidth]{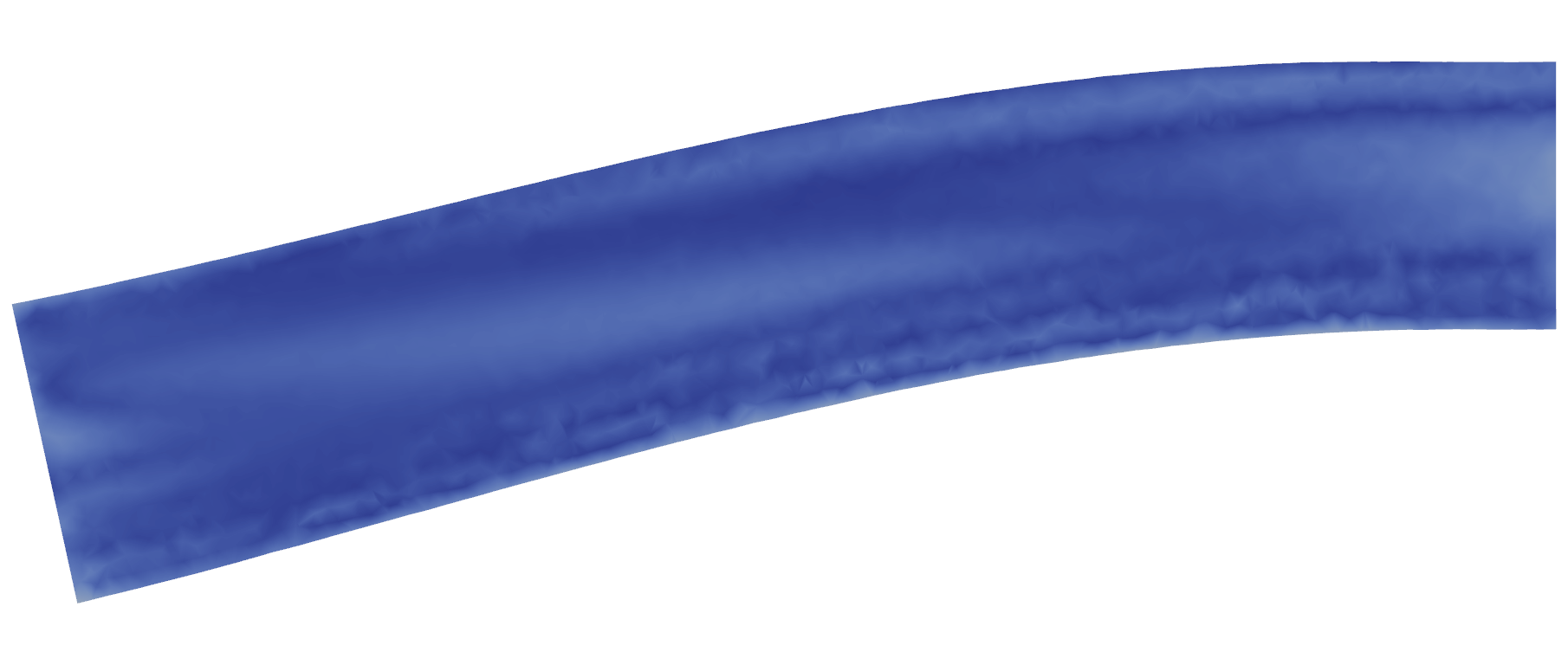} &
     \includegraphics[width=.19\textwidth]{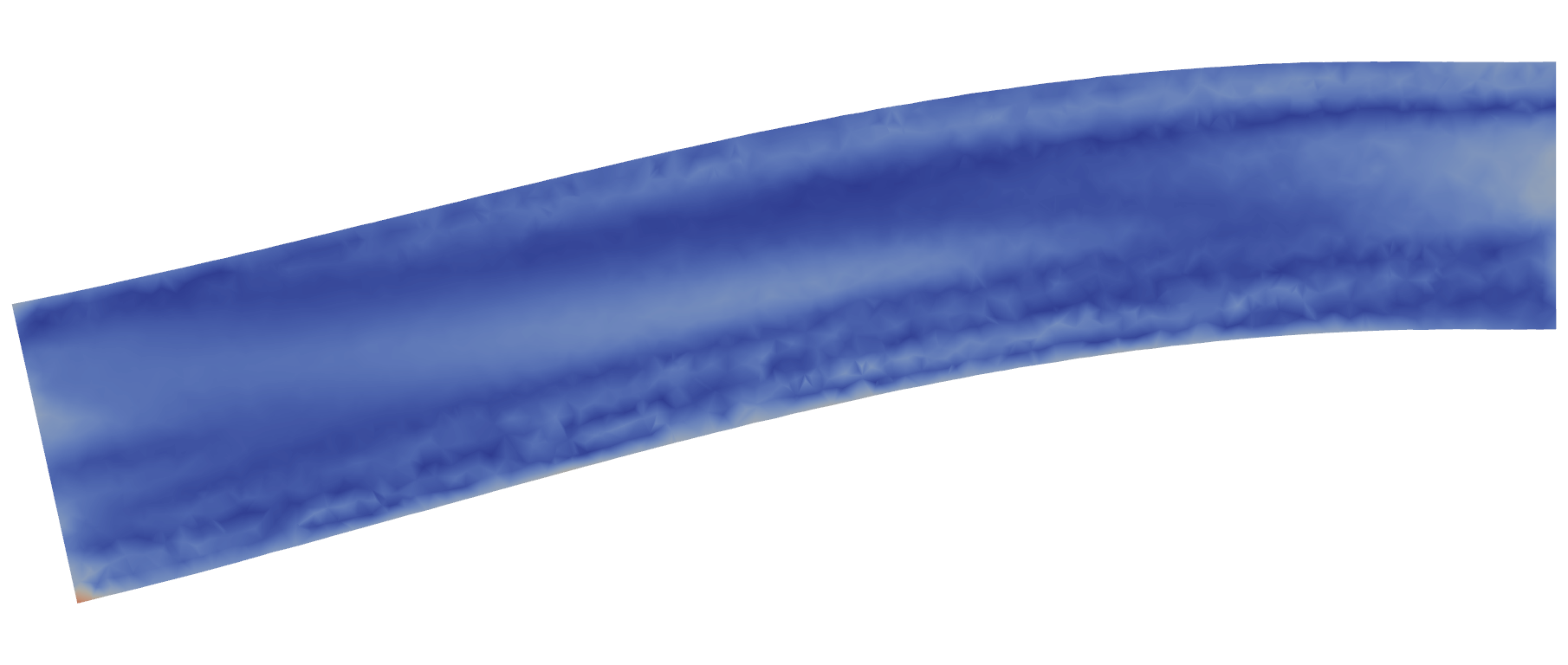} &
     \includegraphics[width=.19\textwidth]{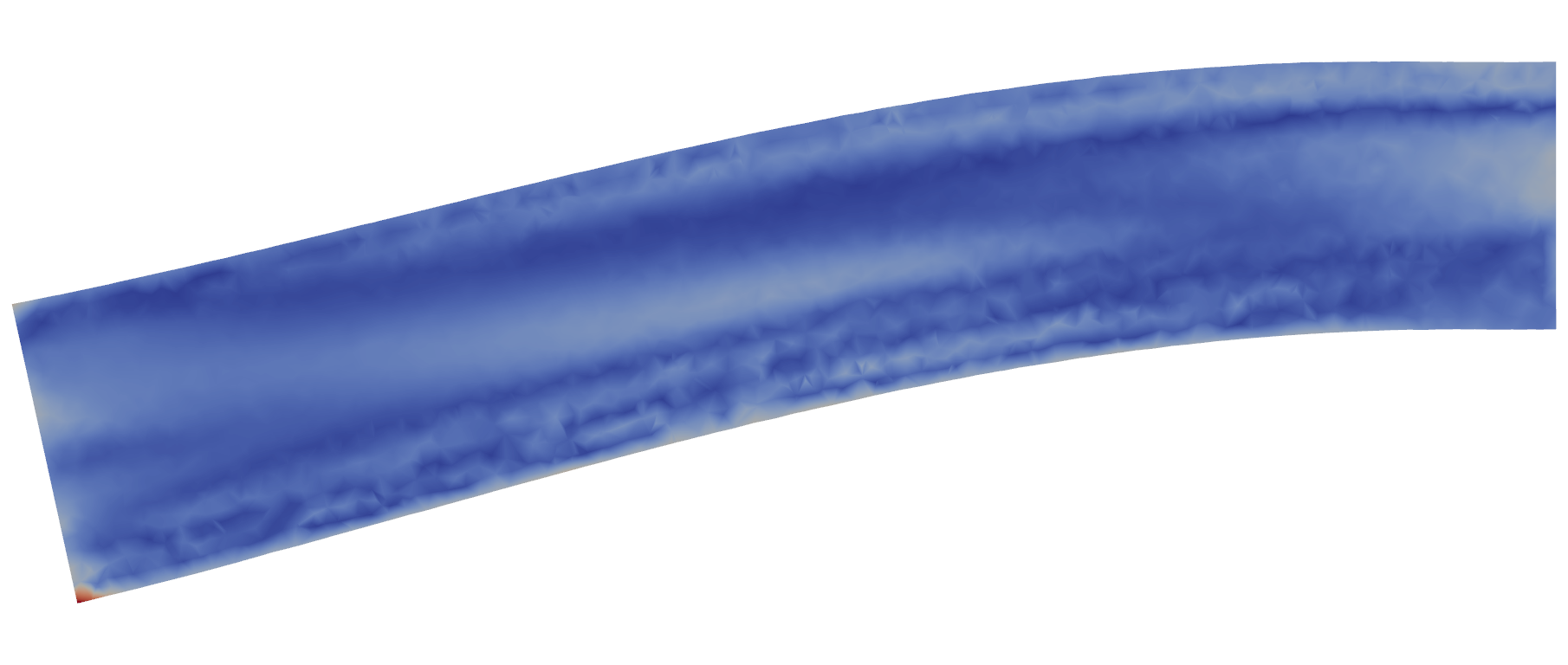} \\
    & \multicolumn{4}{c}{\includegraphics[width=0.5\textwidth]{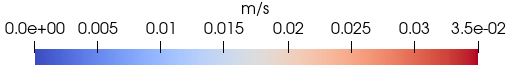}}
\end{tabular}
\caption{Visulatization of the difference between data and computed velocity fields on bent geometry ($h=1.5\text{ mm}$) using MINI element, stabilized MINI element and stabilized $P_1/P_1$ element for multiple values of $\theta$.}
\label{fig:bent_errors}
\end{figure}

\begin{figure}
\centering
\begin{tabular}{c c c c c}
    \rotatebox{90}{\begin{tabular}{c}
         data
    \end{tabular}} &  
    \includegraphics[width=0.19\textwidth]{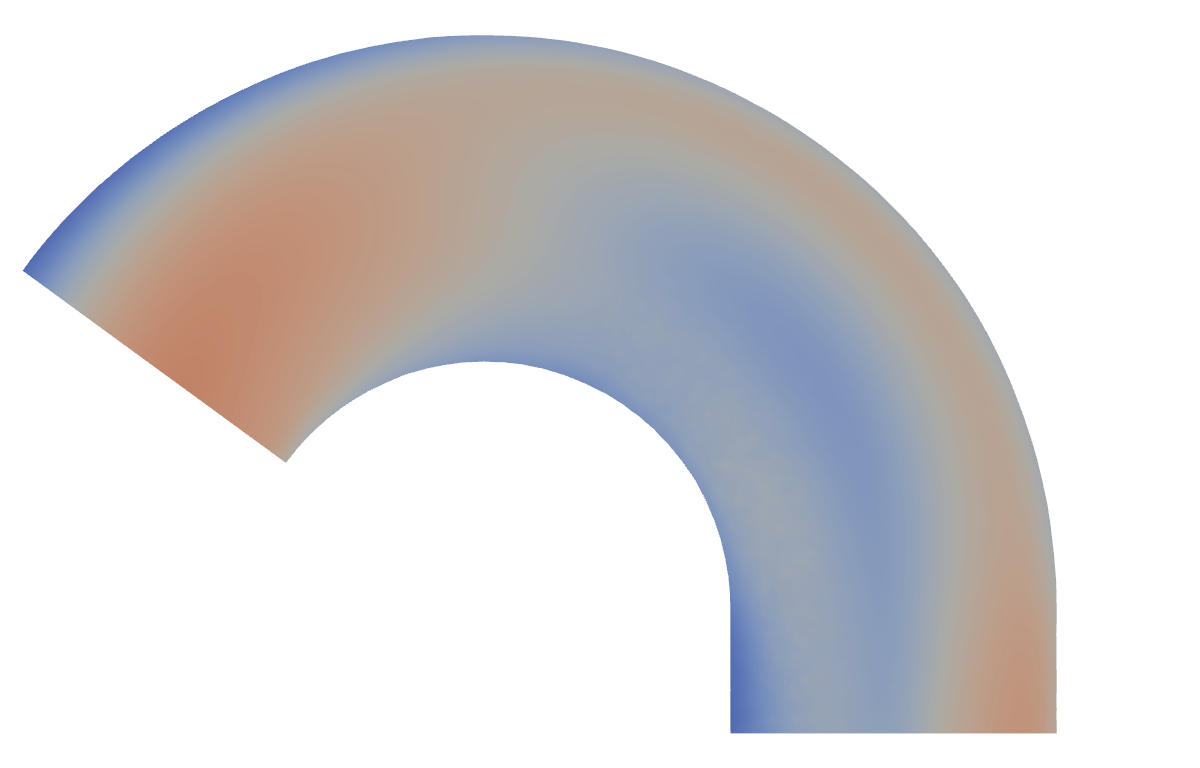} &
    \includegraphics[width=0.19\textwidth]{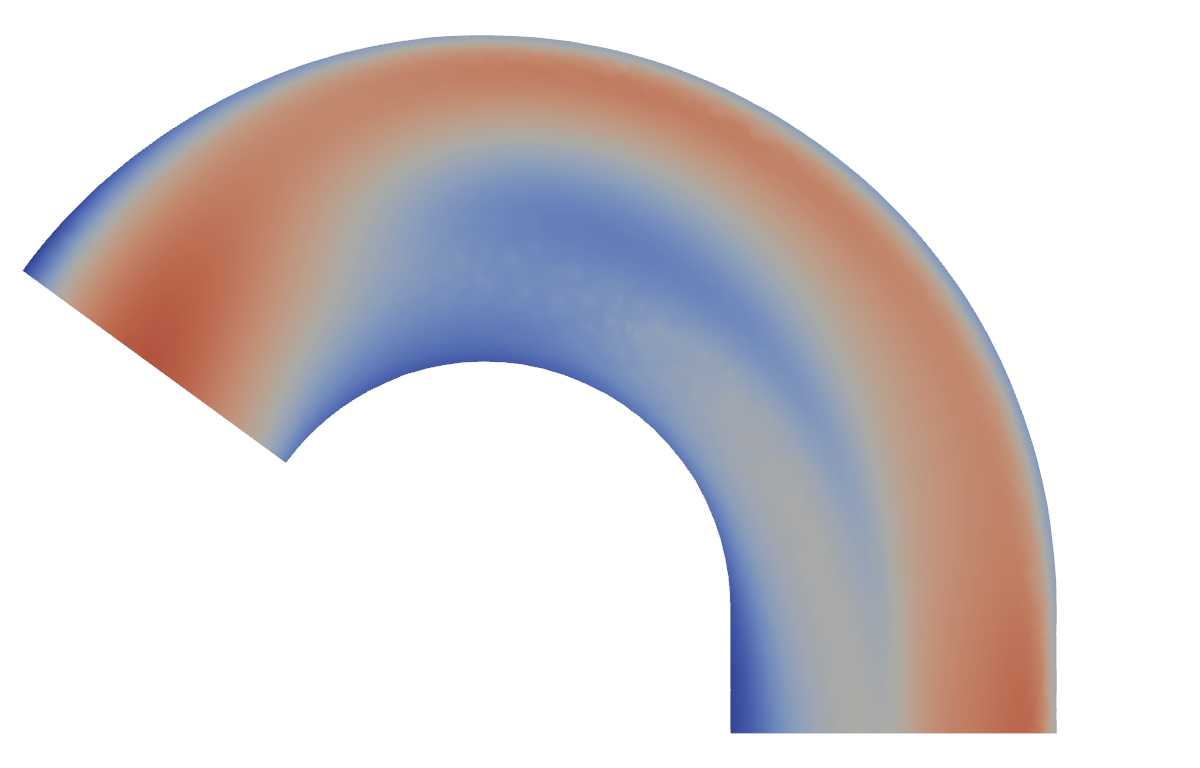} &
    \includegraphics[width=0.19\textwidth]{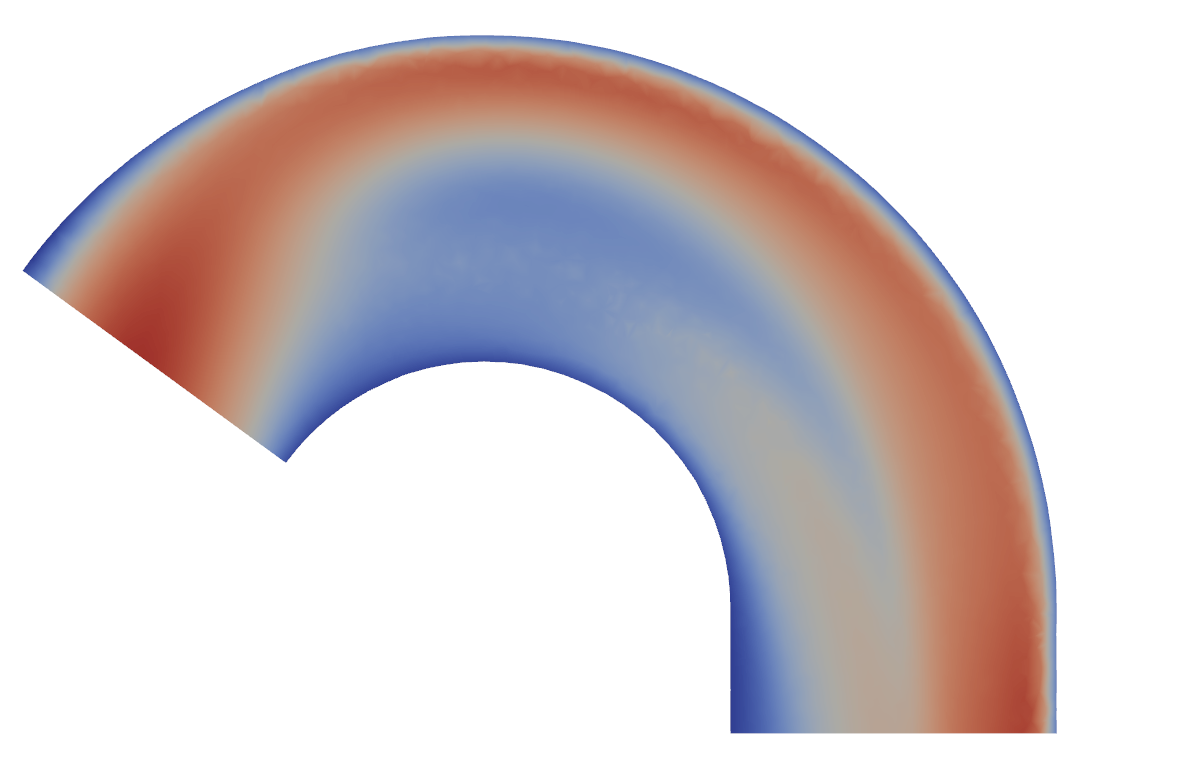} &
    \includegraphics[width=0.19\textwidth]{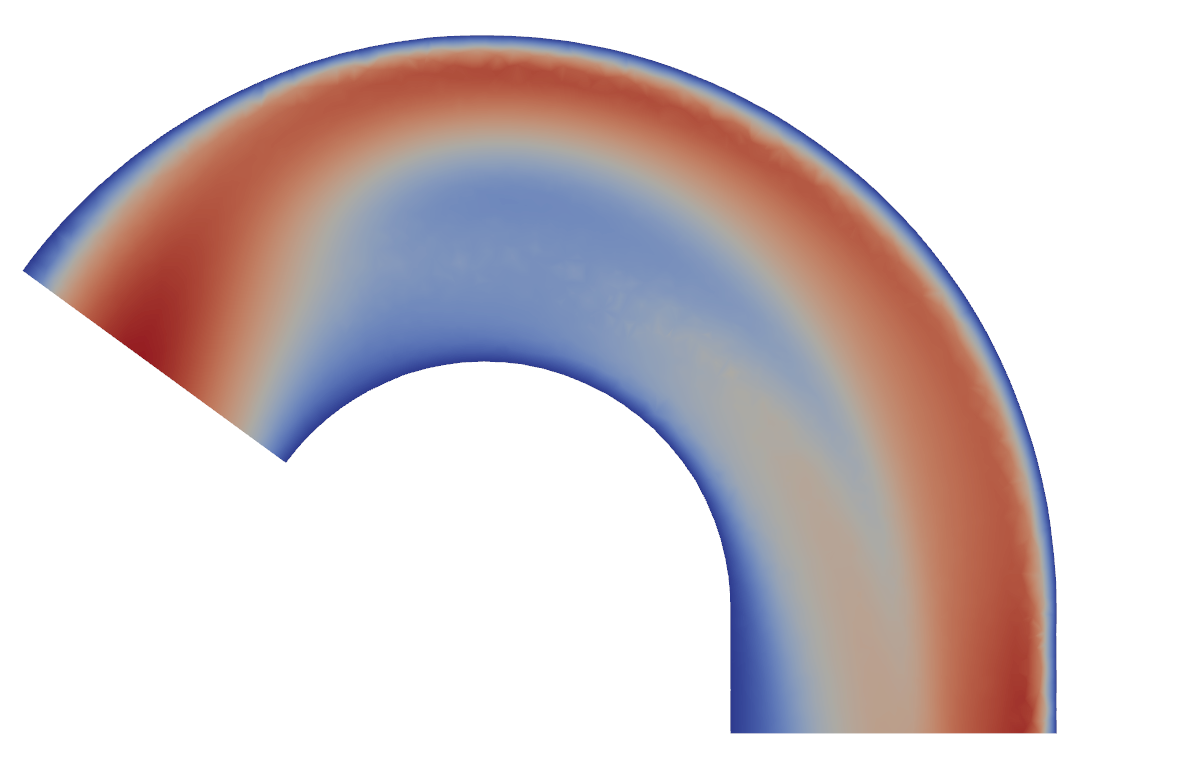} \\
    & $\theta=0.2$ & $\theta=0.5$ & $\theta=0.8$ & $\theta=1$ \\
    \rotatebox{90}{\begin{tabular}{c}
         MINI
    \end{tabular}} & 
    \includegraphics[width=0.19\textwidth]{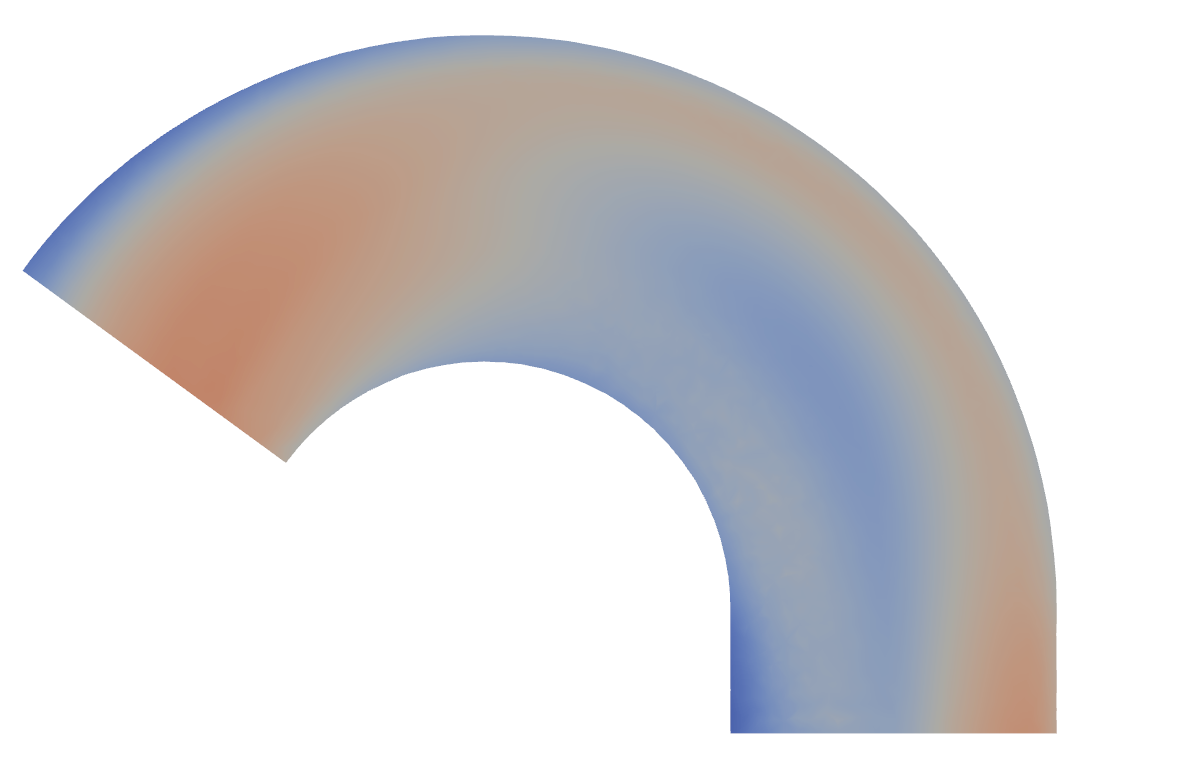} &
    \includegraphics[width=0.19\textwidth]{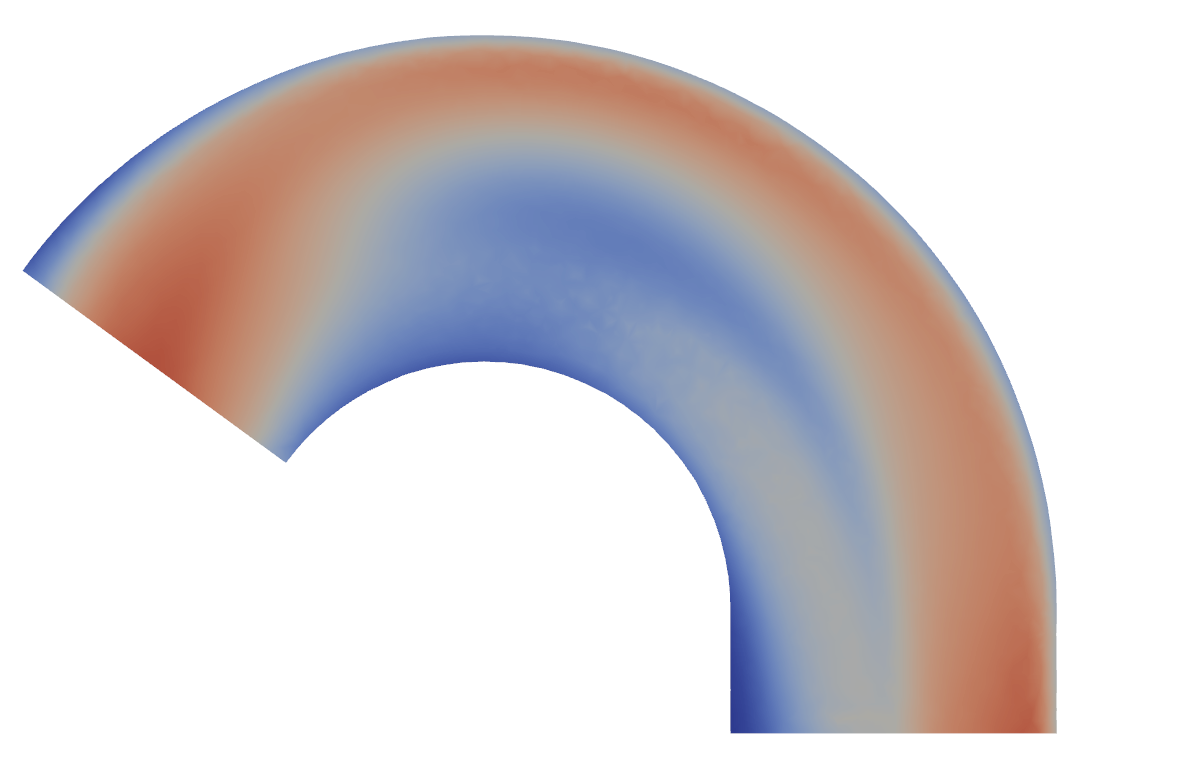} &
    \includegraphics[width=0.19\textwidth]{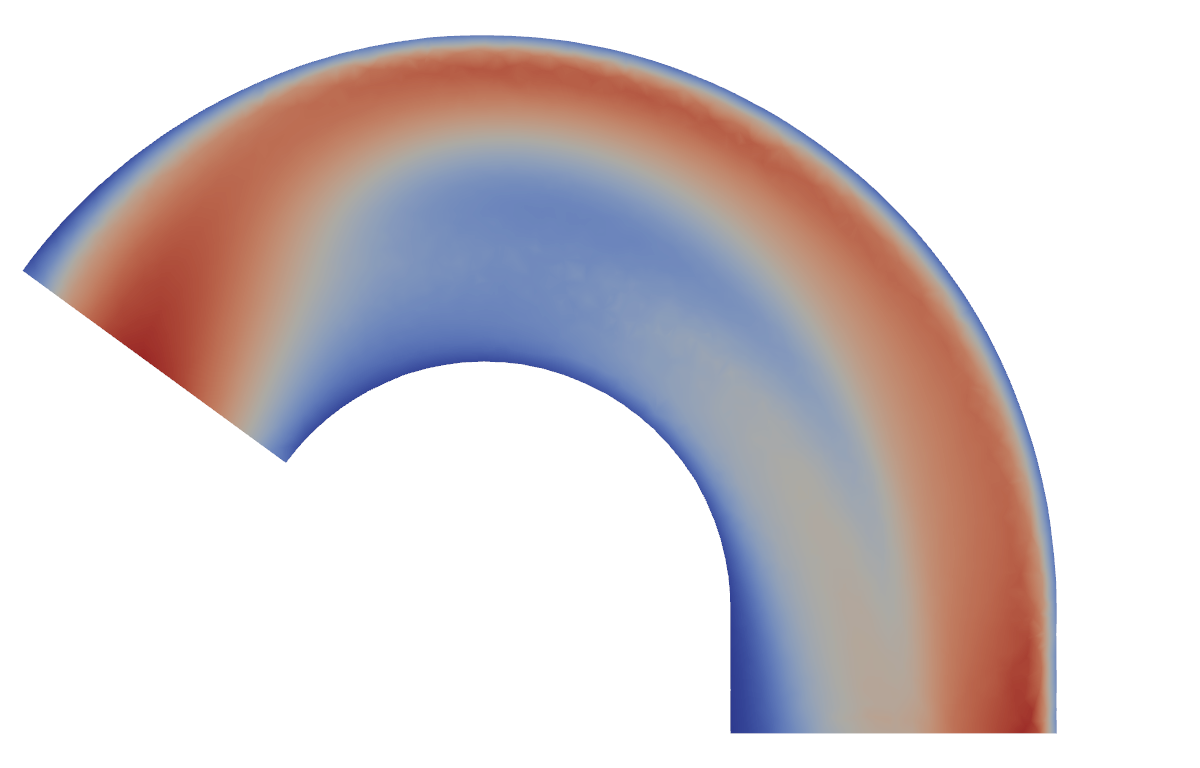} &
    \includegraphics[width=0.19\textwidth]{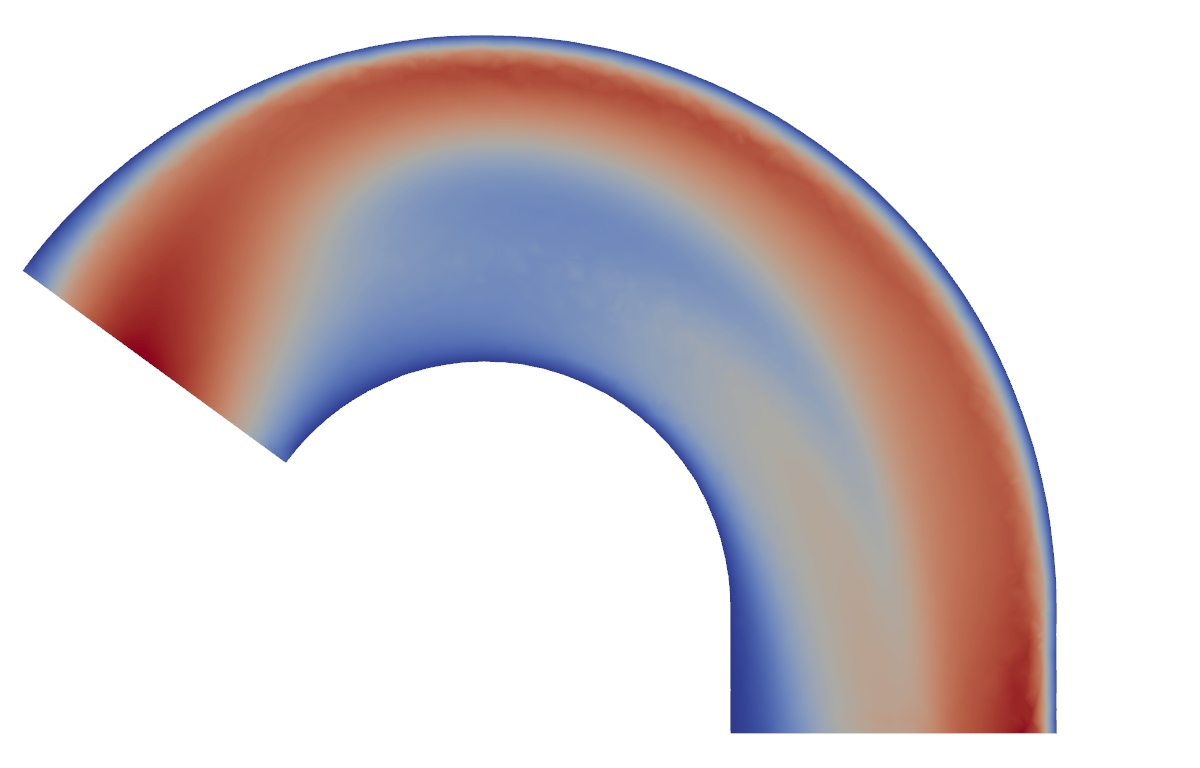} \\
     & $\thet{opt}=0.185$ & $\thet{opt}=0.476$ & $\thet{opt}=0.768$ & $\thet{opt}=0.959$ \\
     & $\mathcal{J}=\rnum{2.475e-05}$ & $\mathcal{J}=\rnum{6.657e-05}$ & $\mathcal{J}=\rnum{0.0001032}$ & $\mathcal{J}=\rnum{0.0001292}$ \\
     & $\mathcal{R}=\rnum{0.0001286}$ & $\mathcal{R}=\rnum{0.0002611}$ & $\mathcal{R}=\rnum{0.0003433}$ & $\mathcal{R}=\rnum{0.0003858}$ \\
     & iterations: 34 & iterations: 32 & iterations: 27 & iterations: 24 \\
    \rotatebox{90}{\begin{tabular}{c}
         MINI \\
         stab.
    \end{tabular}} & 
    \includegraphics[width=0.19\textwidth]{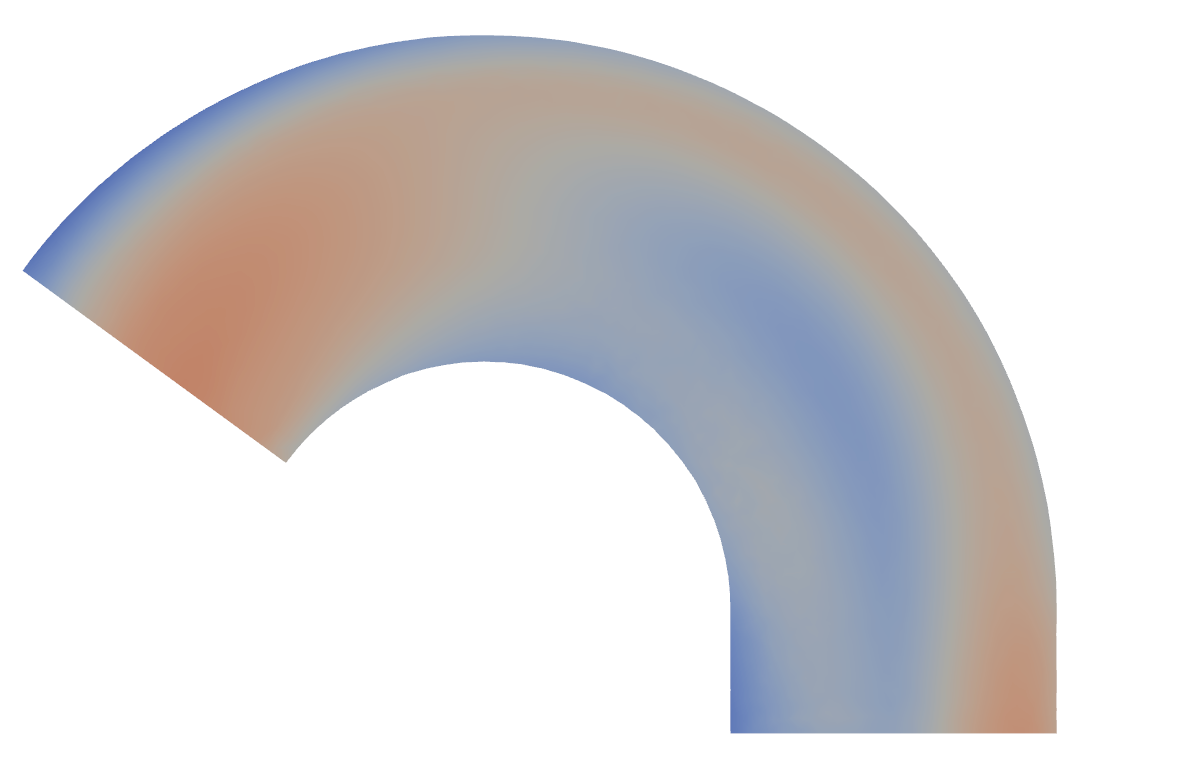} &
    \includegraphics[width=0.19\textwidth]{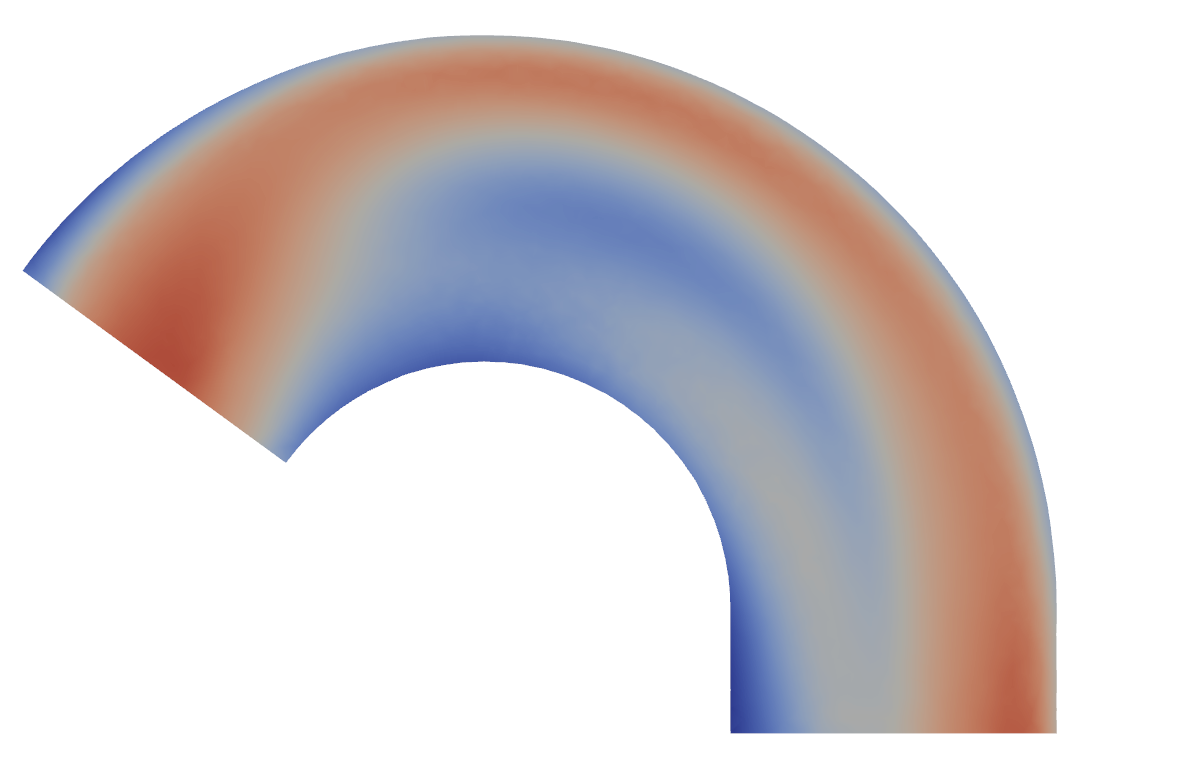} &
    \includegraphics[width=0.19\textwidth]{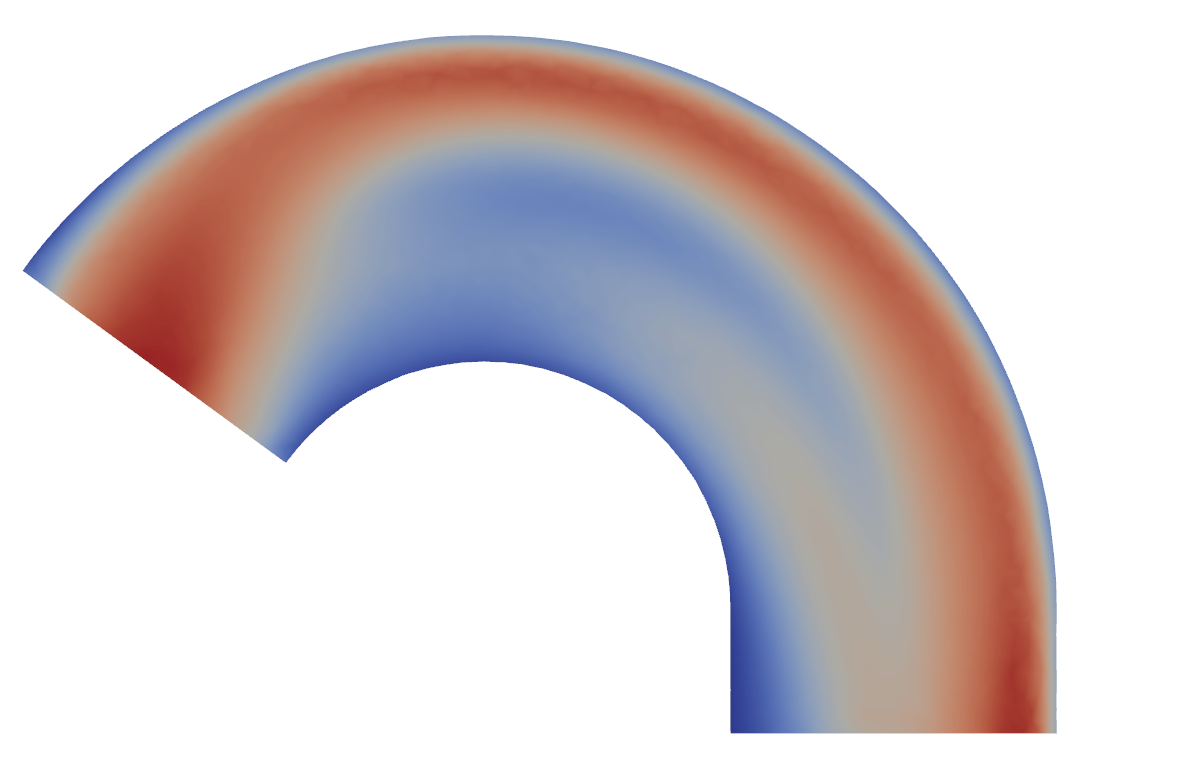} &
    \includegraphics[width=0.19\textwidth]{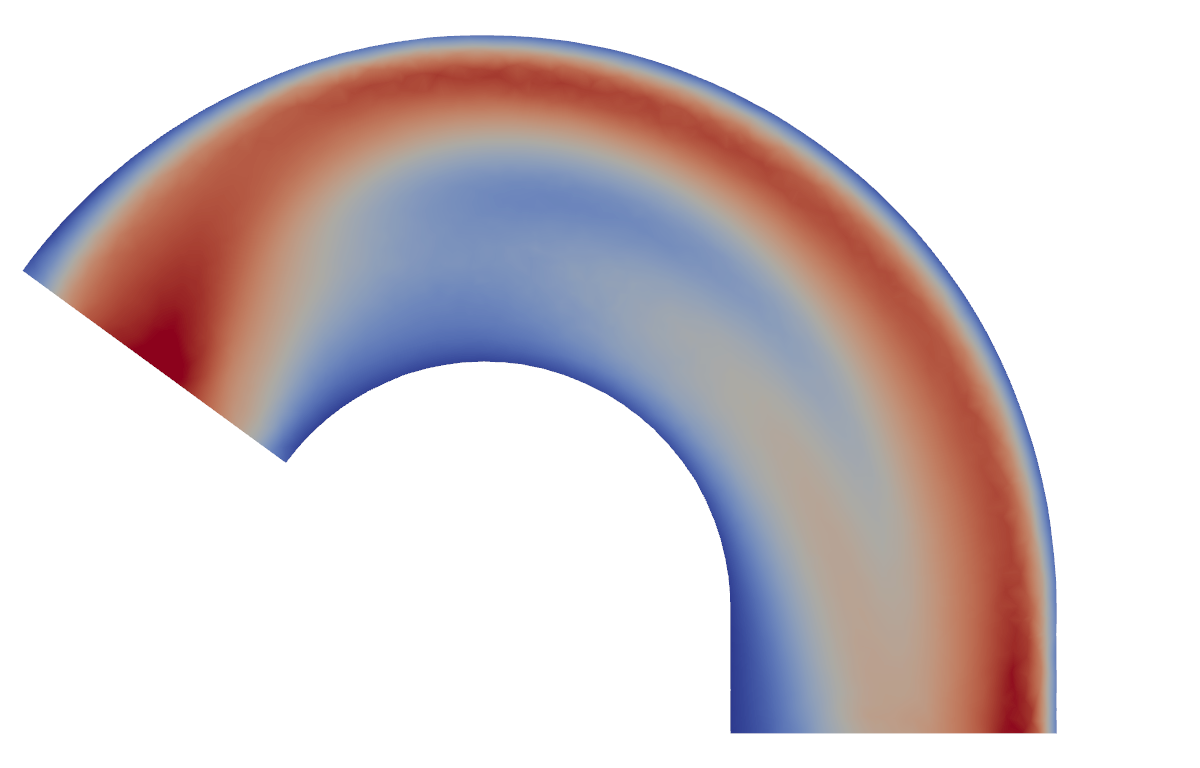} \\
     & $\thet{opt}=0.195$ & $\thet{opt}=0.465$ & $\thet{opt}=0.698$ & $\thet{opt}=0.847$ \\
     & $\mathcal{J}=\rnum{0.0001034}$ & $\mathcal{J}=\rnum{0.0004654}$ & $\mathcal{J}=\rnum{0.0009135}$ & $\mathcal{J}=\rnum{0.001294}$ \\
     & $\mathcal{R}=\rnum{0.0001337}$ & $\mathcal{R}=\rnum{0.0002958}$ & $\mathcal{R}=\rnum{0.000417}$ & $\mathcal{R}=\rnum{0.0004797}$ \\
     & iterations: 34 & iterations: 33 & iterations: 26 & iterations: 22 \\
    \rotatebox{90}{\begin{tabular}{c}
         $P_1/P_1$ \\
         stab.
    \end{tabular}} &  
    \includegraphics[width=0.19\textwidth]{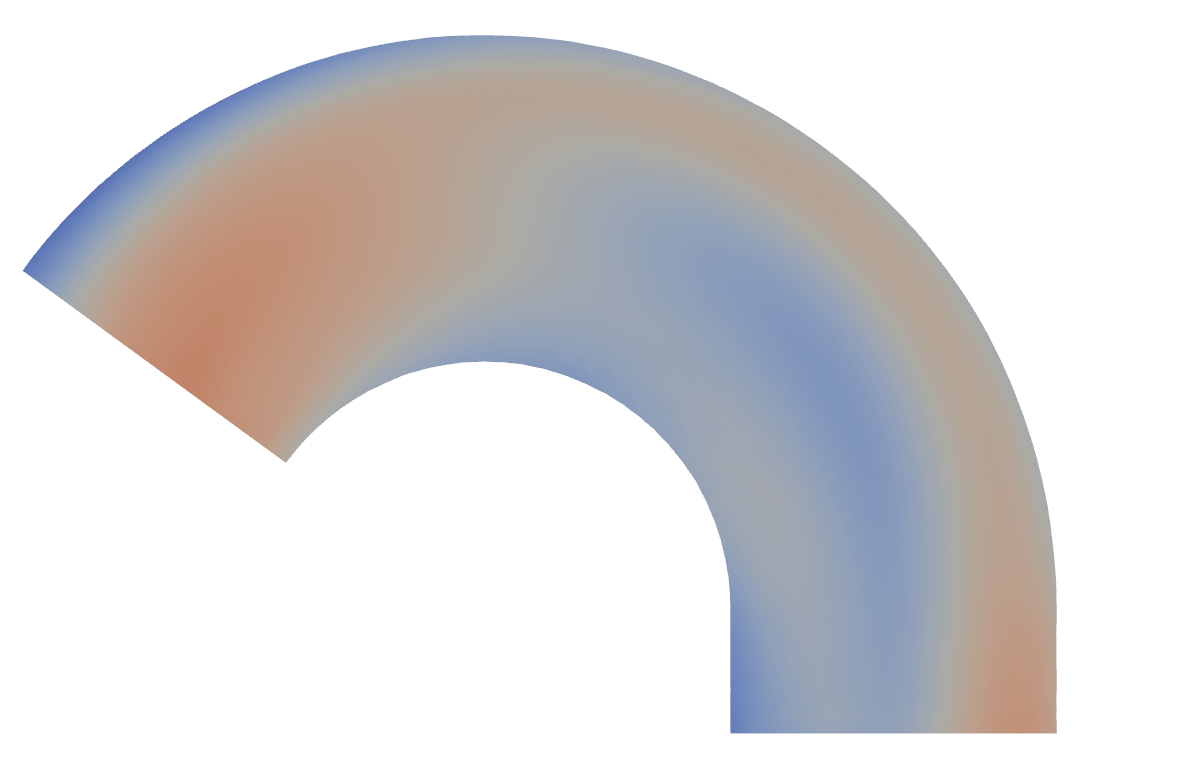} &
    \includegraphics[width=0.19\textwidth]{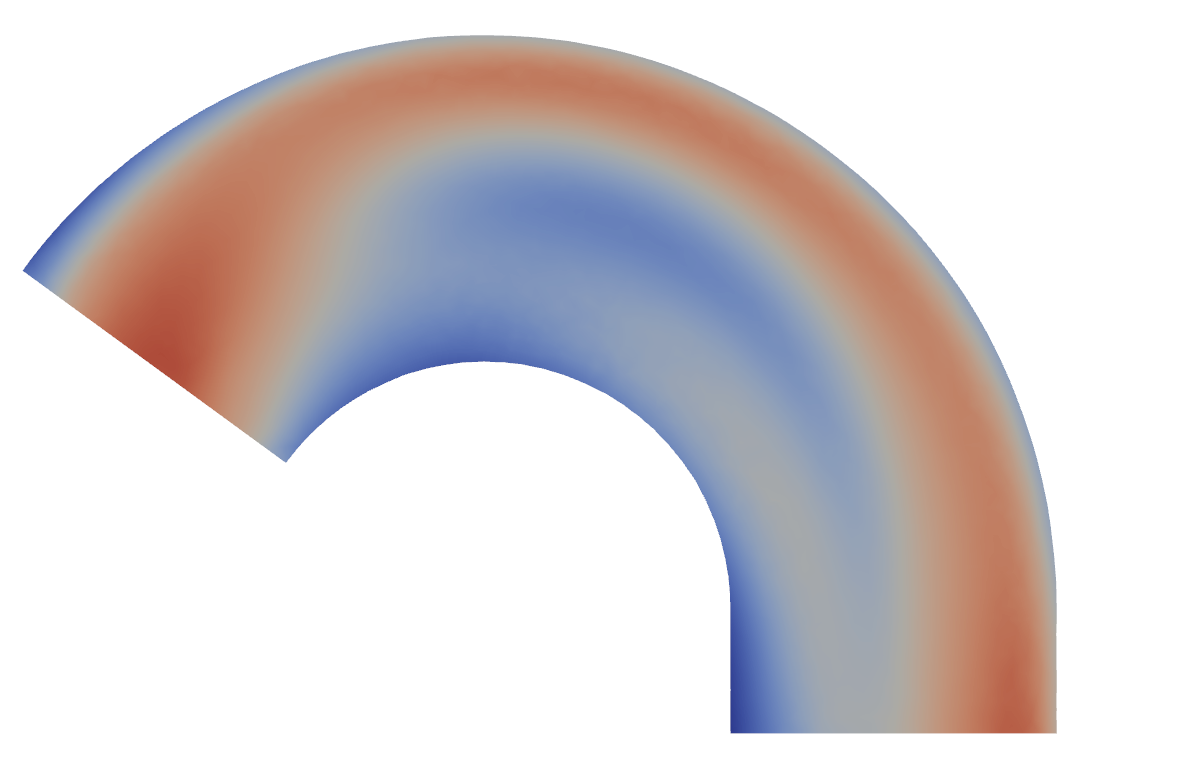} &
    \includegraphics[width=0.19\textwidth]{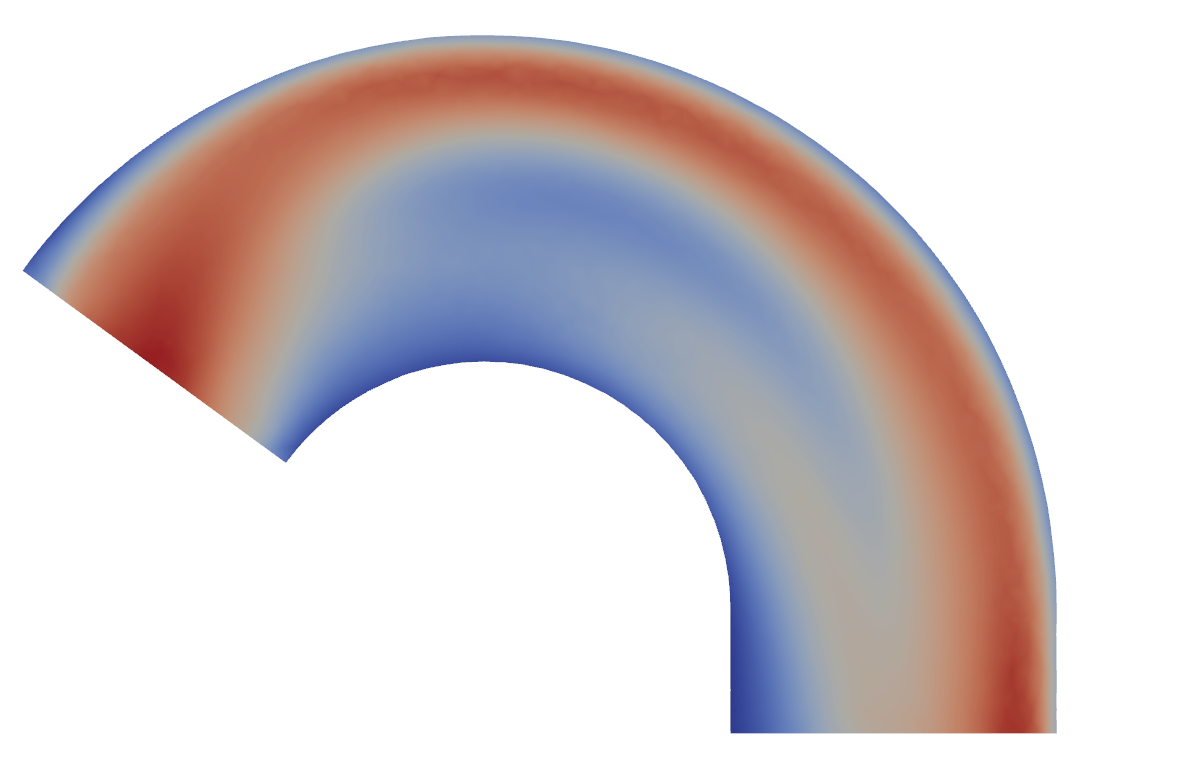} &
    \includegraphics[width=0.19\textwidth]{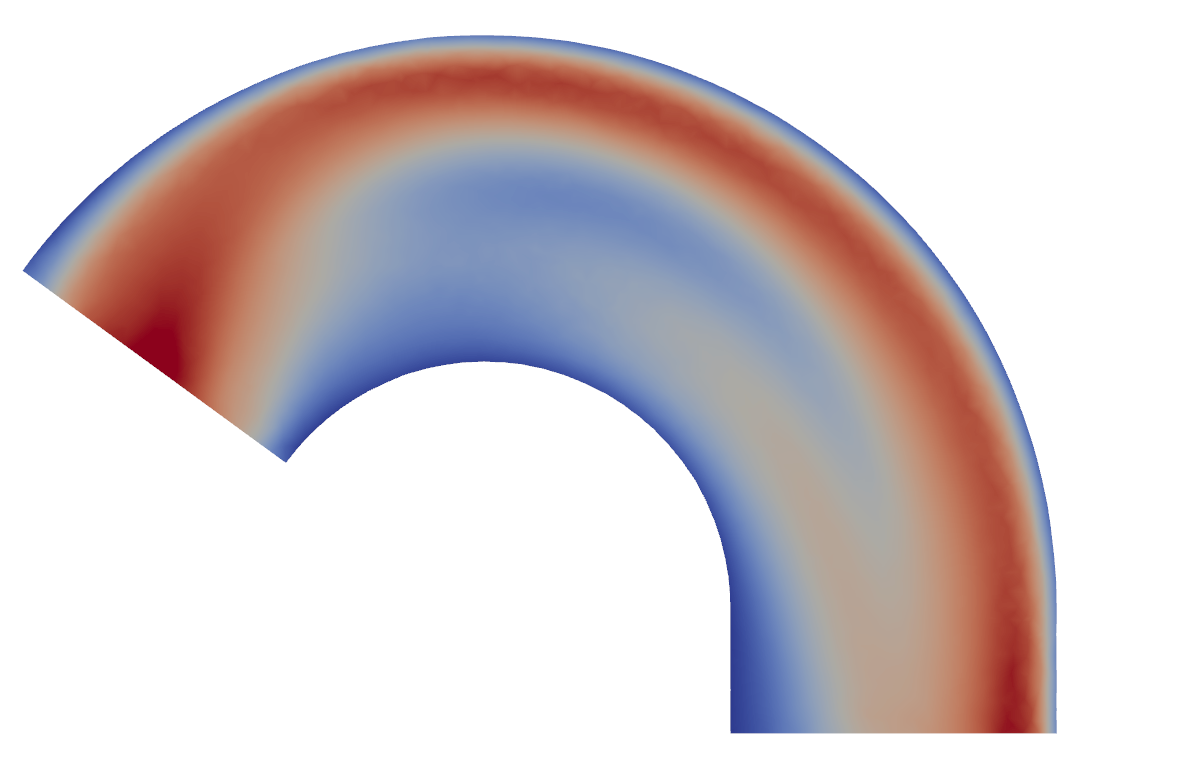} \\
     & $\thet{opt}=0.194$ & $\thet{opt}=0.463$ & $\thet{opt}=0.691$ & $\thet{opt}=0.834$ \\
     & $\mathcal{J}=\rnum{0.0001375}$ & $\mathcal{J}=\rnum{0.0006408}$ & $\mathcal{J}=\rnum{0.001267}$ & $\mathcal{J}=\rnum{0.001807}$ \\
     & $\mathcal{R}=\rnum{0.0001349}$ & $\mathcal{R}=\rnum{0.0003067}$ & $\mathcal{R}=\rnum{0.0004333}$ & $\mathcal{R}=\rnum{0.0005093}$ \\
     & iterations: 30 & iterations: 30 & iterations: 28 & iterations: 21 \\
     & \multicolumn{4}{c}{\includegraphics[width=0.5\textwidth]{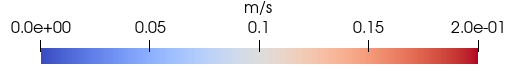}}
\end{tabular}
\caption{Comparison of data without noise with assimilation velocity results in arch geometry with edge length $h=1.5\text{ mm}$ using MINI element, stabilized MINI element ($\alpha_v=0.01, \alpha_p=0$) and stabilized $P_1/P_1$ element ($\alpha_v=\alpha_p=0.01$) for multiple values of $\theta$.}
\label{fig:elem_arch}
\medskip
\centering
\begin{tabular}{c c c c c}
    & $\theta=0.2$ & $\theta=0.5$ & $\theta=0.8$ & $\theta=1.0$ \\ 
     \rotatebox{90}{MINI} &
     \includegraphics[width=.19\textwidth]{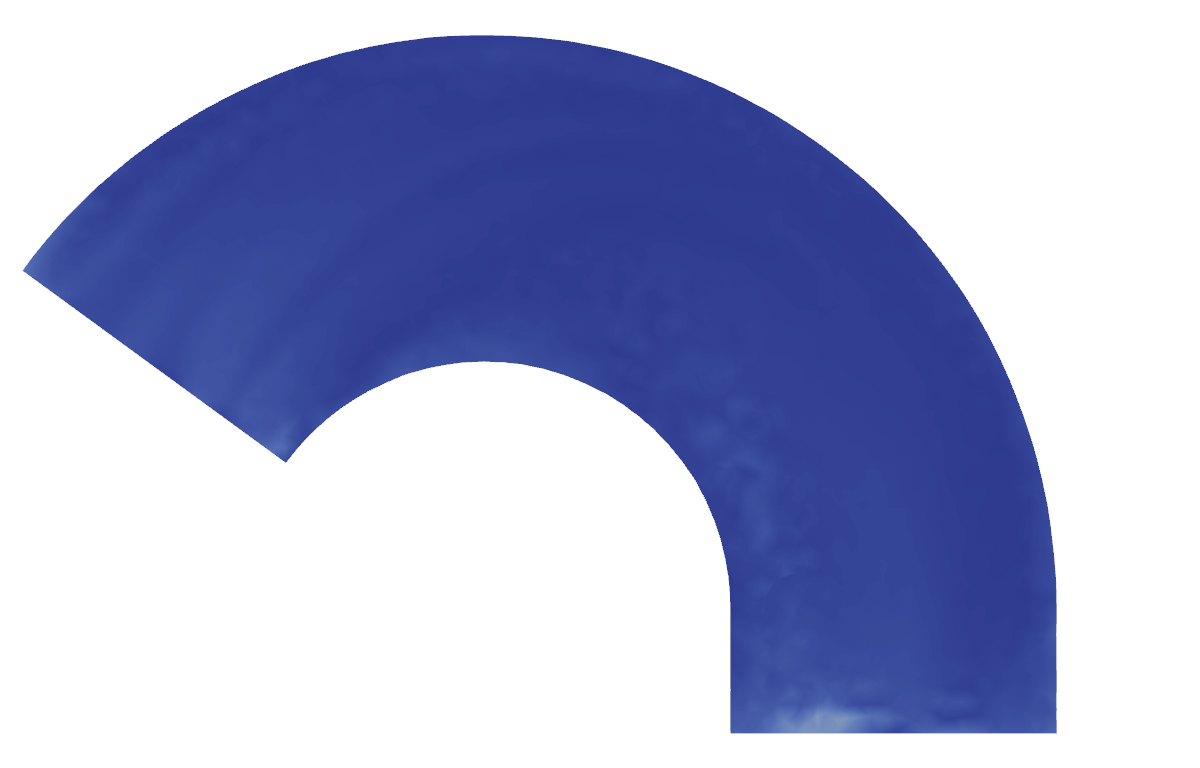} & 
     \includegraphics[width=.19\textwidth]{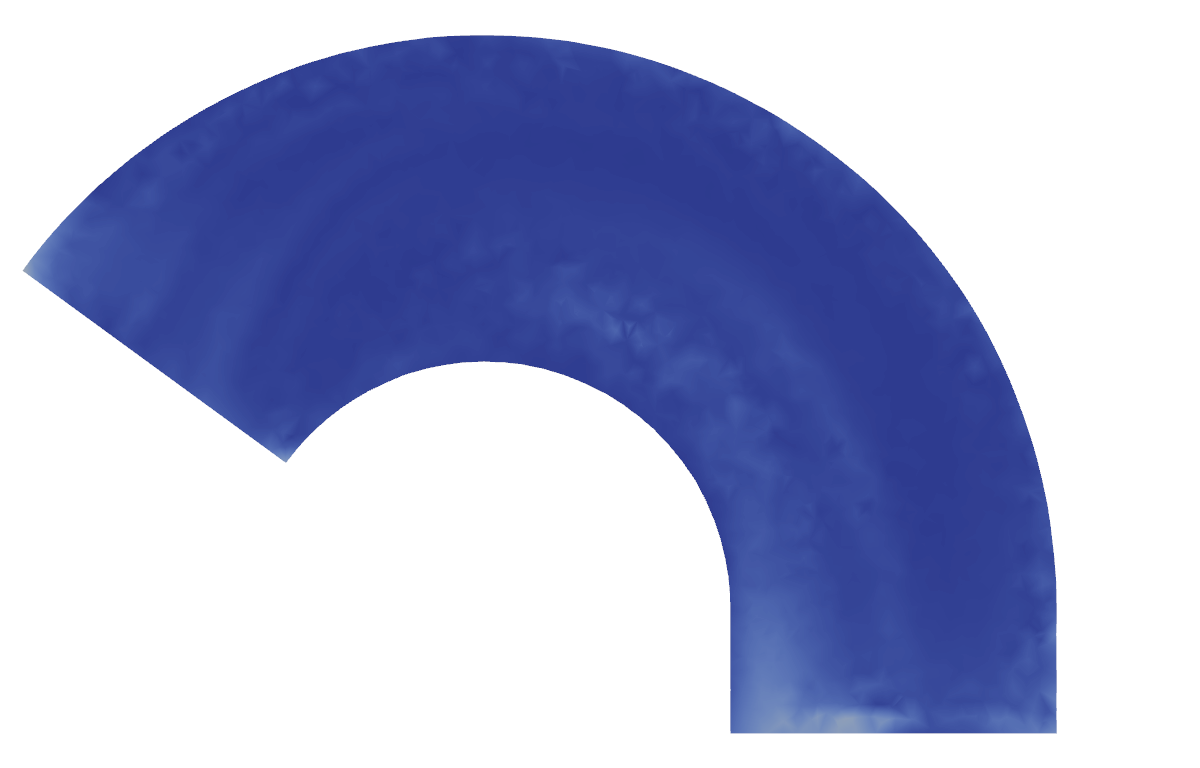} &
     \includegraphics[width=.19\textwidth]{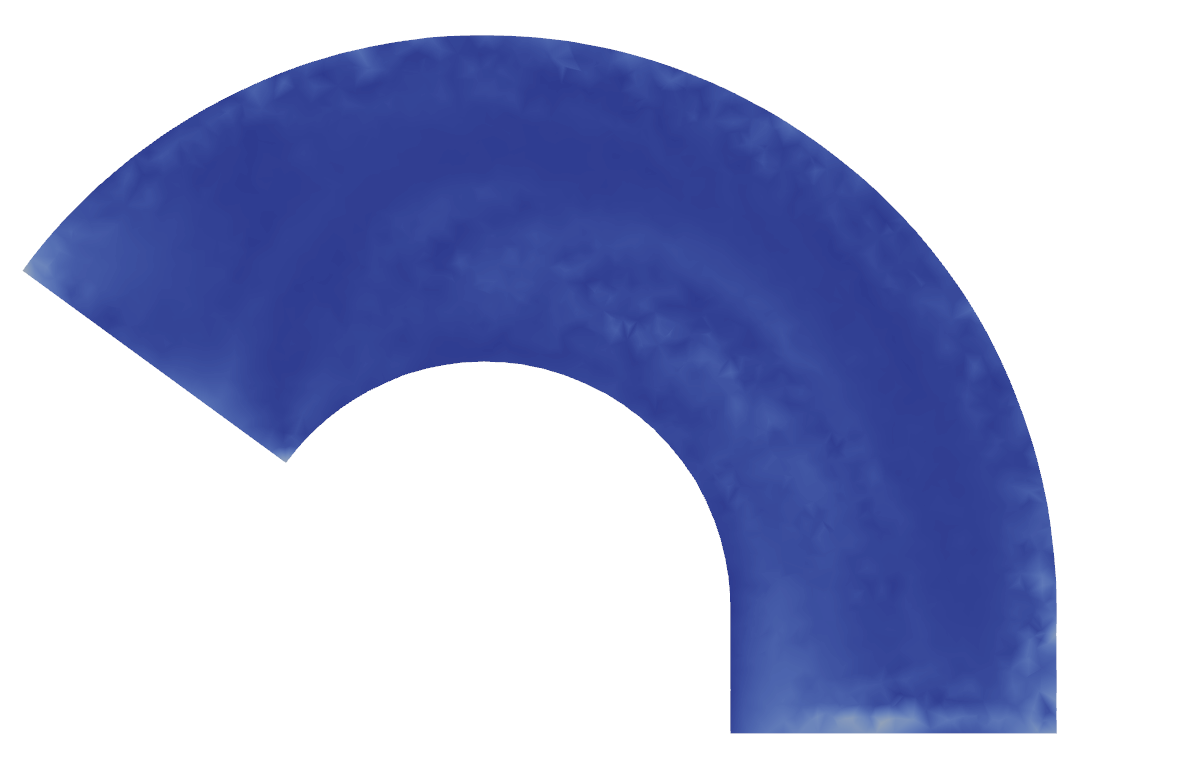} &
     \includegraphics[width=.19\textwidth]{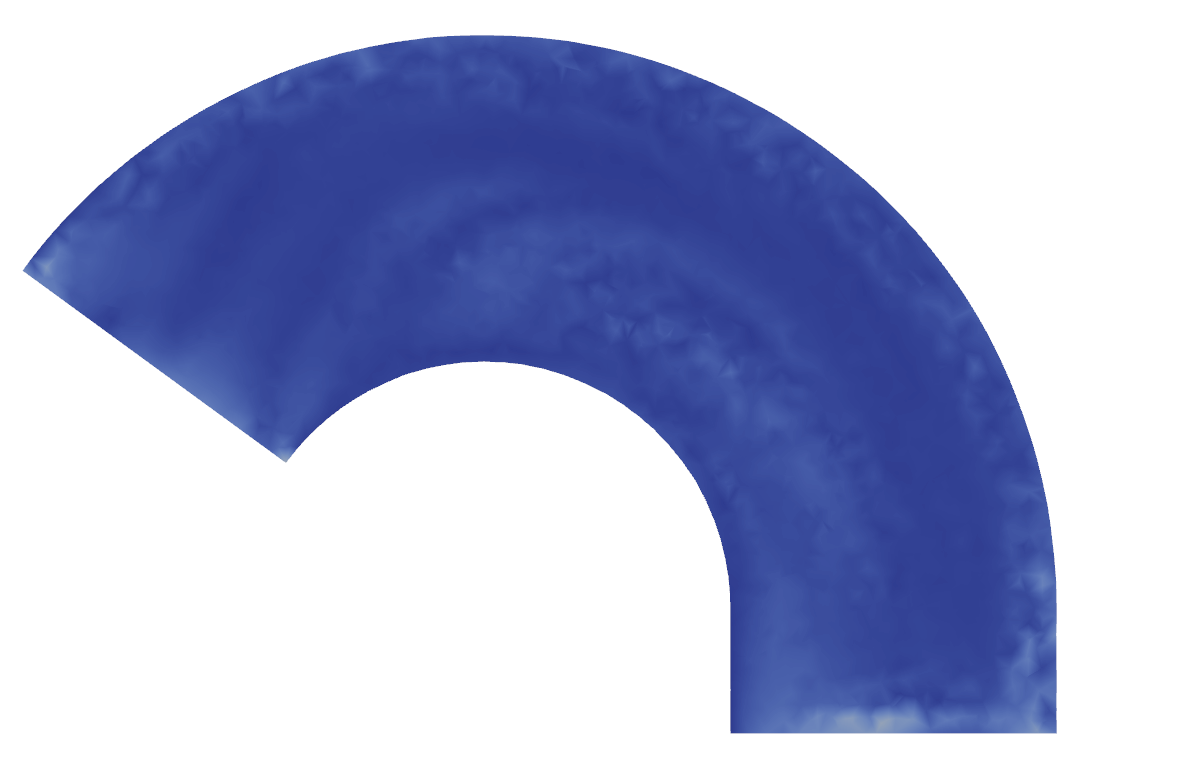} \\
     \rotatebox{90}{\begin{tabular}{c} 
        MINI\\
        stab.
    \end{tabular}}&
     \includegraphics[width=.19\textwidth]{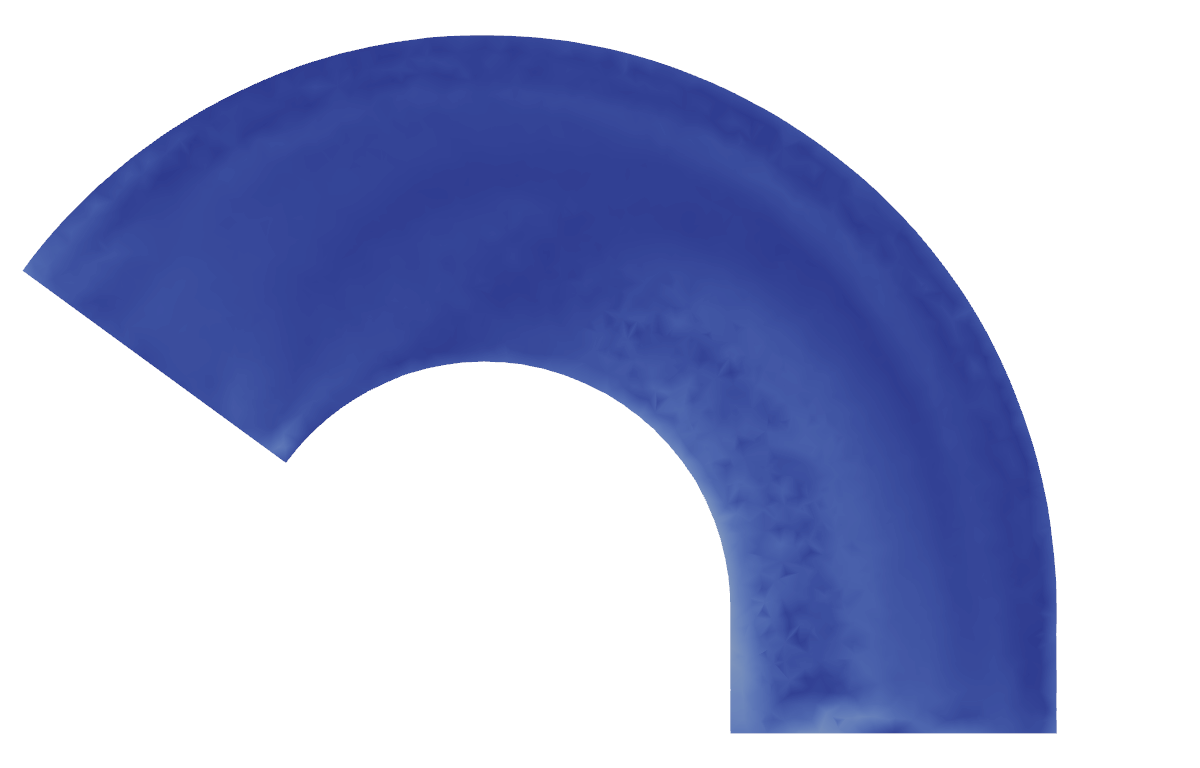} &
     \includegraphics[width=.19\textwidth]{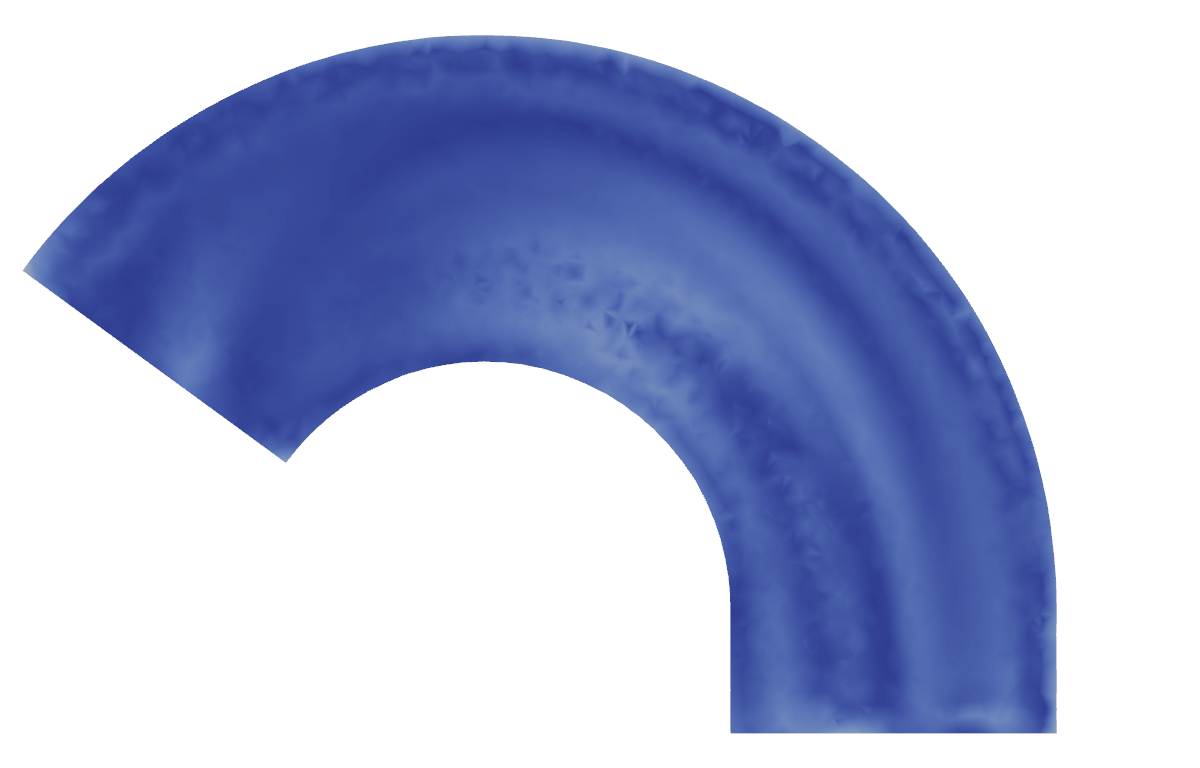} &
     \includegraphics[width=.19\textwidth]{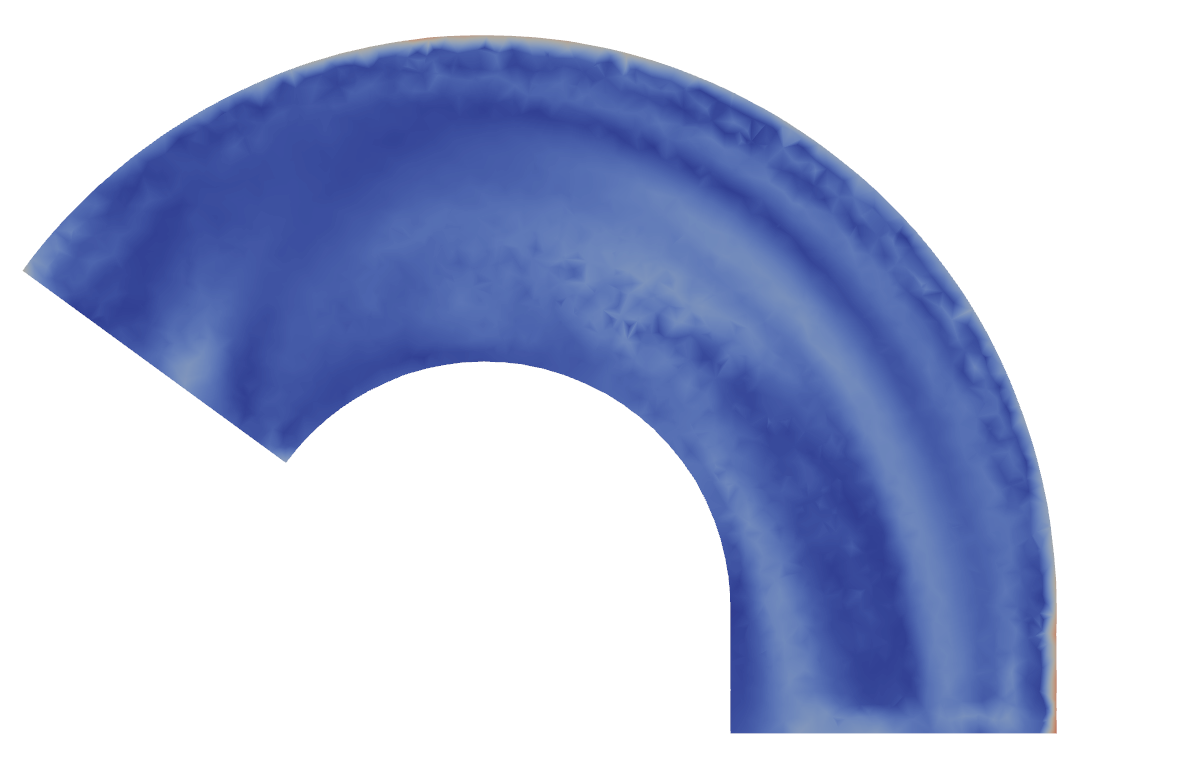} &
     \includegraphics[width=.19\textwidth]{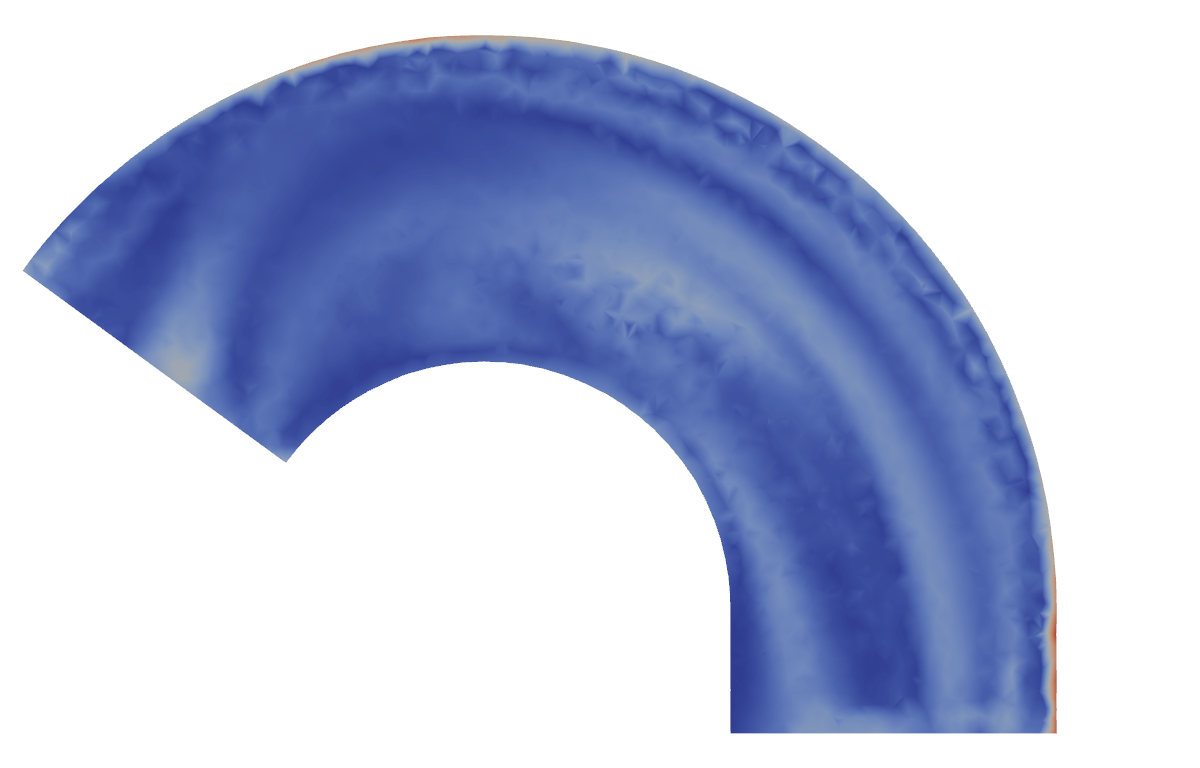} \\
        \rotatebox{90}{\begin{tabular}{c} 
            $P_1/P_1$\\
            stab.
    \end{tabular}}&
     \includegraphics[width=.19\textwidth]{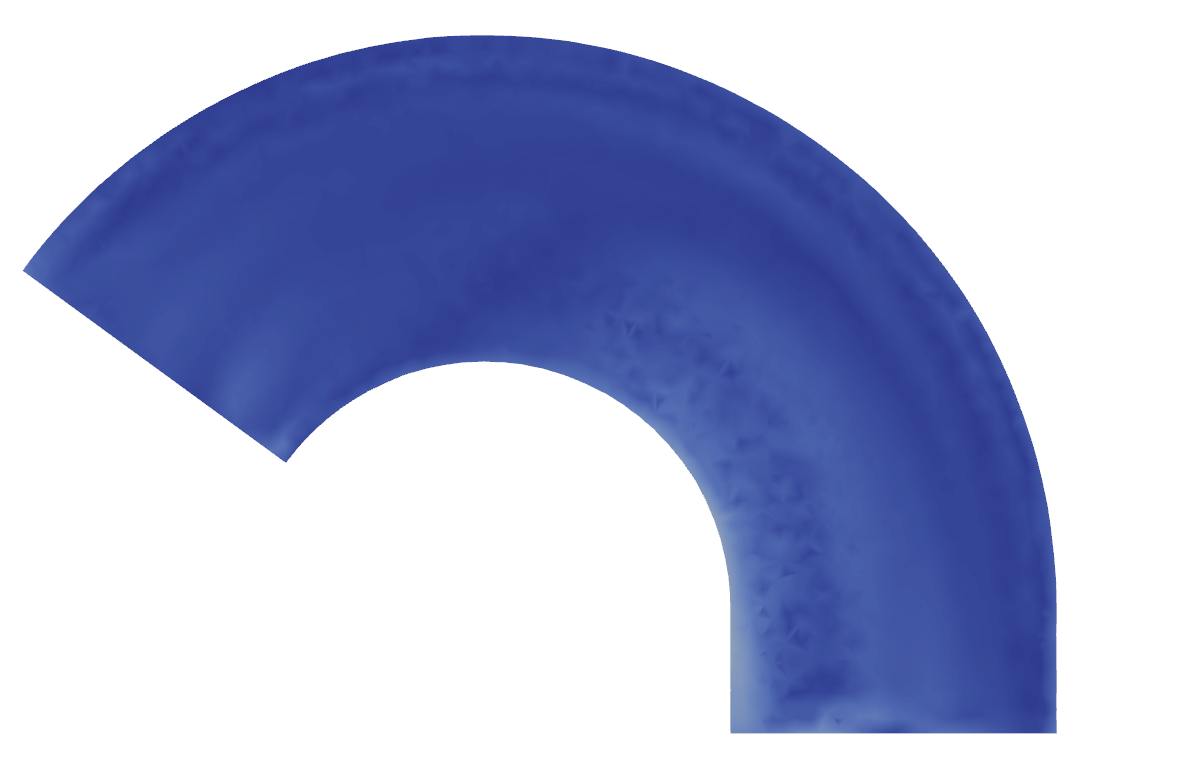} &
     \includegraphics[width=.19\textwidth]{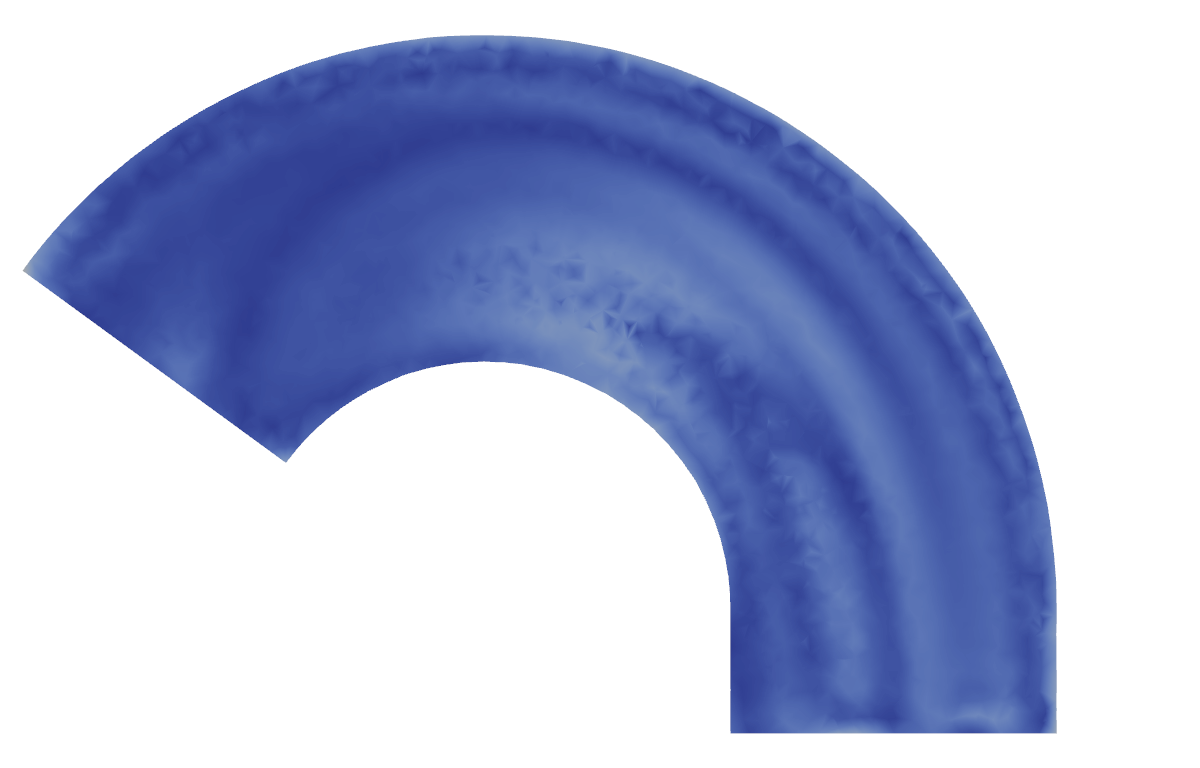} &
     \includegraphics[width=.19\textwidth]{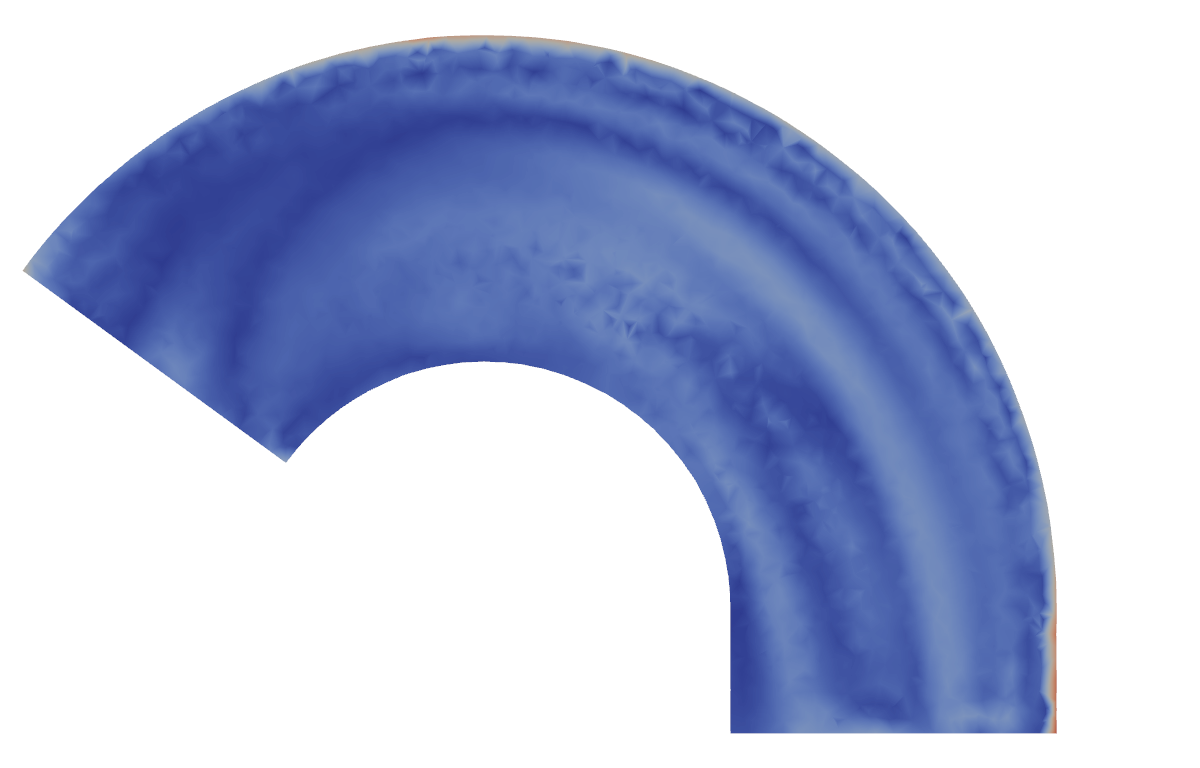} &
     \includegraphics[width=.19\textwidth]{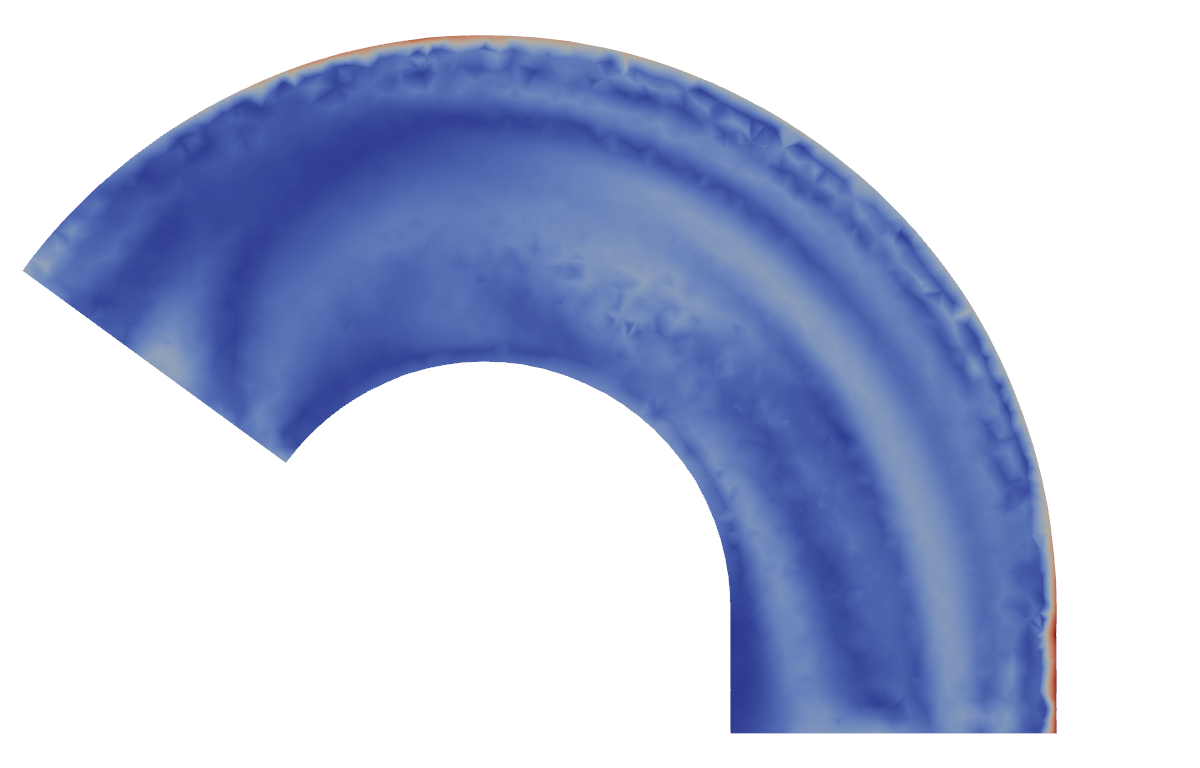} \\
    & \multicolumn{4}{c}{\includegraphics[width=0.5\textwidth]{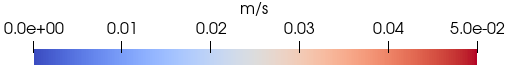}}
\end{tabular}
\caption{Visulatization of the difference between data and computed velocity fields on arch geometry ($h=1.5\text{ mm}$) using MINI element, stabilized MINI element and stabilized $P_1/P_1$ element for multiple values of $\theta$.}
\label{fig:arch_errors}
\end{figure}

The method was tested for both the MINI element and the $P_1/P_1$ element. 
The MINI element was used without and with stabilization using weights  $\alpha_v=0.01$ and $\alpha_p=0$. 
Since the $P_1/P_1$ element is not inf-sup stable by itself, we tested it only with stabilization using weights $\alpha_v = \alpha_p=0.01$. 
The results for bent tube geometry and arch geometry with edge length $h= 1.5\text{ mm}$ are present in Figures \ref{fig:elem_bent} and \ref{fig:elem_arch}, respectively. The errors with respect to the ground truth velocities are shown in Figures \ref{fig:bent_errors} and \ref{fig:arch_errors}.
The results show that the stabilization influences $\thet{opt}$ especially close to no-slip because it introduces additional diffusion. 
However, it must be included with the $P_1/P_1$ element to ensure its stability. 

\subsection{Mesh coarseness}
\begin{figure}
\centering
\begin{tabular}{c c c c c}
    \rotatebox{90}{\begin{tabular}{c}
         data
    \end{tabular}} &  
    \includegraphics[width=0.19\textwidth]{bent_data_200.png} &
    \includegraphics[width=0.19\textwidth]{bent_data_500.png} &
    \includegraphics[width=0.19\textwidth]{bent_data_800.png} &
    \includegraphics[width=0.19\textwidth]{bent_data_1000.png} \\
    & $\theta=0.2$ & $\theta=0.5$ & $\theta=0.8$ & $\theta=1$ \\
    \rotatebox{90}{\begin{tabular}{c}
         coarser\\
         mesh        
    \end{tabular}} 
    &  
    \includegraphics[width=0.19\textwidth]{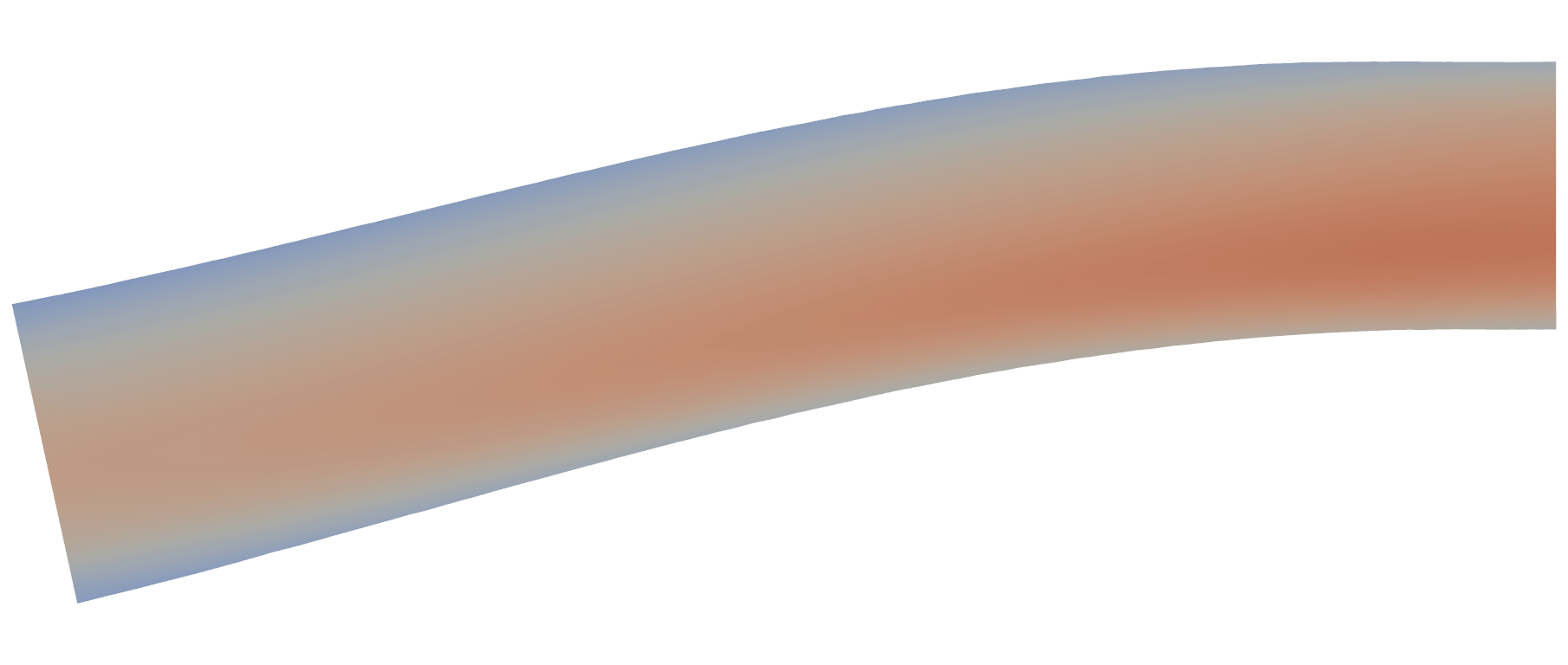} &
    \includegraphics[width=0.19\textwidth]{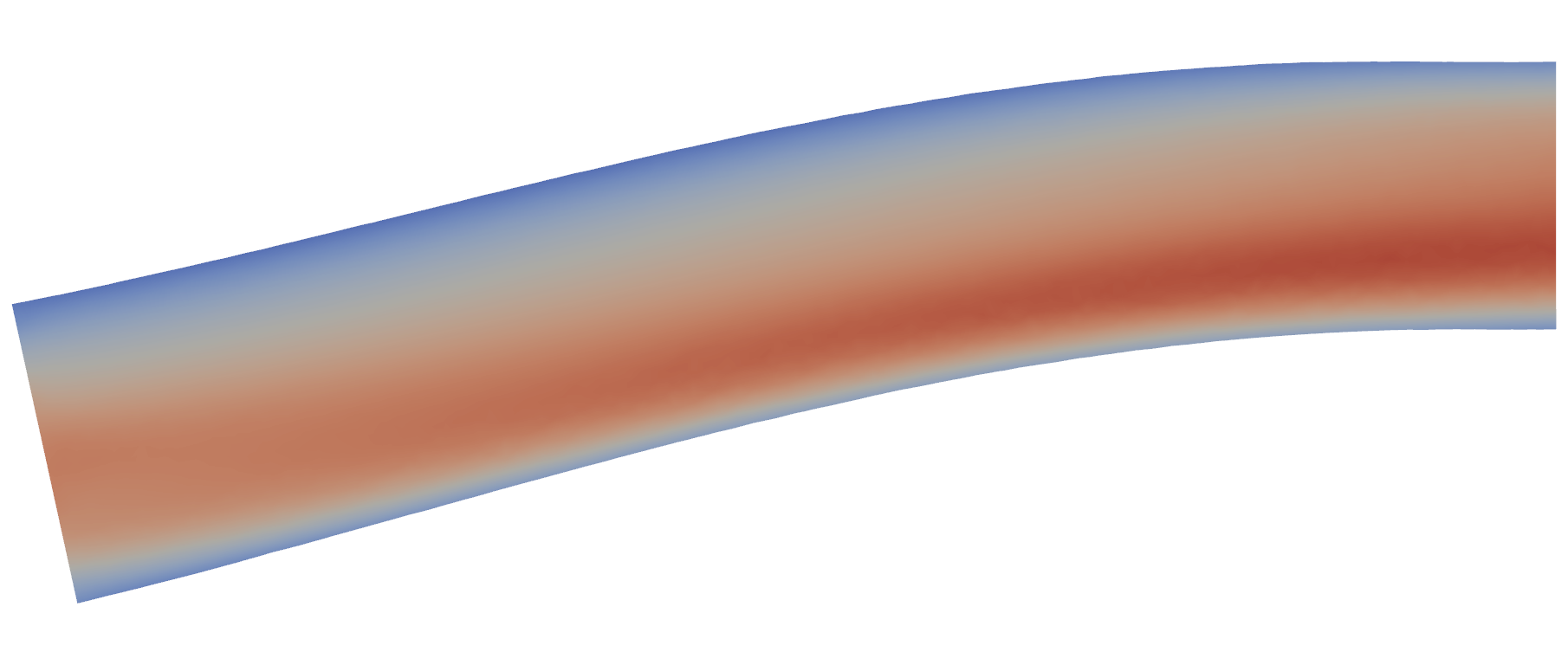} &
    \includegraphics[width=0.19\textwidth]{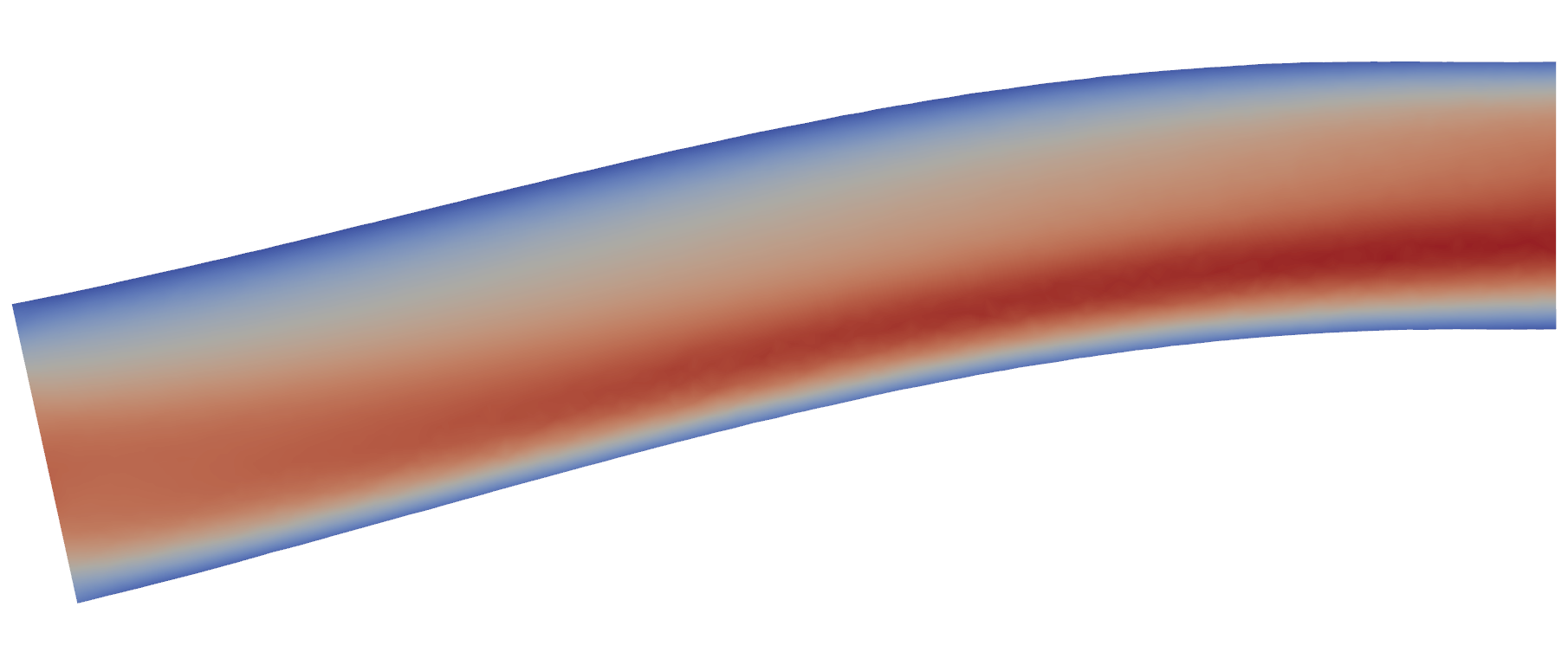} &
    \includegraphics[width=0.19\textwidth]{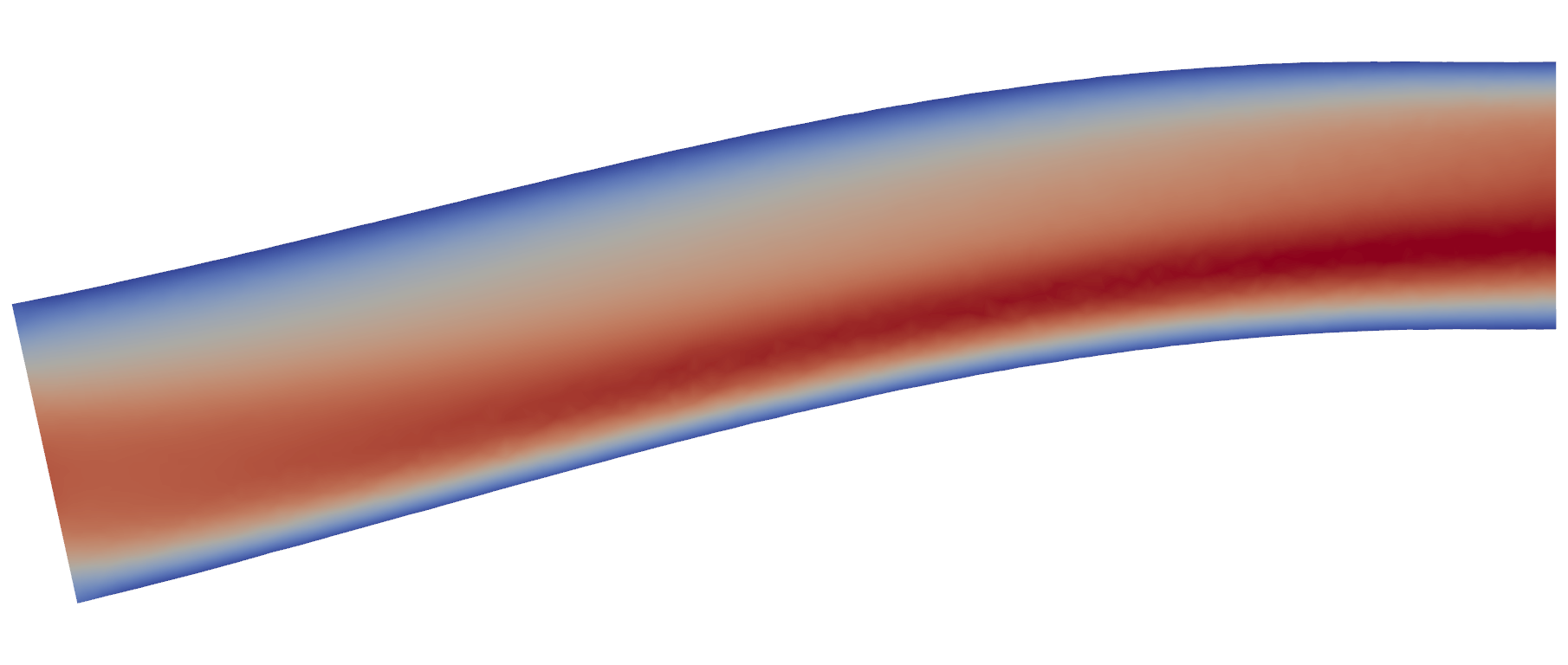} \\
    & $\thet{opt}=0.199$ & $\thet{opt}=0.488$ & $\thet{opt}=0.745$ & $\thet{opt}=0.9$ \\
     & $\mathcal{J}=\rnum{1.606e-05}$ & $\mathcal{J}=\rnum{0.0001248}$ & $\mathcal{J}=\rnum{0.0002822}$ & $\mathcal{J}=\rnum{0.0003827}$ \\
     & $\mathcal{R}=\rnum{2.697e-05}$ & $\mathcal{R}=\rnum{0.0001044}$ & $\mathcal{R}=\rnum{0.000168}$ & $\mathcal{R}=\rnum{0.0001949}$ \\
     & iterations: 25 & iterations: 23 & iterations: 21 & iterations: 21 \\
        \rotatebox{90}{\begin{tabular}{c}
         finer\\
         mesh 
    \end{tabular}} & 
    \includegraphics[width=0.19\textwidth]{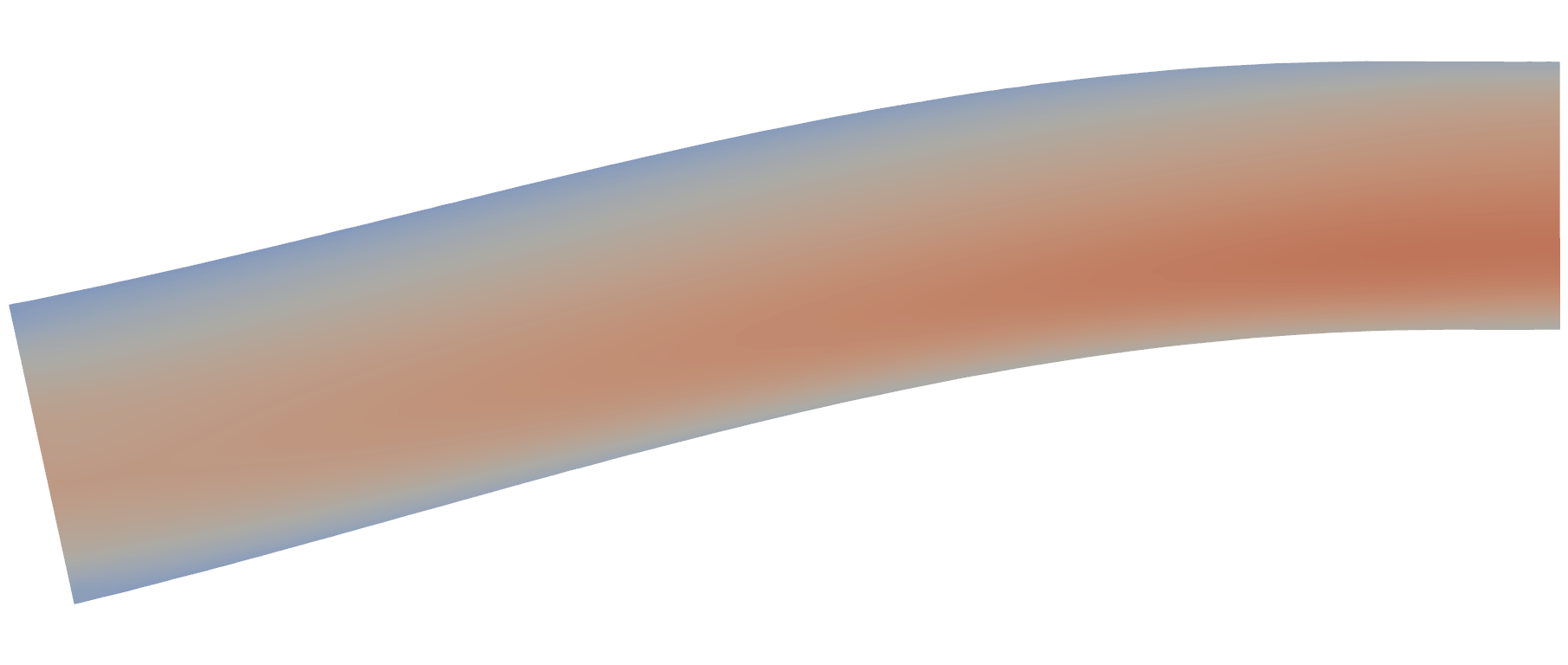} &
    \includegraphics[width=0.19\textwidth]{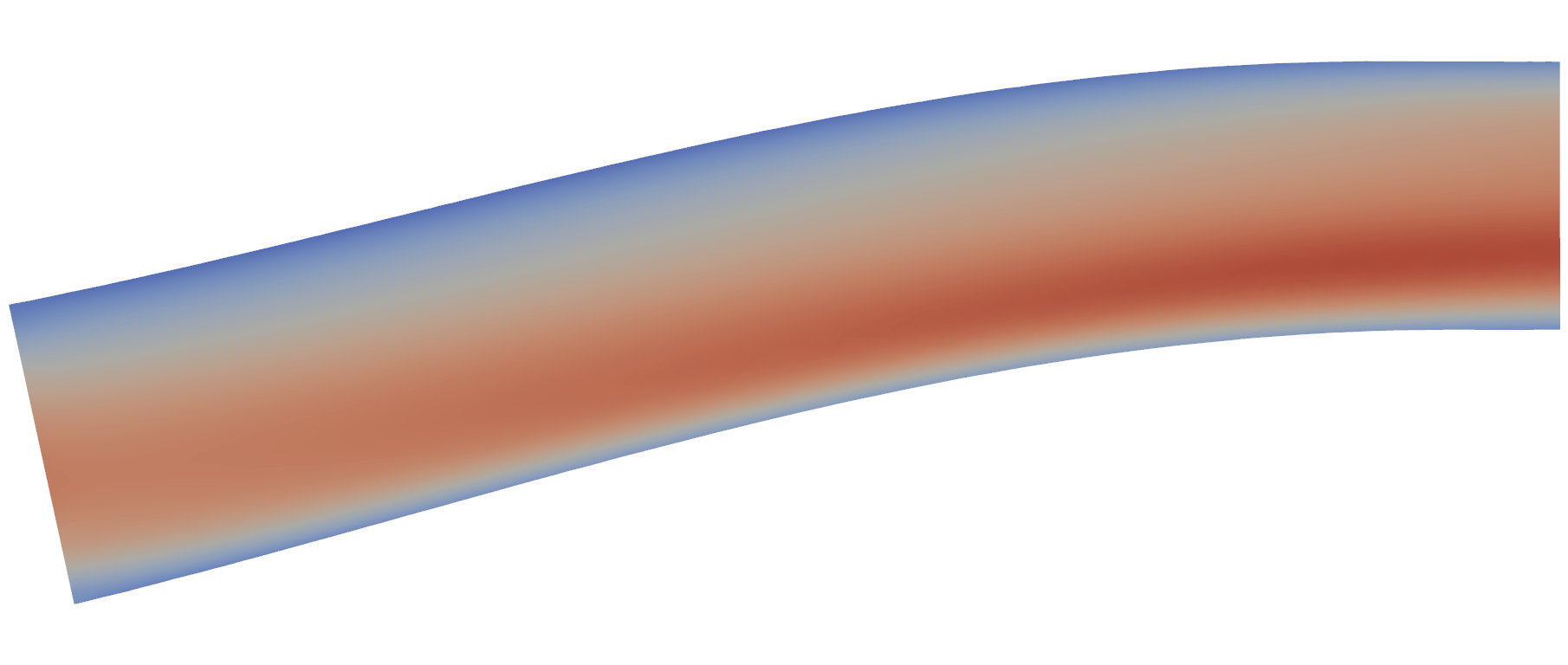} &
    \includegraphics[width=0.19\textwidth]{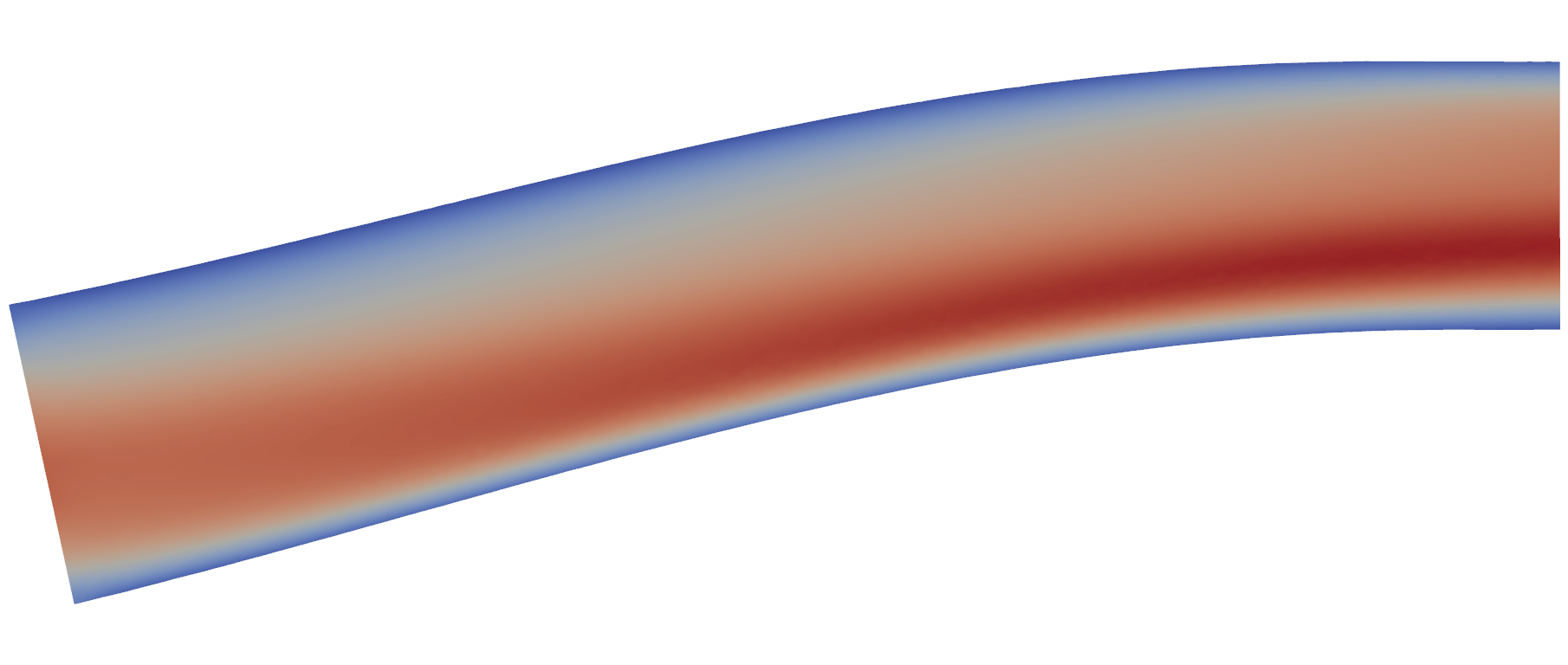} &
    \includegraphics[width=0.19\textwidth]{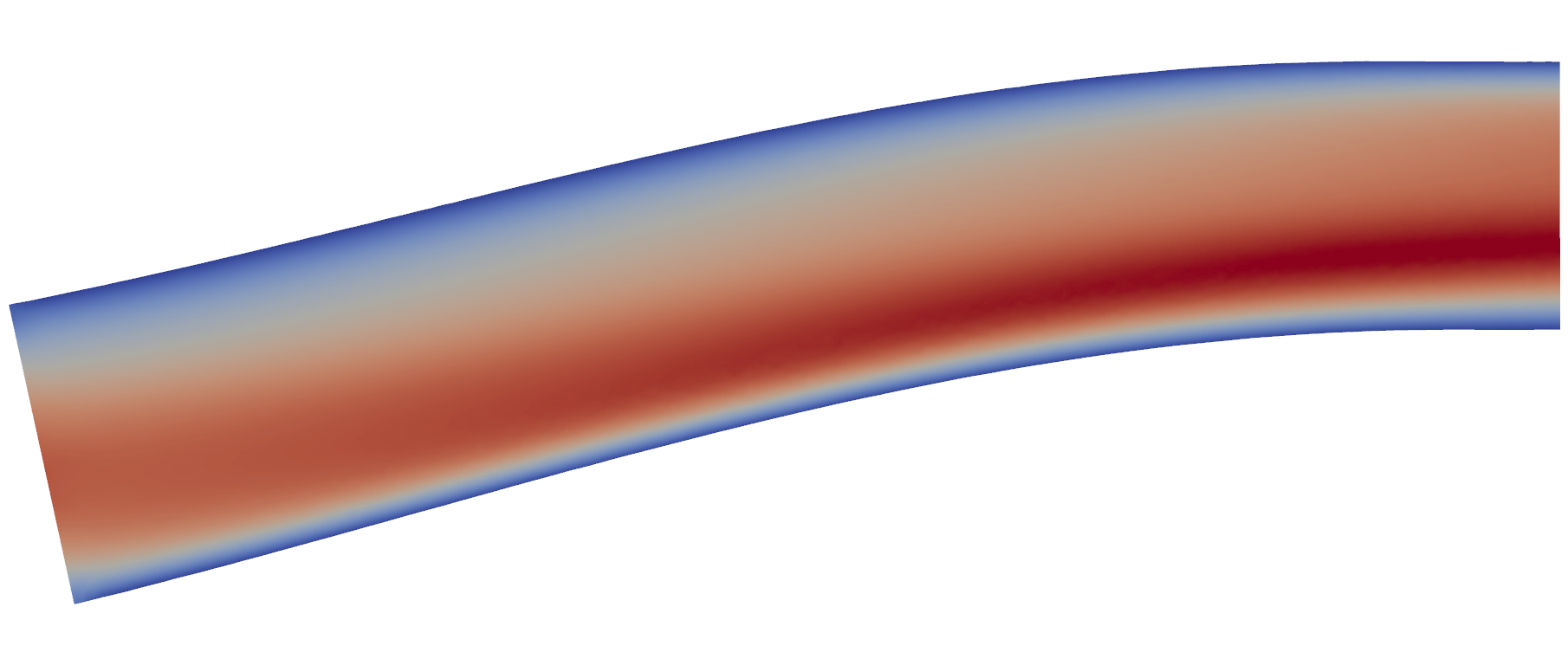} \\
     & $\thet{opt}=0.198$ & $\thet{opt}=0.486$ & $\thet{opt}=0.763$ & $\thet{opt}=0.938$ \\
     & $\mathcal{J}=\rnum{6.128e-06}$ & $\mathcal{J}=\rnum{4.079e-05}$ & $\mathcal{J}=\rnum{7.812e-05}$ & $\mathcal{J}=\rnum{0.0001044}$ \\
     & $\mathcal{R}=\rnum{2.66e-05}$ & $\mathcal{R}=\rnum{9.699e-05}$ & $\mathcal{R}=\rnum{0.0001597}$ & $\mathcal{R}=\rnum{0.0001856}$ \\
     & iterations: 26 & iterations: 25 & iterations: 24 & iterations: 22 \\
     \rotatebox{90}{\begin{tabular}{c}
         finest\\
         mesh
    \end{tabular}} &  
    \includegraphics[width=0.19\textwidth]{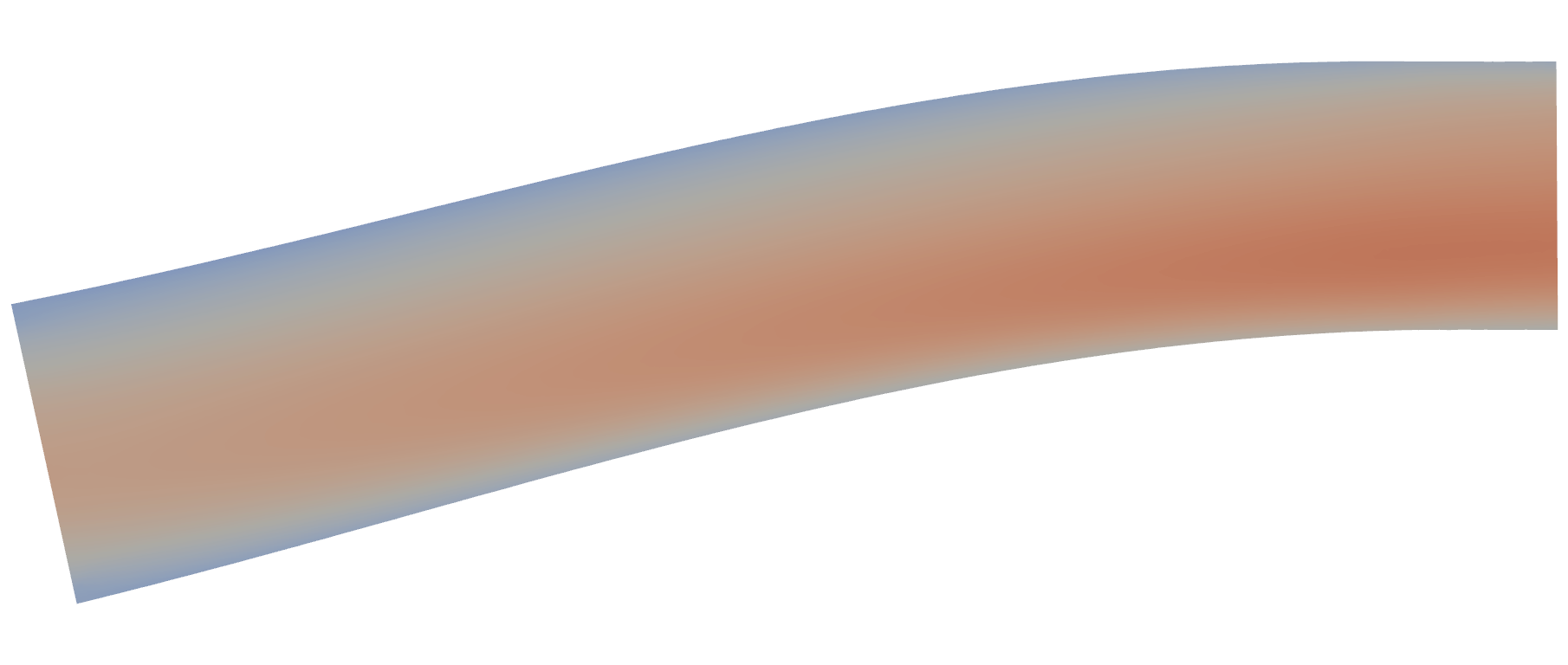} &
    \includegraphics[width=0.19\textwidth]{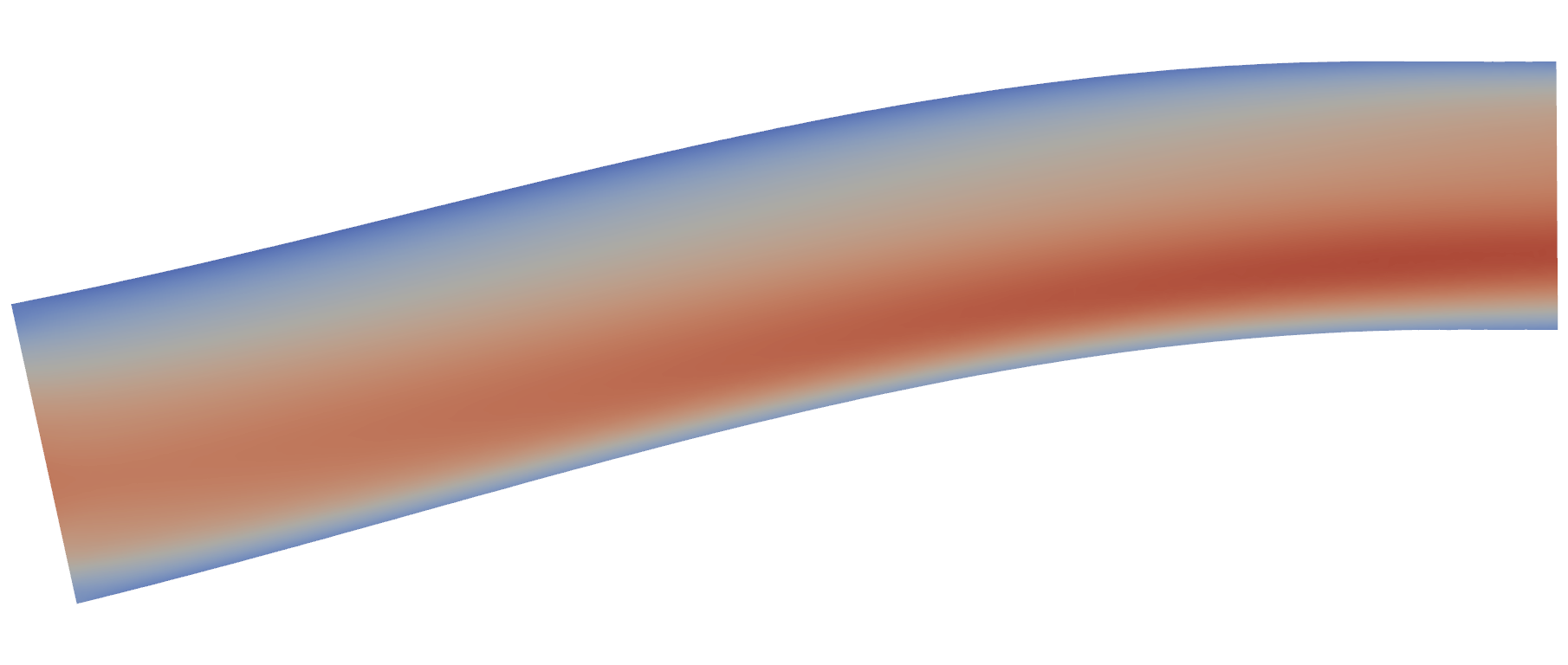} &
    \includegraphics[width=0.19\textwidth]{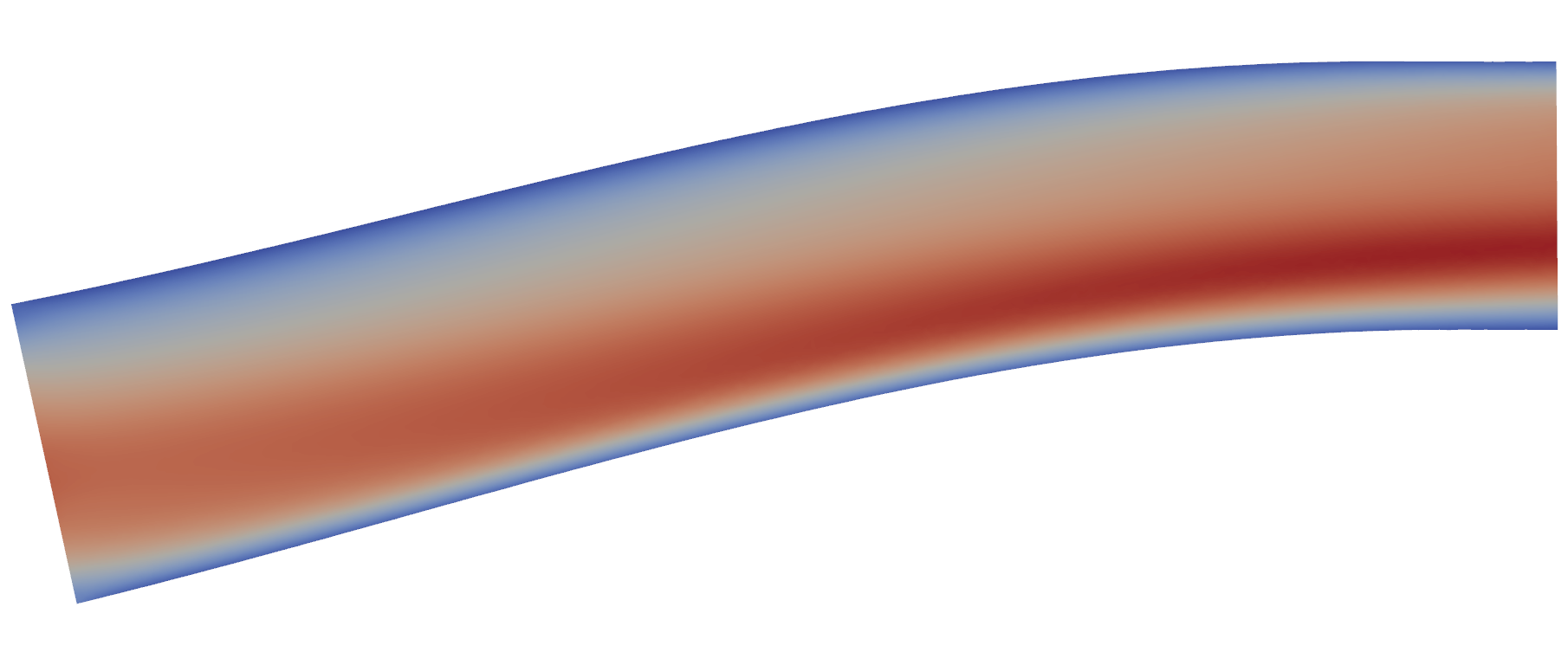} &
    \includegraphics[width=0.19\textwidth]{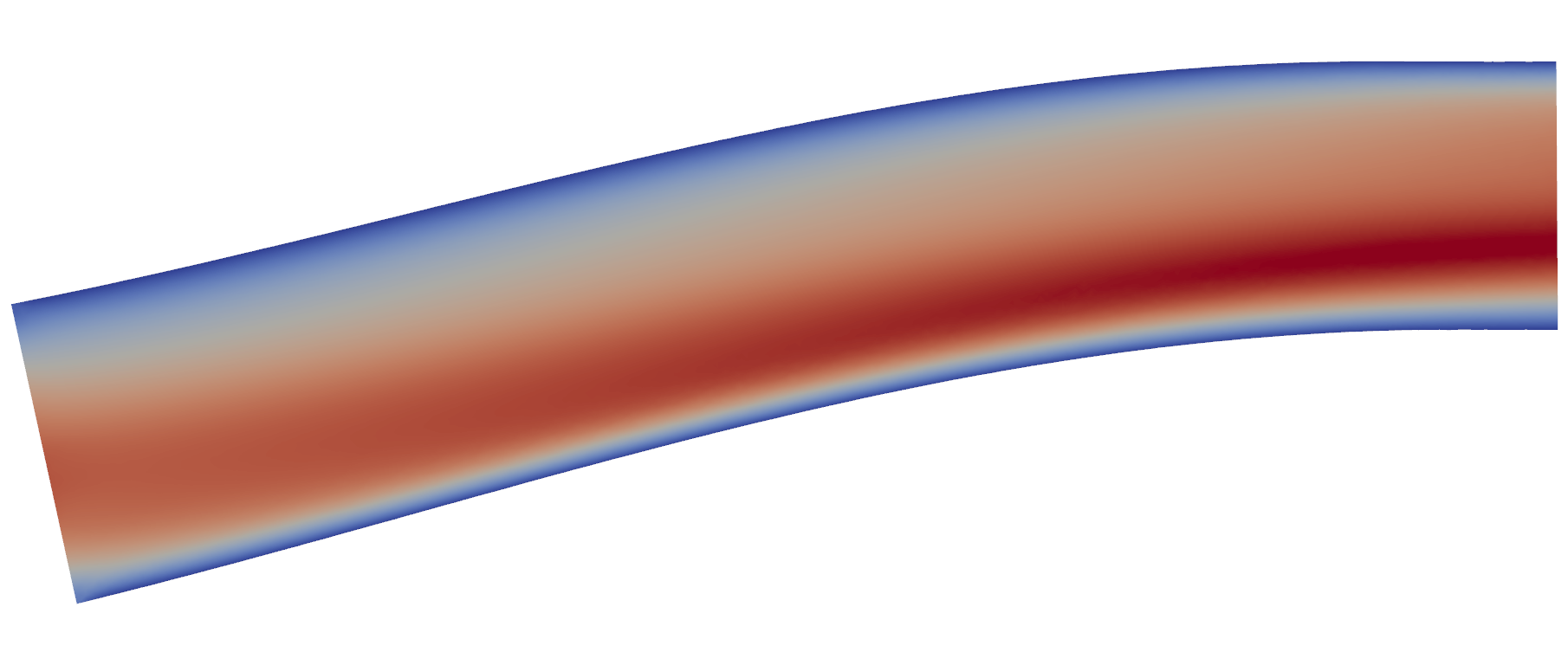} \\
     & $\thet{opt}=0.197$ & $\thet{opt}=0.489$ & $\thet{opt}=0.773$ & $\thet{opt}=0.958$ \\
     & $\mathcal{J}=\rnum{3.442e-06}$ & $\mathcal{J}=\rnum{1.814e-05}$ & $\mathcal{J}=\rnum{3.778e-05}$ & $\mathcal{J}=\rnum{4.916e-05}$ \\
     & $\mathcal{R}=\rnum{2.541e-05}$ & $\mathcal{R}=\rnum{9.646e-05}$ & $\mathcal{R}=\rnum{0.0001593}$ & $\mathcal{R}=\rnum{0.0001855}$ \\
     & iterations: 28 & iterations: 27 & iterations: 25 & iterations: 25 \\  
    & \multicolumn{4}{c}{\includegraphics[width=0.5\textwidth]{scale0.25.png}}
\end{tabular}
\caption{Comparison of data without noise with assimilation velocity results in the bent tube geometry using stabilized $P_1/P_1$ element with $\alpha_p=\alpha_v=0.01$ for the coarser mesh ($h=1.5\text{ mm}$, $\text{number of DOFs}=97,928$), the finer mesh ($h=1\text{ mm}$, $\text{number of DOFs}=322,864$), the finest mesh ($h=0.8\text{ mm}$, $\text{number of DOFs}=623,088$) and multiple values of $\theta$.}
\label{fig:mesh_bent}
\medskip
\centering
\begin{tabular}{c c c c c}
    & $\theta=0.2$ & $\theta=0.5$ & $\theta=0.8$ & $\theta=1$ \\
    \rotatebox{90}{\begin{tabular}{c}
         coarser\\
         mesh
    \end{tabular}} &  
    \includegraphics[width=0.19\textwidth]{bent_err_p1p1_stab0.01_0.01_200.png} &
    \includegraphics[width=0.19\textwidth]{bent_err_p1p1_stab0.01_0.01_500.png} &
    \includegraphics[width=0.19\textwidth]{bent_err_p1p1_stab0.01_0.01_800.png} &
    \includegraphics[width=0.19\textwidth]{bent_err_p1p1_stab0.01_0.01_1000.png} \\
        \rotatebox{90}{\begin{tabular}{c}
         finer\\
         mesh 
    \end{tabular}} & 
    \includegraphics[width=0.19\textwidth]{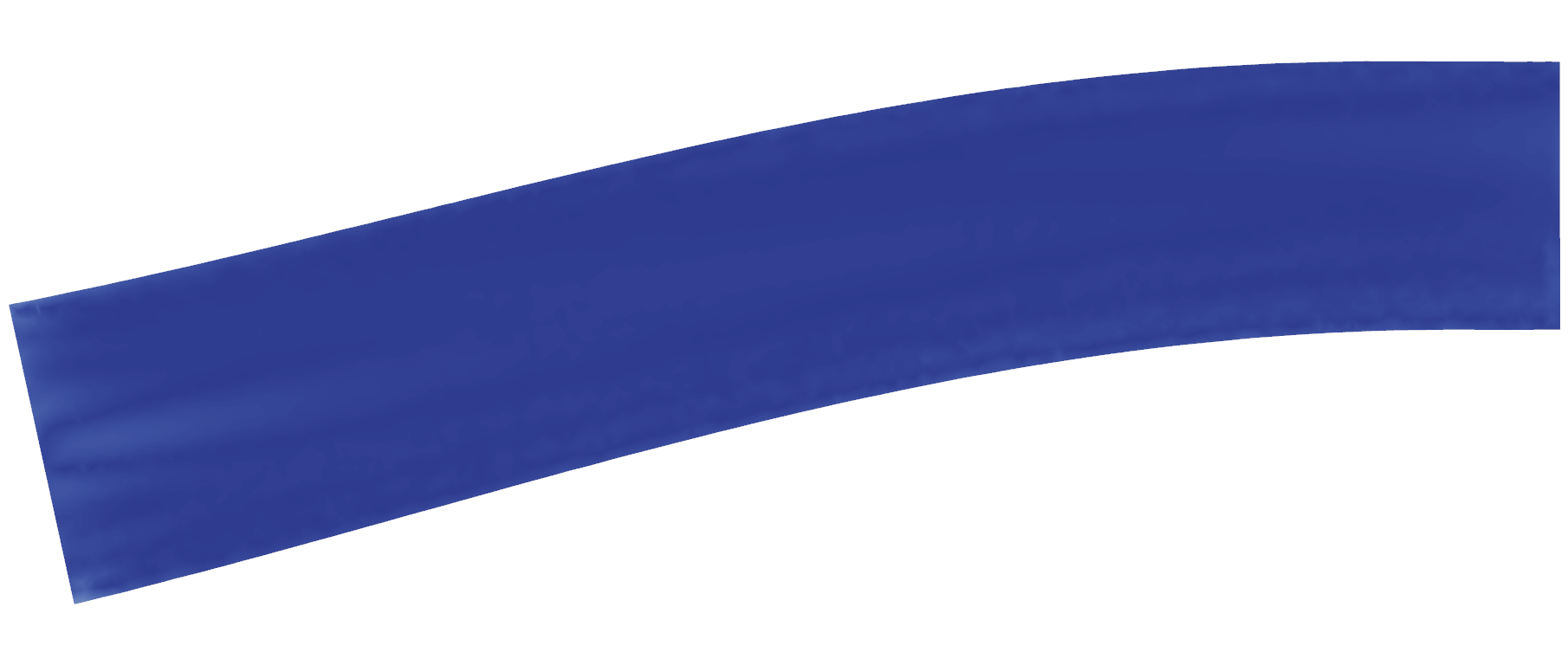} &
    \includegraphics[width=0.19\textwidth]{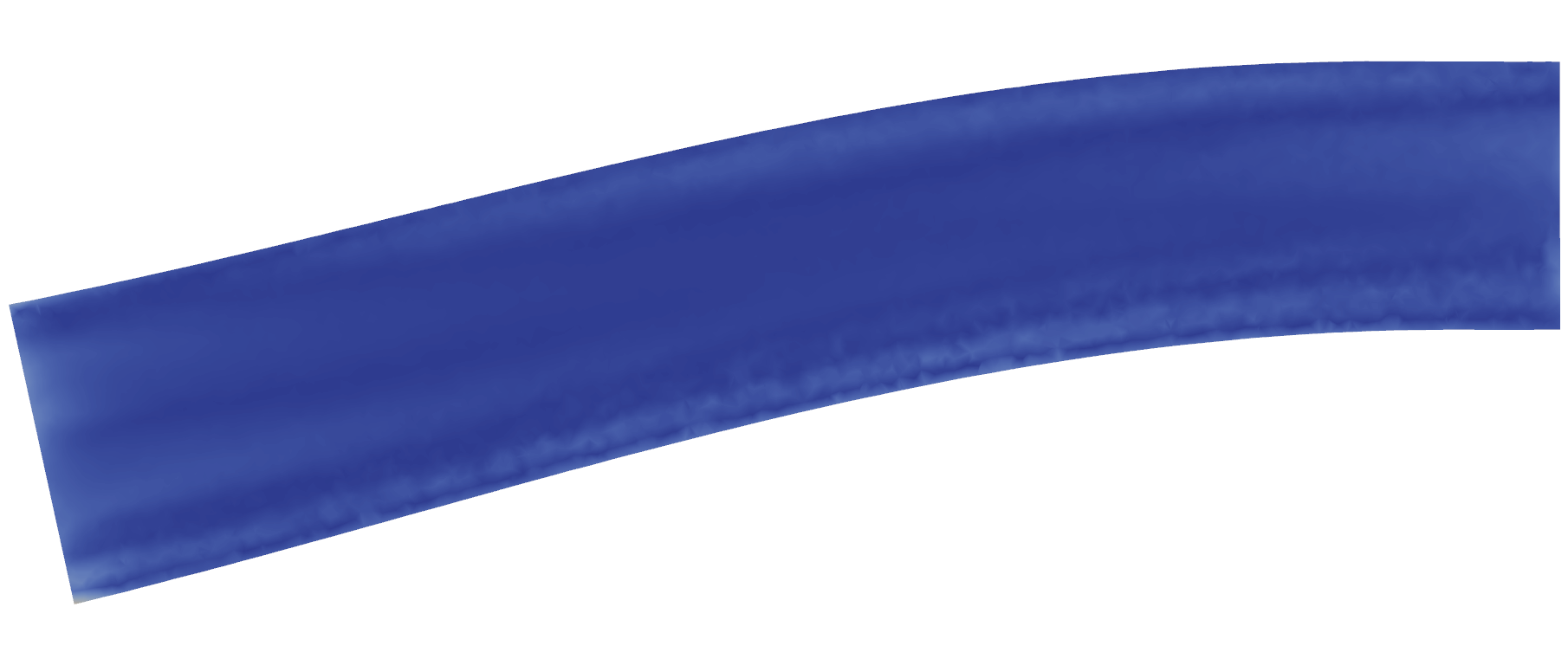} &
    \includegraphics[width=0.19\textwidth]{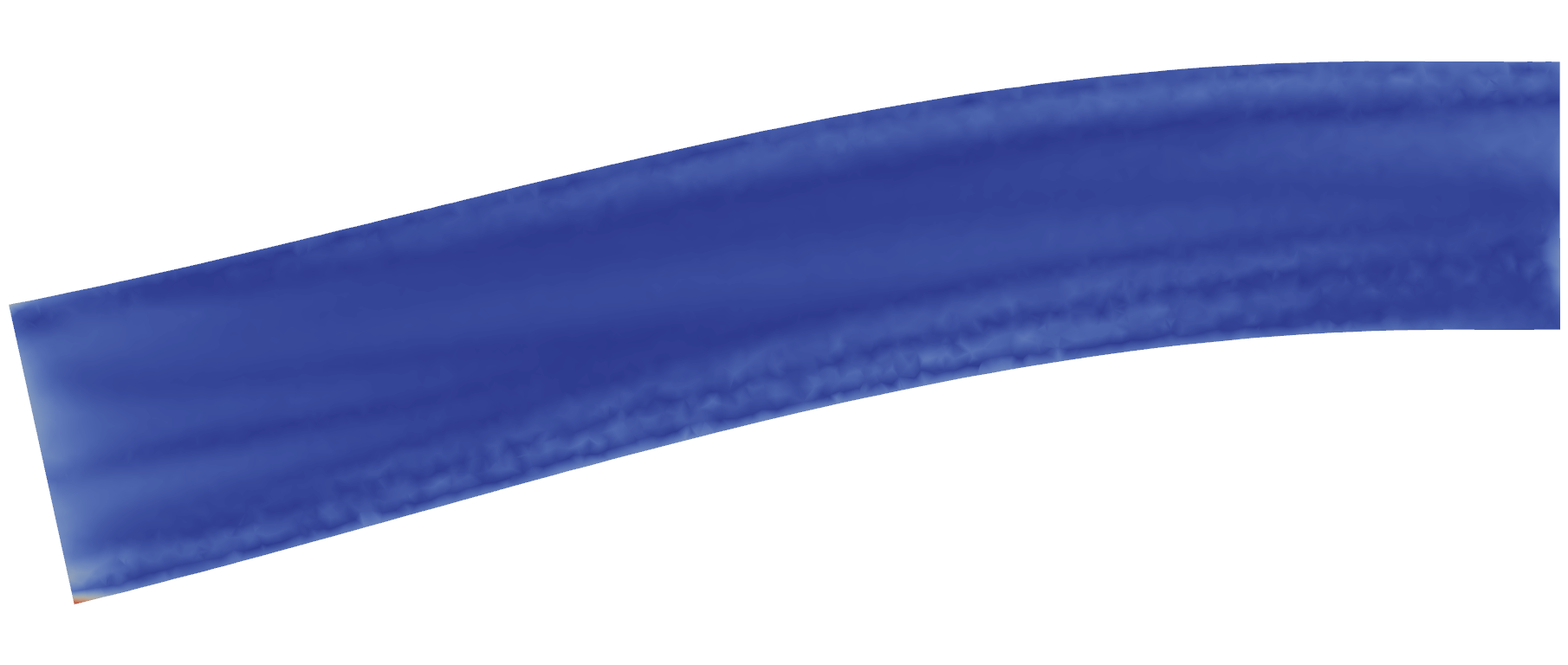} &
    \includegraphics[width=0.19\textwidth]{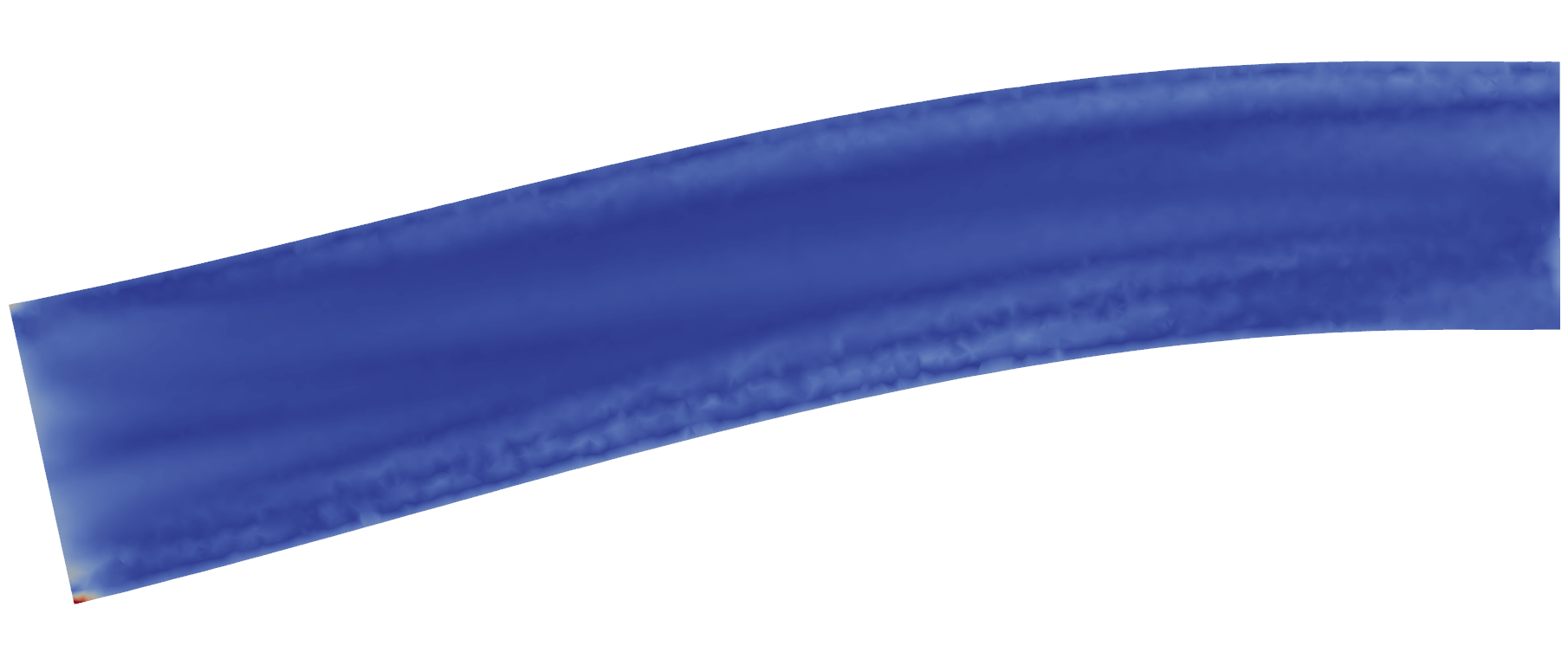} \\
     \rotatebox{90}{\begin{tabular}{c}
         finest\\
         mesh
    \end{tabular}} &  
    \includegraphics[width=0.19\textwidth]{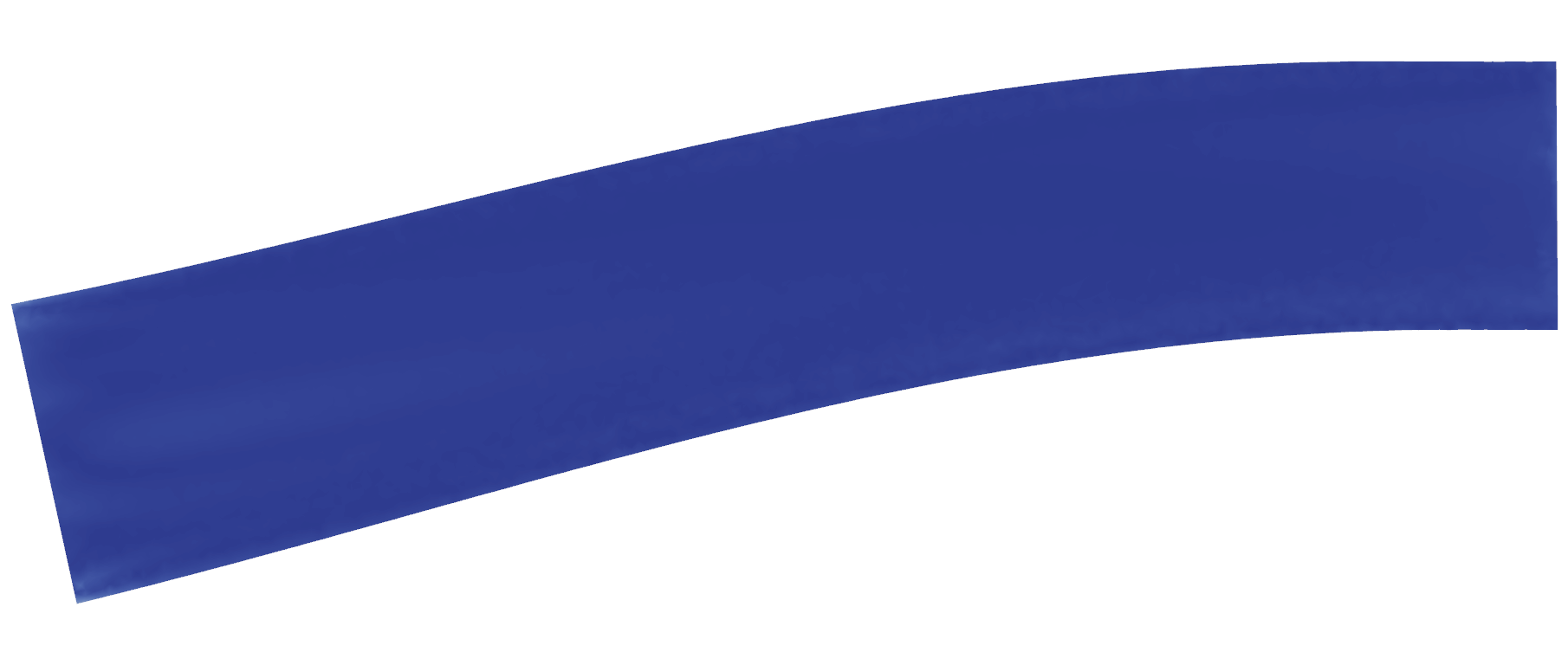} &
    \includegraphics[width=0.19\textwidth]{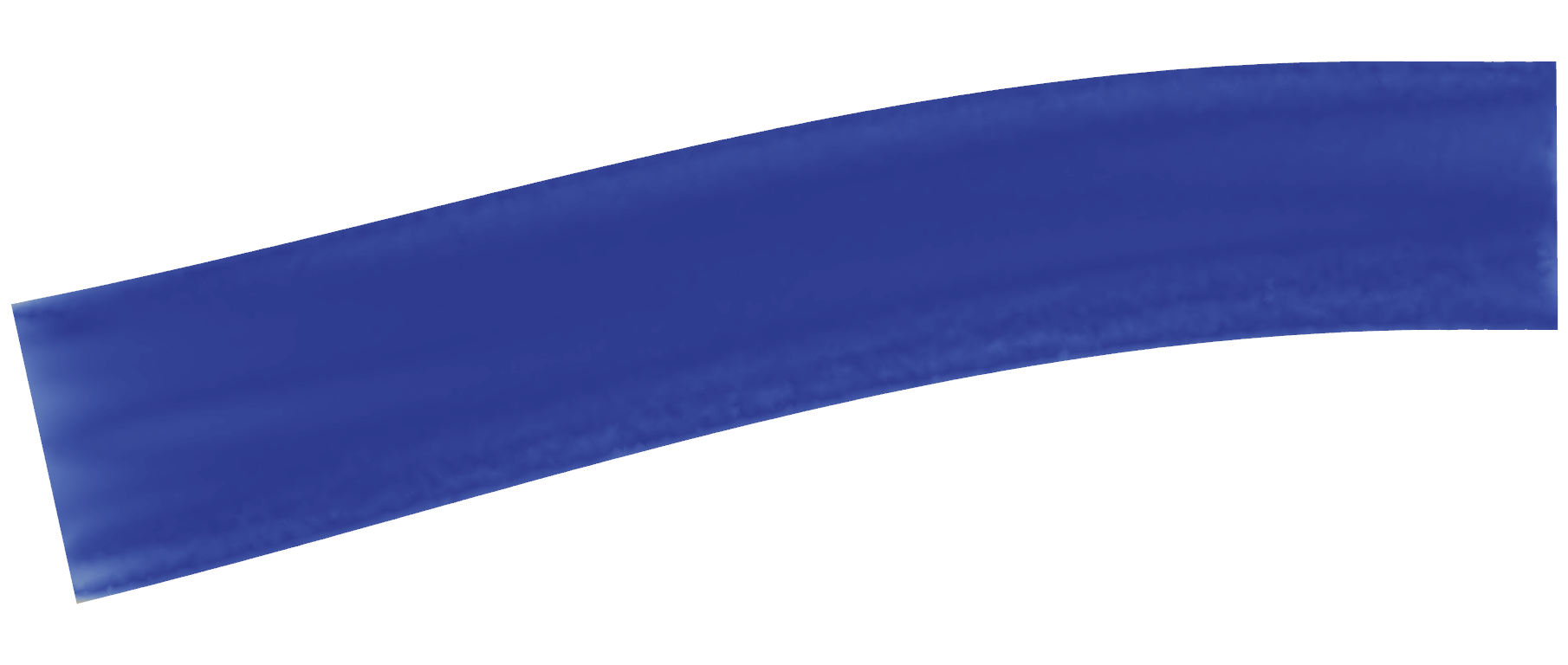} &
    \includegraphics[width=0.19\textwidth]{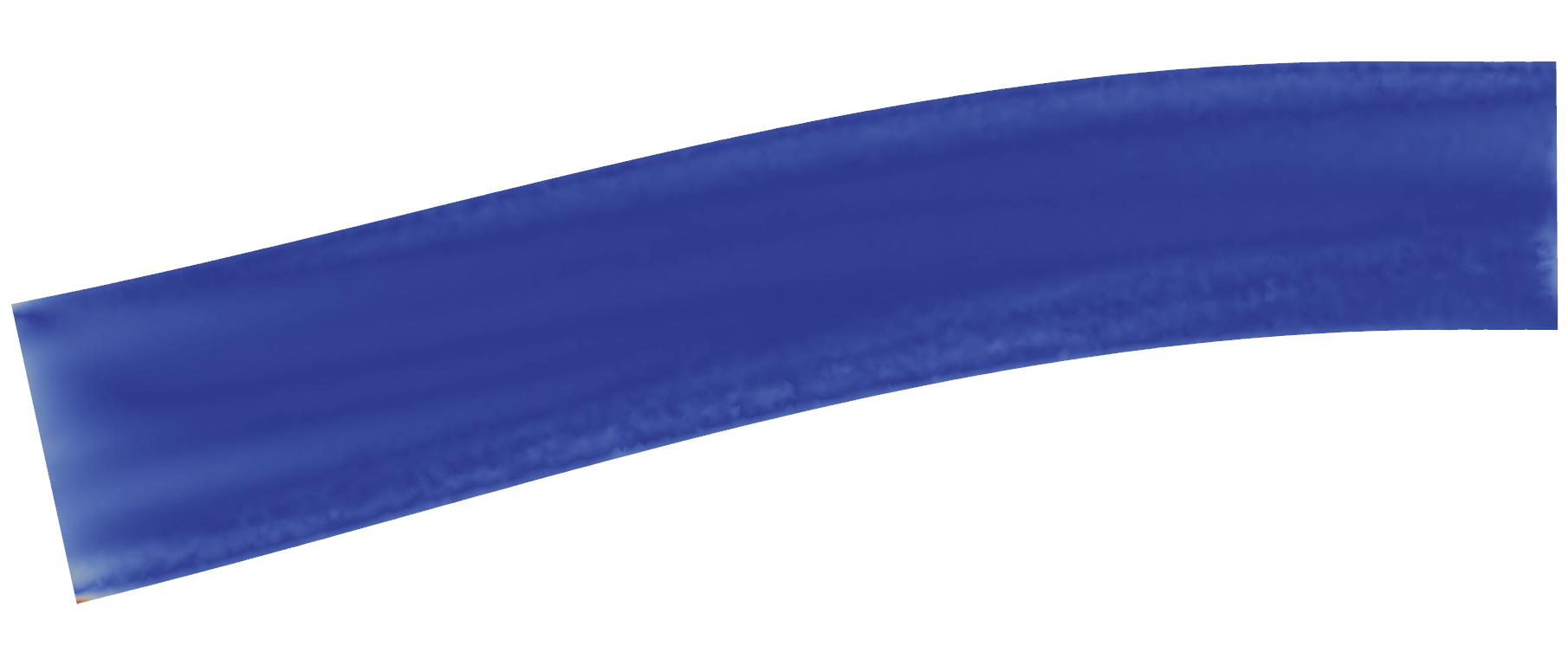} &
    \includegraphics[width=0.19\textwidth]{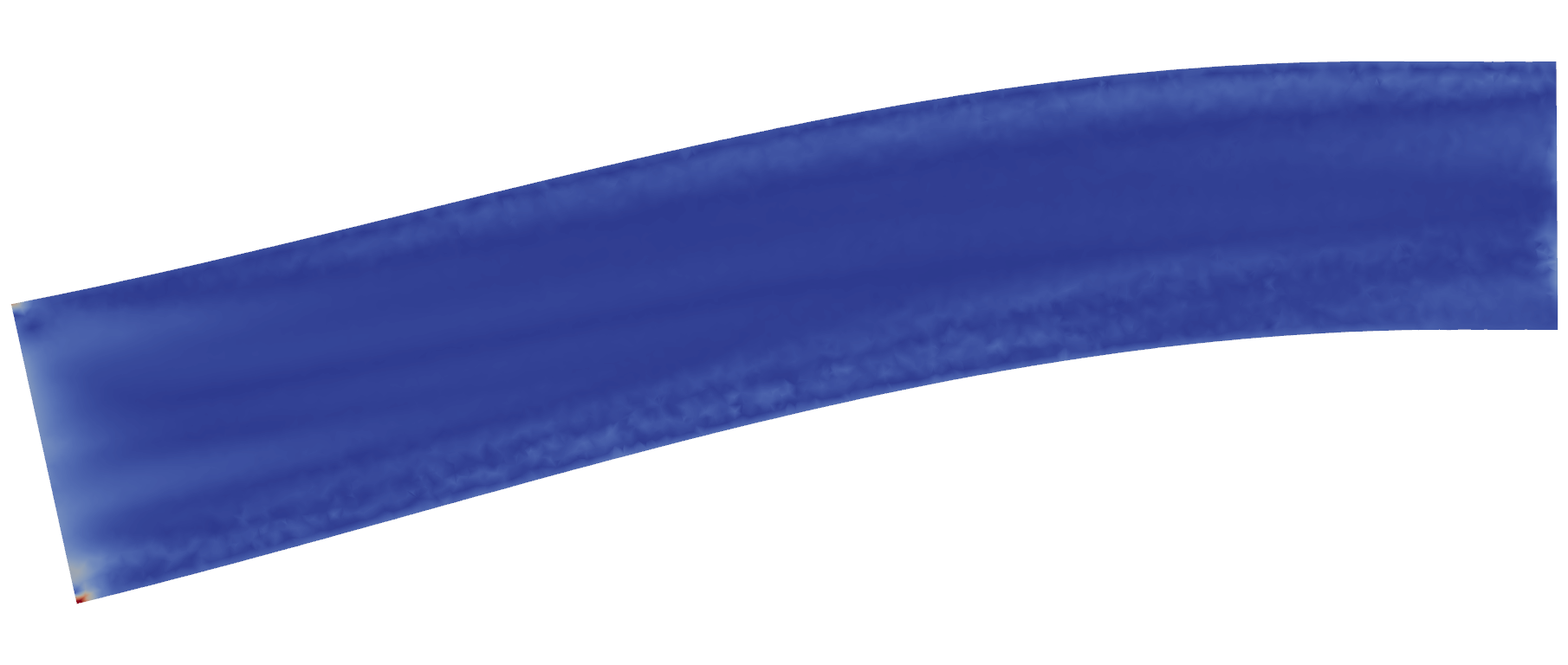} \\
    & \multicolumn{4}{c}{\includegraphics[width=0.5\textwidth]{scale0.035.png}}
\end{tabular}
\caption{Visulatization of the difference between data and computed velocity fields on bent geometry using stabilized $P_1/P_1$ element with $\alpha_p=\alpha_v=0.01$ for the coarser mesh ($h=1.5\text{ mm}$, $\text{number of DOFs}=97,928$), the finer mesh ($h=1\text{ mm}$, $\text{number of DOFs}=322,864$), the finest mesh ($h=0.8\text{ mm}$, $\text{number of DOFs}=623,088$) and multiple values of $\theta$.}
\label{fig:mesh_bent_error}
\end{figure}

\begin{figure}
\centering
\begin{tabular}{c c c c c}
    \rotatebox{90}{\begin{tabular}{c}
         data
    \end{tabular}} &  
    \includegraphics[width=0.19\textwidth]{arch_data_200.png} &
    \includegraphics[width=0.19\textwidth]{arch_data_500.png} &
    \includegraphics[width=0.19\textwidth]{arch_data_800.png} &
    \includegraphics[width=0.19\textwidth]{arch_data_1000.png} \\
    & $\theta=0.2$ & $\theta=0.5$ & $\theta=0.8$ & $\theta=1$ \\
    \rotatebox{90}{\begin{tabular}{c}
         coarser \\
         mesh
    \end{tabular}} &  
    \includegraphics[width=0.19\textwidth]{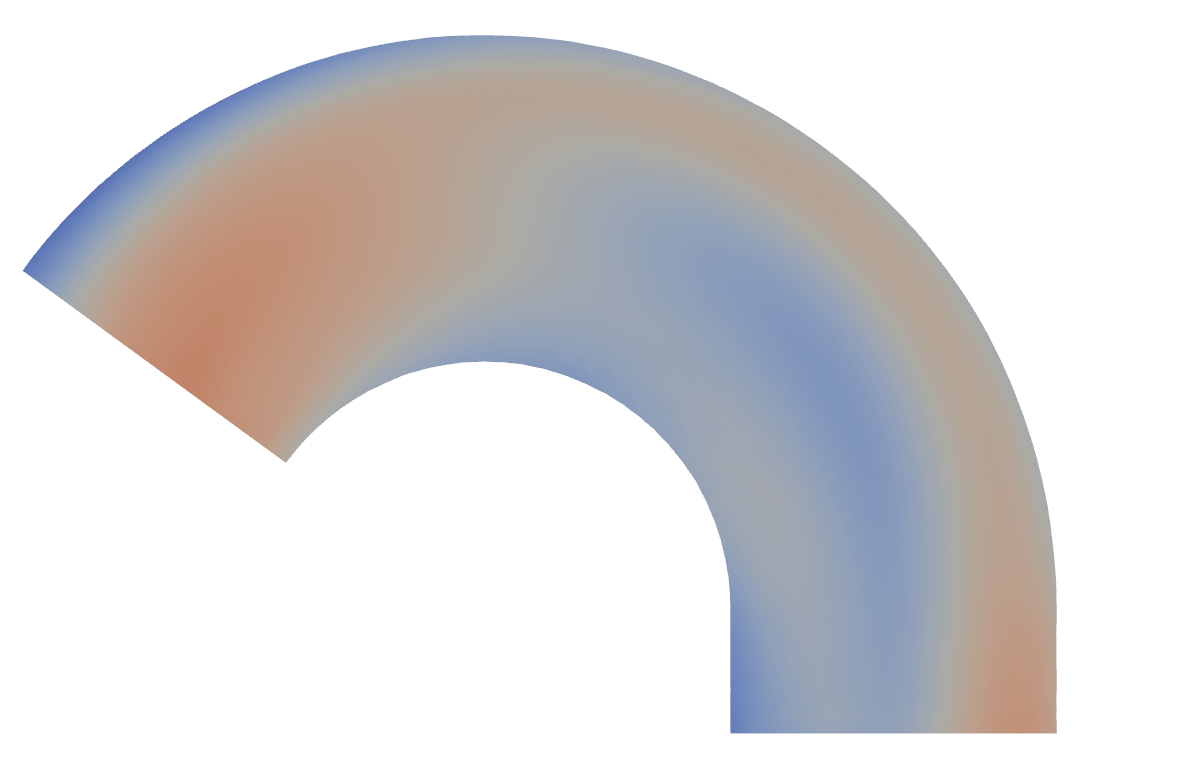} &
    \includegraphics[width=0.19\textwidth]{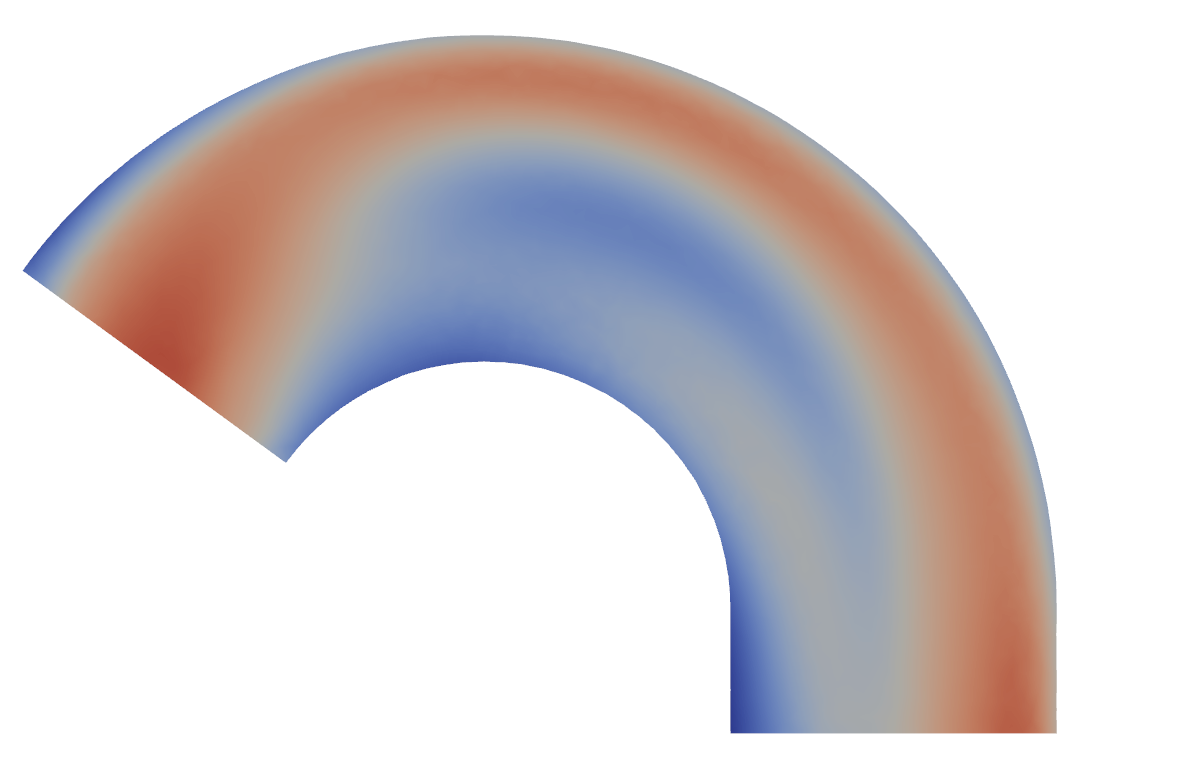} &
    \includegraphics[width=0.19\textwidth]{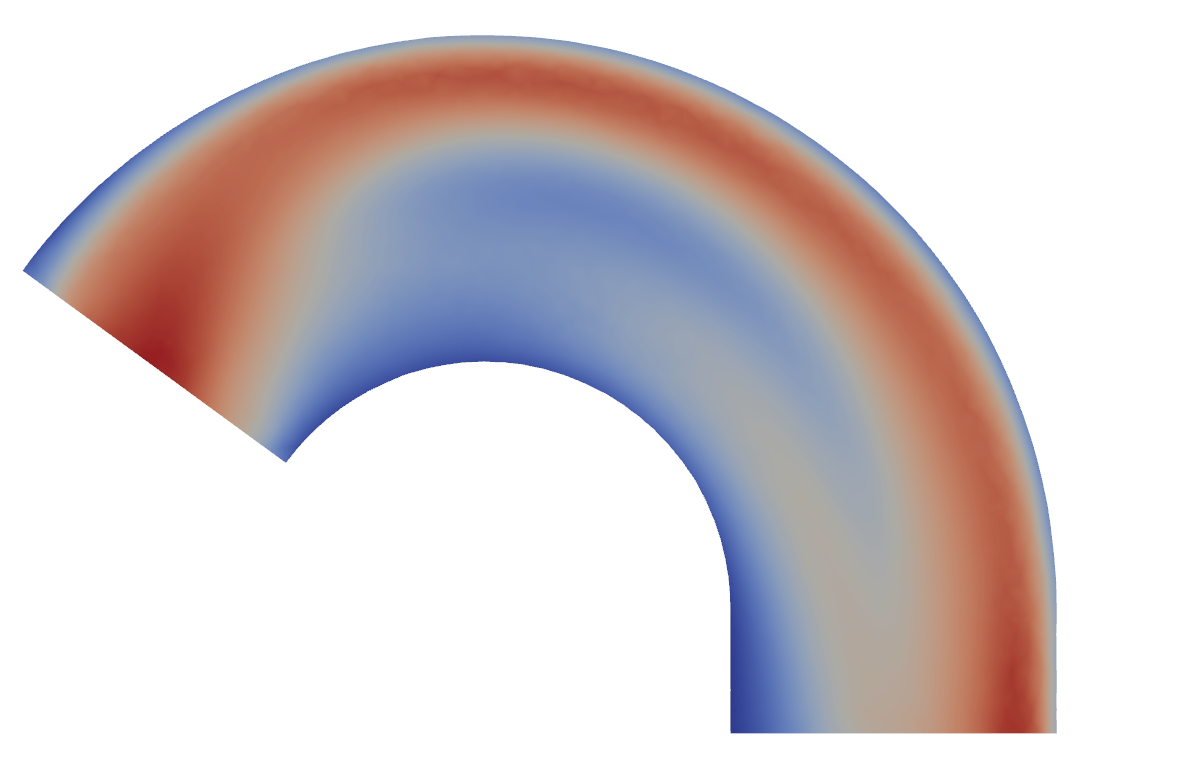} &
    \includegraphics[width=0.19\textwidth]{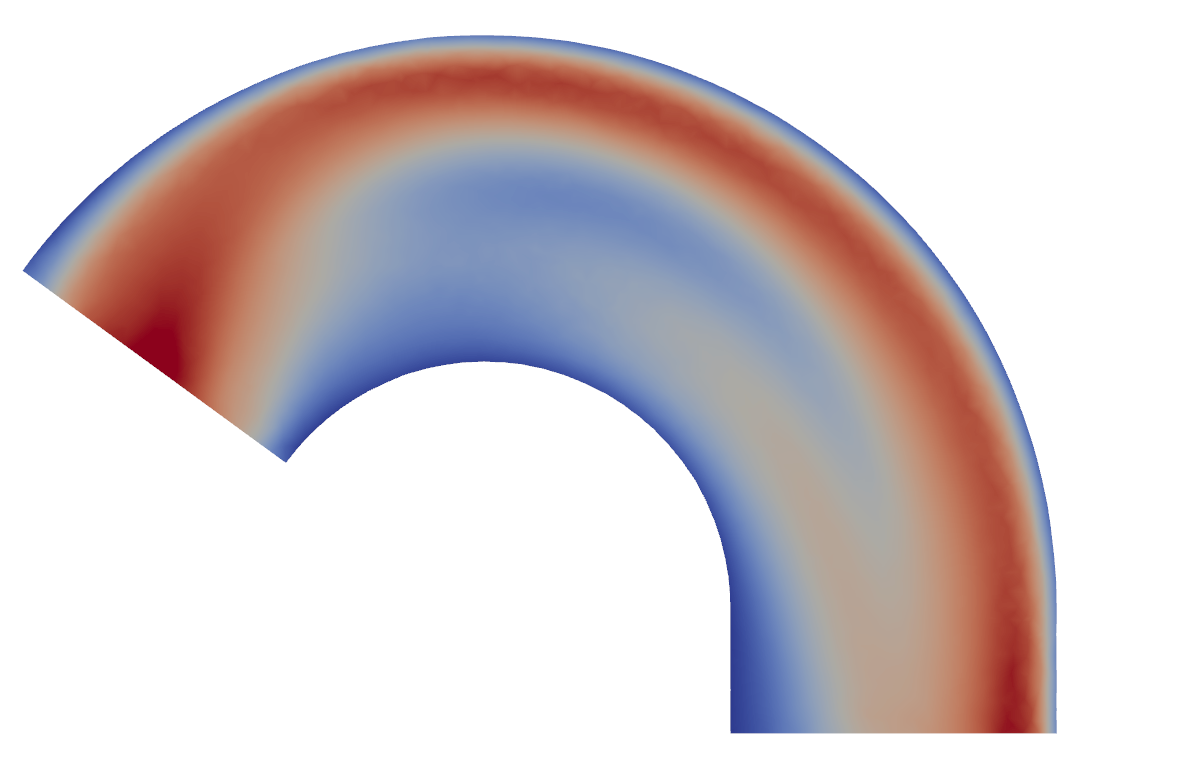} \\
     & $\thet{opt}=0.194$ & $\thet{opt}=0.463$ & $\thet{opt}=0.691$ & $\thet{opt}=0.834$ \\
     & $\mathcal{J}=\rnum{0.0001375}$ & $\mathcal{J}=\rnum{0.0006408}$ & $\mathcal{J}=\rnum{0.001267}$ & $\mathcal{J}=\rnum{0.001807}$ \\
     & $\mathcal{R}=\rnum{0.0001349}$ & $\mathcal{R}=\rnum{0.0003067}$ & $\mathcal{R}=\rnum{0.0004333}$ & $\mathcal{R}=\rnum{0.0005093}$ \\
     & iterations: 30 & iterations: 30 & iterations: 28 & iterations: 21 \\
    \rotatebox{90}{\begin{tabular}{c}
         finer\\
         mesh
    \end{tabular}} & 
    \includegraphics[width=0.19\textwidth]{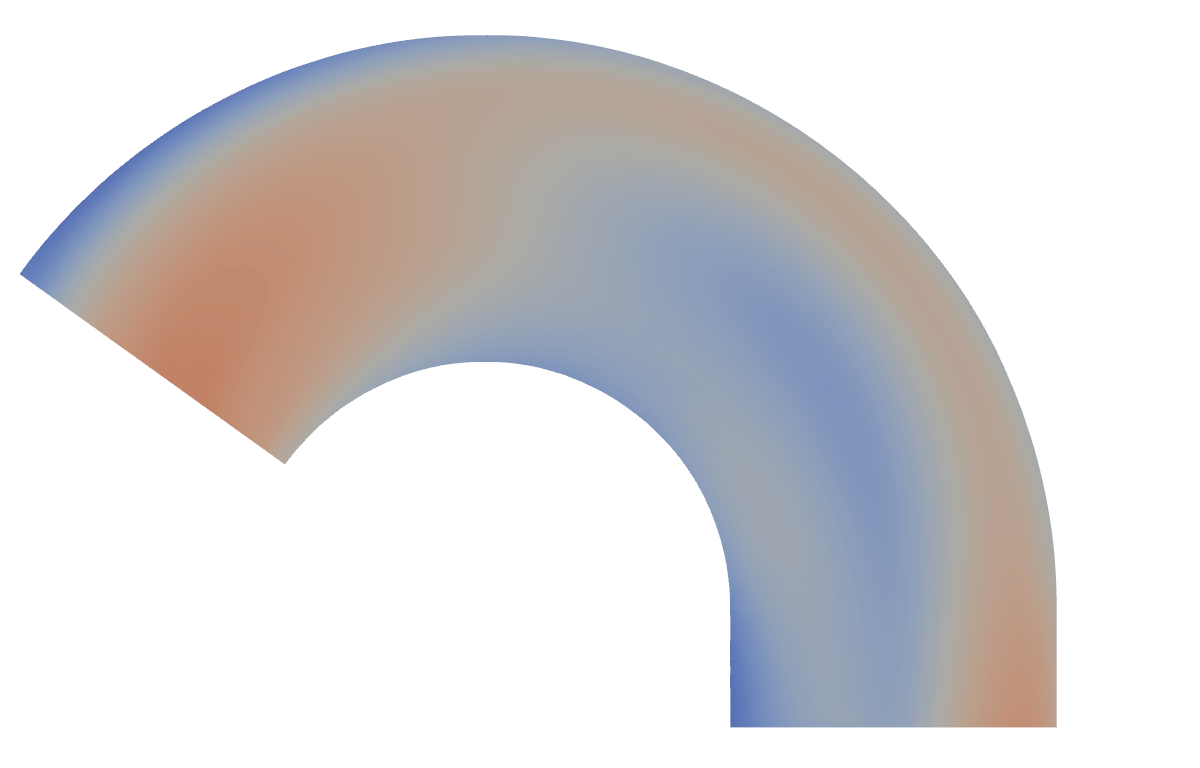} &
    \includegraphics[width=0.19\textwidth]{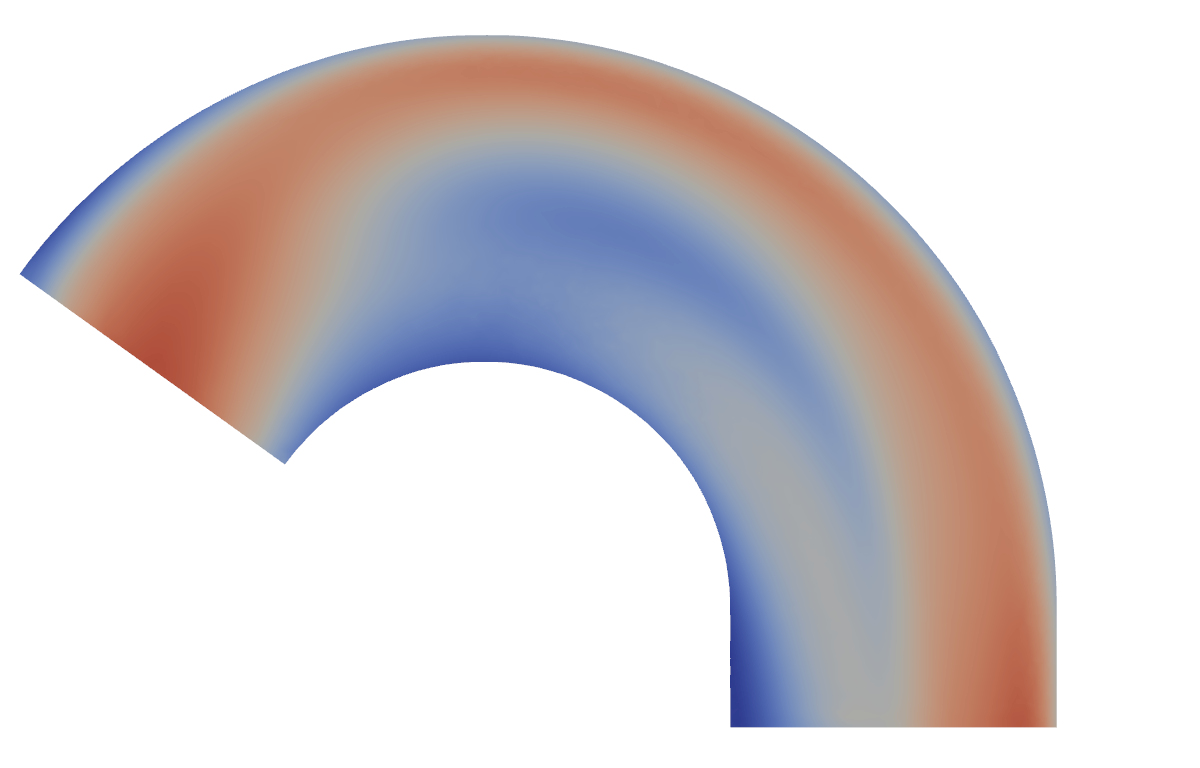} &
    \includegraphics[width=0.19\textwidth]{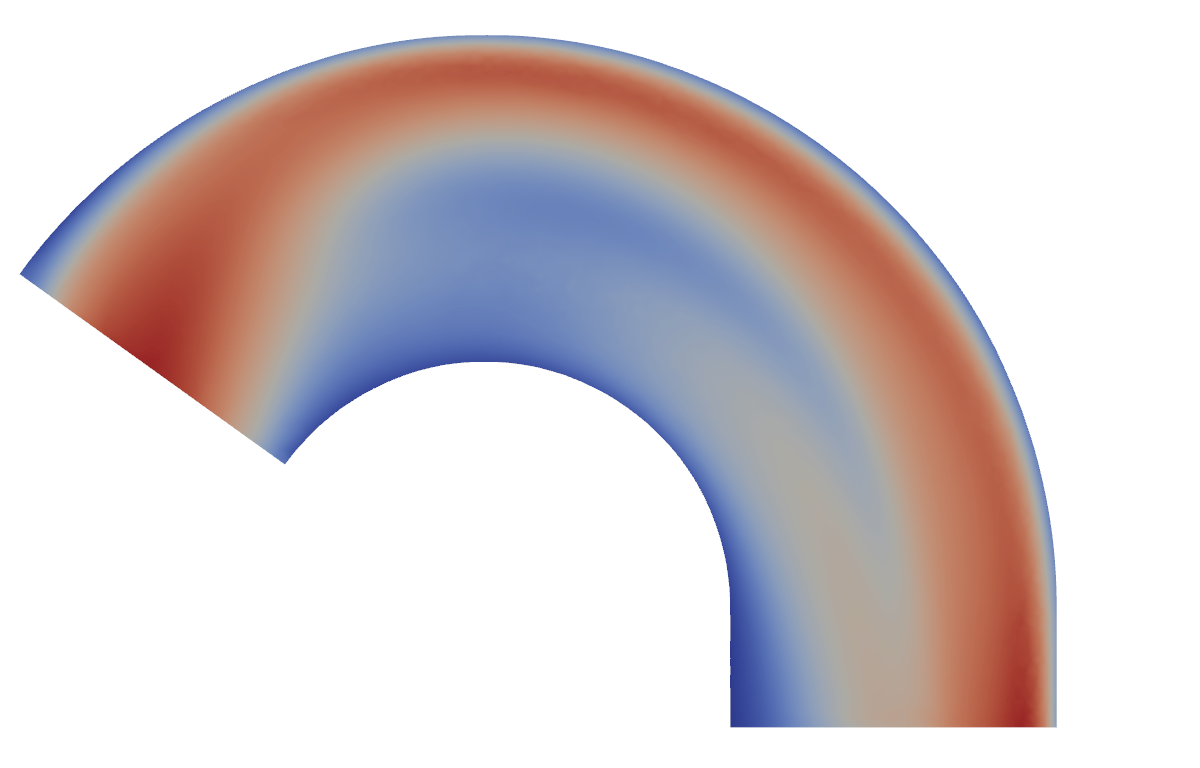} &
    \includegraphics[width=0.19\textwidth]{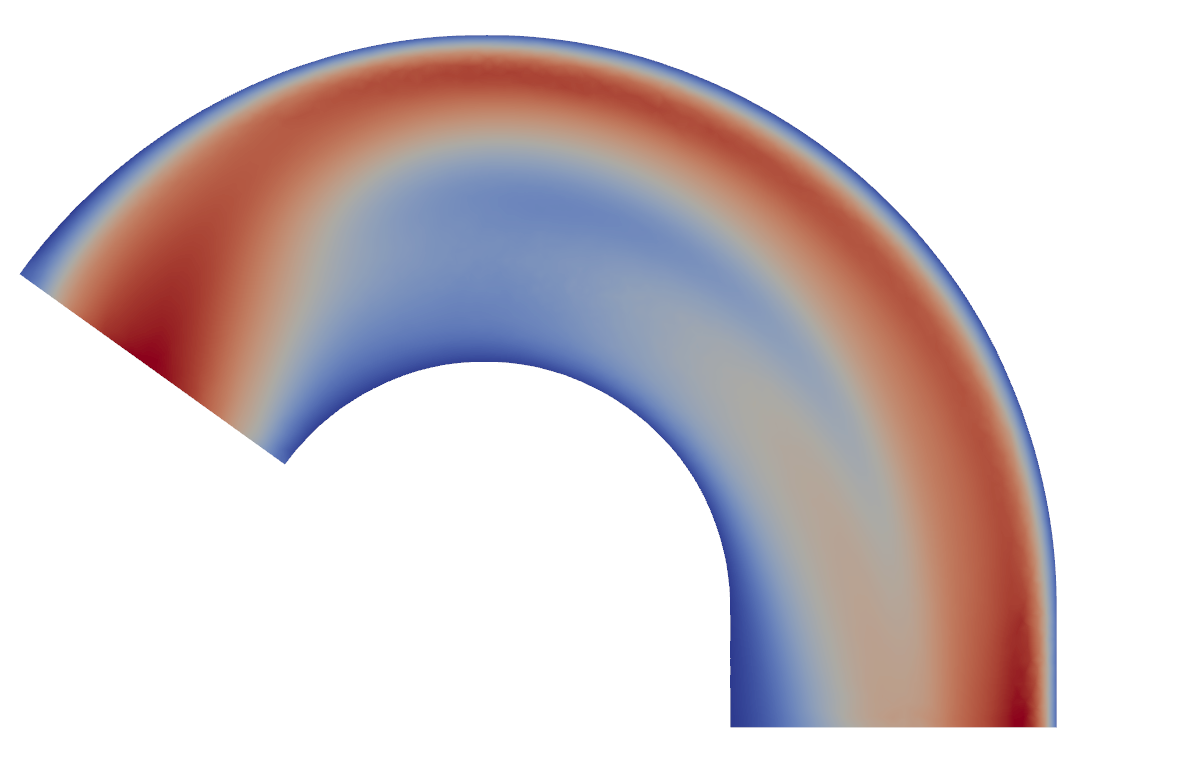} \\
     & $\thet{opt}=0.192$ & $\thet{opt}=0.473$ & $\thet{opt}=0.732$ & $\thet{opt}=0.899$ \\
     & $\mathcal{J}=\rnum{4.637e-05}$ & $\mathcal{J}=\rnum{0.0001957}$ & $\mathcal{J}=\rnum{0.0003714}$ & $\mathcal{J}=\rnum{0.0005119}$ \\
     & $\mathcal{R}=\rnum{0.0001327}$ & $\mathcal{R}=\rnum{0.0002781}$ & $\mathcal{R}=\rnum{0.0003695}$ & $\mathcal{R}=\rnum{0.0004181}$ \\
     & iterations: 33 & iterations: 33 & iterations: 31 & iterations: 29 \\
    \rotatebox{90}{\begin{tabular}{c}
         finest\\
         mesh
    \end{tabular}} & 
    \includegraphics[width=0.19\textwidth]{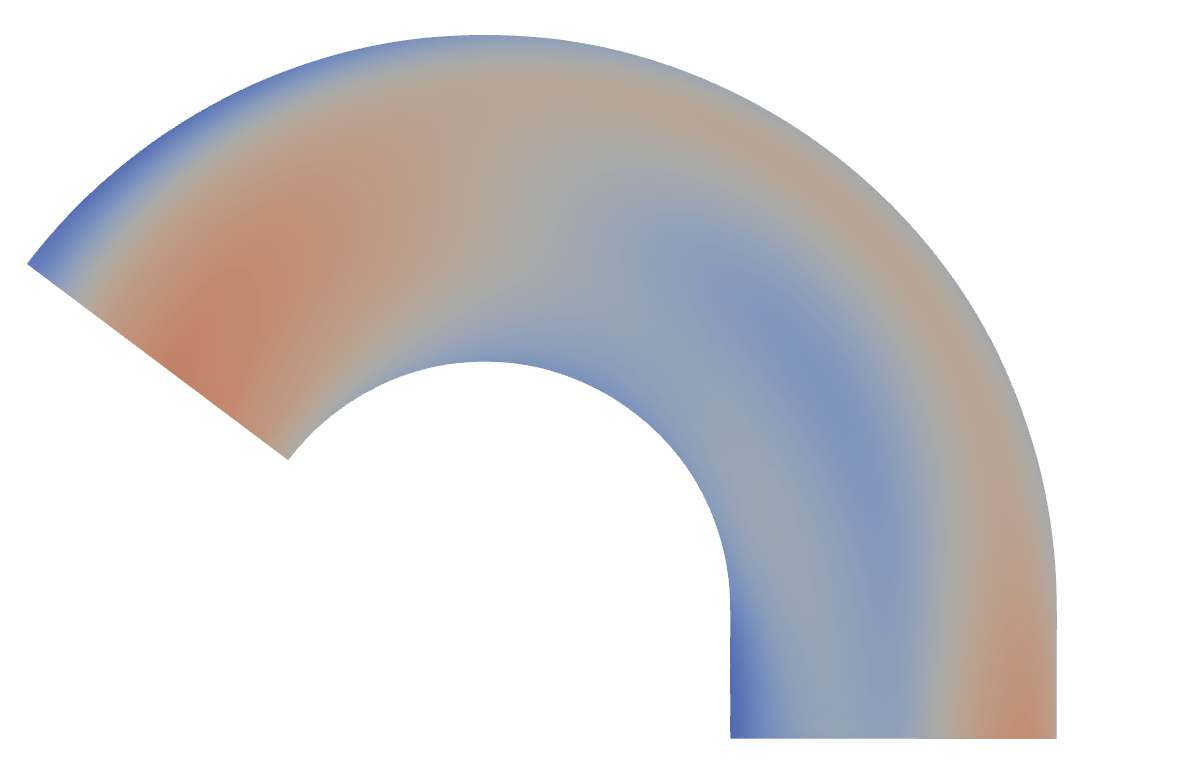} &
    \includegraphics[width=0.19\textwidth]{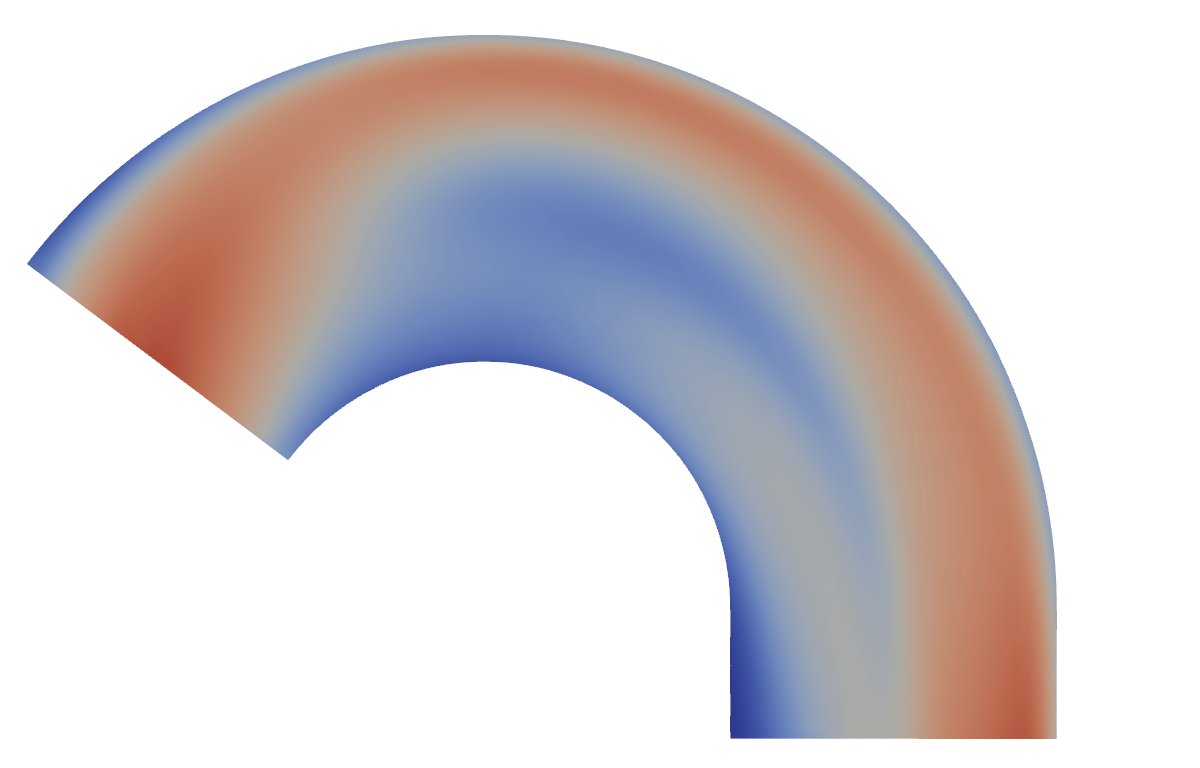} &
    \includegraphics[width=0.19\textwidth]{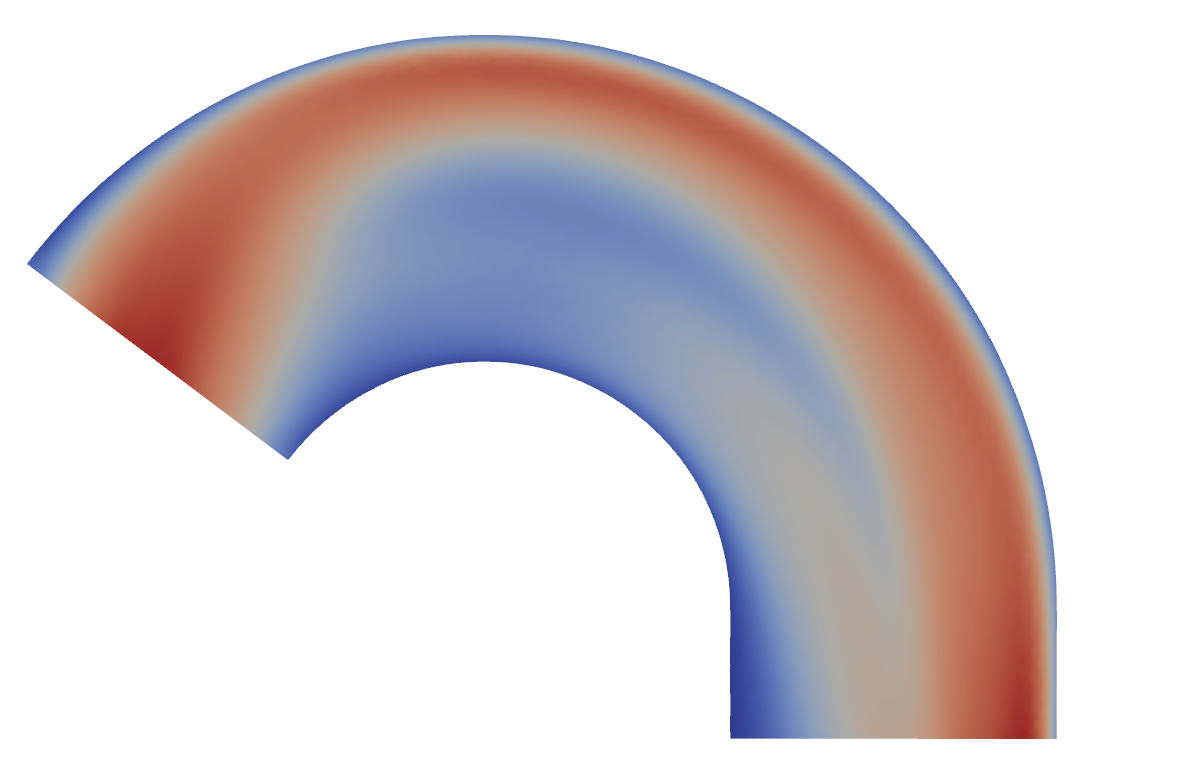} &
    \includegraphics[width=0.19\textwidth]{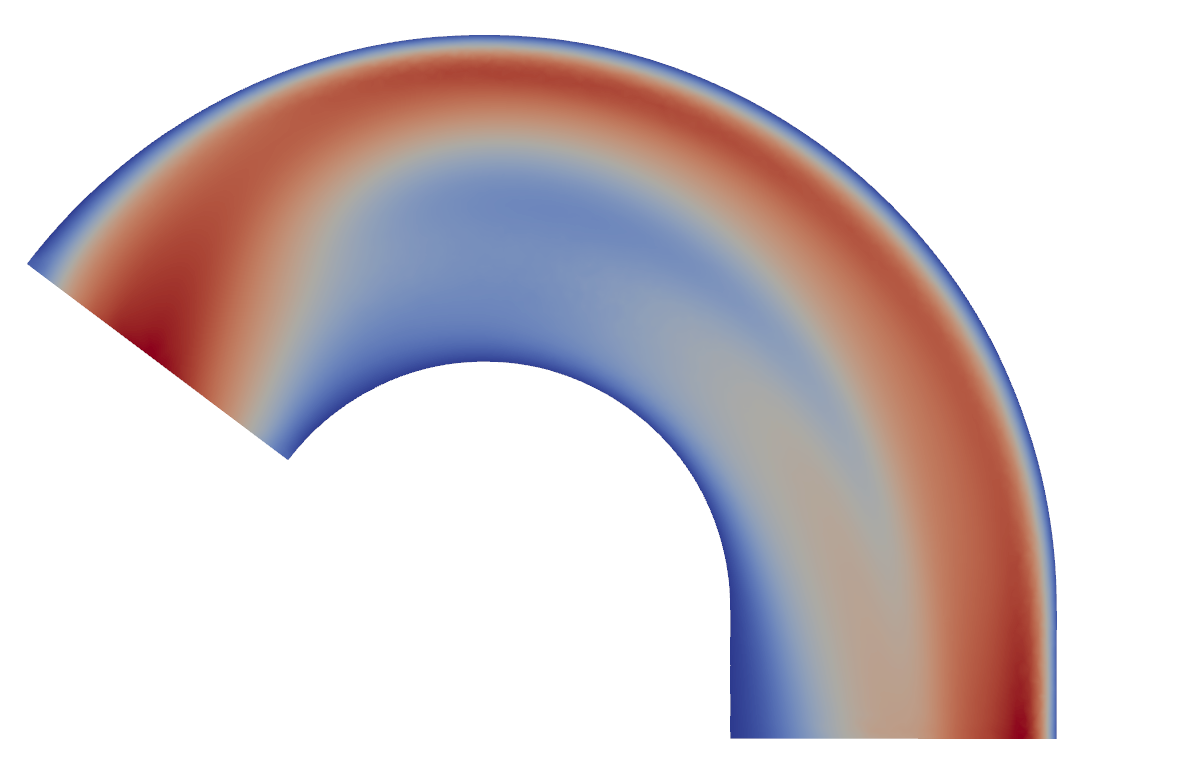} \\
     & $\thet{opt}=0.195$ & $\thet{opt}=0.476$ & $\thet{opt}=0.754$ & $\thet{opt}=0.933$ \\
     & $\mathcal{J}=\rnum{2.757e-05}$ & $\mathcal{J}=\rnum{0.0001003}$ & $\mathcal{J}=\rnum{0.00018}$ & $\mathcal{J}=\rnum{0.0002381}$ \\ 
     & $\mathcal{R}=\rnum{0.0001329}$ & $\mathcal{R}=\rnum{0.0002816}$ & $\mathcal{R}=\rnum{0.0003727}$ & $\mathcal{R}=\rnum{0.0004148}$ \\
     & iterations: 34 & iterations: 31 & iterations: 28 & iterations: 29 \\
    & \multicolumn{4}{c}{\includegraphics[width=0.5\textwidth]{scale0.2.png}}
\end{tabular}
\caption{Comparison of data without noise with assimilation velocity results in the arch geometry using stabilized $P_1/P_1$ element with $\alpha_p=\alpha_v=0.01$ for the coarser mesh ($h=1.5\text{ mm}$, $\text{number of DOFs}=106,052$), the finer mesh ($h=1\text{ mm}$, $\text{number of DOFs}=358,216$), the finest mesh ($h=0.8\text{ mm}$, $\text{number of DOFs}=658,980$) and multiple values of $\theta$.}
\label{fig:mesh_arch}
\centering
\medskip
\begin{tabular}{c c c c c}
    & $\theta=0.2$ & $\theta=0.5$ & $\theta=0.8$ & $\theta=1$ \\
    \rotatebox{90}{\begin{tabular}{c}
         coarser\\
         mesh
    \end{tabular}} &  
    \includegraphics[width=0.19\textwidth]{arch_err_p1p1_stab0.01_0.01_200.png} &
    \includegraphics[width=0.19\textwidth]{arch_err_p1p1_stab0.01_0.01_500.png} &
    \includegraphics[width=0.19\textwidth]{arch_err_p1p1_stab0.01_0.01_800.png} &
    \includegraphics[width=0.19\textwidth]{arch_err_p1p1_stab0.01_0.01_1000.png} \\
        \rotatebox{90}{\begin{tabular}{c}
         finer\\
         mesh 
    \end{tabular}} & 
    \includegraphics[width=0.19\textwidth]{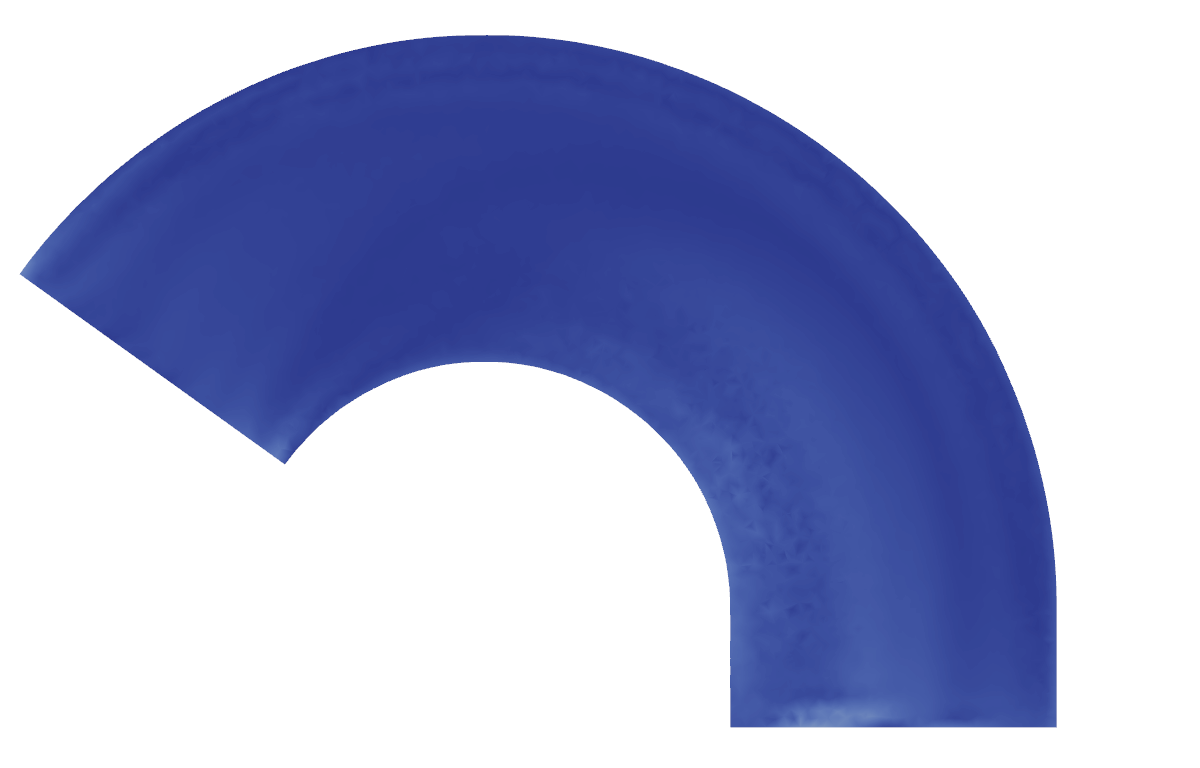} &
    \includegraphics[width=0.19\textwidth]{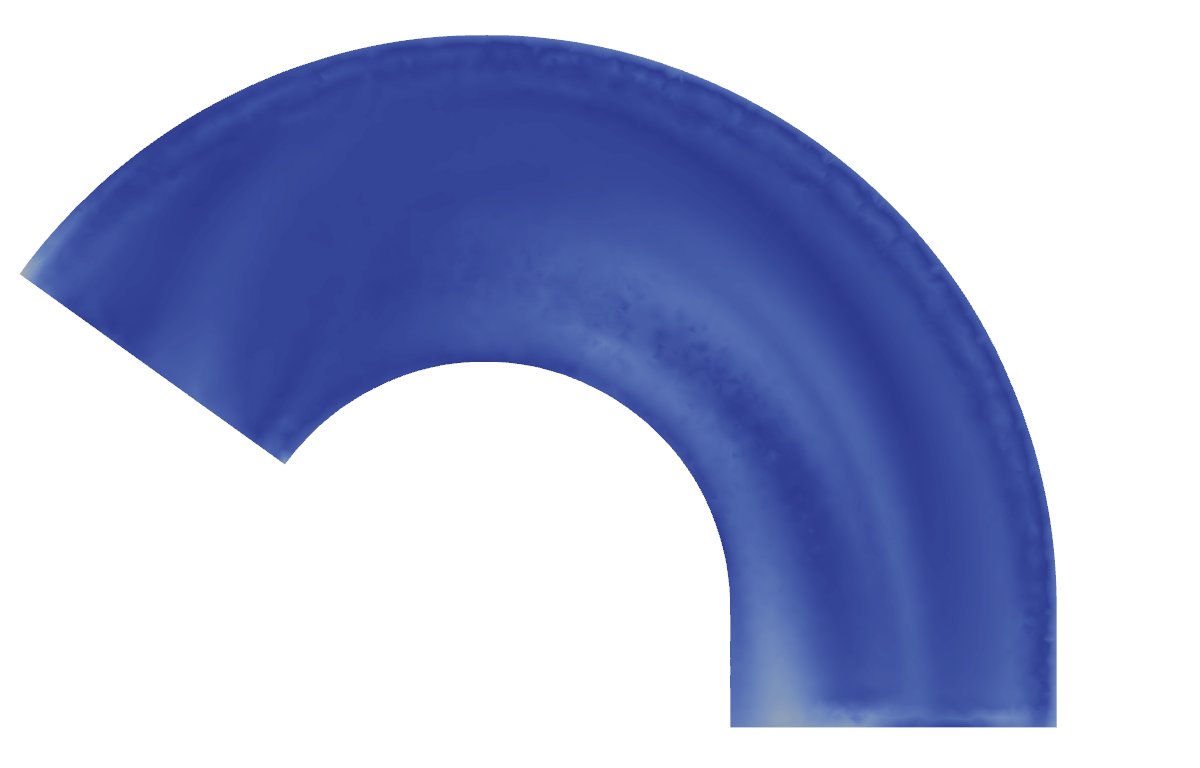} &
    \includegraphics[width=0.19\textwidth]{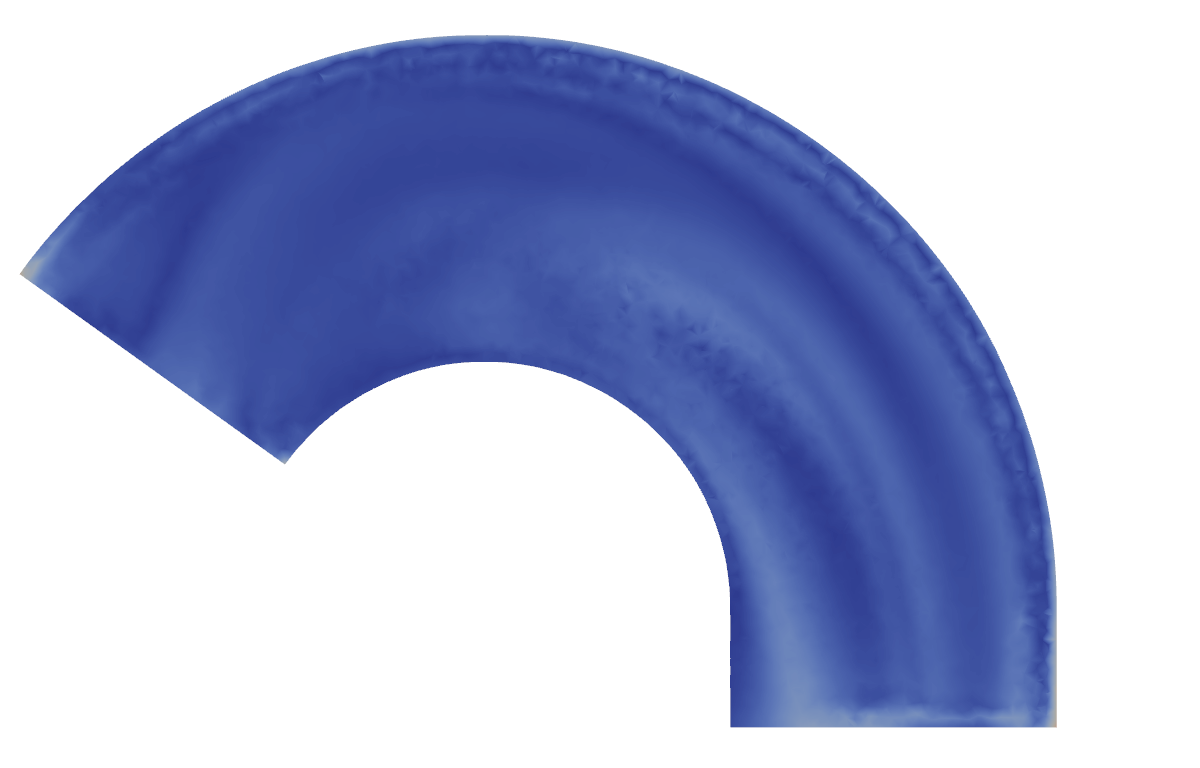} &
    \includegraphics[width=0.19\textwidth]{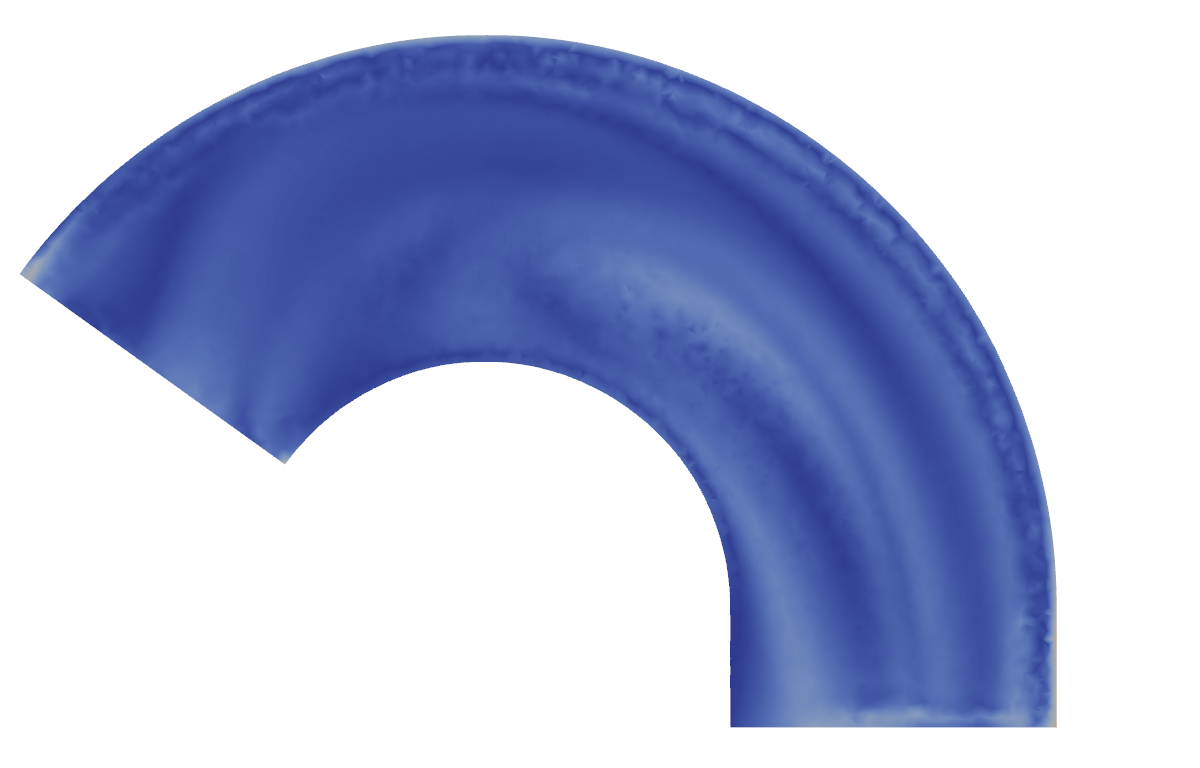} \\
     \rotatebox{90}{\begin{tabular}{c}
         finest\\
         mesh
    \end{tabular}} &  
    \includegraphics[width=0.19\textwidth]{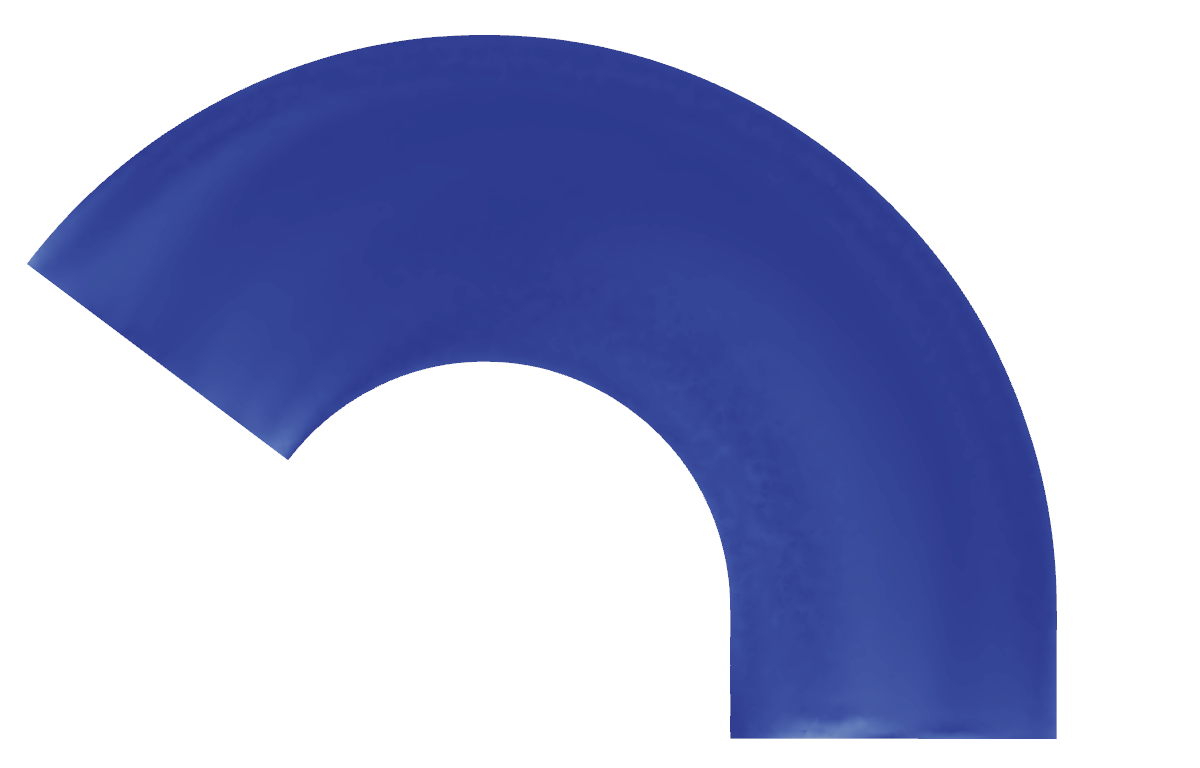} &
    \includegraphics[width=0.19\textwidth]{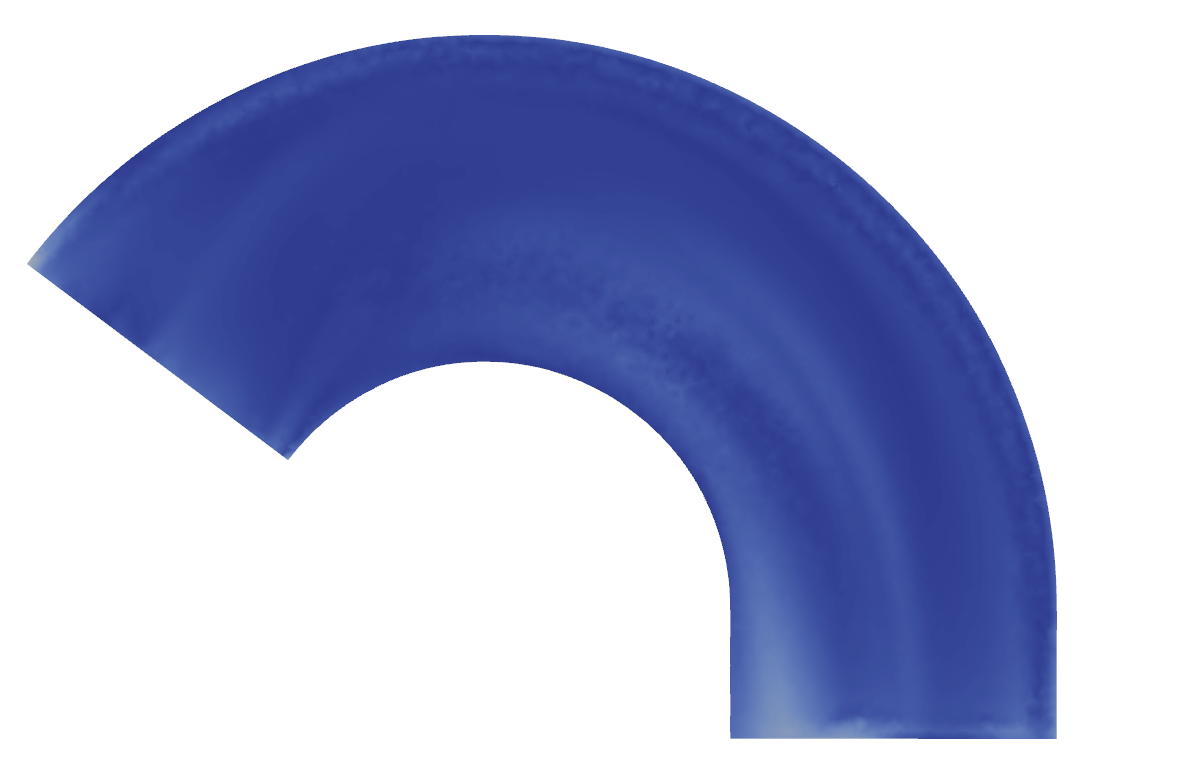} &
    \includegraphics[width=0.19\textwidth]{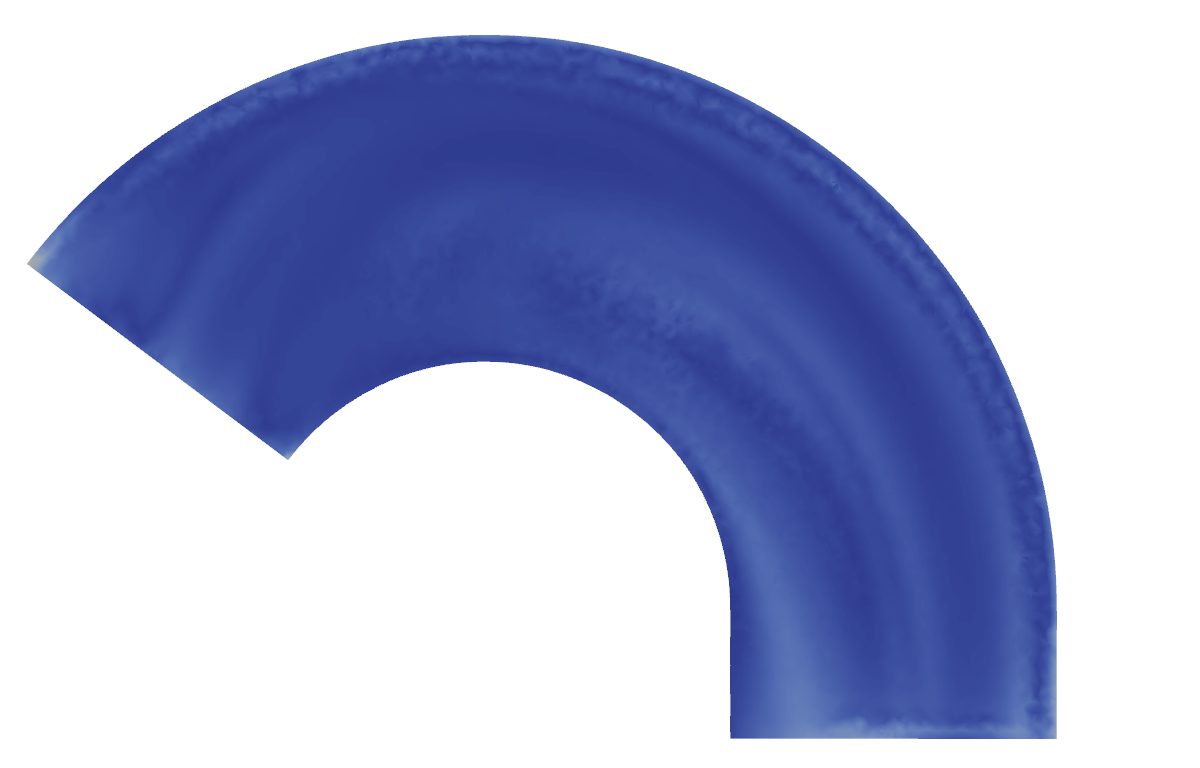} &
    \includegraphics[width=0.19\textwidth]{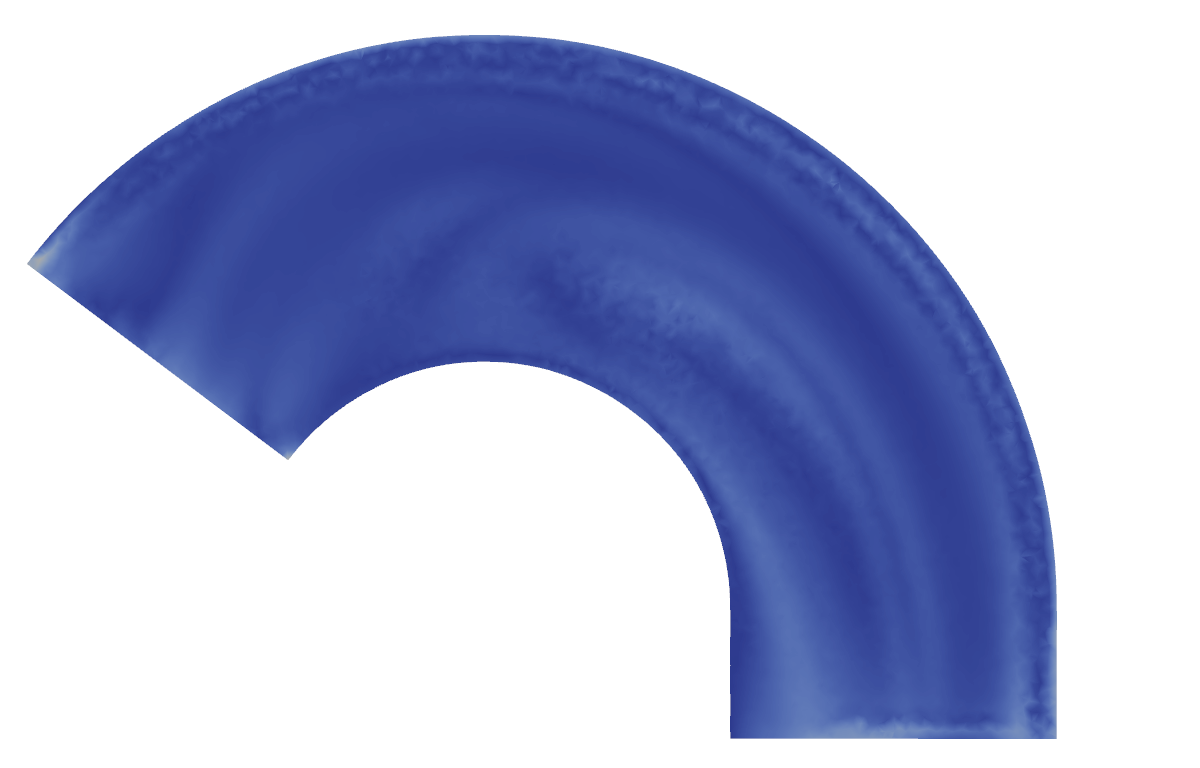} \\
    & \multicolumn{4}{c}{\includegraphics[width=0.5\textwidth]{scale0.05.png}}
\end{tabular}
\caption{Visulatization of the difference between data and computed velocity fields on arch geometry using stabilized $P_1/P_1$ element with $\alpha_p=\alpha_v=0.01$ for the coarser mesh ($h=1.5\text{ mm}$, $\text{number of DOFs}=106,052$), the finer mesh ($h=1\text{ mm}$, $\text{number of DOFs}=358,216$), the finest mesh ($h=0.8\text{ mm}$, $\text{number of DOFs}=658,980$) and multiple values of $\theta$.}
\label{fig:mesh_arch_error}
\end{figure}

Three meshes with different levels of refinement were used to assess the convergence of results computed using stabilized $P_1/P_1$ element with $\alpha_v=\alpha_p=0.01$.  
The edge lengths of the meshes were selected as $h=1.5\text{ mm}$, $h=1\text{ mm}$ and $h=0.8\text{ mm}$. 
Therefore, the finest mesh has approximately the same number of degrees of freedom for the $P_1/P_1$ element as the coarsest mesh using the MINI element.
The comparisons of the results are shown in Figures \ref{fig:mesh_bent} and \ref{fig:mesh_arch}, and the errors with respect to the ground truth velocities are shown in Figures \ref{fig:mesh_bent_error} and \ref{fig:mesh_arch_error}. 
We can see that the velocity field and the slip parameter $\thet{opt}$ get closer to the ground truth with the increasing number of degrees of freedom.
Of course, it is important to point out that the problems involving finer meshes are much more computationally expensive and often more prone to divergence of the nonlinear solver.

\subsection{Amount of noise}
\begin{figure}
\centering
\begin{tabular}{c c c c c}
     & $\text{SNR}=\infty$ & $\text{SNR}=2$ & $\text{SNR}=1$ & $\text{SNR}=0.5$\\
    \rotatebox{90}{\begin{tabular}{c}
         data \\
         $\theta=0.5$
    \end{tabular}} &  
    \includegraphics[width=.205\textwidth]{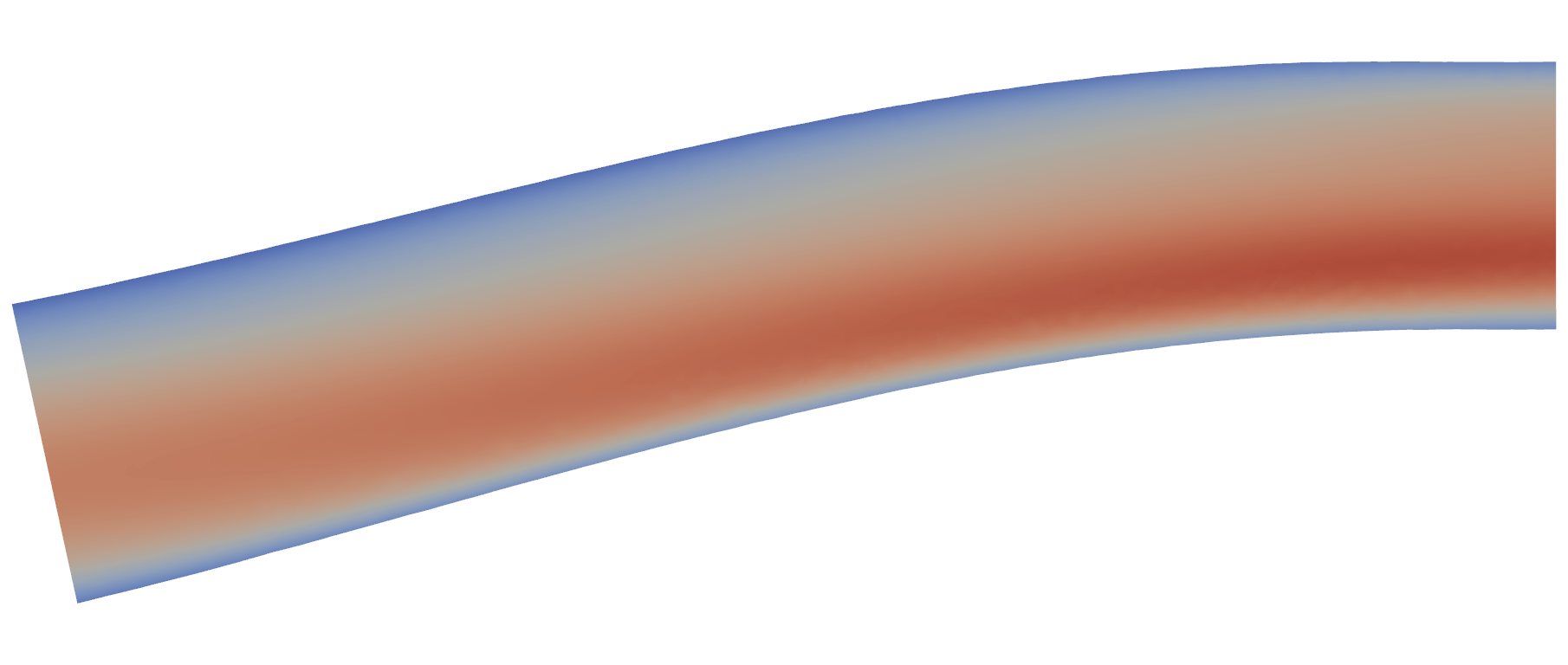} &
    \includegraphics[width=.205\textwidth]{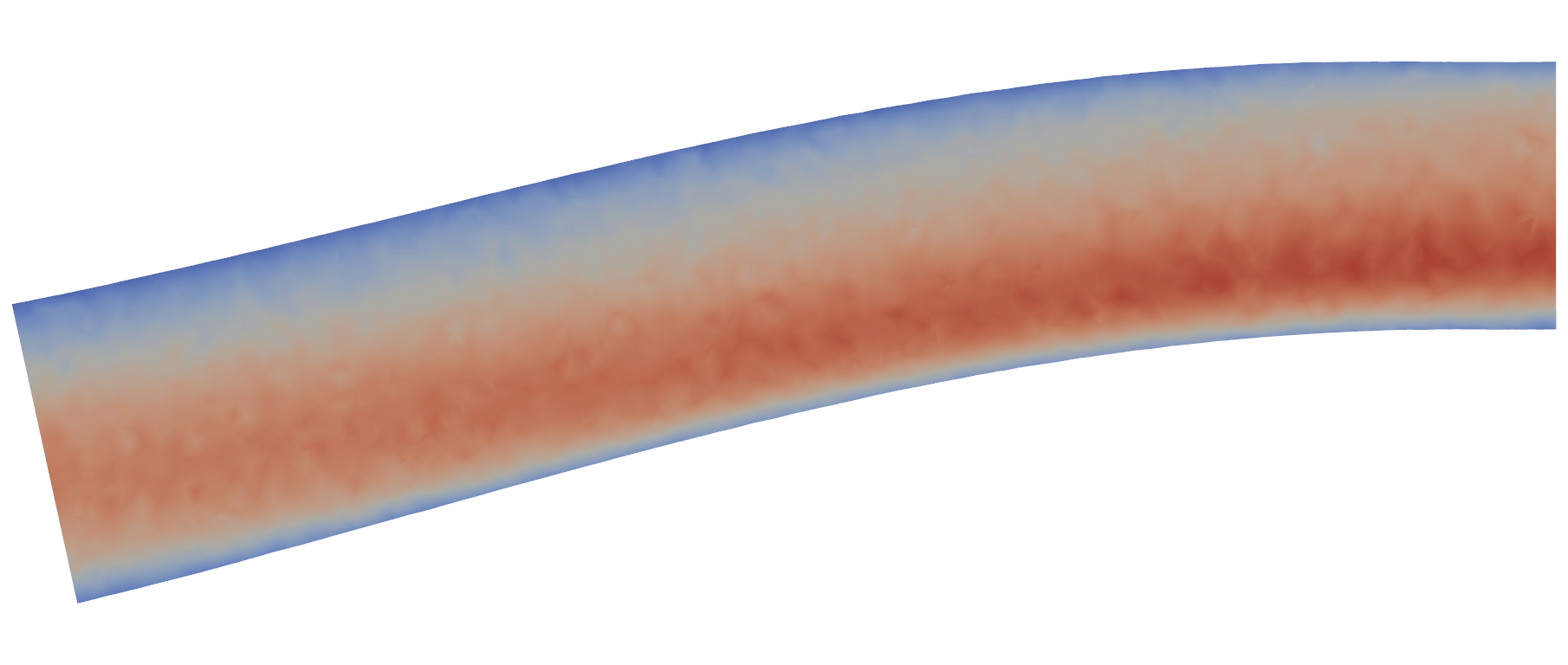} &
    \includegraphics[width=.205\textwidth]{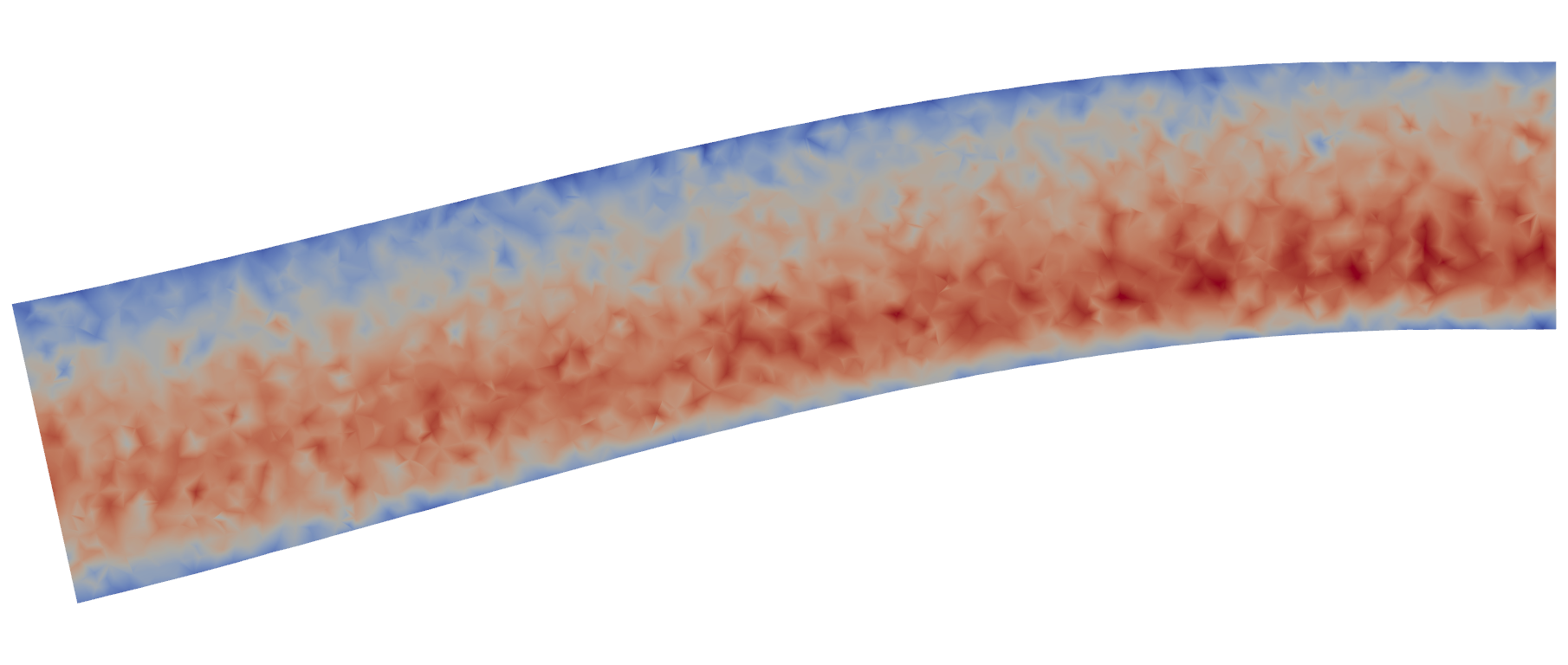} &
    \includegraphics[width=.205\textwidth]{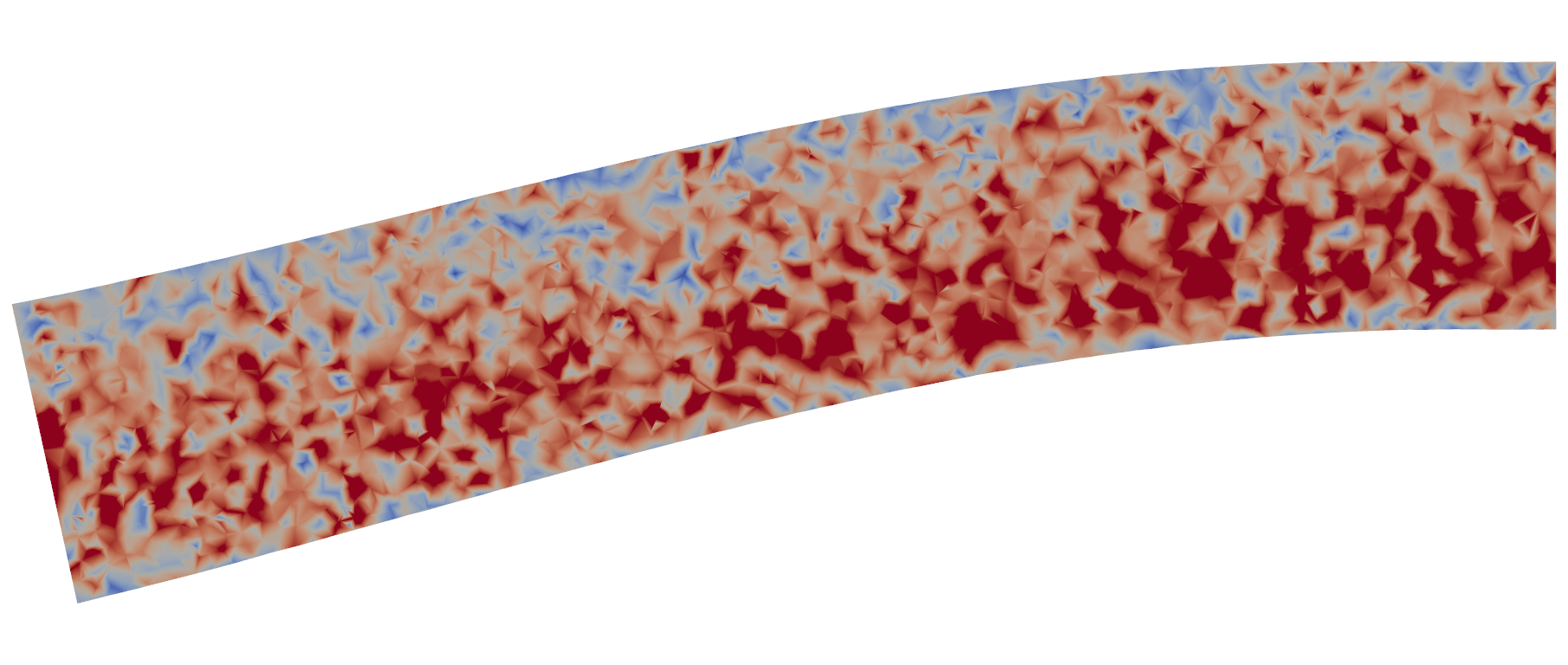} \\
    \rotatebox{90}{\begin{tabular}{c}
         assimilation \\
         $\theta=0.5$
    \end{tabular}} & 
    \includegraphics[width=.205\textwidth]{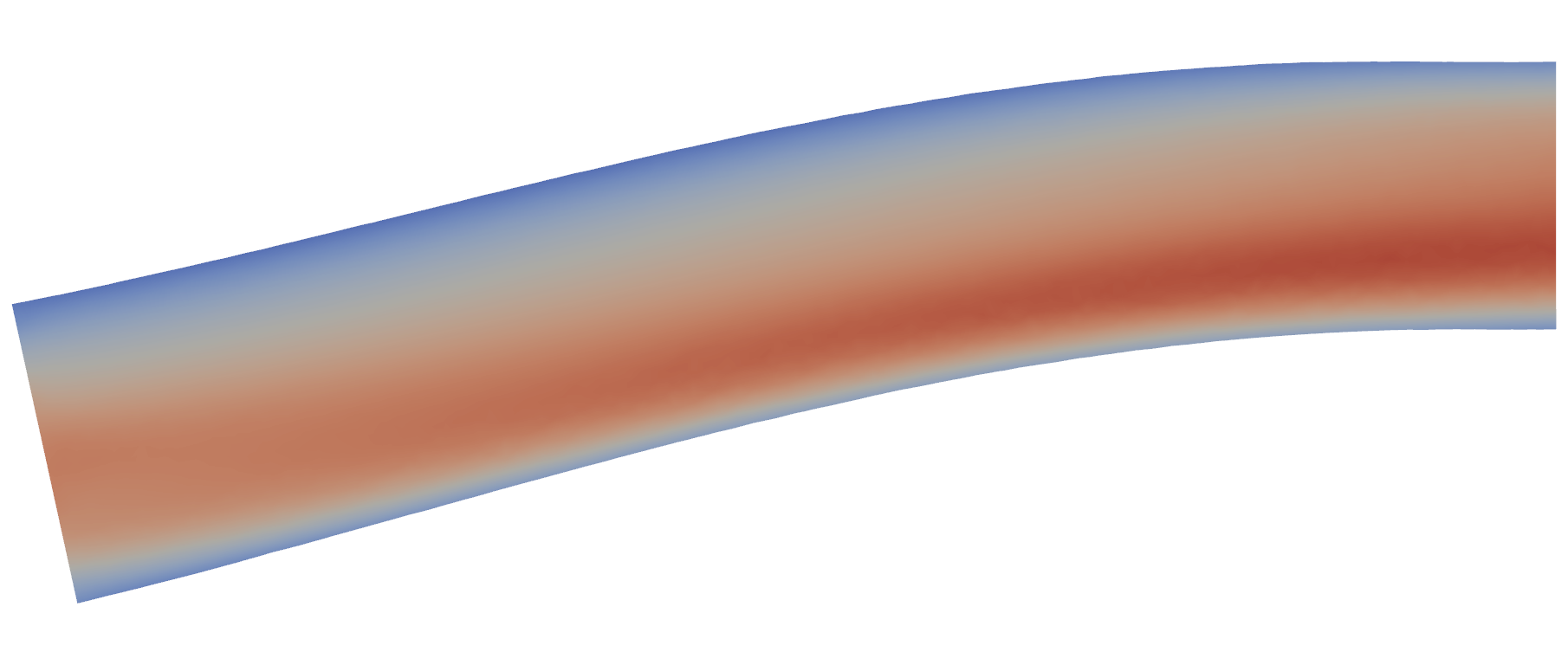} &
    \includegraphics[width=.205\textwidth]{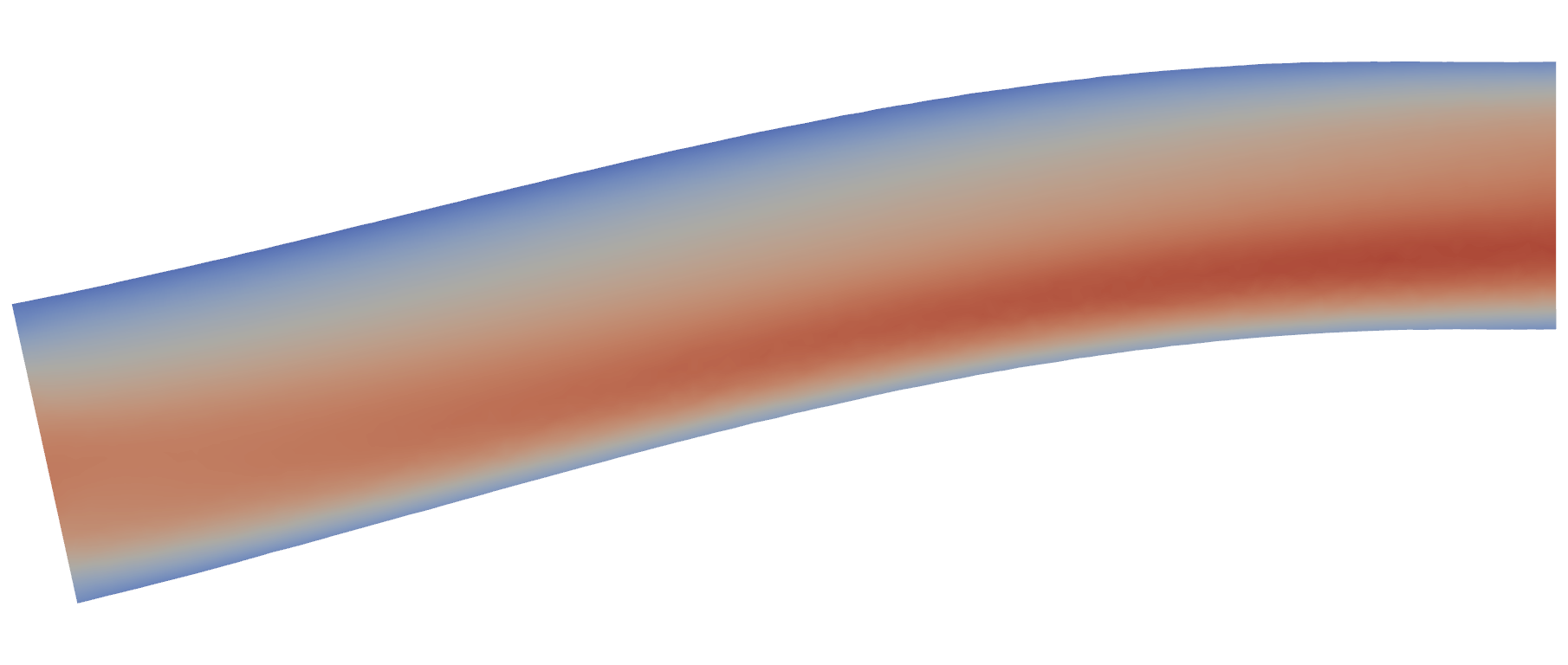} &
    \includegraphics[width=.205\textwidth]{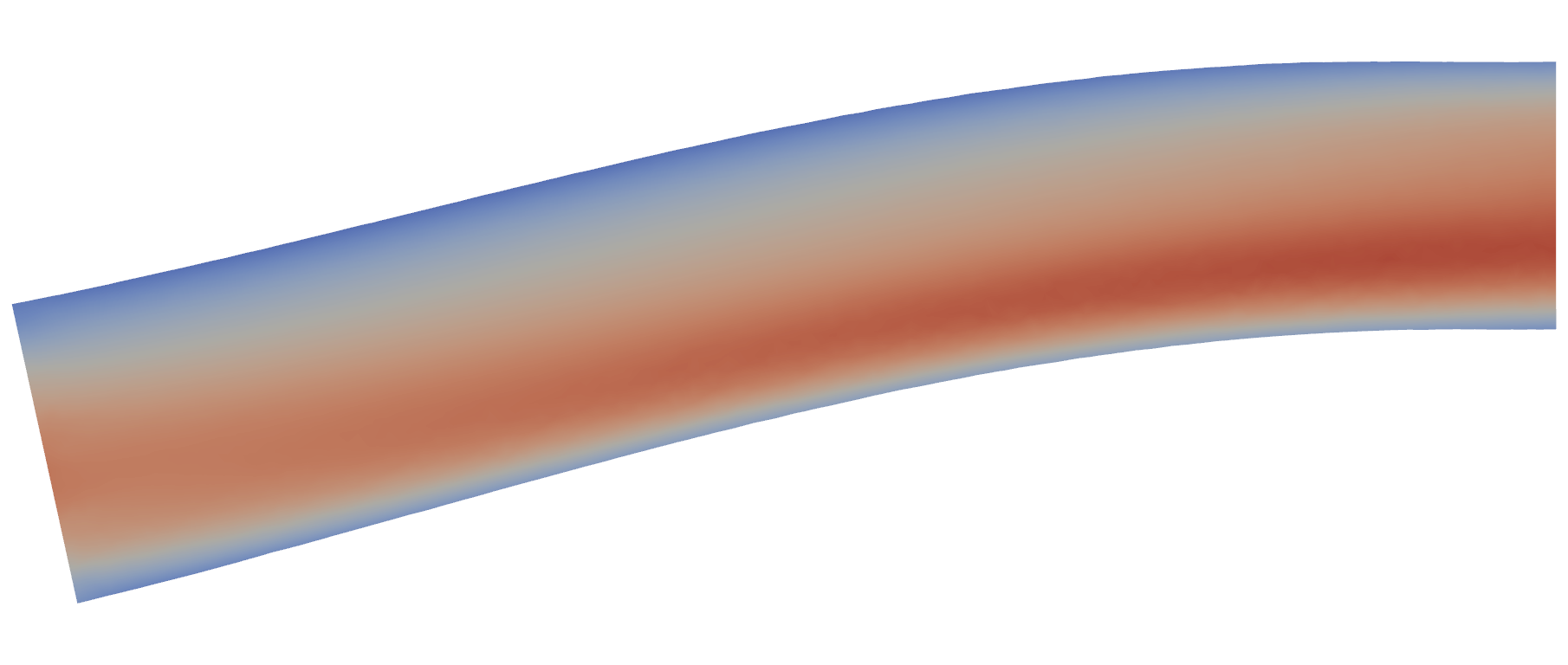} &
    \includegraphics[width=.205\textwidth]{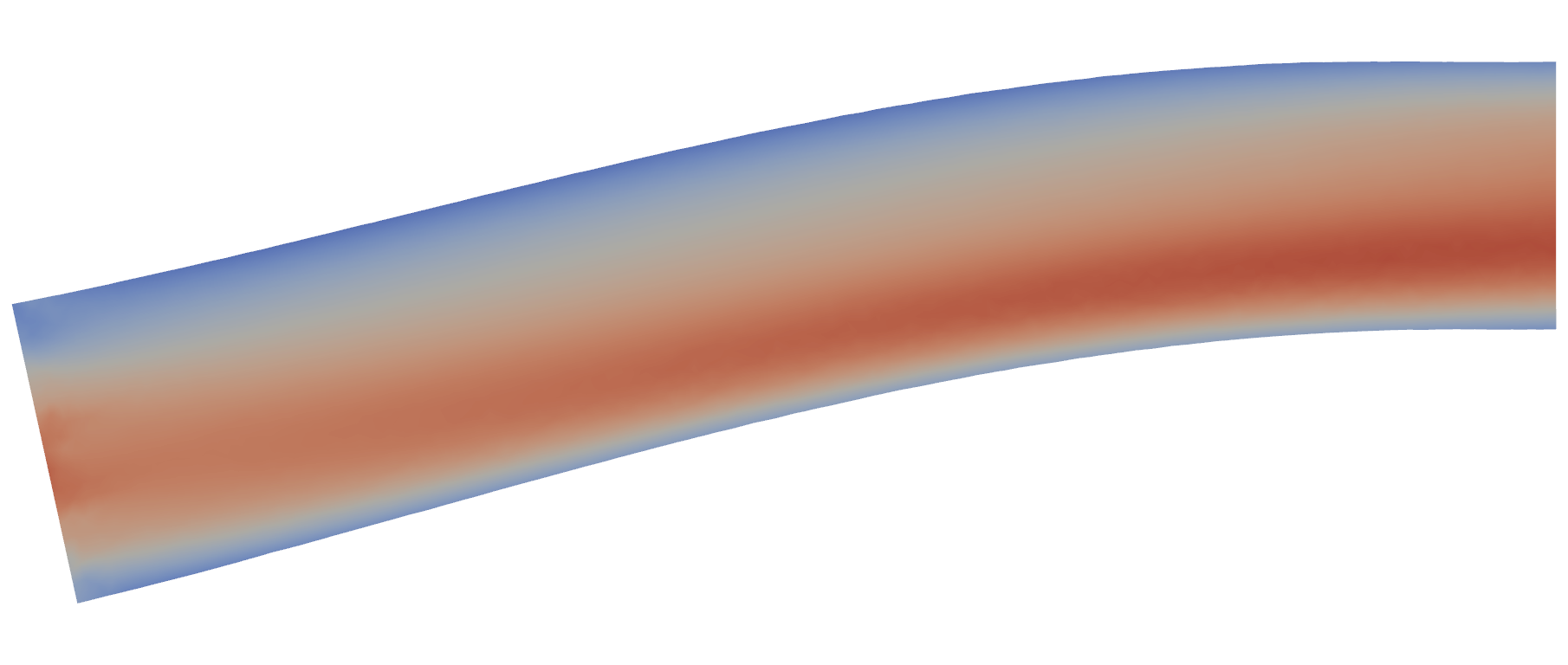} \\
     & $\thet{opt}=0.488$ & $\thet{opt}=0.488$ & $\thet{opt}=0.483$ & $\thet{opt}=0.469$ \\
     & $\mathcal{J}=\rnum{0.0001248}$ & $\mathcal{J}=\rnum{0.0004456}$ & $\mathcal{J}=\rnum{0.00528}$ & $\mathcal{J}=\rnum{0.0826}$ \\
     & $\mathcal{R}=\rnum{0.0001044}$ & $\mathcal{R}=\rnum{0.0001044}$ & $\mathcal{R}=\rnum{0.0001061}$ & $\mathcal{R}=\rnum{0.0002052}$ \\
     & iterations: 23 & iterations: 24 & iterations: 29 & iterations: 32 \\[5mm]
    \rotatebox{90}{\begin{tabular}{c}
         data \\
         $\theta=0.8$
    \end{tabular}} &  
    \includegraphics[width=.205\textwidth]{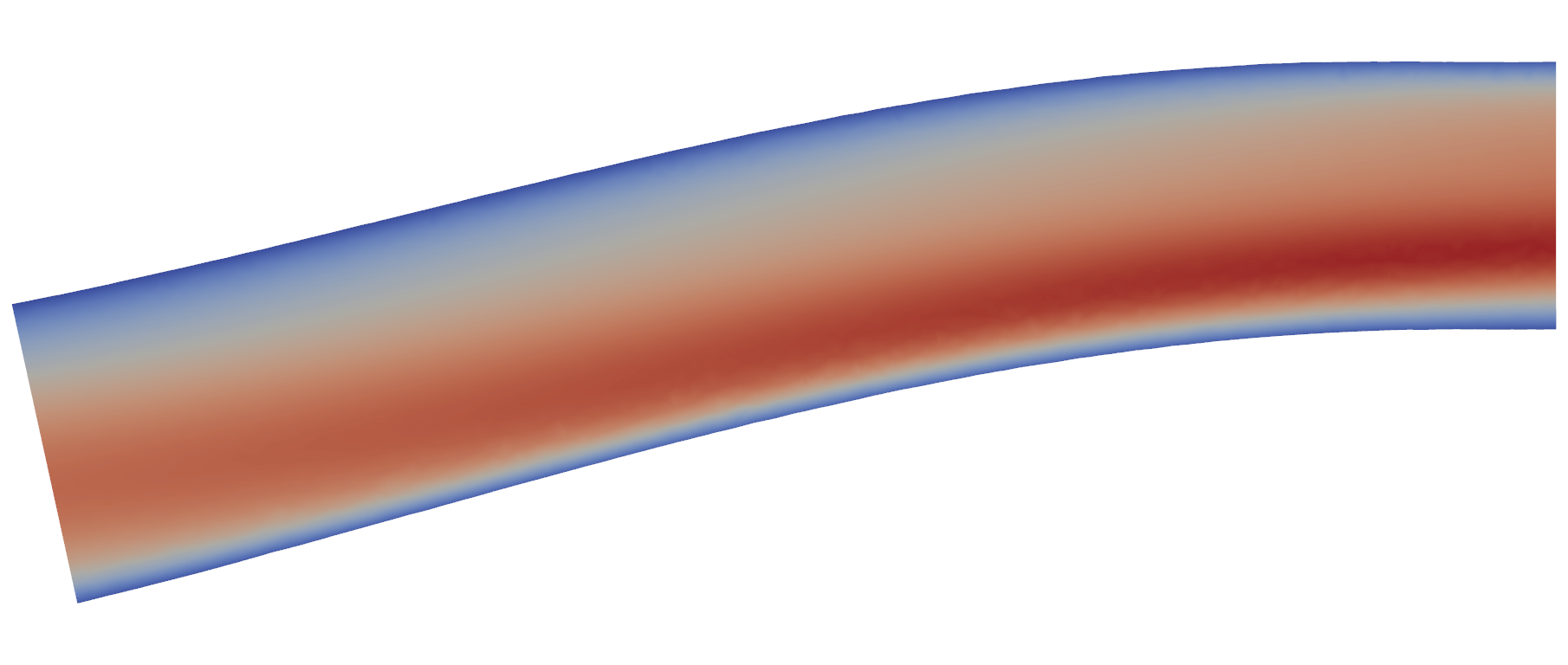} &
    \includegraphics[width=.205\textwidth]{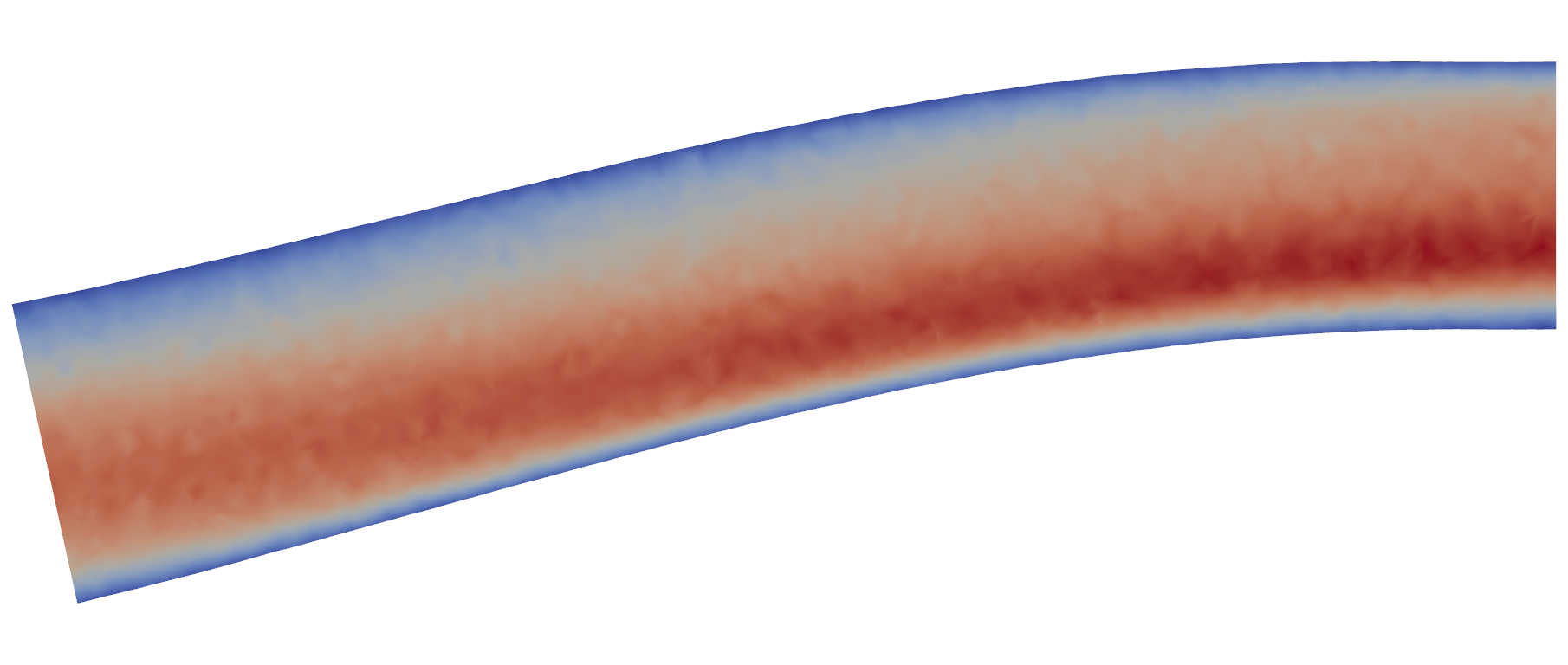} &
    \includegraphics[width=.205\textwidth]{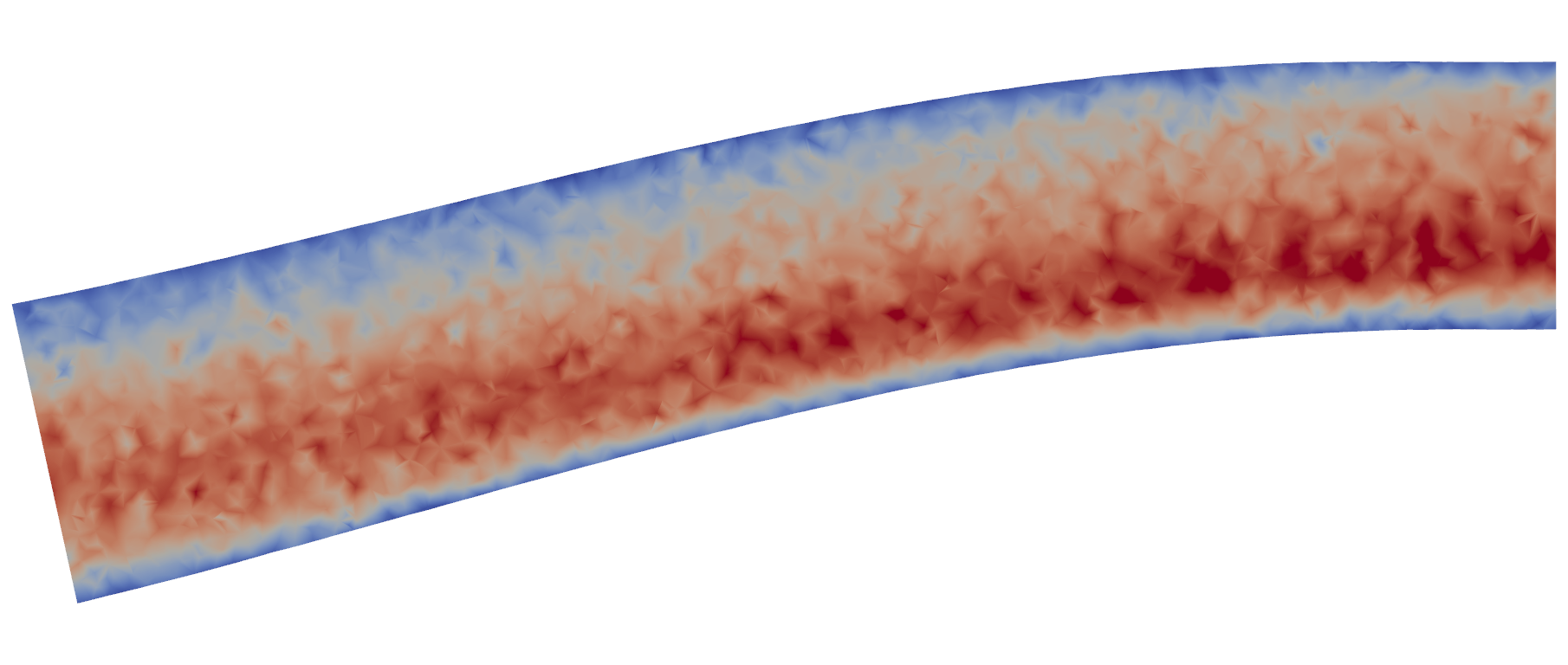} &
    \includegraphics[width=.205\textwidth]{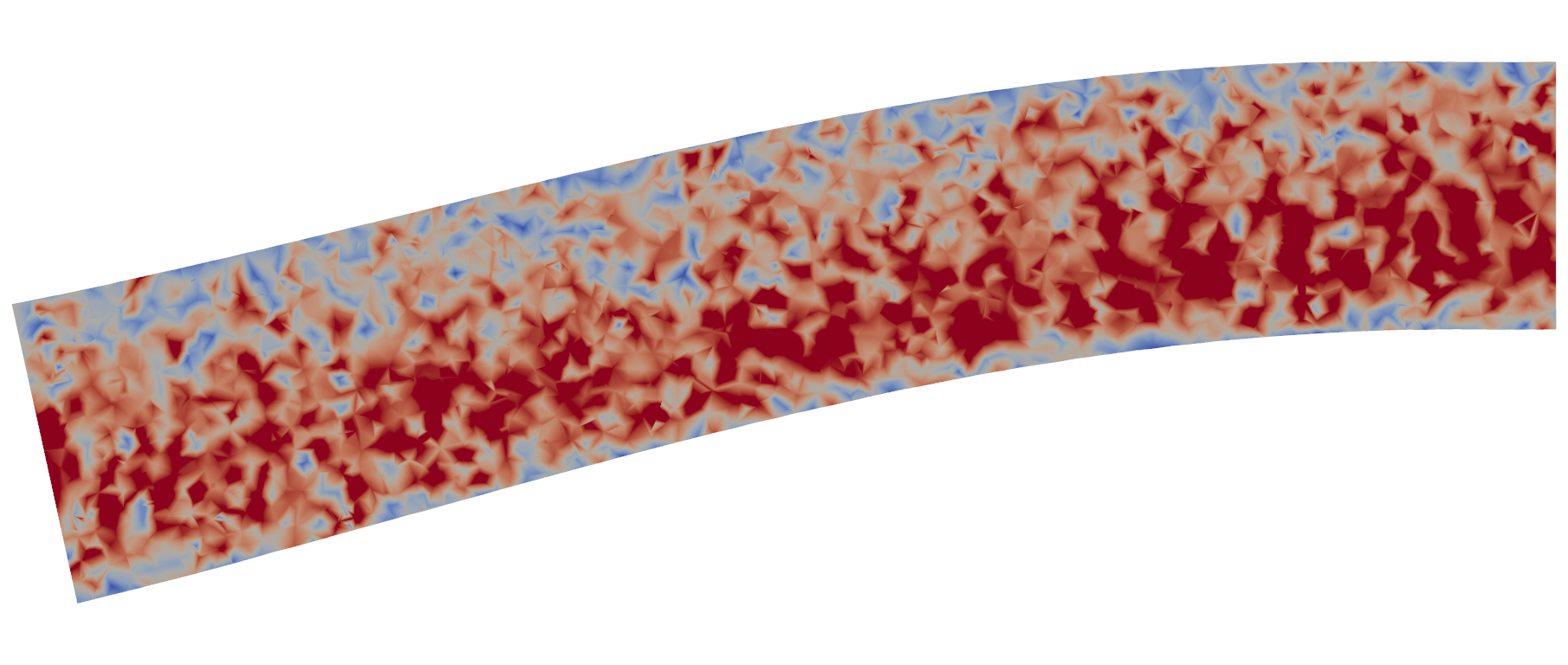} \\
    \rotatebox{90}{\begin{tabular}{c}
         assimilation \\
         $\theta=0.8$
    \end{tabular}} & 
    \includegraphics[width=.205\textwidth]{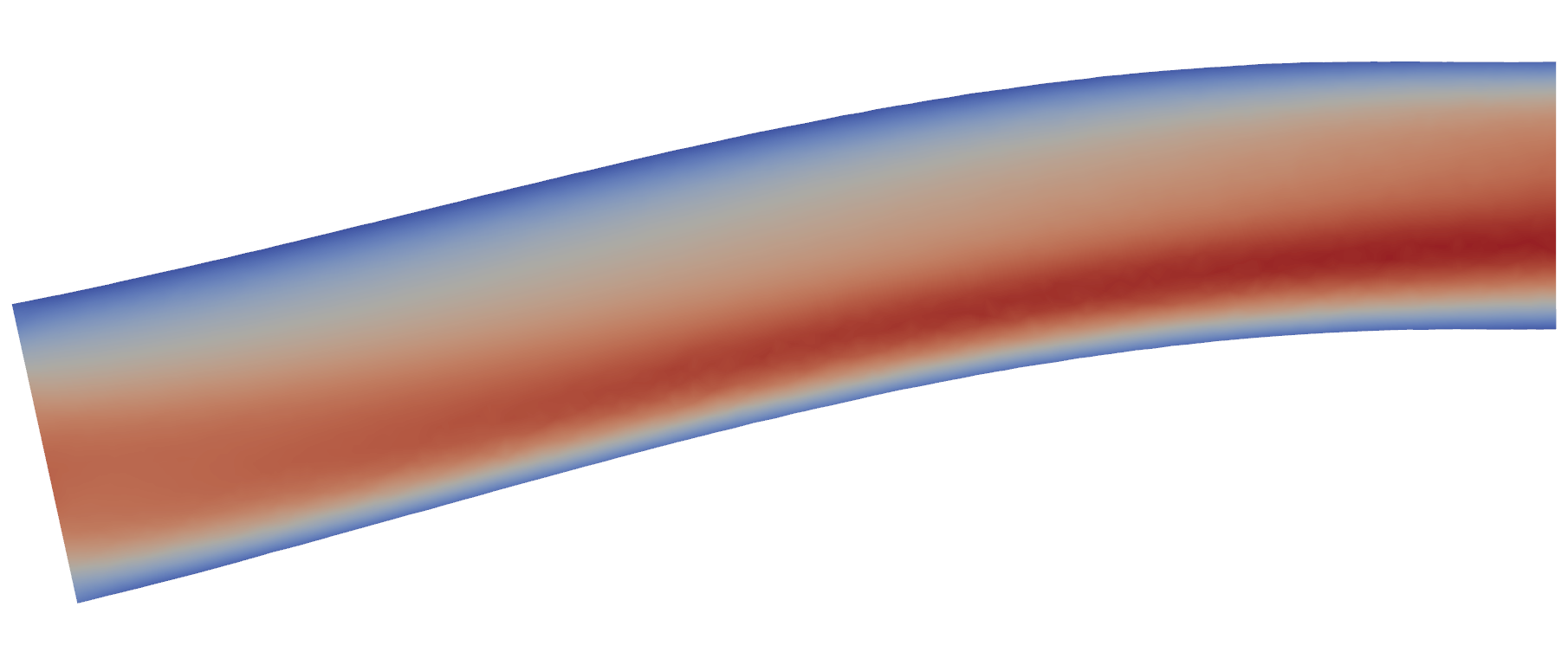}  &
    \includegraphics[width=.205\textwidth]{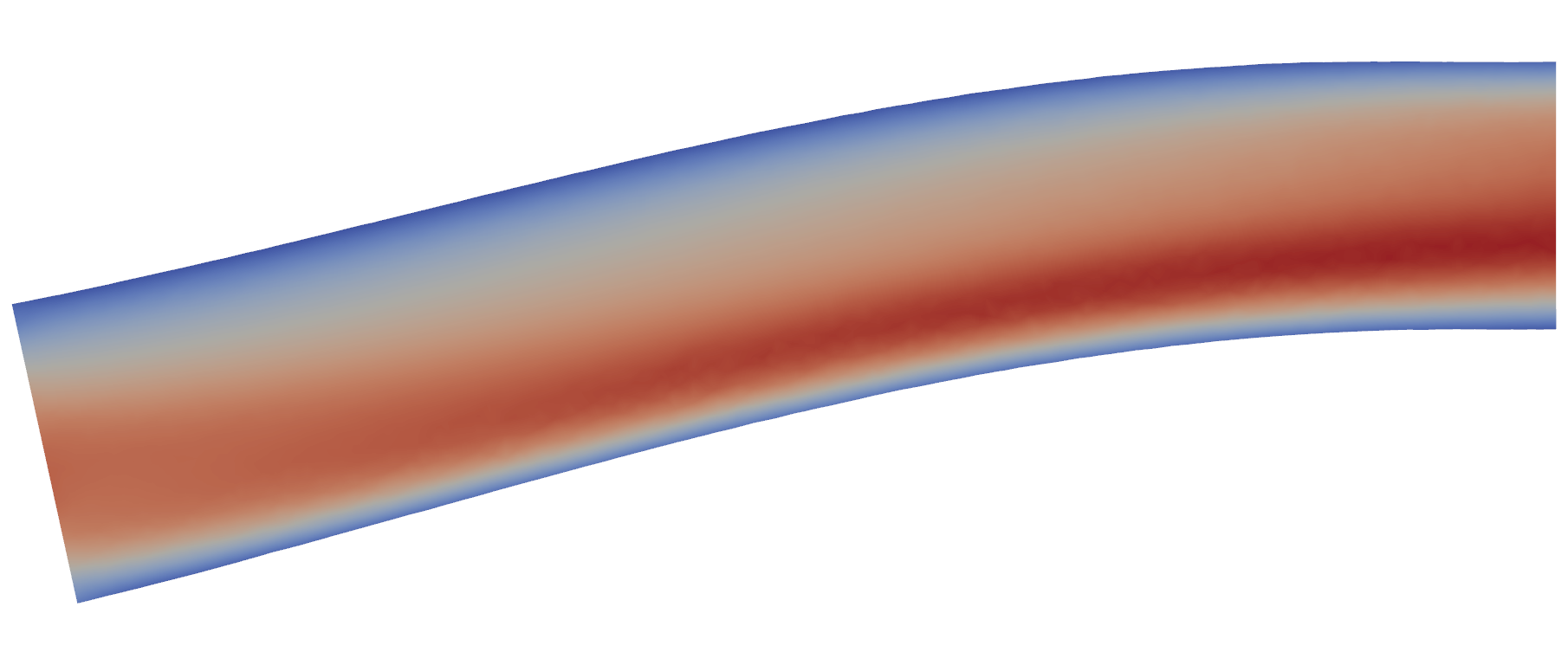} &
    \includegraphics[width=.205\textwidth]{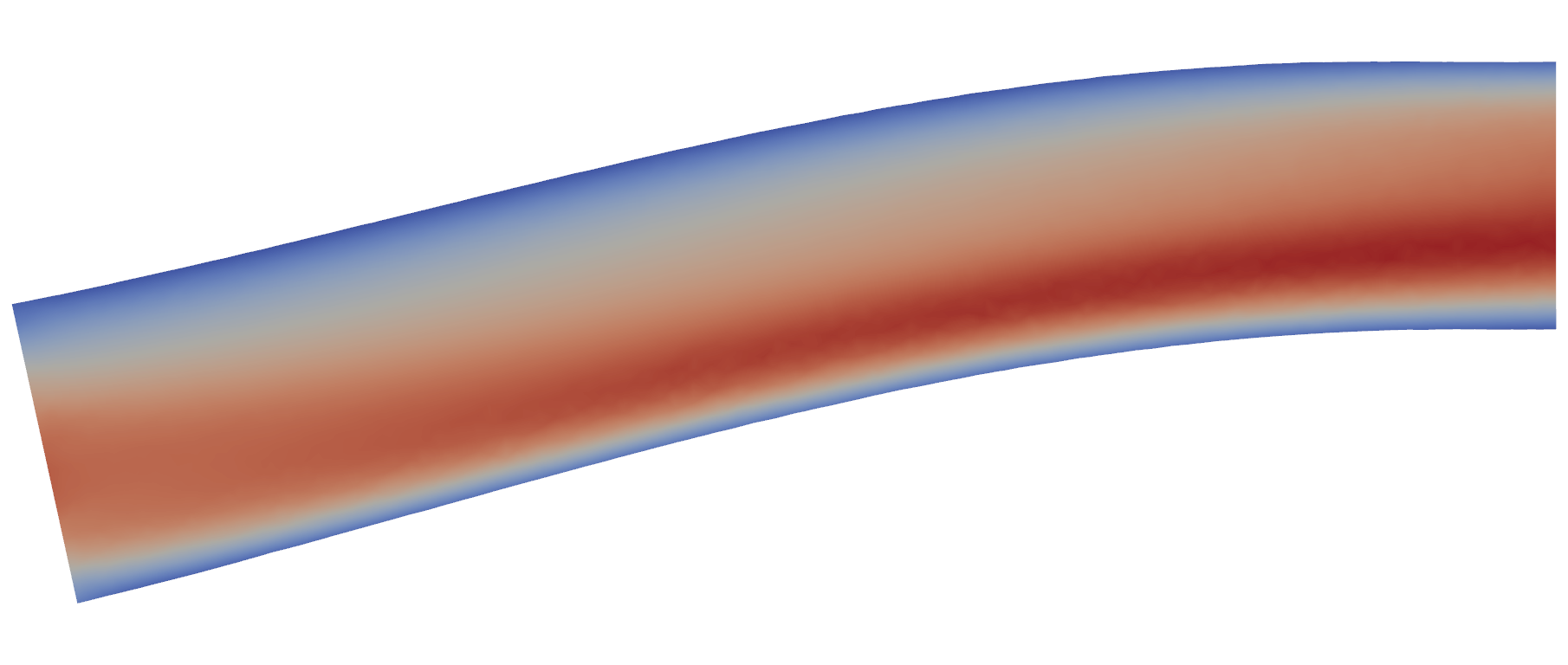} &
    \includegraphics[width=.205\textwidth]{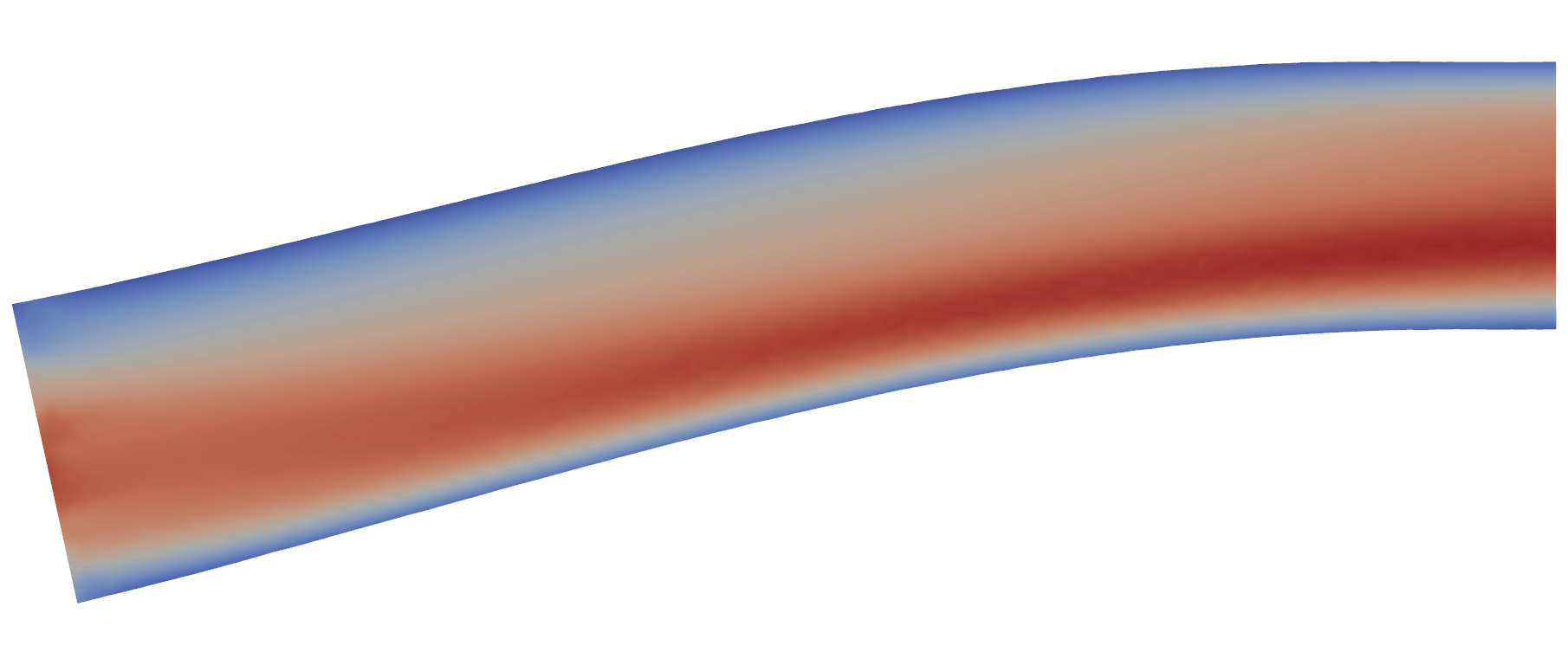} \\
     & $\thet{opt}=0.745$ & $\thet{opt}=0.743$ & $\thet{opt}=0.74$ & $\thet{opt}=0.733$ \\
     & $\mathcal{J}=\rnum{0.0002822}$ & $\mathcal{J}=\rnum{0.000604}$ & $\mathcal{J}=\rnum{0.005439}$ & $\mathcal{J}=\rnum{0.08277}$ \\
     & $\mathcal{R}=\rnum{0.000168}$ & $\mathcal{R}=\rnum{0.0001676}$ & $\mathcal{R}=\rnum{0.0001747}$ & $\mathcal{R}=\rnum{0.0002847}$ \\
     & iterations: 21 & iterations: 24 & iterations: 22 & iterations: 22 \\
     & \multicolumn{4}{c}{\includegraphics[width=0.5\textwidth]{scale0.25.png}}
\end{tabular}
\caption{Comparison of data with various amounts of noise with assimilation velocity results on the bent tube geometry with edge length $h=1.5\text{ mm}$ using stabilized $P_1/P_1$ element with $\alpha_v=\alpha_p=0.01$ for $\theta=0.5$ and $\theta=0.8$.}
\label{fig:noise_bent}
\end{figure}

\begin{figure}
\centering
\begin{tabular}{c c c c c}
     & $\text{SNR}=\infty$ & $\text{SNR}=2$ & $\text{SNR}=1$ & $\text{SNR}=0.5$\\
    \rotatebox{90}{\begin{tabular}{c}
         data \\
         $\theta=0.5$
    \end{tabular}} &  
    \includegraphics[width=.205\textwidth]{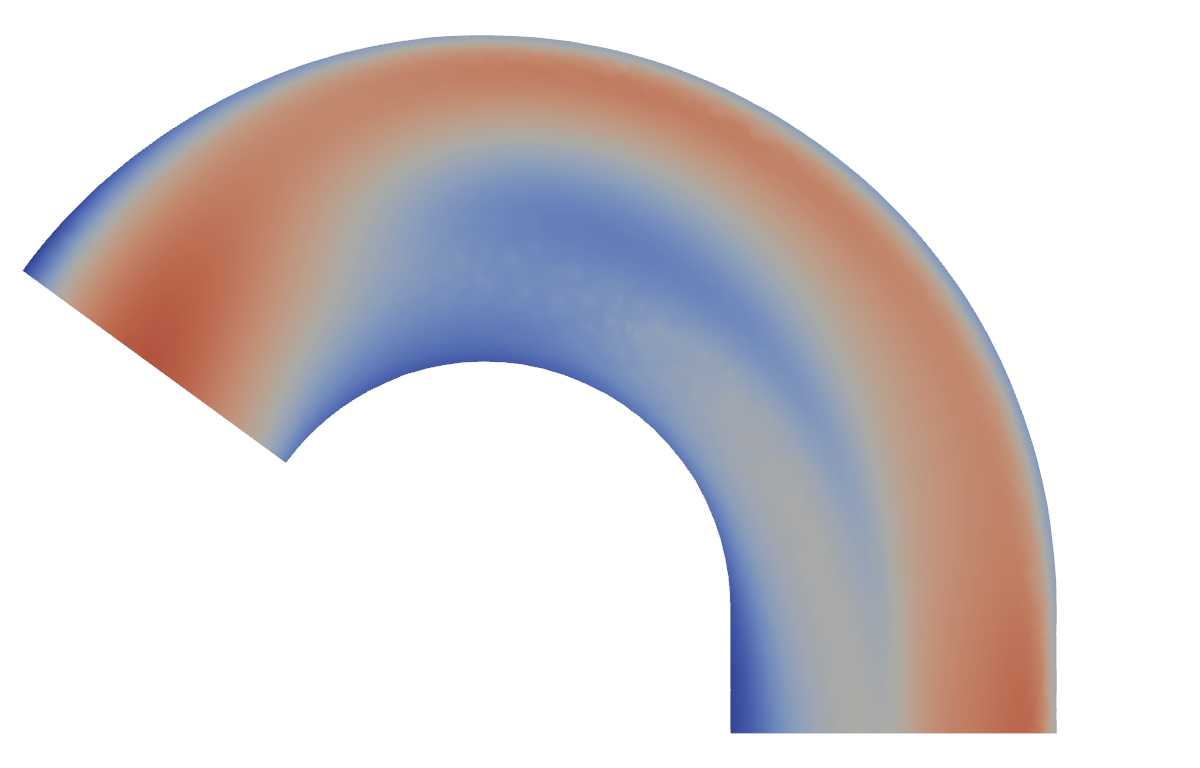} &
    \includegraphics[width=.205\textwidth]{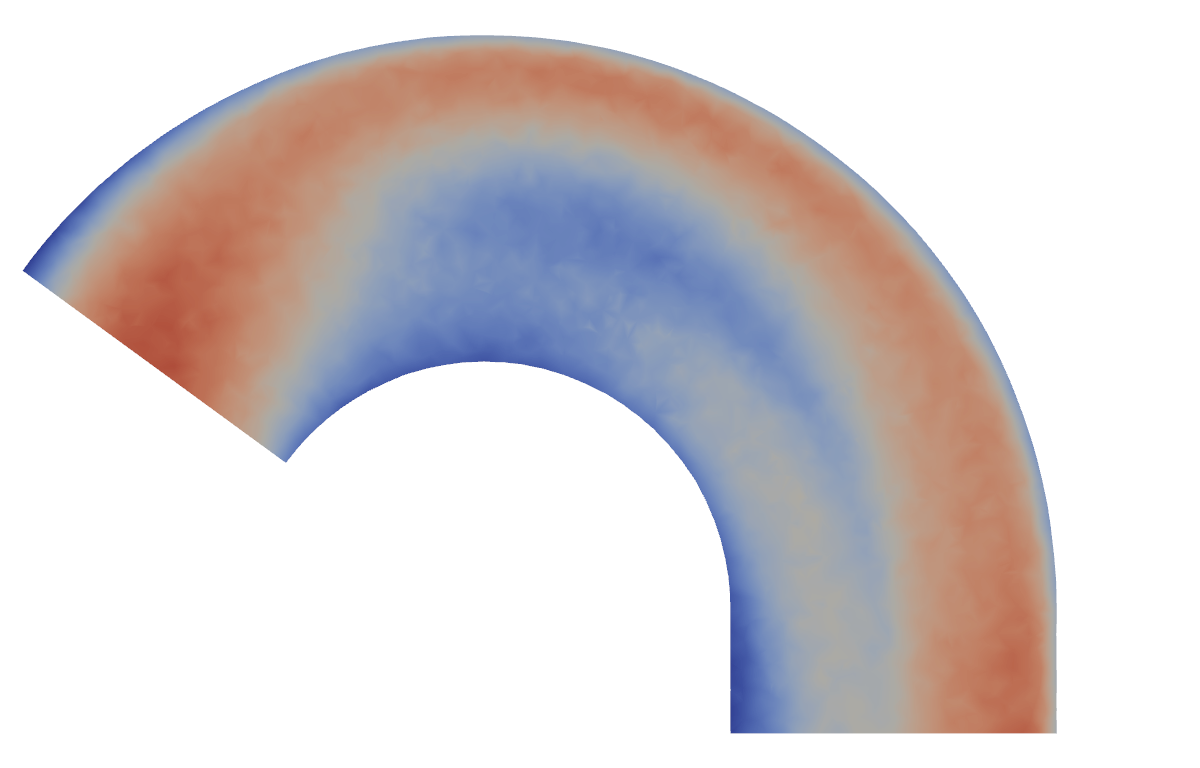} &
    \includegraphics[width=.205\textwidth]{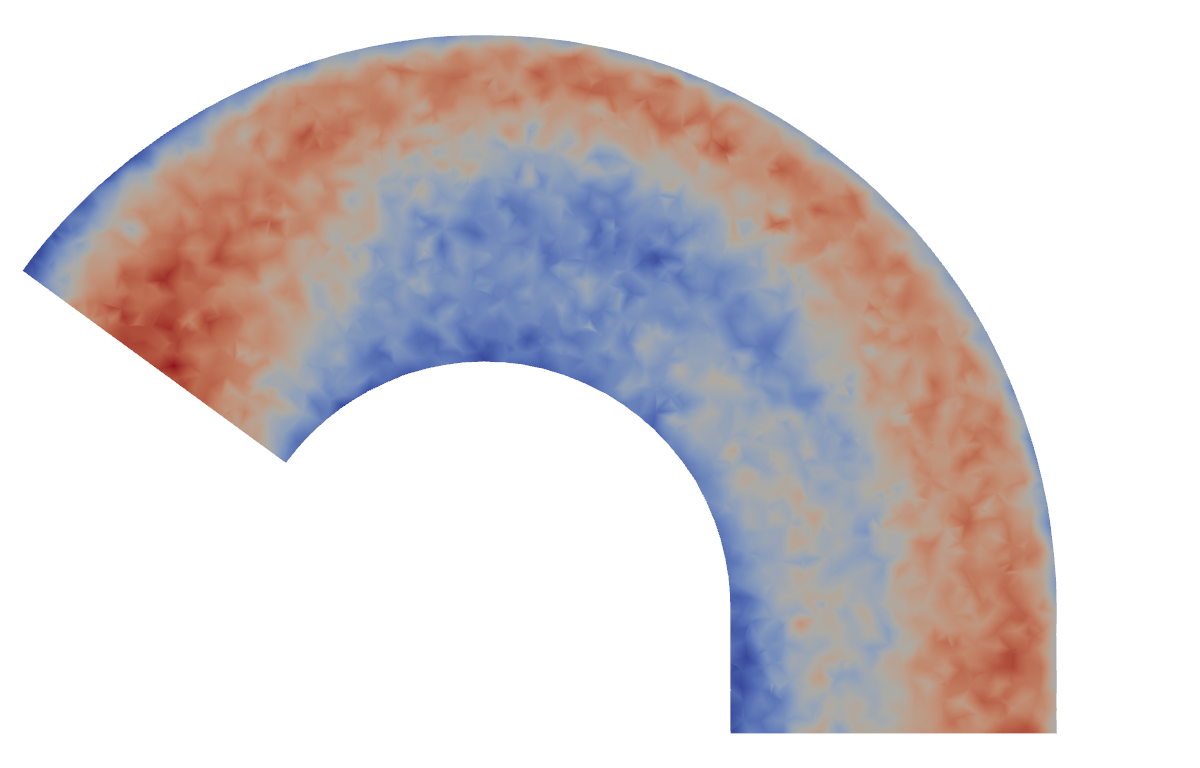} &
    \includegraphics[width=.205\textwidth]{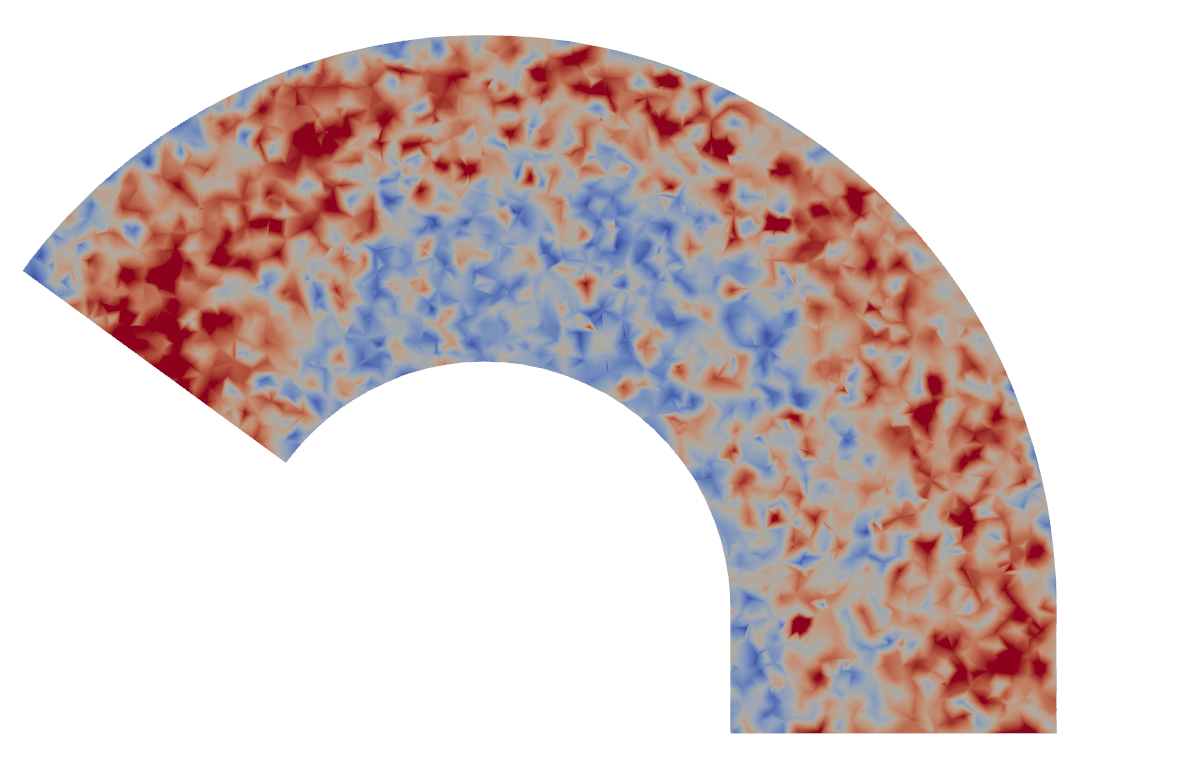} \\
    \rotatebox{90}{\begin{tabular}{c}
         assimilation \\
         $\theta=0.5$
    \end{tabular}} & 
    \includegraphics[width=.205\textwidth]{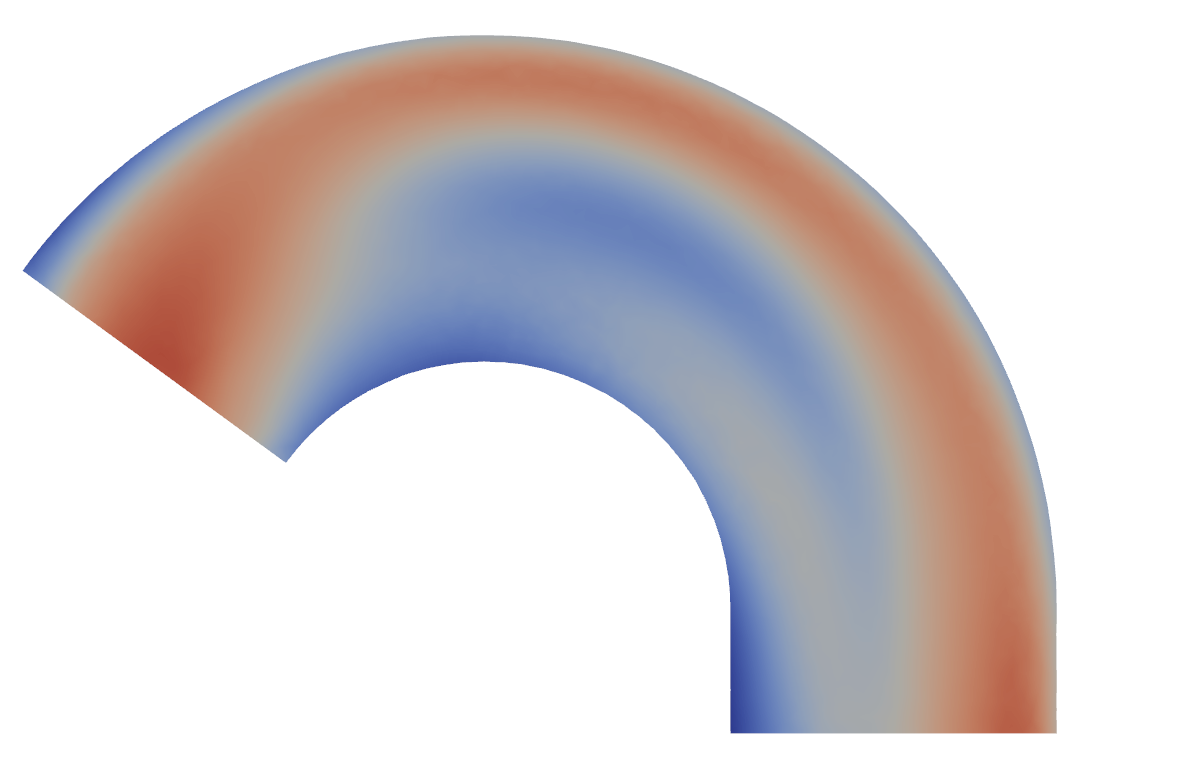} &
    \includegraphics[width=.205\textwidth]{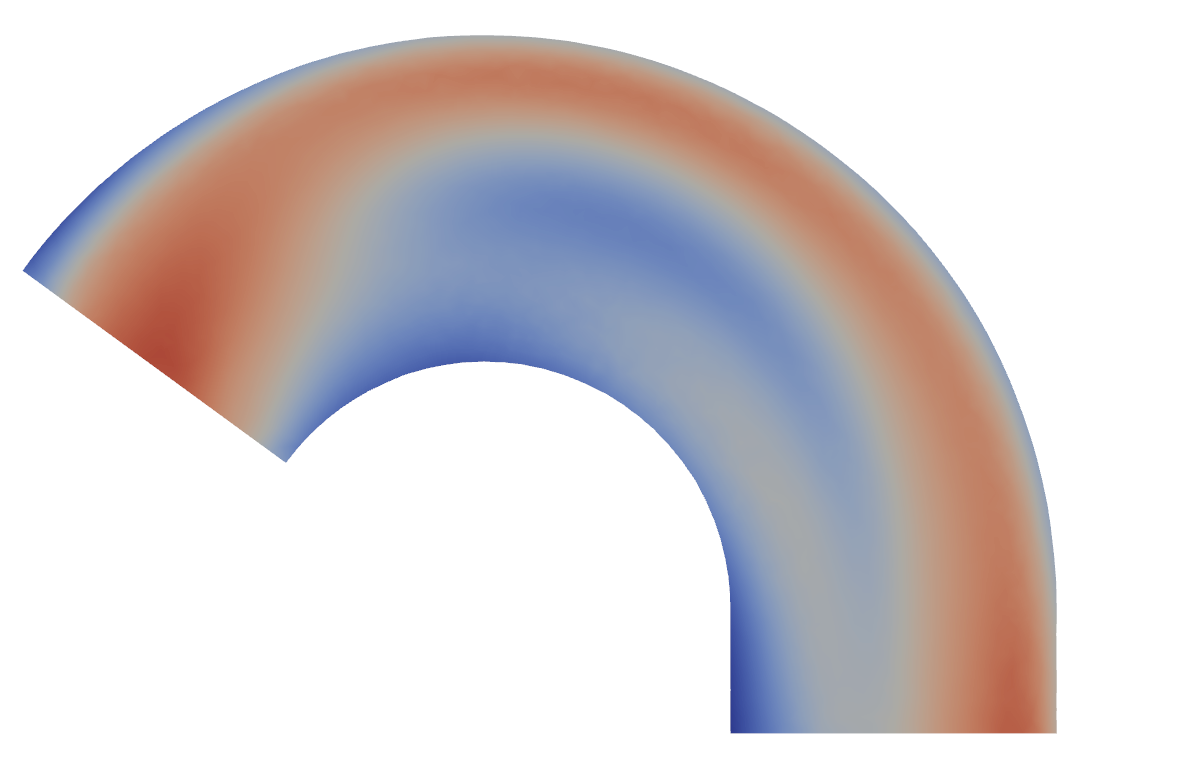} &
    \includegraphics[width=.205\textwidth]{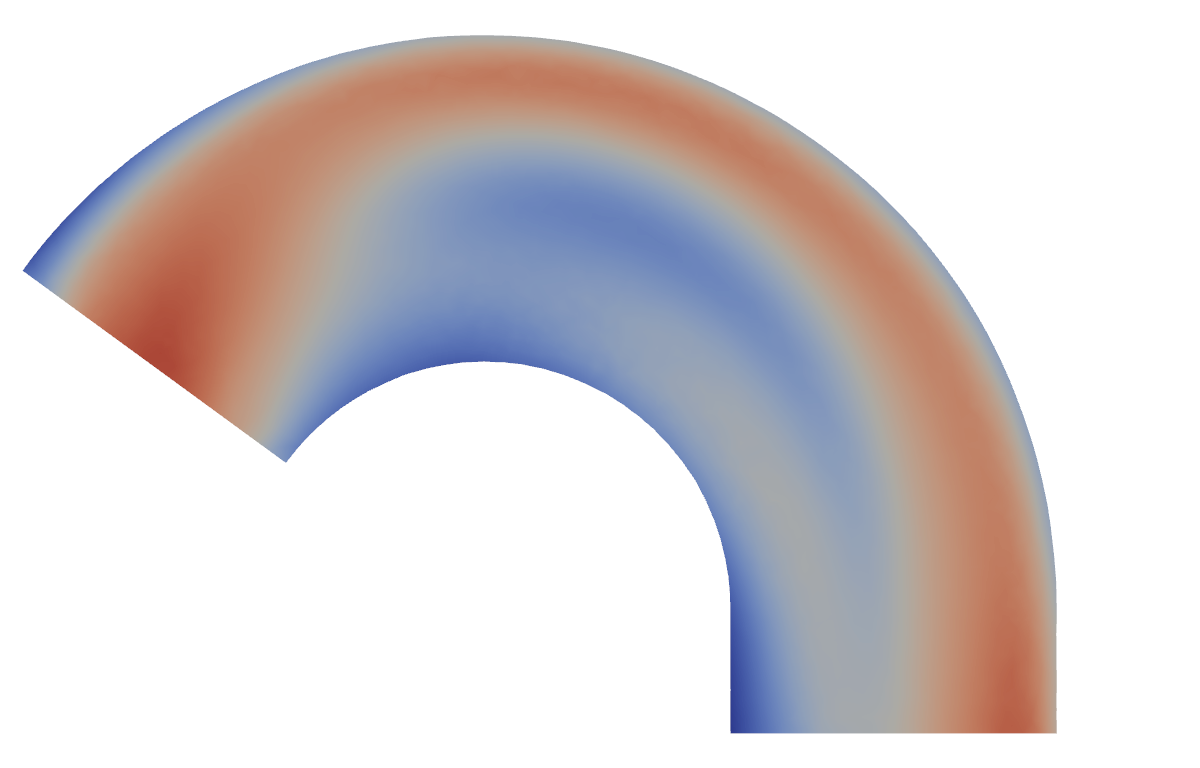} &
    \includegraphics[width=.205\textwidth]{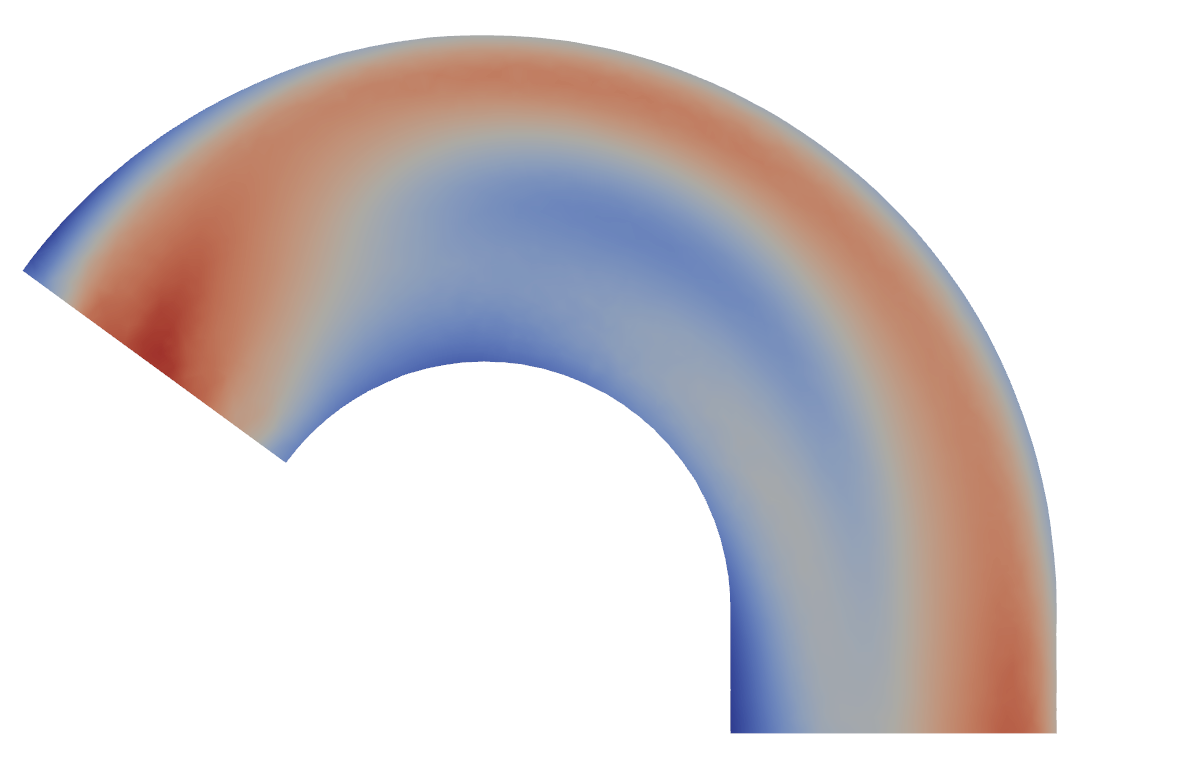} \\
     & $\thet{opt}=0.488$ & $\thet{opt}=0.488$ & $\thet{opt}=0.483$ & $\thet{opt}=0.469$ \\
     & $\mathcal{J}=\rnum{0.0001248}$ & $\mathcal{J}=\rnum{0.0004456}$ & $\mathcal{J}=\rnum{0.00528}$ & $\mathcal{J}=\rnum{0.0826}$ \\
     & $\mathcal{R}=\rnum{0.0001044}$ & $\mathcal{R}=\rnum{0.0001044}$ & $\mathcal{R}=\rnum{0.0001061}$ & $\mathcal{R}=\rnum{0.0002052}$ \\
     & iterations: 23 & iterations: 24 & iterations: 29 & iterations: 32 \\[5mm]
    \rotatebox{90}{\begin{tabular}{c}
         data \\
         $\theta=0.8$
    \end{tabular}} &  
    \includegraphics[width=.205\textwidth]{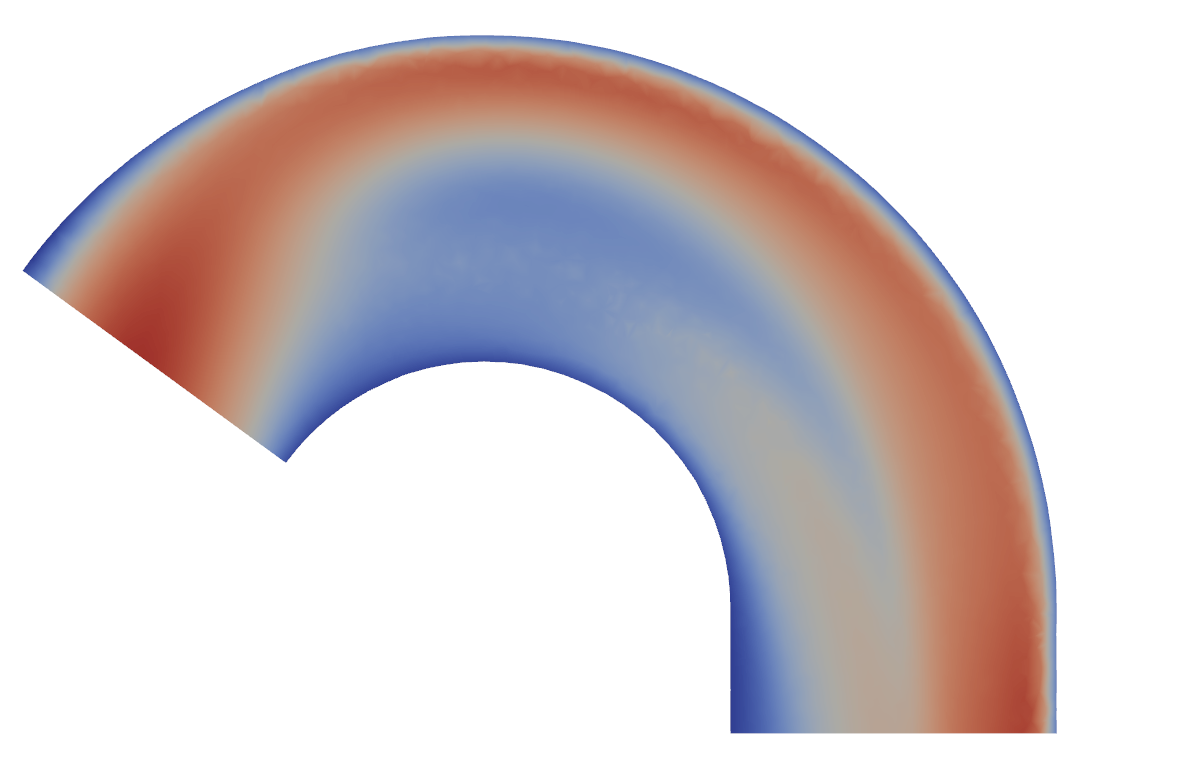} &
    \includegraphics[width=.205\textwidth]{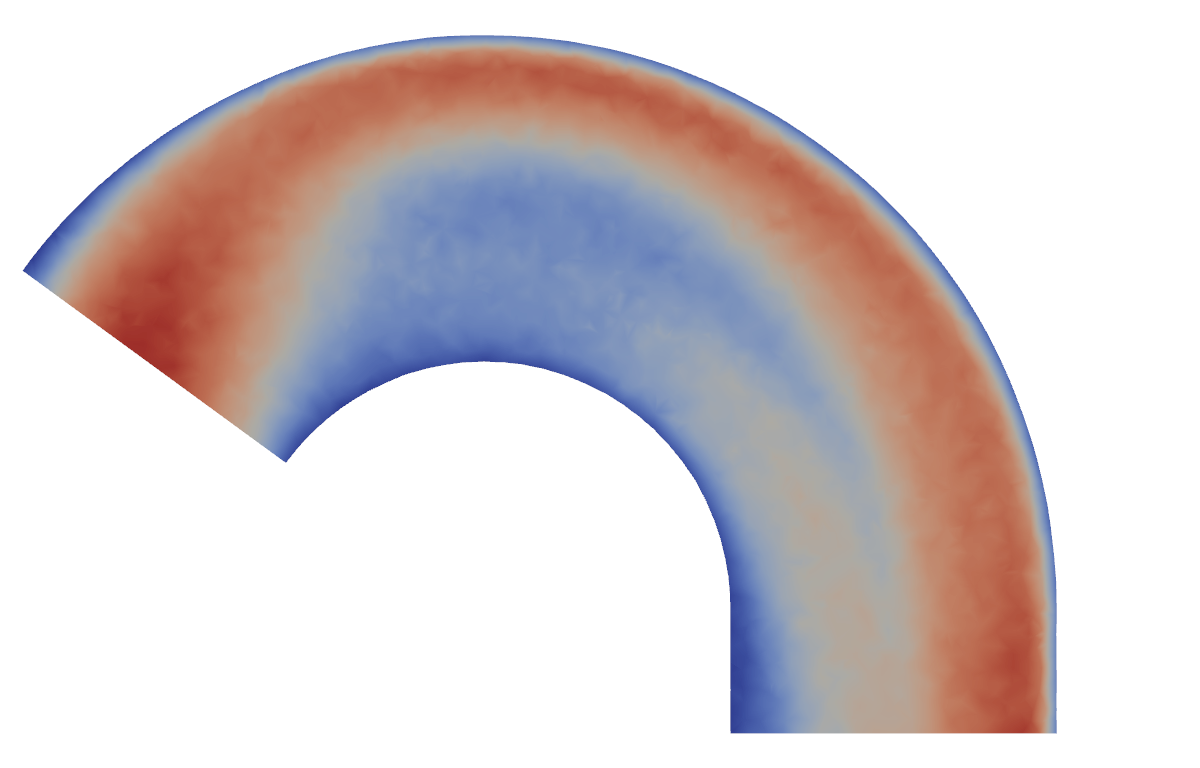} &
    \includegraphics[width=.205\textwidth]{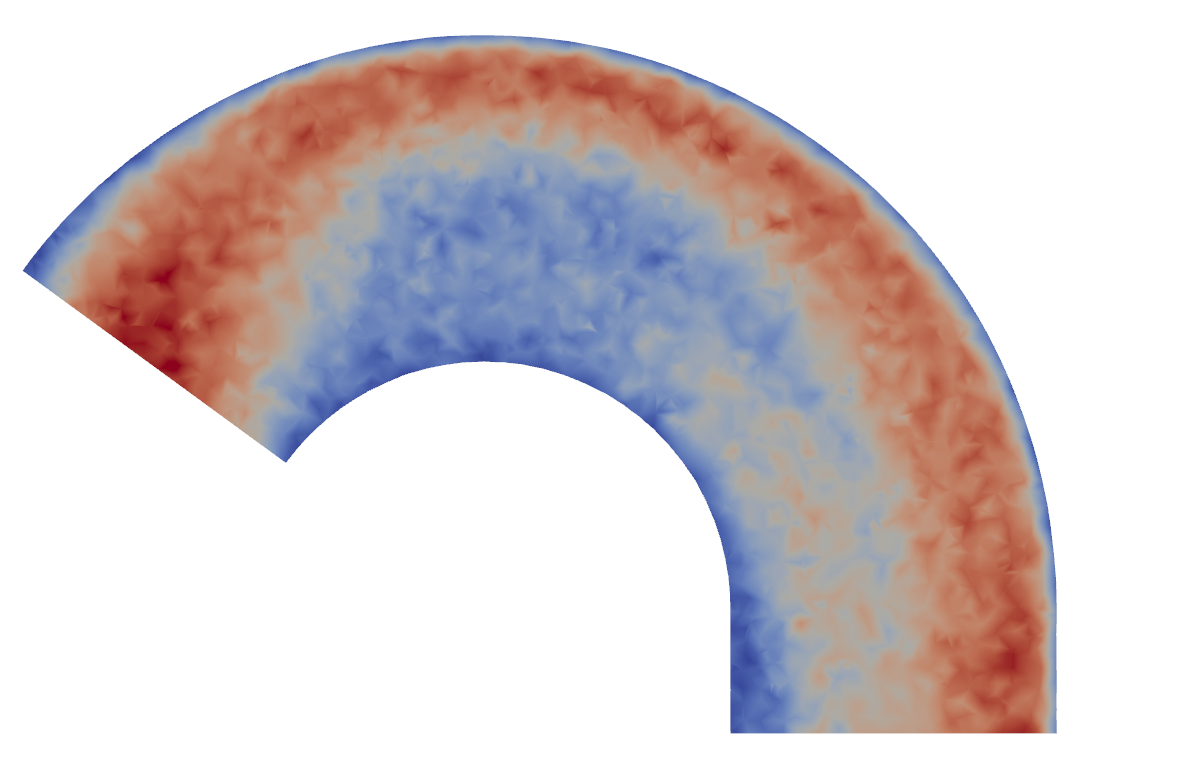} &
    \includegraphics[width=.205\textwidth]{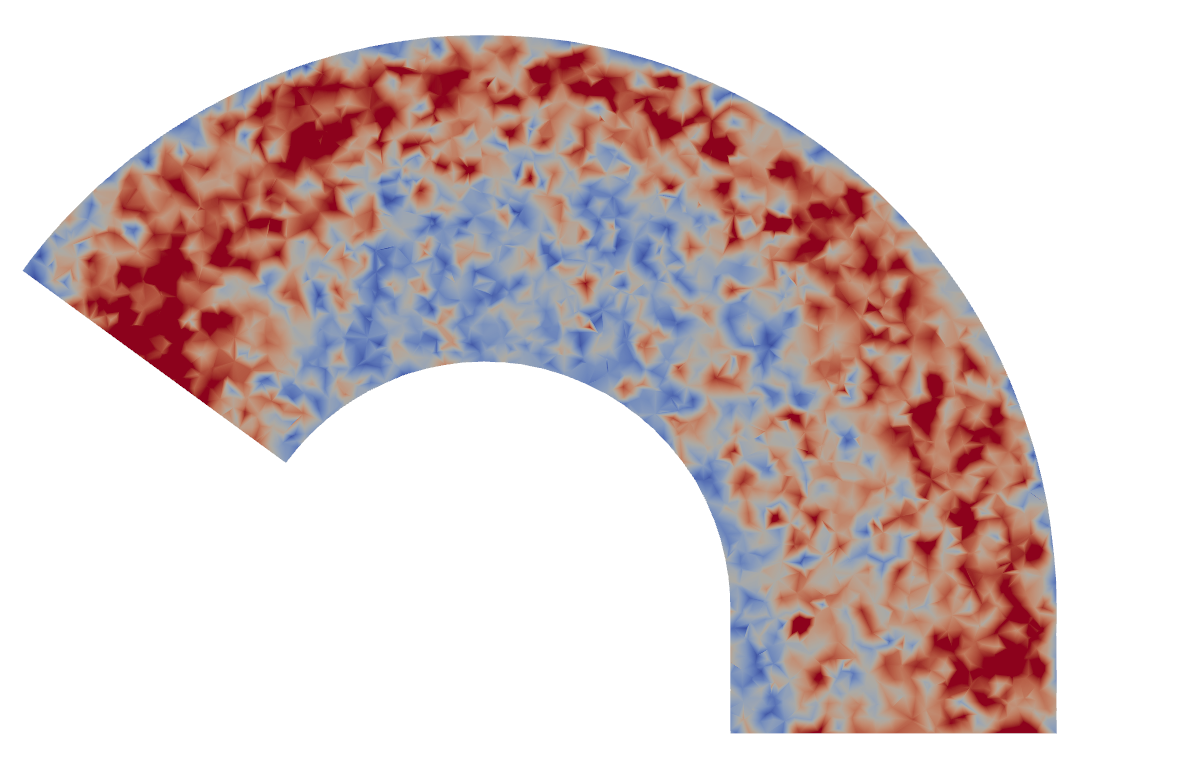} \\
    \rotatebox{90}{\begin{tabular}{c}
         assimilation \\
         $\theta=0.8$
    \end{tabular}} & 
    \includegraphics[width=.205\textwidth]{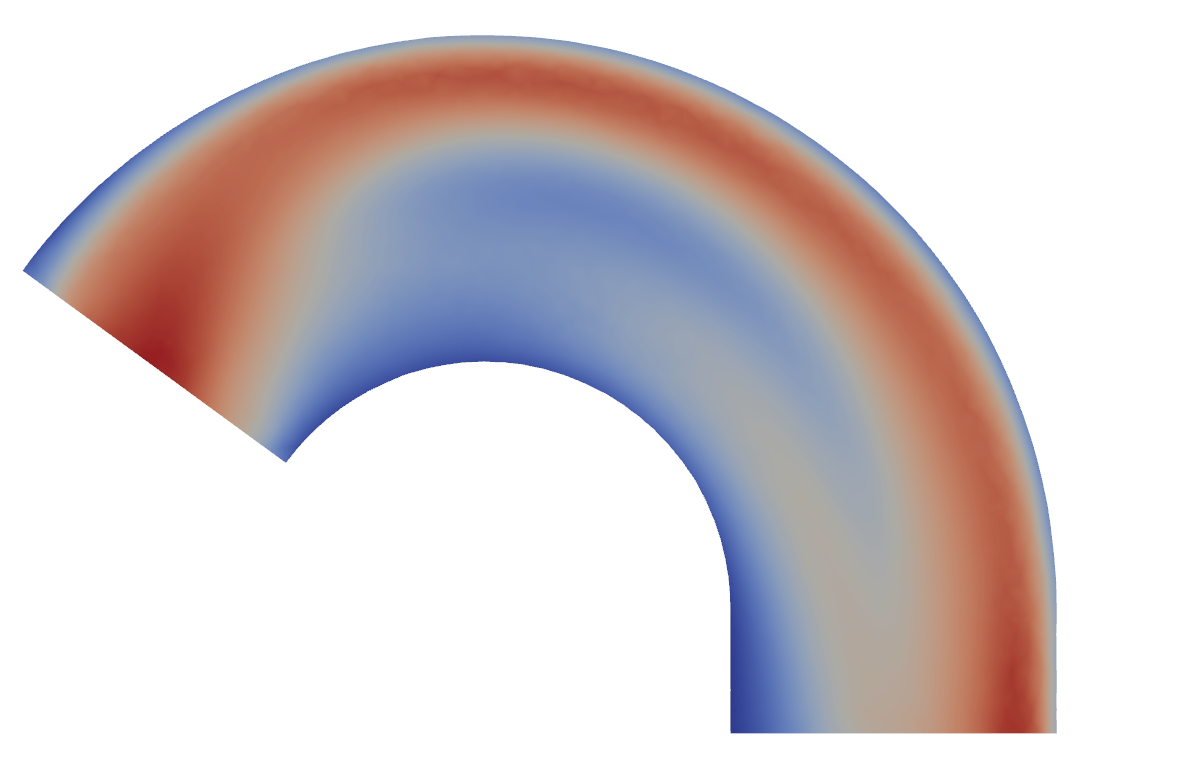}  &
    \includegraphics[width=.205\textwidth]{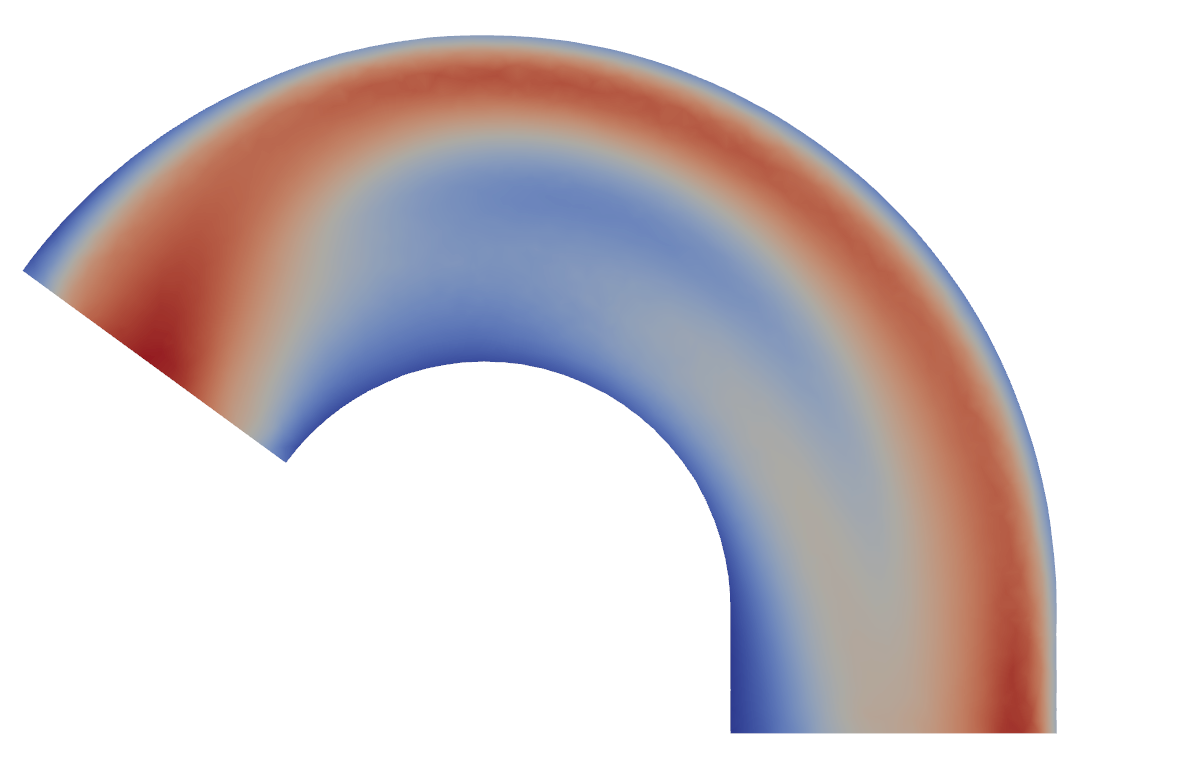} &
    \includegraphics[width=.205\textwidth]{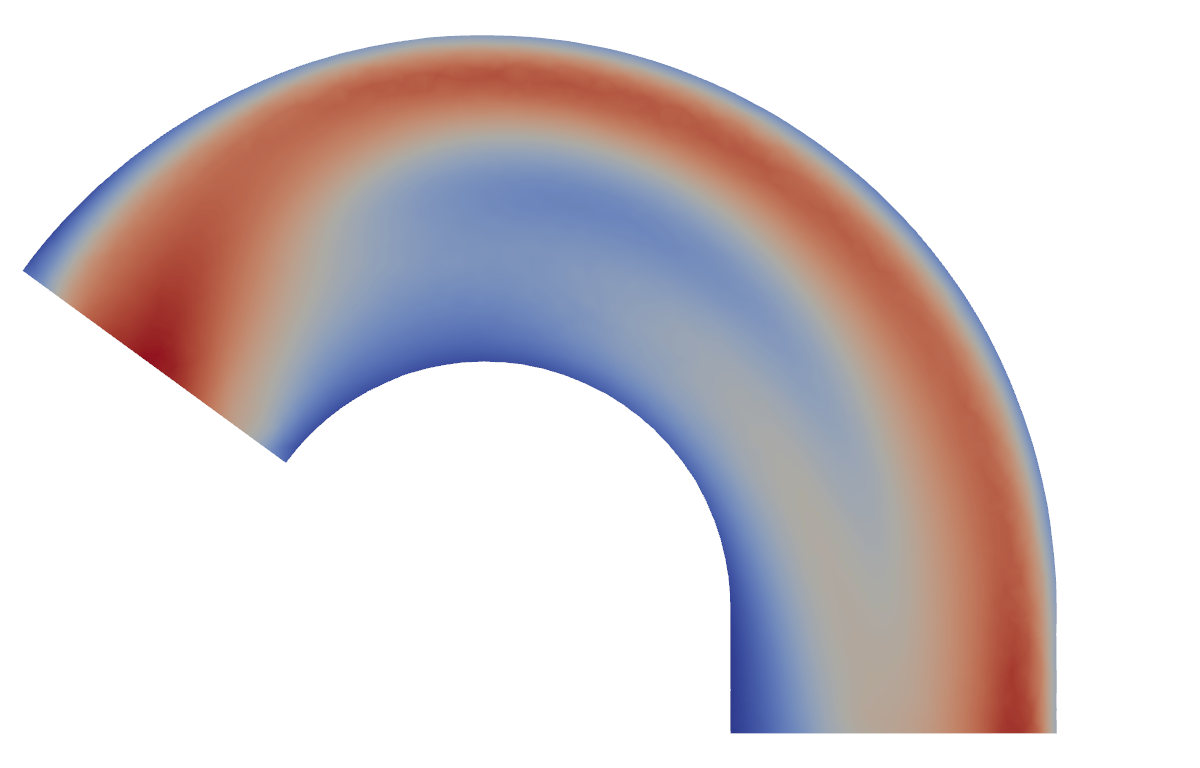} &
    \includegraphics[width=.205\textwidth]{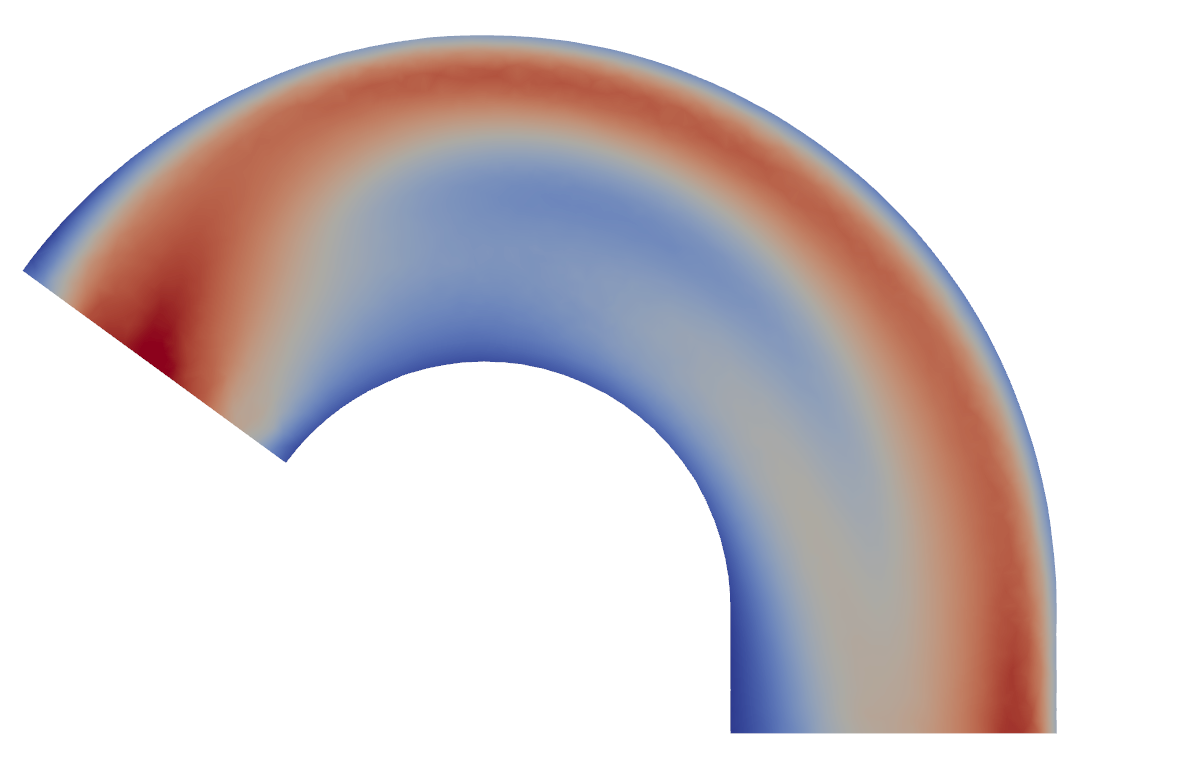} \\
     & $\thet{opt}=0.691$ & $\thet{opt}=0.691$ & $\thet{opt}=0.691$ & $\thet{opt}=0.692$ \\
     & $\mathcal{J}=\rnum{0.001267}$ & $\mathcal{J}=\rnum{0.001419}$ & $\mathcal{J}=\rnum{0.003605}$ & $\mathcal{J}=\rnum{0.03835}$ \\
     & $\mathcal{R}=\rnum{0.0004333}$ & $\mathcal{R}=\rnum{0.0004344}$ & $\mathcal{R}=\rnum{0.0004435}$ & $\mathcal{R}=\rnum{0.0005083}$ \\
     & iterations: 28 & iterations: 28 & iterations: 26 & iterations: 27 \\
     & \multicolumn{4}{c}{\includegraphics[width=0.5\textwidth]{scale0.2.png}}
\end{tabular}
\caption{Comparison of data with various amounts of noise with assimilation velocity results on the arch geometry with edge length $h=1.5\text{ mm}$ using stabilized $P_1/P_1$ element with $\alpha_v=\alpha_p=0.01$ for $\theta=0.5$ and $\theta=0.8$.}
\label{fig:noise_arch}
\end{figure}

The robustness of the method with respect to the signal-to-noise ratio ($\text{SNR}$) was tested for various values of the slip parameter $\theta$. In Figures \ref{fig:noise_bent} and \ref{fig:noise_arch}, we present the comparison of the results for $\theta=0.5$ and $\theta=0.8$.
The value of $\text{SNR}$ ranges from $\infty$ to $0.5$, corresponding to no noise and noise with a maximum twice the size of maximal velocity, respectively.
We observed that the amount of Gaussian noise does not have a significant effect on the reconstructed velocity field and value of the slip parameter $\thet{opt}$. This was true also for all the elements and values of $\theta$ we have tested. 

\subsection{Effect of regularization}\label{subsec: reg}
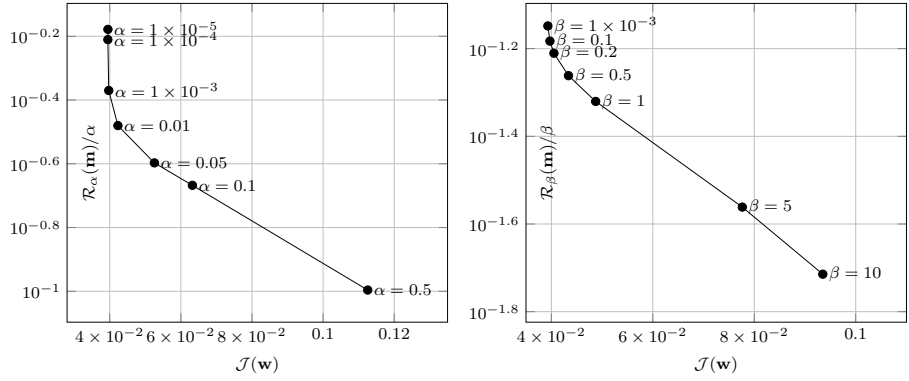
\begin{figure}
    \centering

\pgfplotstableread[col sep=comma]{
error, regularization, parameter, theta
0.11254481521603739,0.10090057673921393,0.5,0.343
0.06324253506909096,0.21499685313717964,0.1,0.425
0.05256027976233574,0.2528147564396192,0.05,0.443
0.04229653205638495,0.33097181950703897,0.01,0.455
0.039719629689376514,0.4264783160187117,0.001,0.463
0.03947179036988822,0.6159182946819455,0.0001,0.463
0.03948744576948983,0.663263952141454,1e-05,0.461
}\lcurvetablealpha

\pgfplotstableread[col sep=comma]{
error, regularization, parameter, theta
0.09351015103261677,0.019303309214796307,10.0,0.405
0.07763971268609125,0.027452627727569543,5.0,0.41
0.04873535817186532,0.0478185153930546,1.0,0.441
0.04337548573669464,0.05474593850966846,0.5,0.453
0.04050910536317977,0.061592643179035594,0.2,0.457
0.039719629689376514,0.06563723399563695,0.1,0.463
0.039310806629592326,0.07108199243149306,0.001,0.467
}\lcurvetablebeta
    
\begin{tikzpicture}[scale=0.78]
\begin{axis}[    
    xlabel={$\small \mathcal{J}(\mathbf{w})$},
    ylabel={$\tiny \mathcal{R_{\alpha}}(\mathbf{m})/\alpha$},
    y label style={at={(0.1,0.5)}},
    width=8cm, height=7cm,
    xmin=0.028, xmax=0.135,
    ymin=0.08, ymax=0.8,
    scaled x ticks=false,
    ymode=log,
    grid=both,
]
\addplot [ 
    point meta = explicit symbolic, 
    nodes near coords, 
    every node near coord/.append style={anchor=west},
    visualization depends on={meta \as \mytag}, 
    nodes near coords={$\tiny \alpha=\num{\mytag}$},
    mark=*] 
    table [x=error, y=regularization, meta=parameter] {\lcurvetablealpha} ;
\end{axis}
\end{tikzpicture}
\hfill
\begin{tikzpicture}[scale=0.78]
\begin{axis}[    
    xlabel={$\small \mathcal{J}(\mathbf{w})$},
    ylabel={$\tiny \mathcal{R_{\beta}}(\mathbf{m})/\beta$},
    y label style={at={(0.1,0.5)}},
    width=8cm, height=7cm,
    xmin=0.035, xmax=0.11,
    ymin=0.015, ymax=0.08,
    scaled x ticks=false,
    ymode=log,
    grid=both,
]
\addplot [ 
    point meta = explicit symbolic, 
    nodes near coords, 
    every node near coord/.append style={anchor=west},
    visualization depends on={meta \as \mytag}, 
    nodes near coords={$\tiny \beta=\num{\mytag}$},
    mark=*
    ] 
    table [x=error, y=regularization, meta=parameter] {\lcurvetablebeta} ;
\end{axis}
\end{tikzpicture}
    
    \caption{The graphs show the relations between the Tikhonov regularization and the distance of the velocity from the noisy data for varying regularization weights in the log-log scale. In both cases, the other regularization weight was kept constant at the value we selected as optimal, i.e. $\alpha=0.001, \beta=0.1$. This type of graph is sometimes called the L-curve. 
    }
    \label{fig:lcurves}
\end{figure}

\begin{figure}
\centering
\begin{tabular}{c c c c c}
     \rotatebox{90}{data} & 
     \includegraphics[width=.2\textwidth]{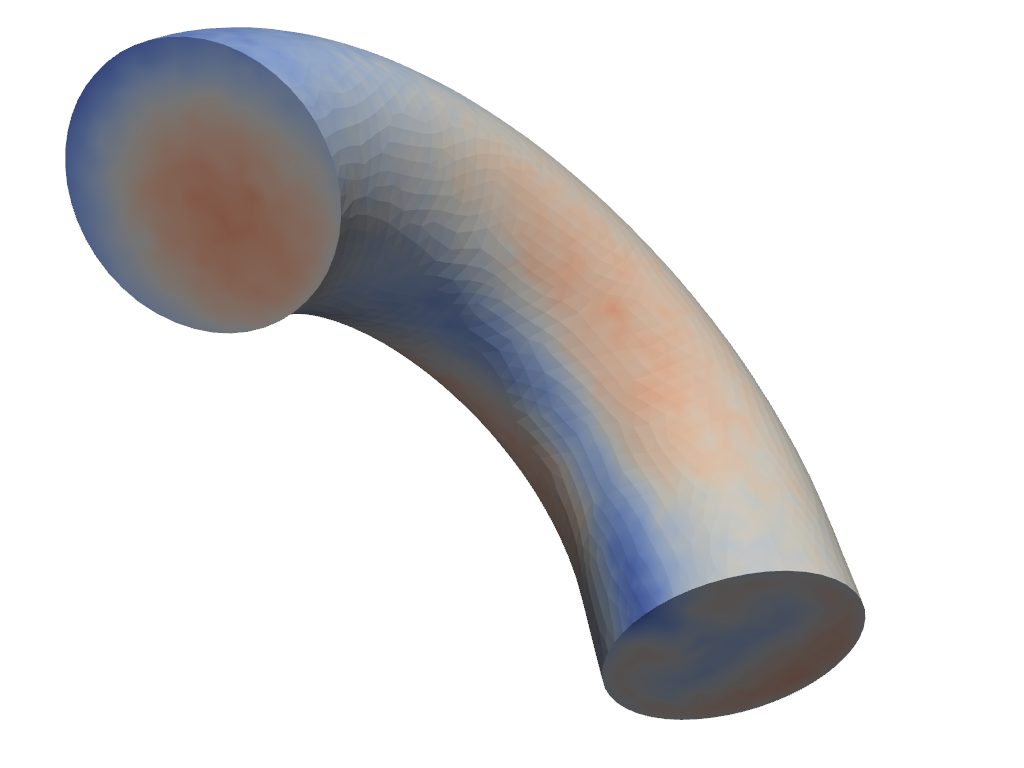} &
     \includegraphics[width=.2\textwidth]{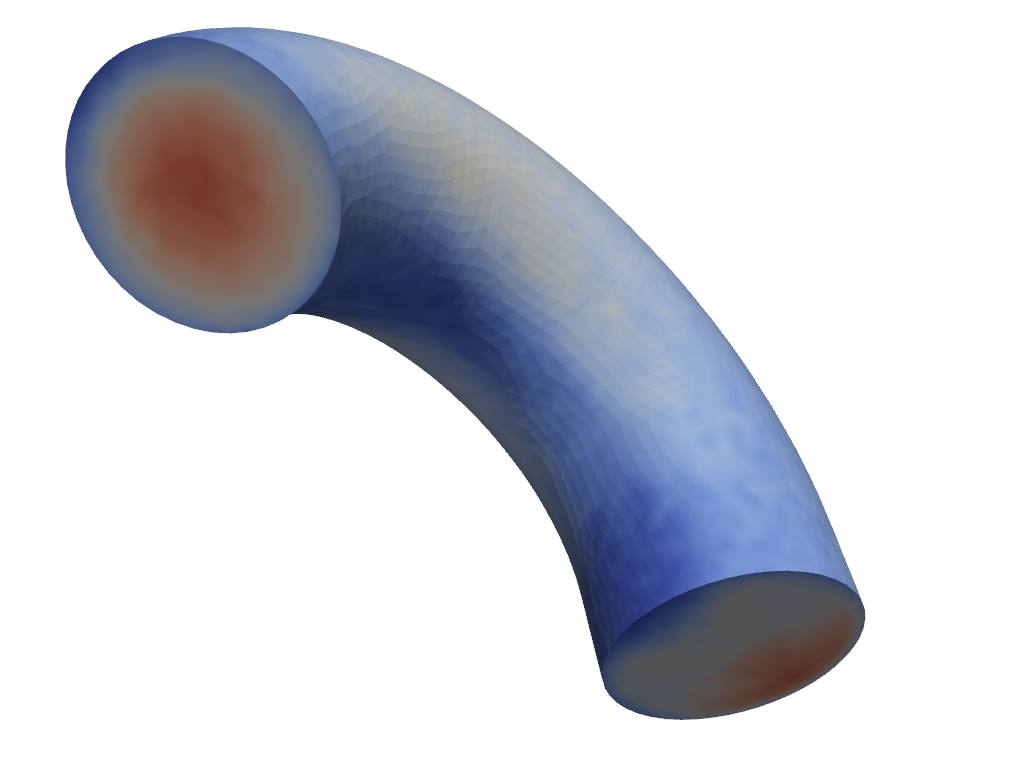} &
     \includegraphics[width=.2\textwidth]{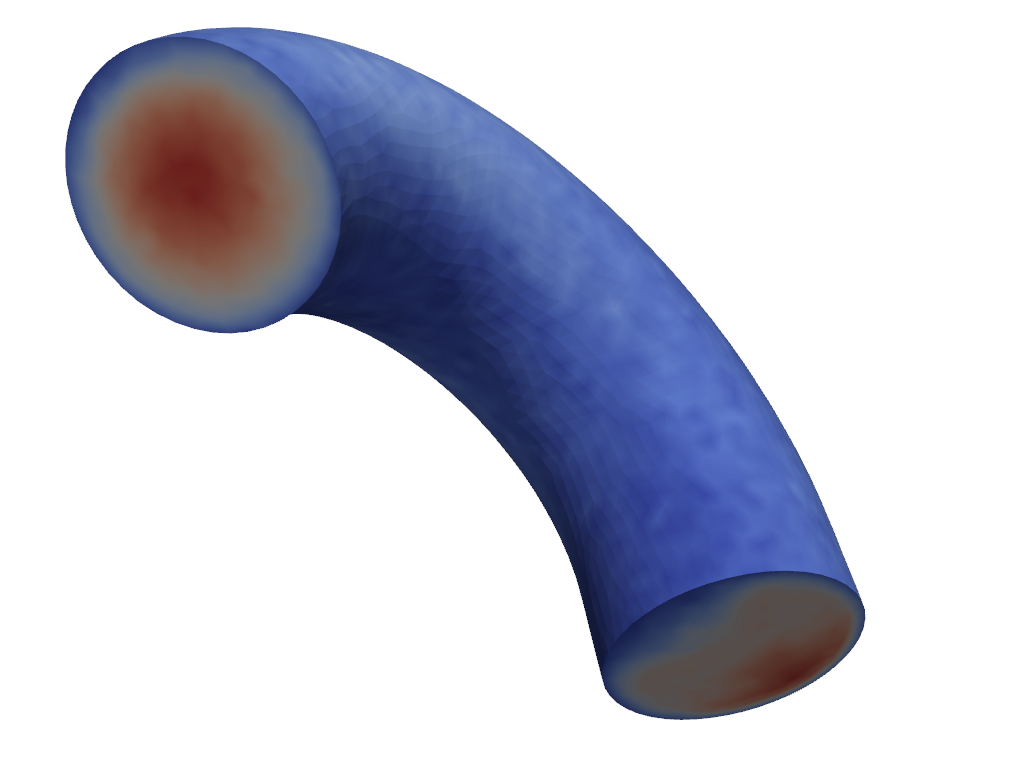} &
     \includegraphics[width=.2\textwidth]{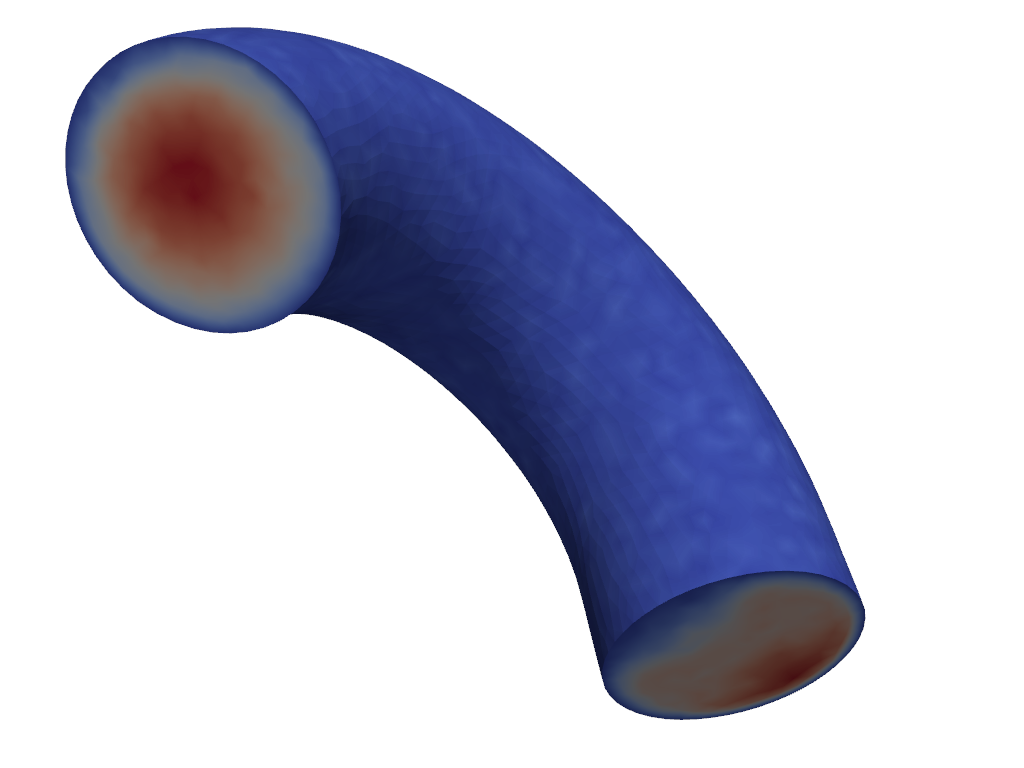} \\
    & $\theta=0.2$ & $\theta=0.5$ & $\theta=0.8$ & $\theta=1.0$ \\ 
     \rotatebox{90}{\begin{tabular}{c} 
     $\alpha=10^{-5}$\\ $\beta=10^{-3}$
     \end{tabular}} &
     \includegraphics[width=.2\textwidth]{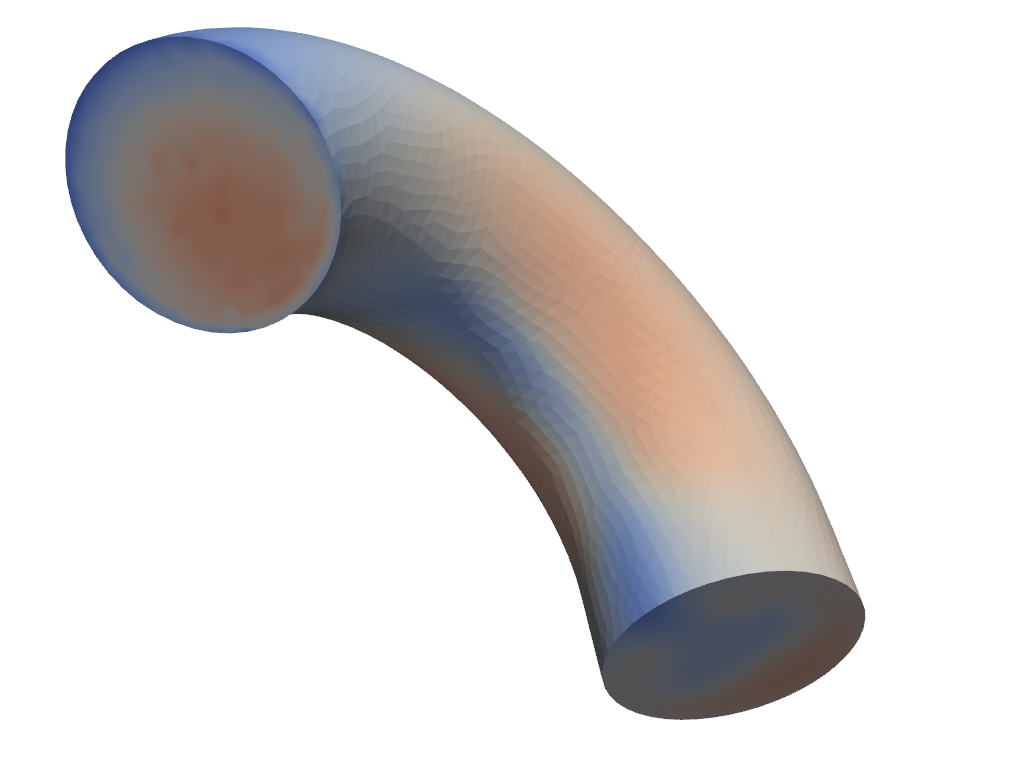} & 
     \includegraphics[width=.2\textwidth]{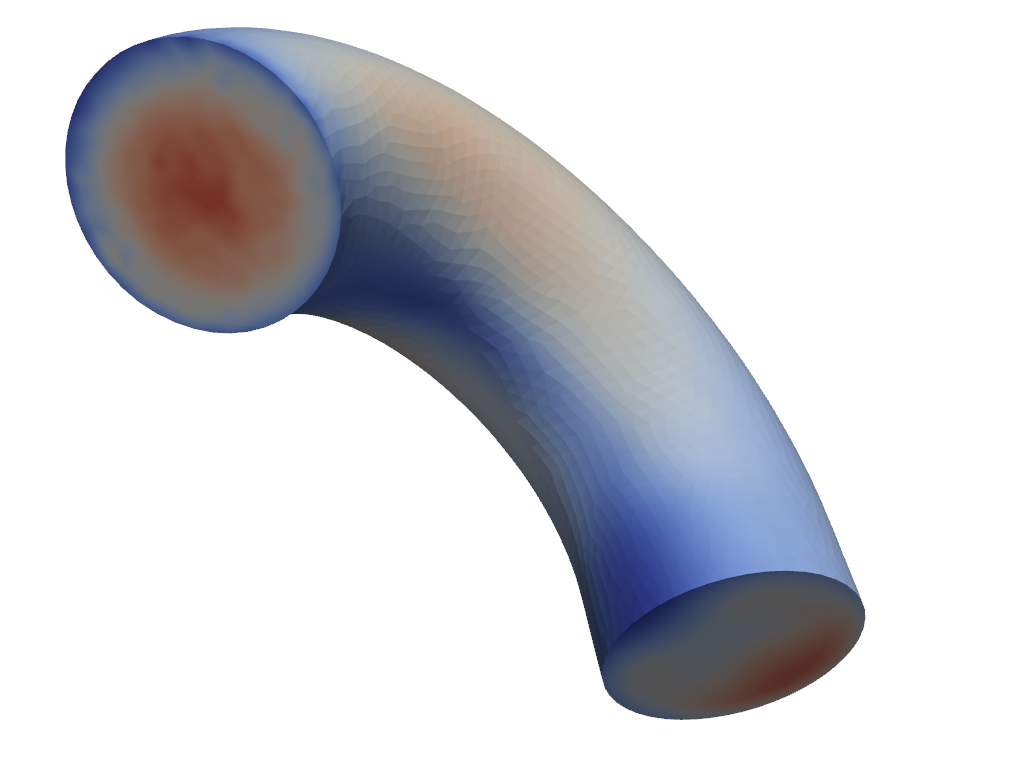} &
     \includegraphics[width=.2\textwidth]{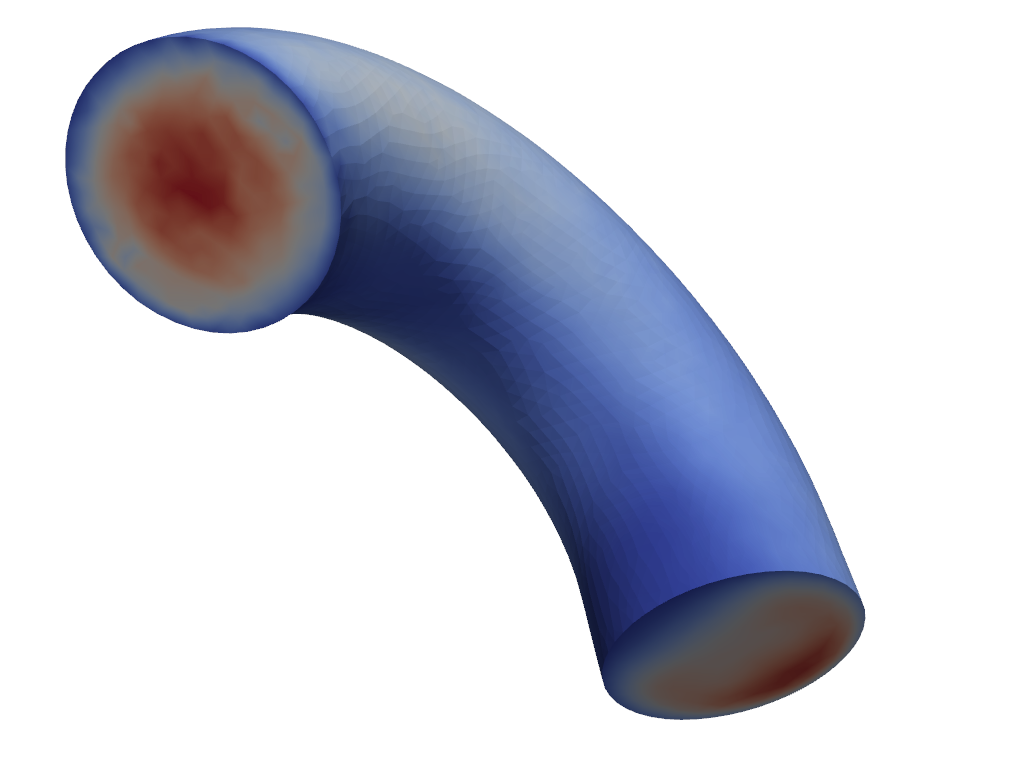} &
     \includegraphics[width=.2\textwidth]{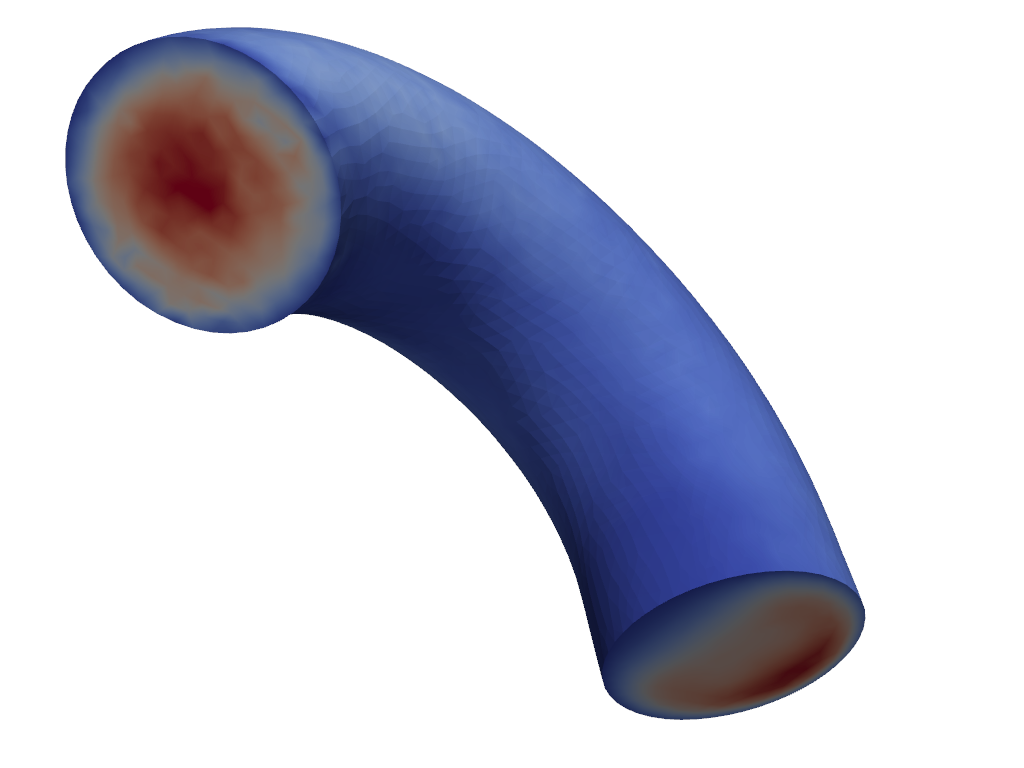} \\
     & $\thet{opt}=0.202$ & $\thet{opt}=0.467$ & $\thet{opt}=0.698$ & $\thet{opt}=0.841$ \\
     & $\mathcal{J}=\rnum{0.000279}$ & $\mathcal{J}=\rnum{0.0007639}$ & $\mathcal{J}=\rnum{0.001371}$ & $\mathcal{J}=\rnum{0.001883}$ \\
     & $\mathcal{R}=\rnum{2.314e-06}$ & $\mathcal{R}=\rnum{4.434e-06}$ & $\mathcal{R}=\rnum{6.161e-06}$ & $\mathcal{R}=\rnum{7.464e-06}$ \\
     & iterations: 34 & iterations: 30 & iterations: 27 & iterations: 34 \\
    \rotatebox{90}{\begin{tabular}{c} 
    $\alpha=10^{-3}$\\ $\beta=0.1$
    \end{tabular}}&
     \includegraphics[width=.2\textwidth]{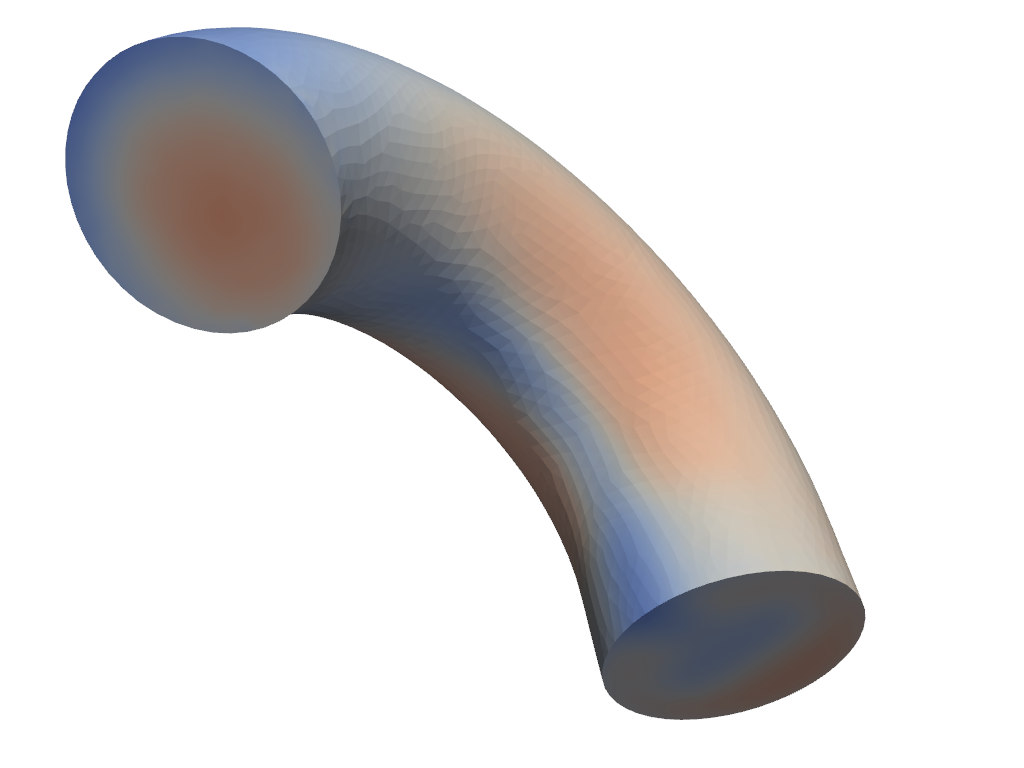} &
     \includegraphics[width=.2\textwidth]{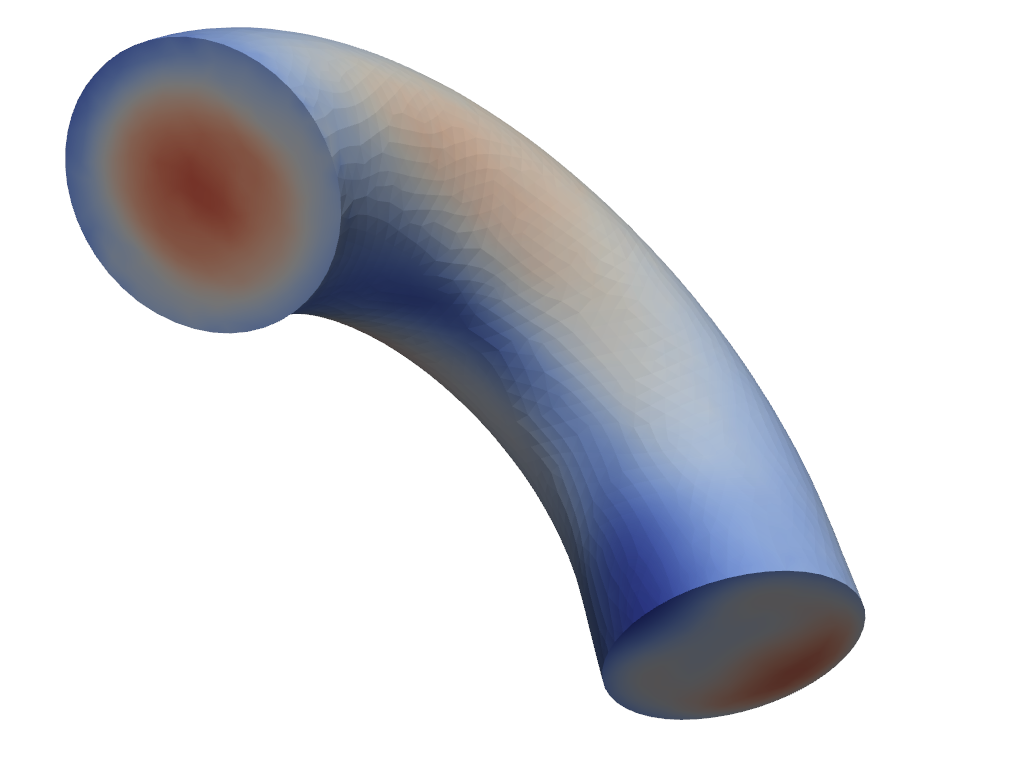} &
     \includegraphics[width=.2\textwidth]{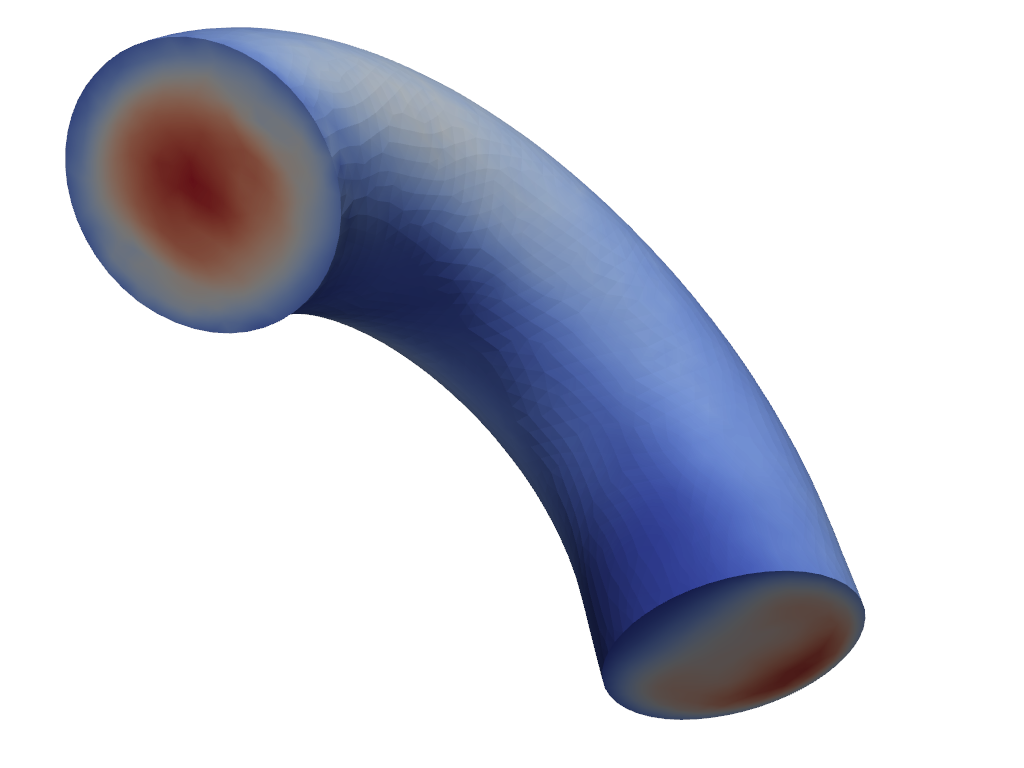} &
     \includegraphics[width=.2\textwidth]{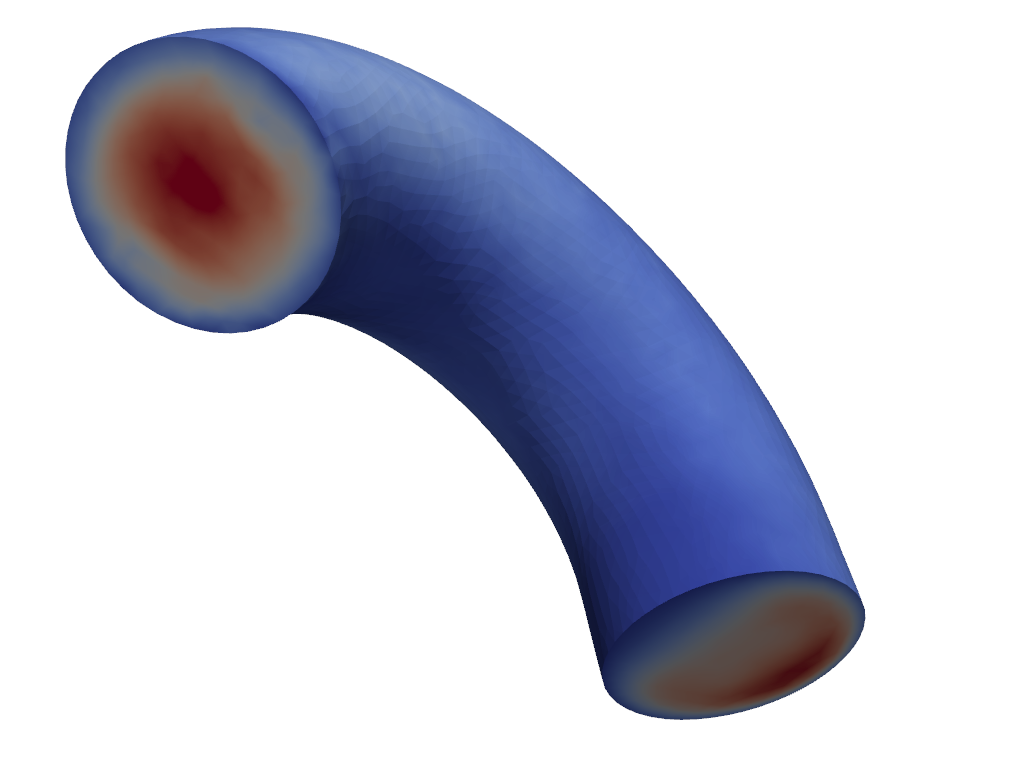} \\
     & $\thet{opt}=0.194$ & $\thet{opt}=0.463$ & $\thet{opt}=0.691$ & $\thet{opt}=0.833$ \\
     & $\mathcal{J}=\rnum{0.0002823}$ & $\mathcal{J}=\rnum{0.0007888}$ & $\mathcal{J}=\rnum{0.001419}$ & $\mathcal{J}=\rnum{0.001959}$ \\
     & $\mathcal{R}=\rnum{0.0001363}$ & $\mathcal{R}=\rnum{0.0003064}$ & $\mathcal{R}=\rnum{0.0004344}$ & $\mathcal{R}=\rnum{0.0005097}$ \\
     & iterations: 29 & iterations: 29 & iterations: 28 & iterations: 21 \\    
     \rotatebox{90}{\begin{tabular}{c}
     $\alpha=0.1$\\ $\beta=10.0$
     \end{tabular}} &
     \includegraphics[width=.2\textwidth]{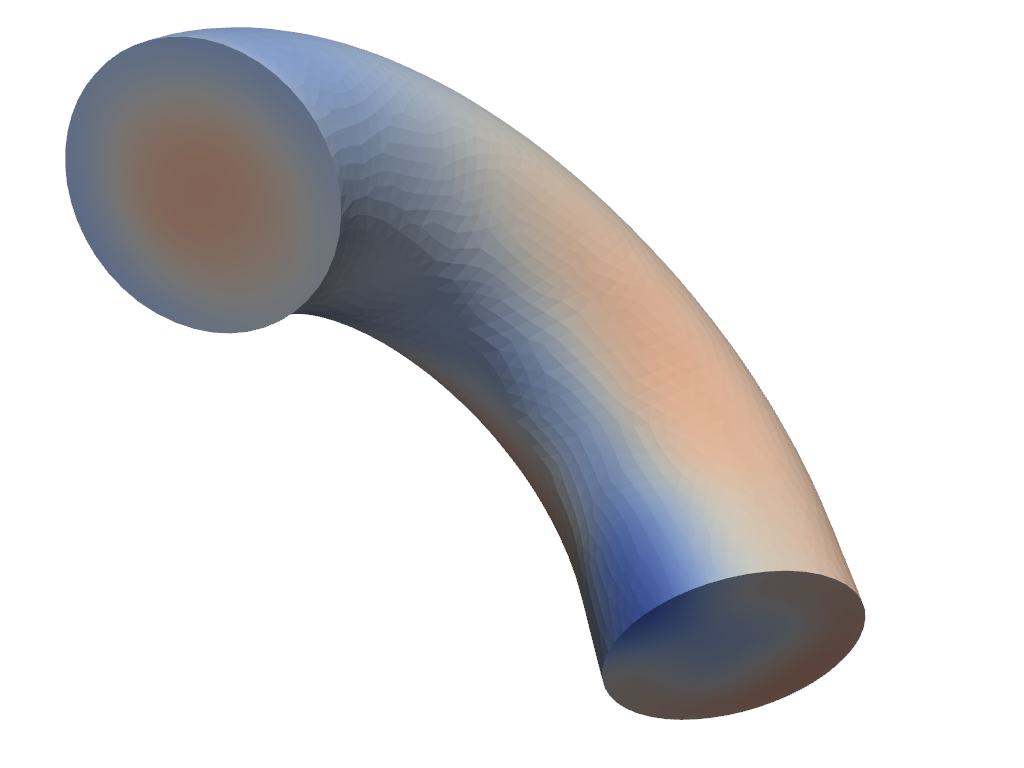} &
     \includegraphics[width=.2\textwidth]{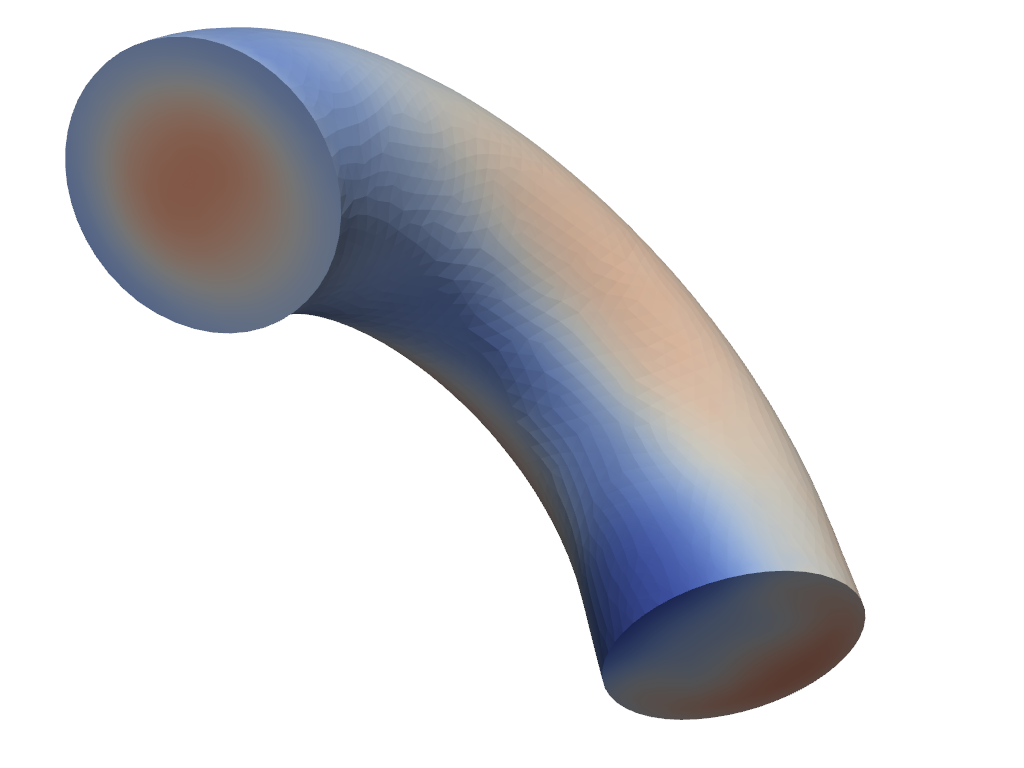} &
     \includegraphics[width=.2\textwidth]{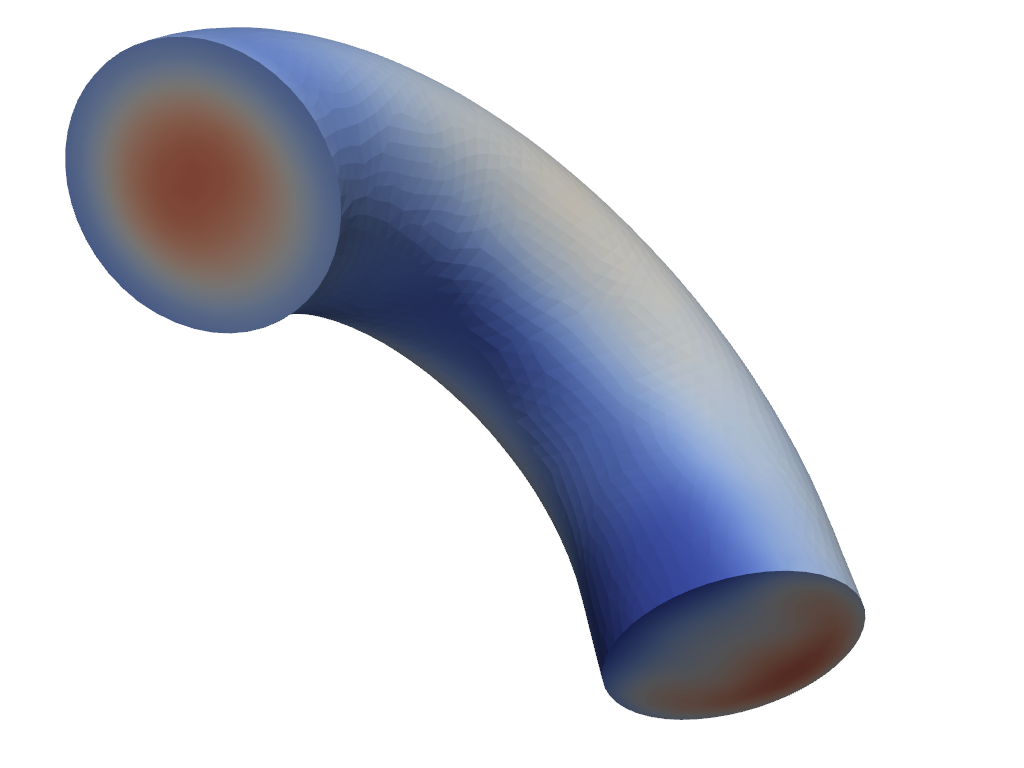} &
     \includegraphics[width=.2\textwidth]{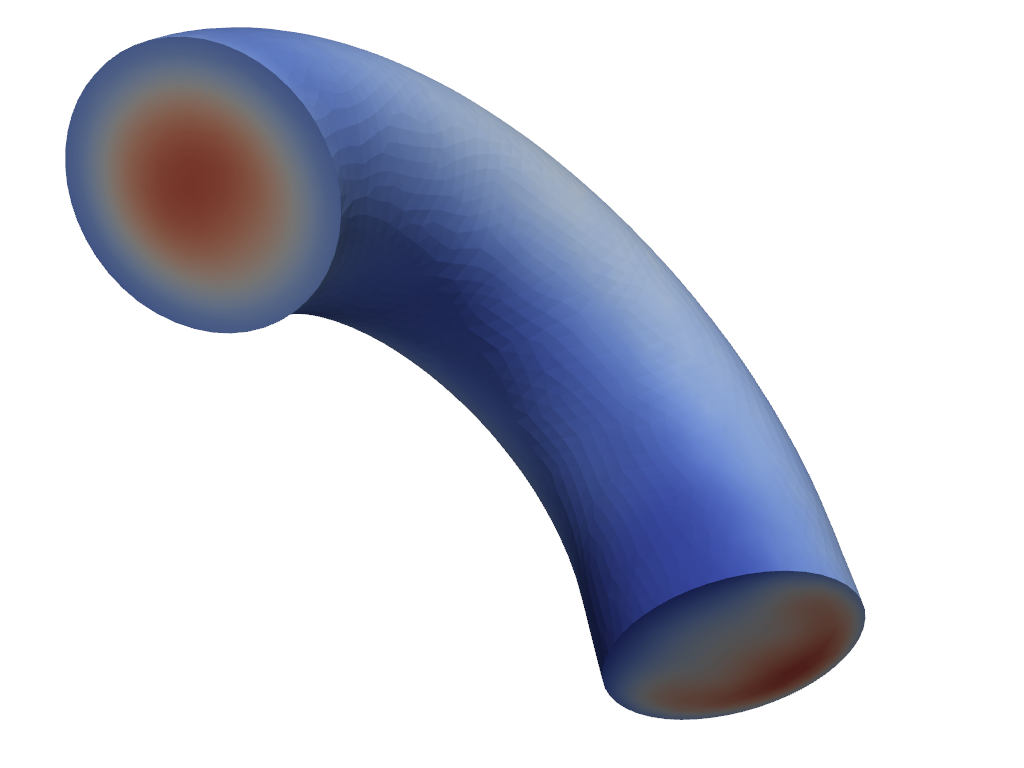} \\
     & $\thet{opt}=0.138$ & $\thet{opt}=0.243$ & $\thet{opt}=0.475$ & $\thet{opt}=0.643$ \\
     & $\mathcal{J}=\rnum{0.002412}$ & $\mathcal{J}=\rnum{0.006166}$ & $\mathcal{J}=\rnum{0.00756}$ & $\mathcal{J}=\rnum{0.008243}$ \\
     & $\mathcal{R}=\rnum{0.002205}$ & $\mathcal{R}=\rnum{0.003722}$ & $\mathcal{R}=\rnum{0.006002}$ & $\mathcal{R}=\rnum{0.00762}$ \\
     & iterations: 11 & iterations: 10 & iterations: 10 & iterations: 10 \\ 
     & \multicolumn{4}{c}{\includegraphics[width=0.5\textwidth]{scale0.2.png}}
\end{tabular}
\caption{Comparison of noisy data (SNR = 2.0) with assimilation velocity results computed using stabilized $P_1/P_1$ element with $\alpha_v=\alpha_p=0.01$ on the arch tube geometry with edge length $h=1.5\text{ mm}$ with various amounts of regularization for multiple values of $\theta$.}
\label{fig:regularization}
\end{figure}

The regularization weights in the error functional \eqref{error functional} can significantly impact the performance and results of the assimilation. 
In Figure \ref{fig:lcurves}, we plotted the L-curve for each regularization weight we used to determine which regularization weights have the best trade-off between the accuracy and the problem stability.
The curves show the relation between $\mathcal{J}(\mathbf{v})$ and $\mathcal{R}(\mathbf{m})$ for varying values of regularization weights in the log-log scale. 
Ideally, the plotted quantities should be balanced to avoid over- and under-regularization.
This should be possible to achieve by picking the highest regularization weights for which $\mathcal{J}(\mathbf{v})$ is still reasonably small. 
This approach for selecting the regulation weights was inspired by a similar problem in discrete inverse problems, see for example \cite{Hansen2010}.
Figure \ref{fig:regularization} shows how the regularization affects the velocity field. 
If the regularization weights $\alpha$ and $\beta$ are not high enough, the inlet velocity profile is not smooth. 
On the other hand, if the regularization weights are too high, the inlet profile does not agree with the data at all, and $\thet{opt}$ is reconstructed poorly.

\subsection{Higher velocities}
\begin{figure}
\centering
\begin{tabular}{c c c c c}
     \rotatebox{90}{\begin{tabular}{c} 
    data\\ $V=0.25$
    \end{tabular}} & 
     \includegraphics[width=.2\textwidth]{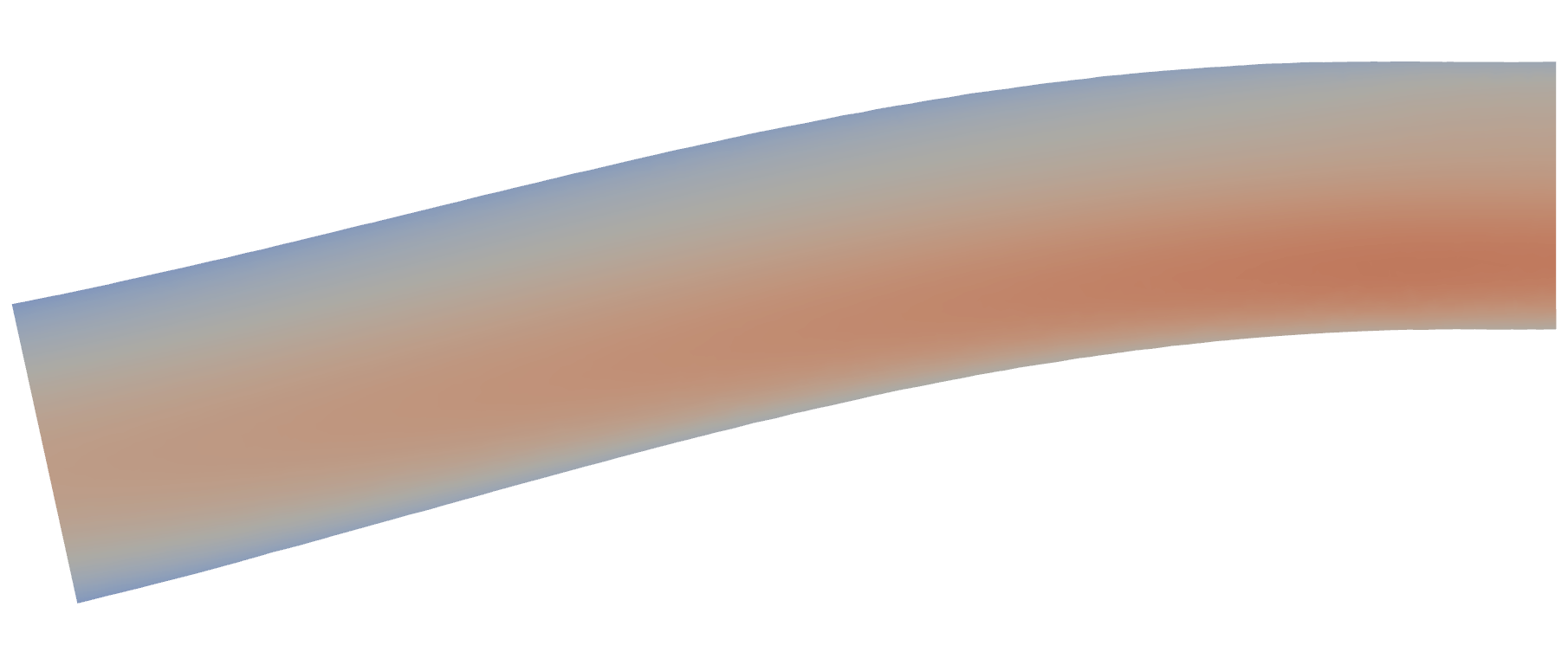} &
     \includegraphics[width=.2\textwidth]{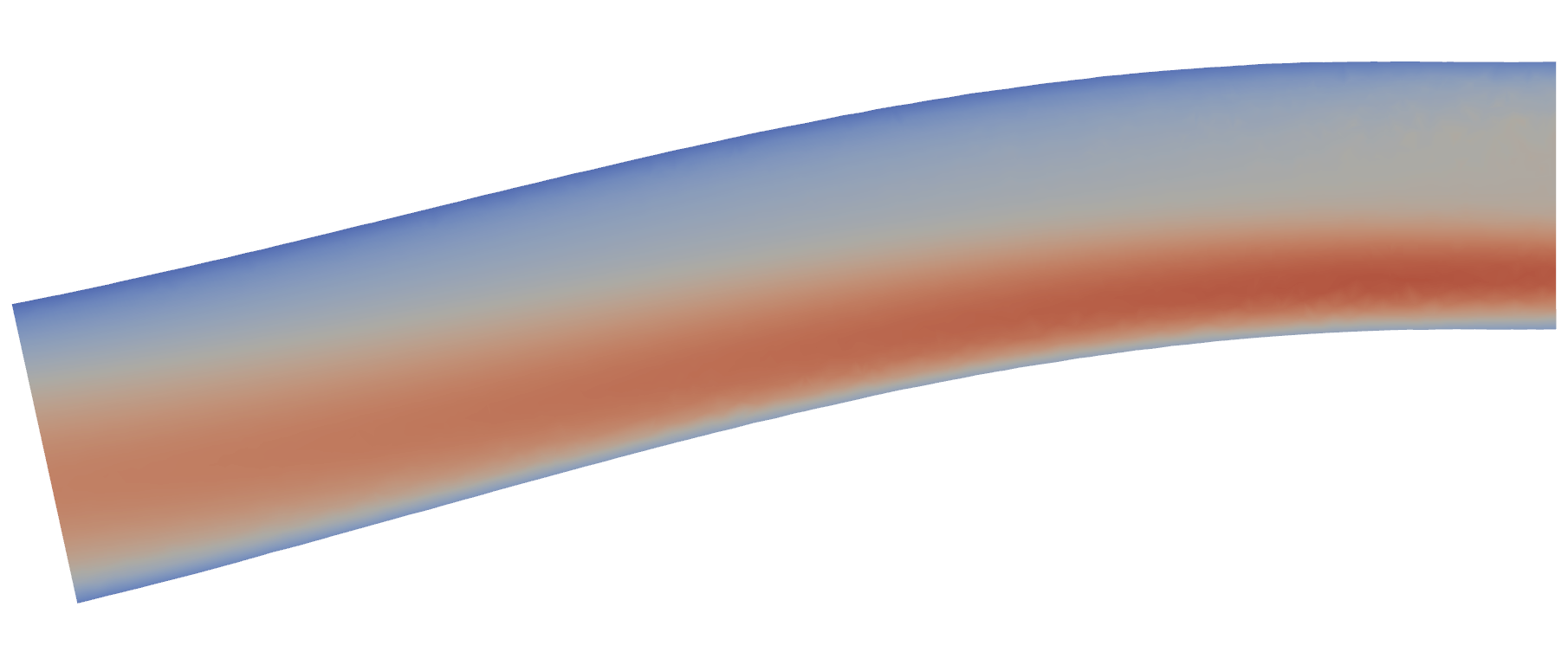} &
     \includegraphics[width=.2\textwidth]{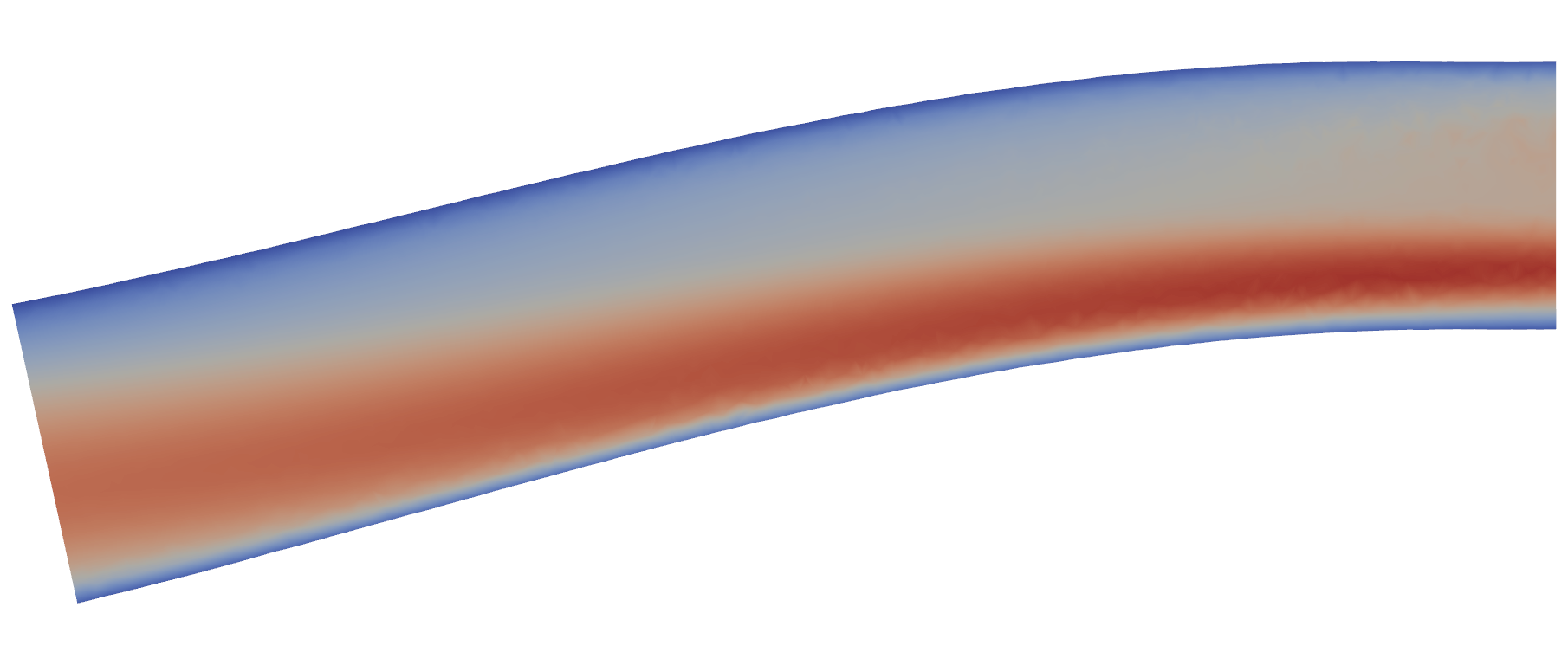} &
     \includegraphics[width=.2\textwidth]{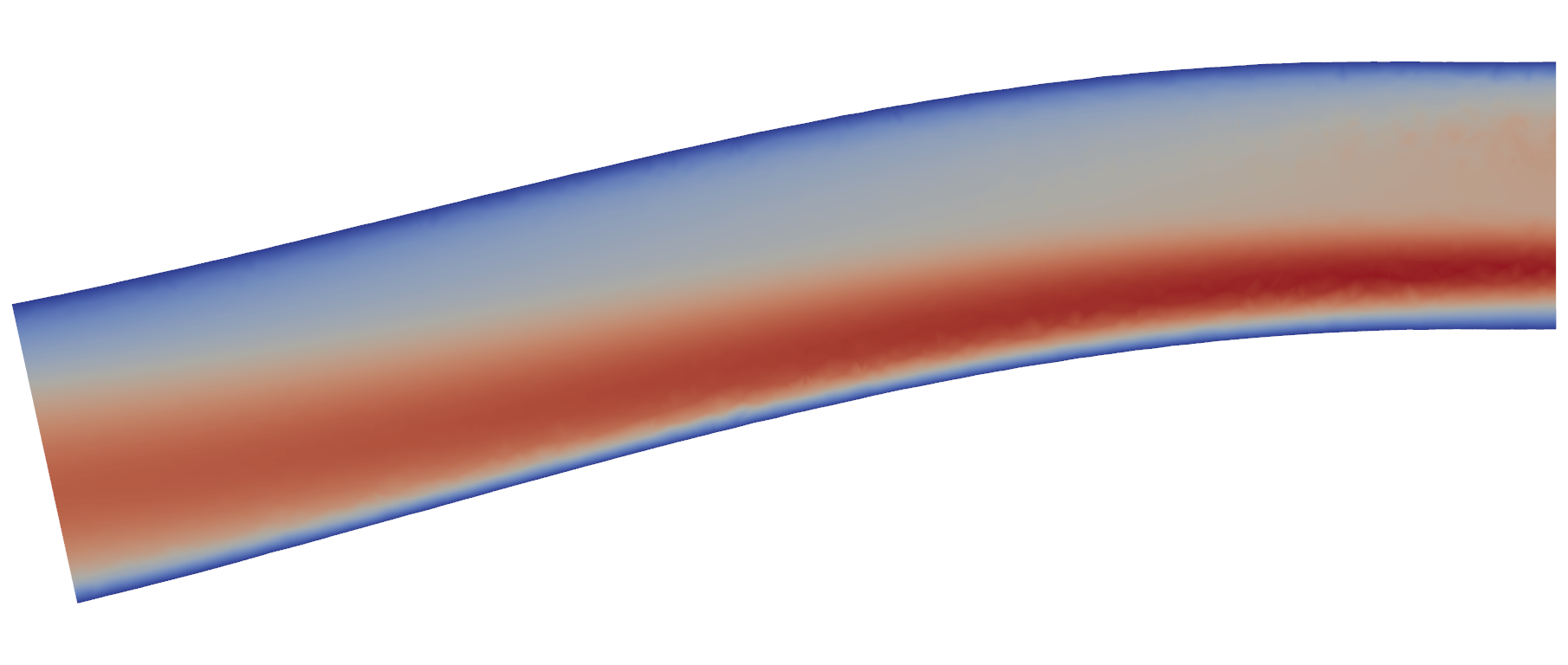} \\
    & $\theta=0.2$ & $\theta=0.5$ & $\theta=0.8$ & $\theta=1.0$ \\ 
     \rotatebox{90}{\begin{tabular}{c} 
     assimilation\\ $V=0.25$
     \end{tabular}} &
     \includegraphics[width=.2\textwidth]{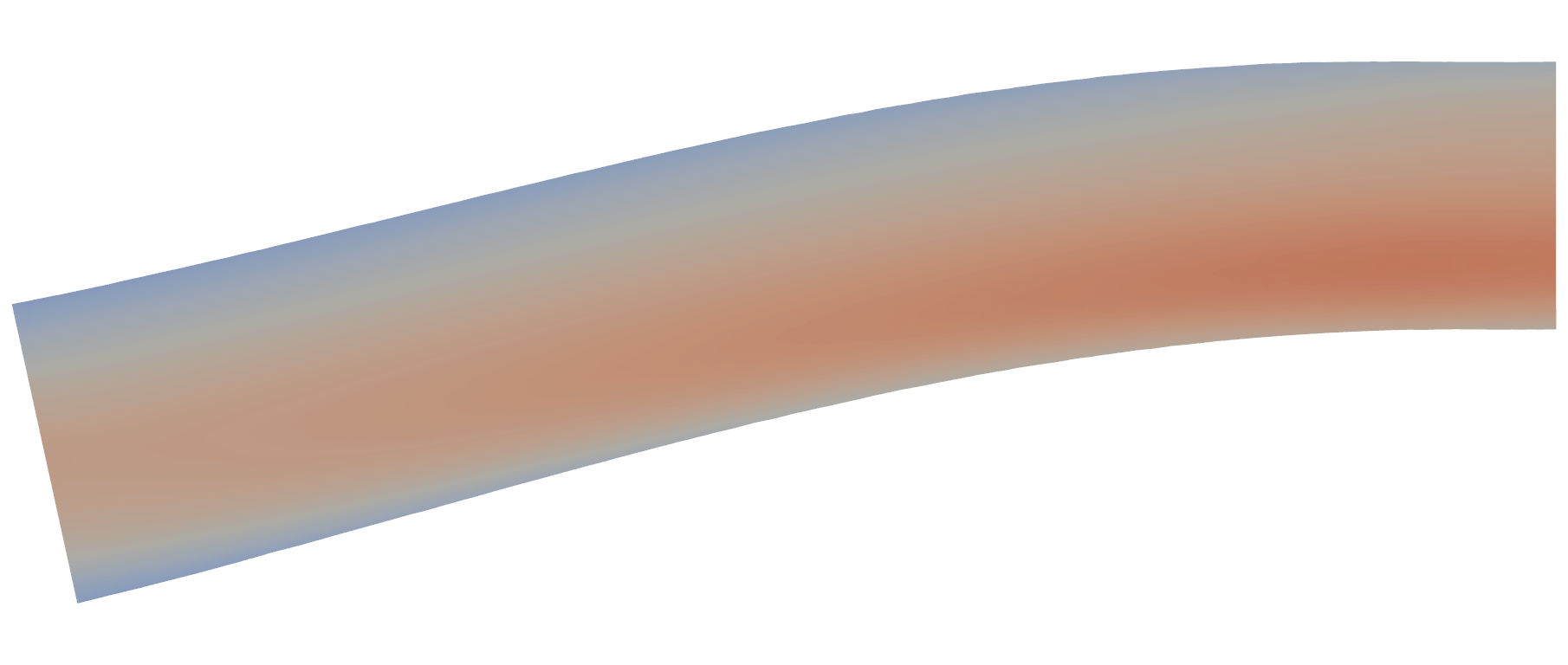} & 
     \includegraphics[width=.2\textwidth]{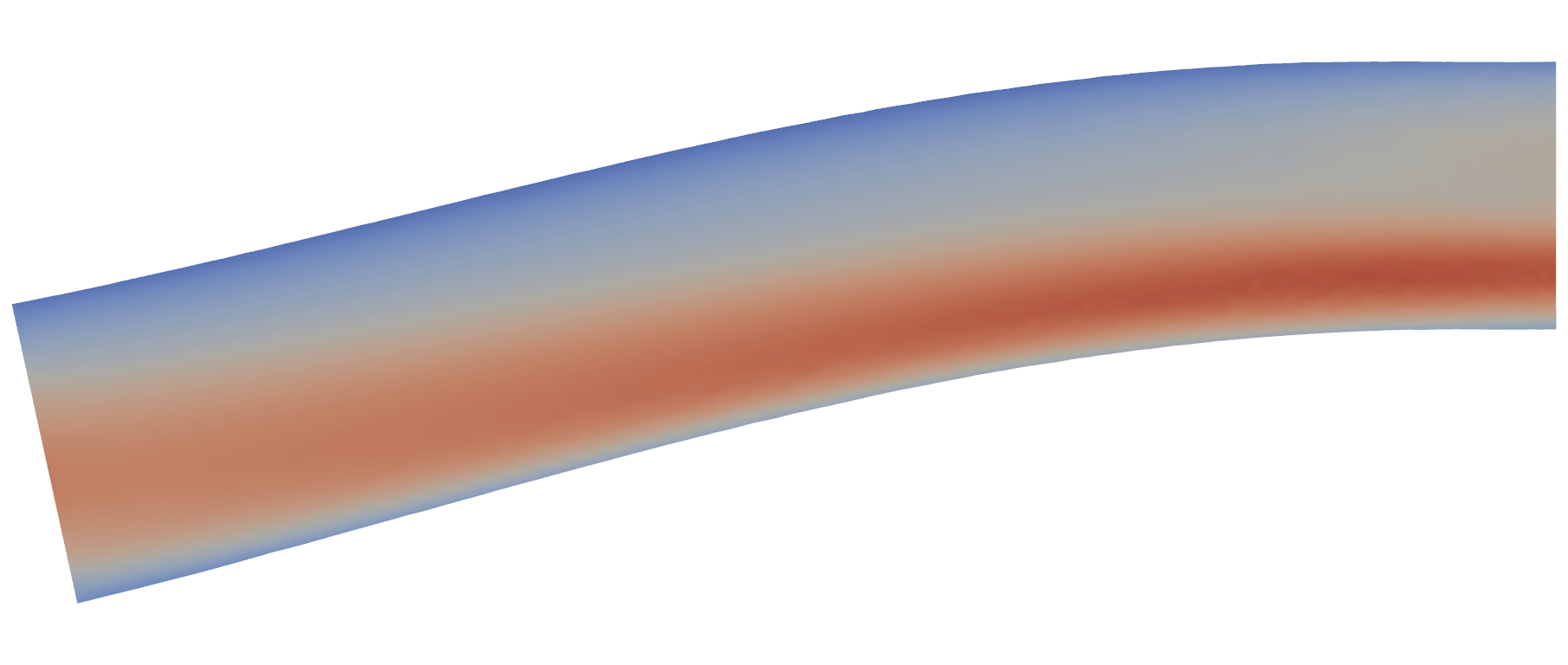} &
     \includegraphics[width=.2\textwidth]{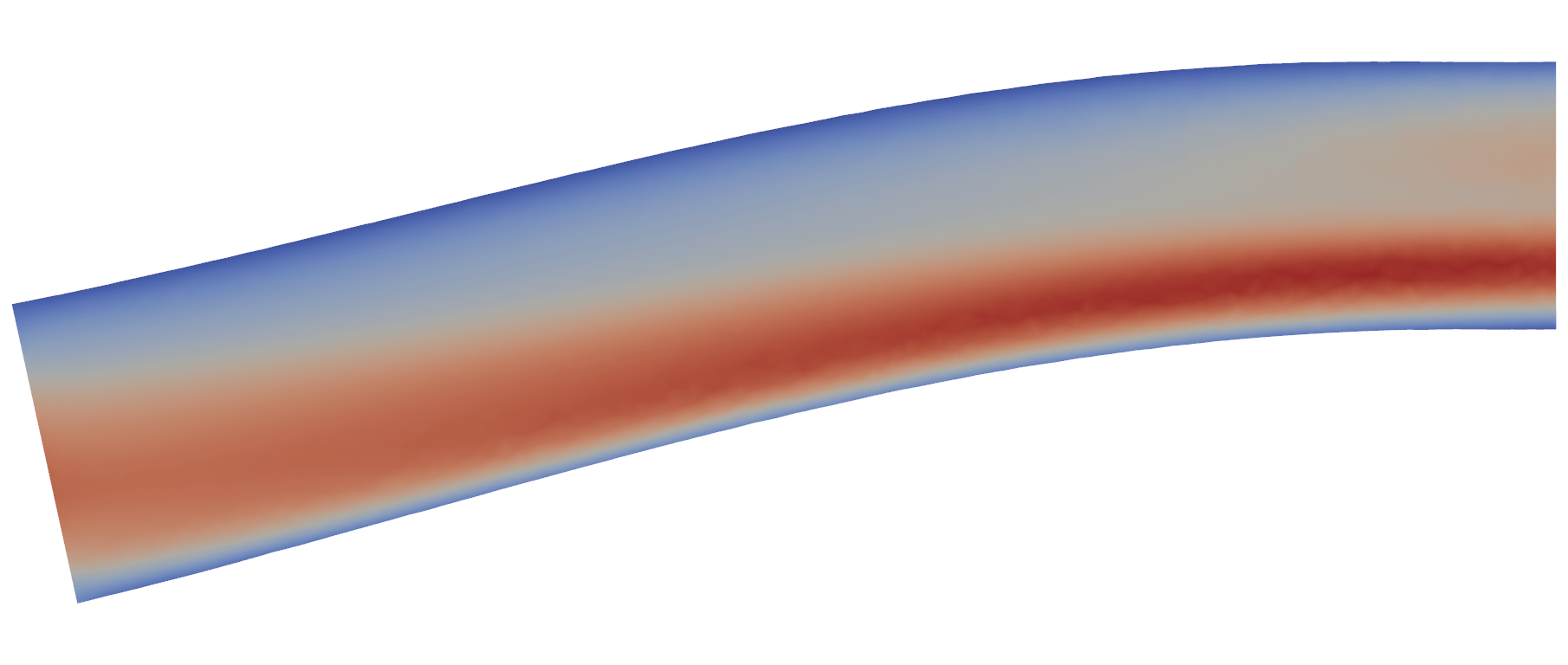} &
     \includegraphics[width=.2\textwidth]{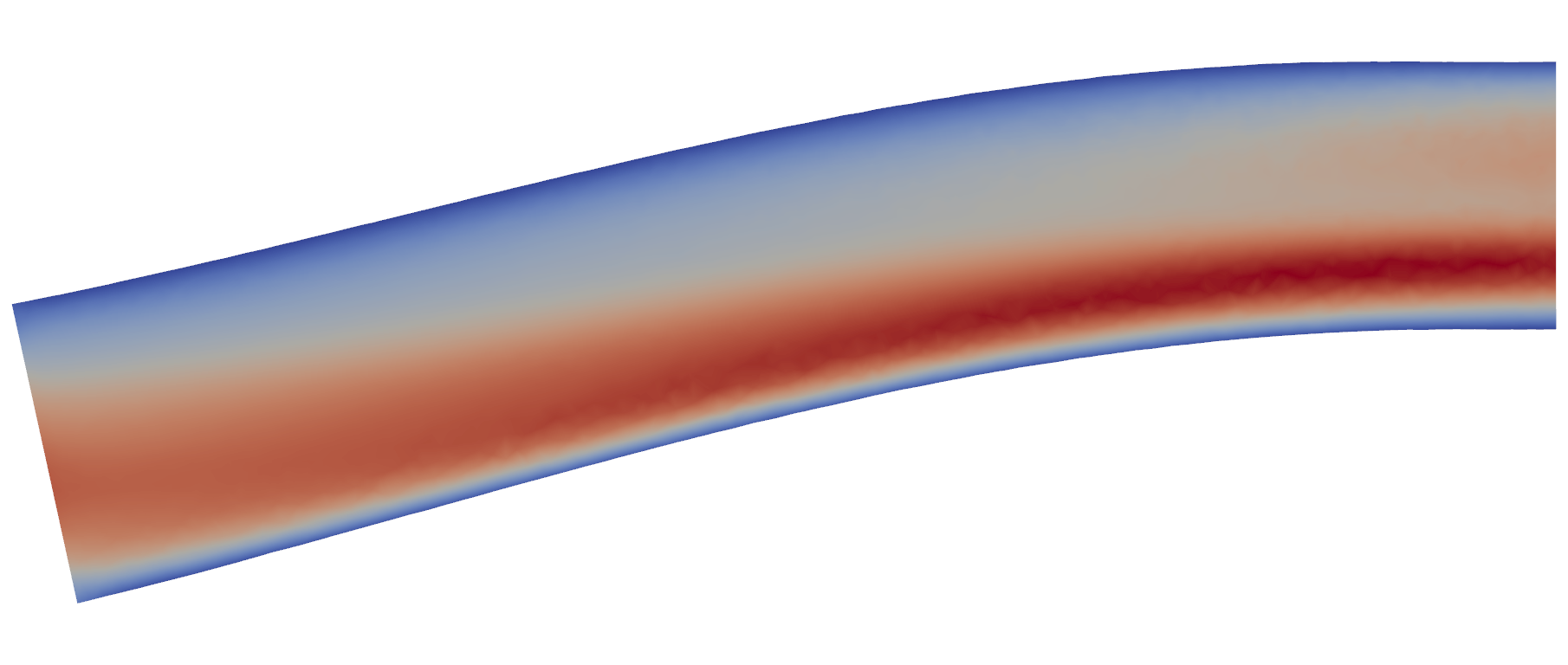} \\
     & $\thet{opt}=0.2$ & $\thet{opt}=0.506$ & $\thet{opt}=0.758$ & $\thet{opt}=0.908$ \\
     & $\mathcal{J}=\rnum{8.463e-05}$ & $\mathcal{J}=\rnum{0.000888}$ & $\mathcal{J}=\rnum{0.002564}$ & $\mathcal{J}=\rnum{0.003796}$ \\
     & $\mathcal{R}=\rnum{0.0001669}$ & $\mathcal{R}=\rnum{0.0007297}$ & $\mathcal{R}=\rnum{0.001222}$ & $\mathcal{R}=\rnum{0.001439}$ \\
     & iterations: 41 & iterations: 50 & iterations: 44 & iterations: 43 \\
     & \multicolumn{4}{c}{\includegraphics[width=0.5\textwidth]{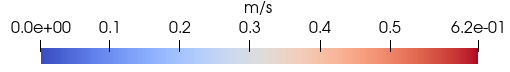}}\\[5mm]
    \rotatebox{90}{\begin{tabular}{c} 
    data\\ $V=0.8$
    \end{tabular}}&
     \includegraphics[width=.2\textwidth]{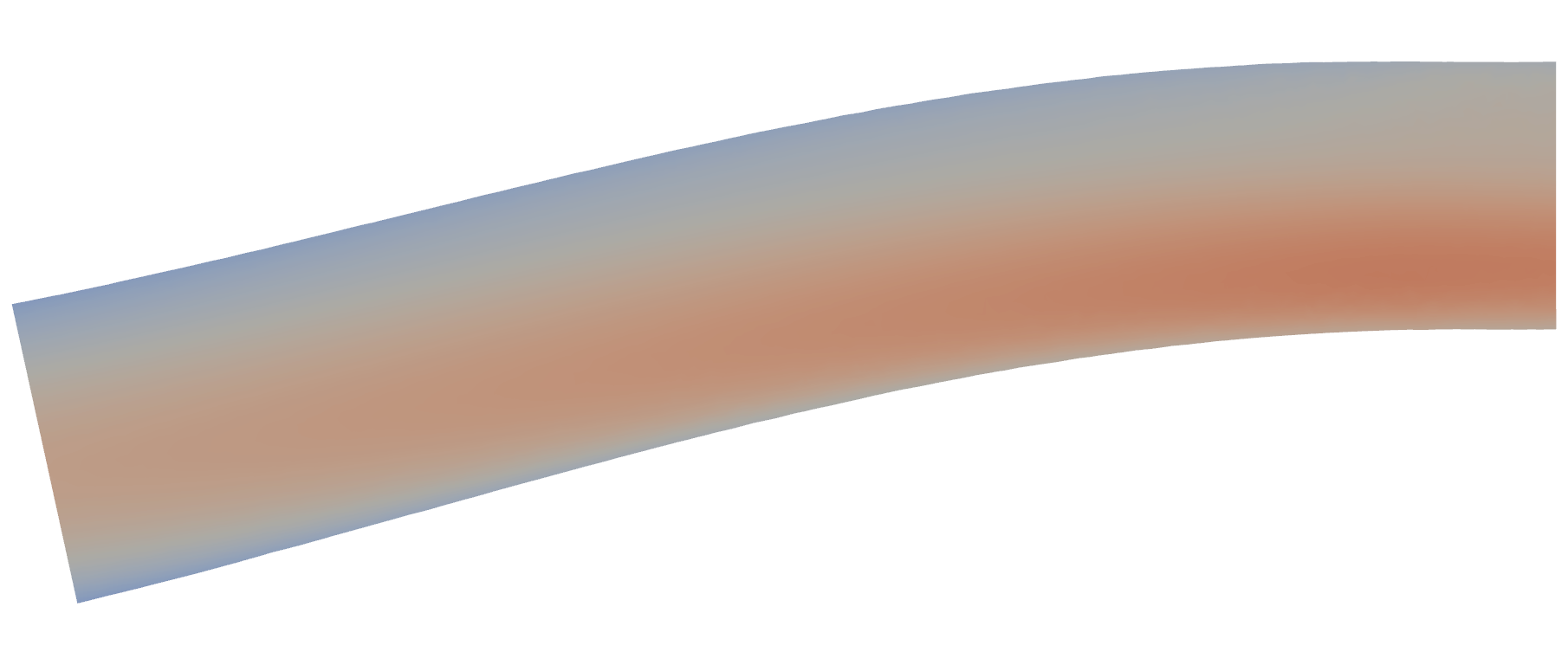} &
     \includegraphics[width=.2\textwidth]{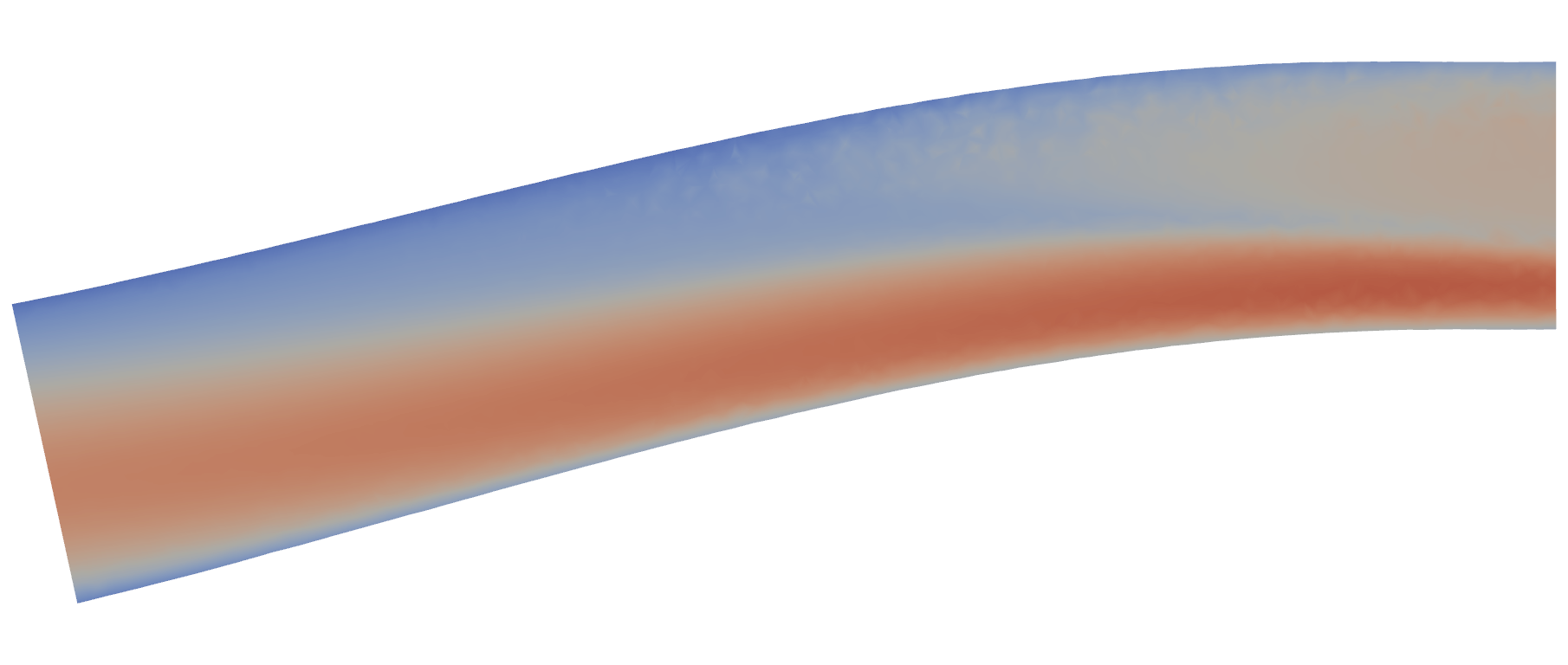} &
     \includegraphics[width=.2\textwidth]{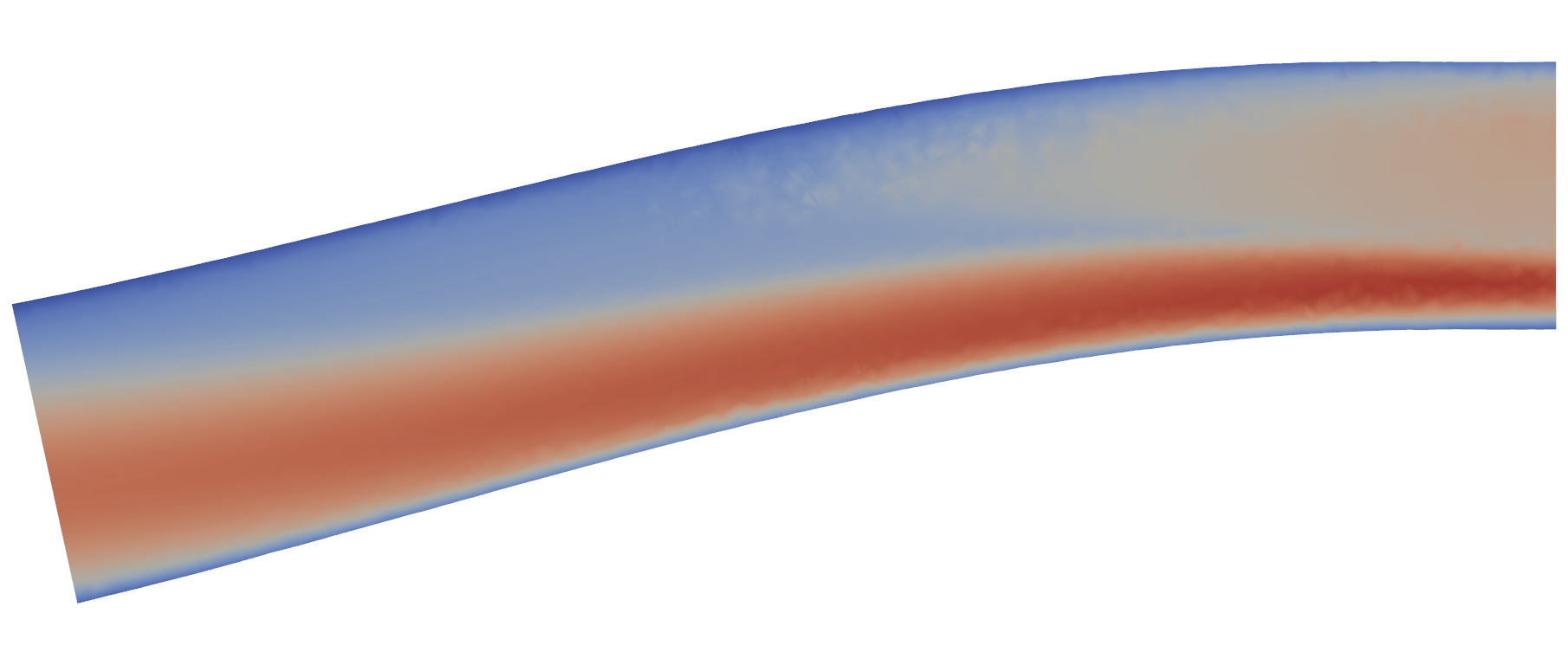} &
     \includegraphics[width=.2\textwidth]{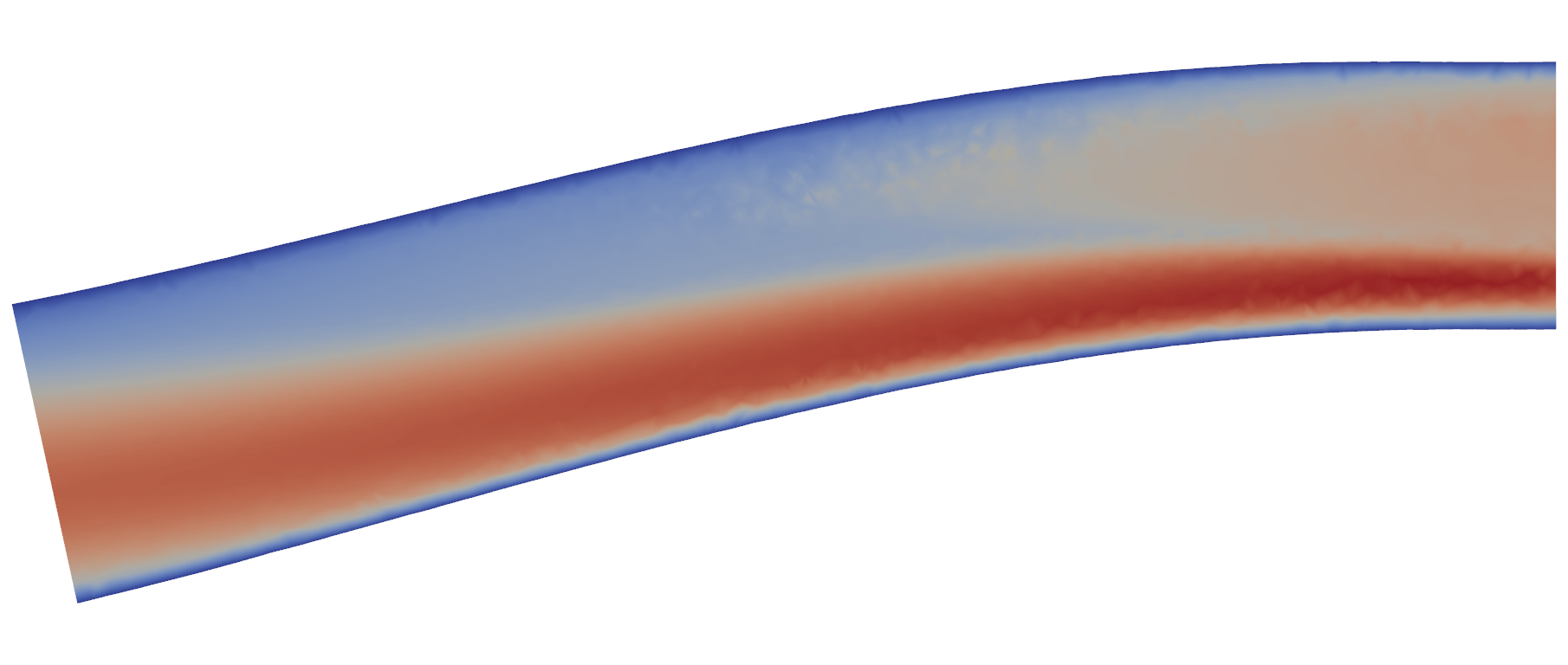} \\
     & $\theta=0.2$ & $\theta=0.5$ & $\theta=0.8$ & $\theta=1.0$ \\ 
     \rotatebox{90}{\begin{tabular}{c}
     assimilation\\ $V=0.8$
     \end{tabular}} &
     \includegraphics[width=.2\textwidth]{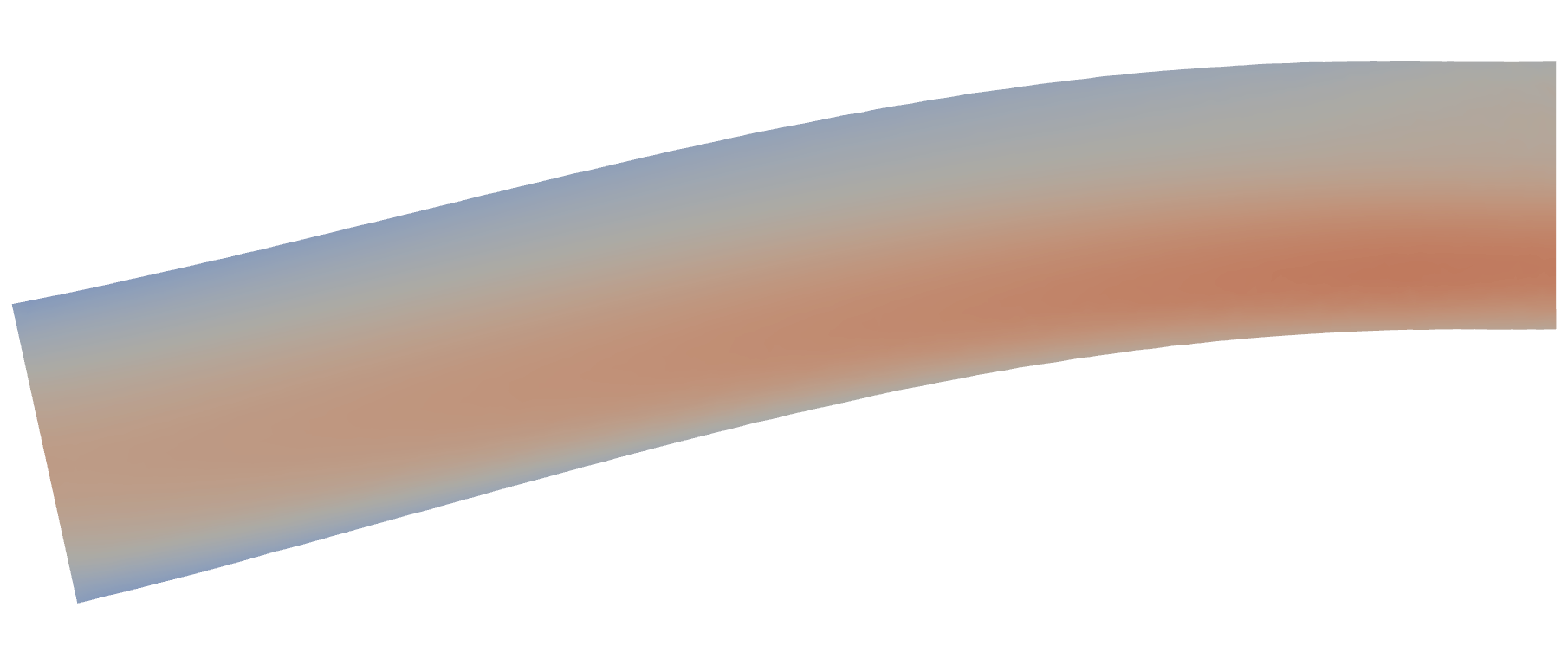} &
     \includegraphics[width=.2\textwidth]{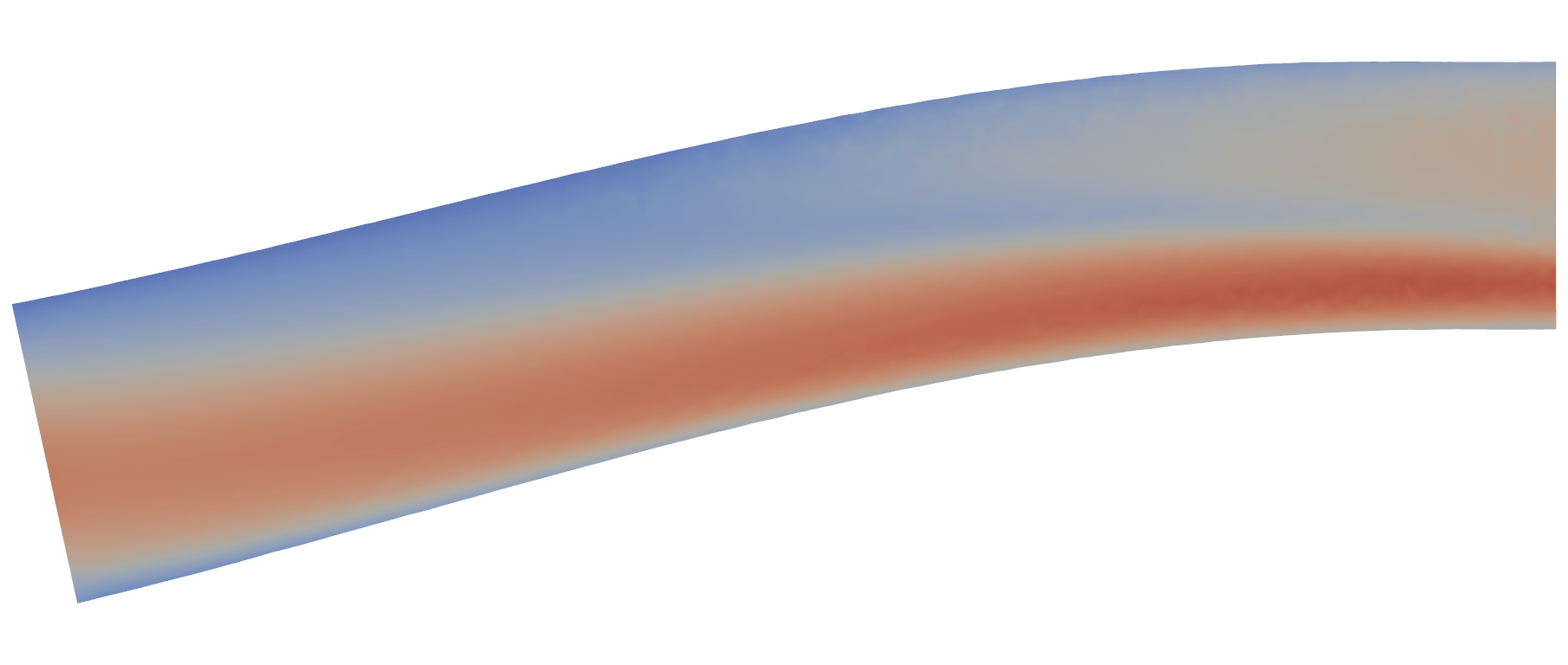} &
     \includegraphics[width=.2\textwidth]{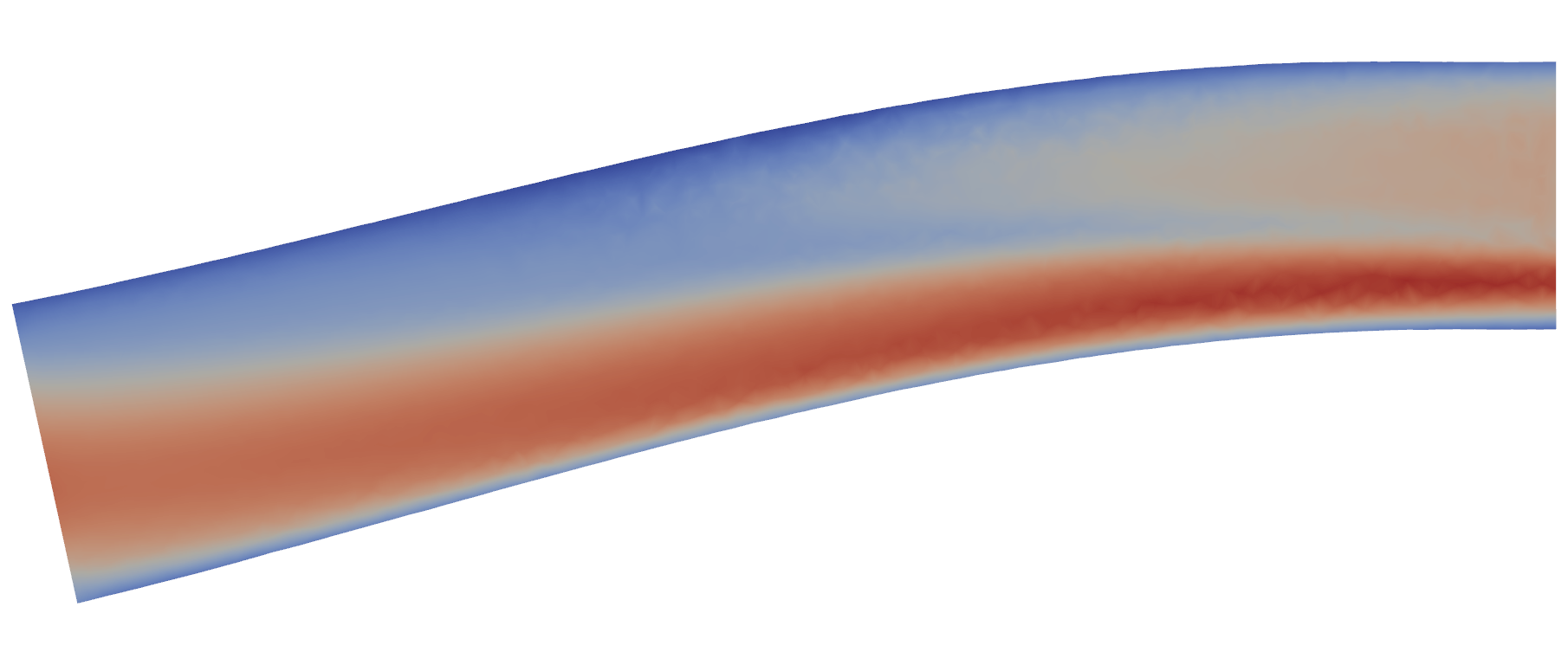} &
     \includegraphics[width=.2\textwidth]{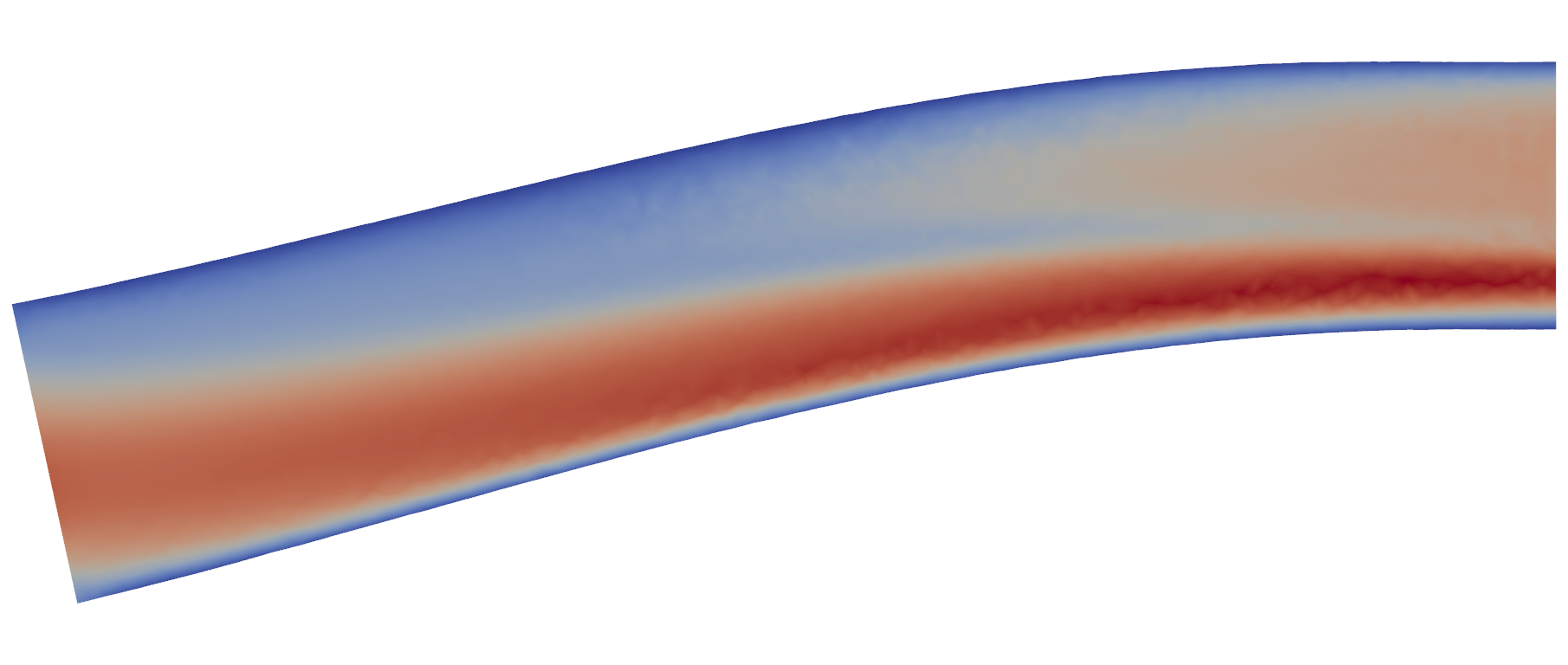} \\
     & $\thet{opt}=0.171$ & $\thet{opt}=0.527$ & $\thet{opt}=0.811$ & $\thet{opt}=0.956$ \\
     & $\mathcal{J}=\rnum{0.0006342}$ & $\mathcal{J}=\rnum{0.005556}$ & $\mathcal{J}=\rnum{0.0173}$ & $\mathcal{J}=\rnum{0.03159}$ \\
     & $\mathcal{R}=\rnum{0.00148}$ & $\mathcal{R}=\rnum{0.008317}$ & $\mathcal{R}=\rnum{0.01468}$ & $\mathcal{R}=\rnum{0.01668}$ \\
     & iterations: 62 & iterations: 95 & iterations: 132 & iterations: 146 \\
     & \multicolumn{4}{c}{\includegraphics[width=0.5\textwidth]{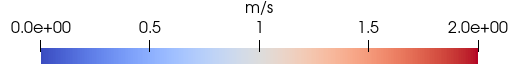}}
\end{tabular}
\caption{Comparison of data without noise with assimilation velocity results for higher velocities using the stabilized $P_1/P_1$ element with $\alpha_v=\alpha_p=0.01$ on the bent tube geometry for multiple values of $\theta$.}
\label{fig:velocities_bent}
\end{figure}

\begin{figure}
\centering
\begin{tabular}{c c c c c}
     \rotatebox{90}{\begin{tabular}{c} 
    data\\ $V=0.25$
    \end{tabular}} & 
     \includegraphics[width=.2\textwidth]{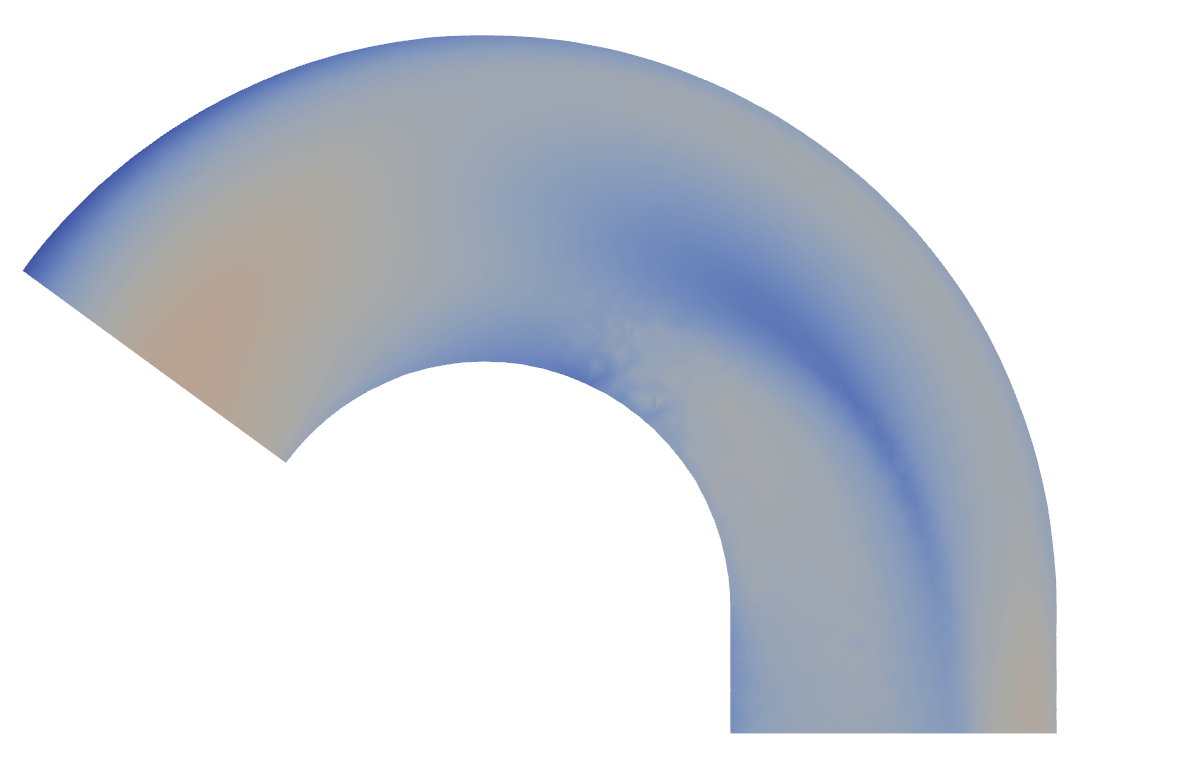} &
     \includegraphics[width=.2\textwidth]{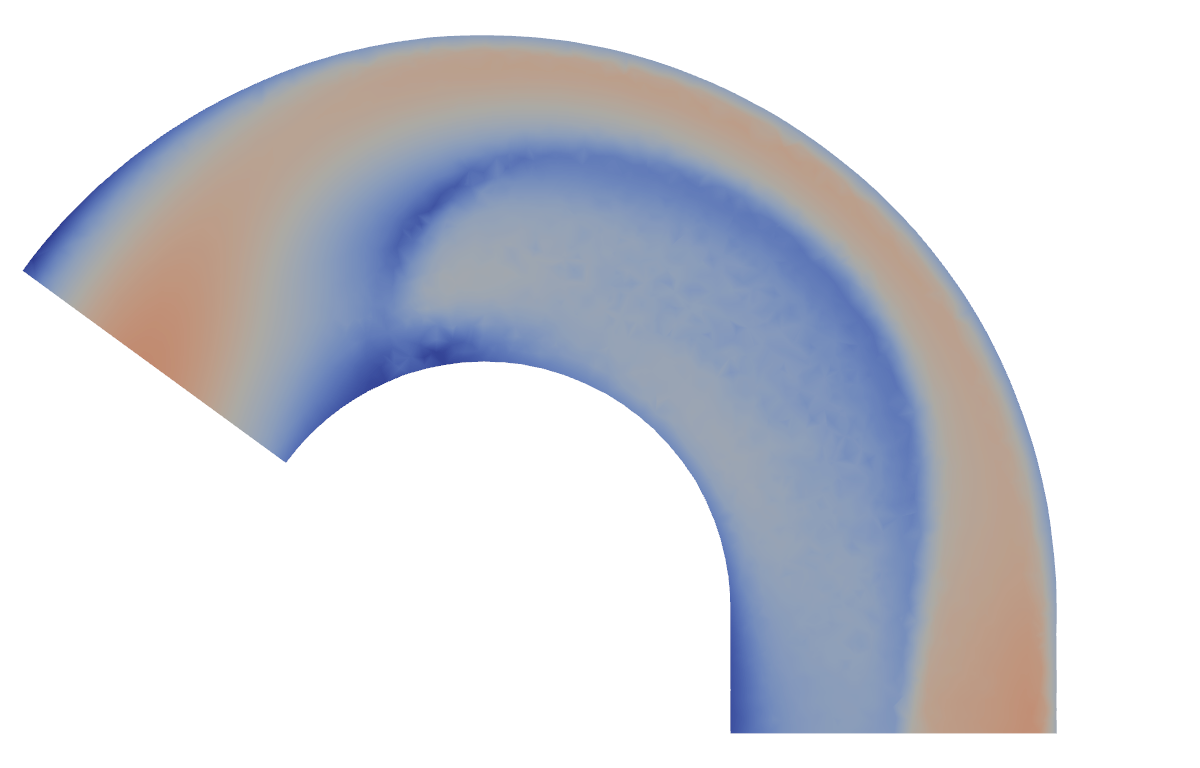} &
     \includegraphics[width=.2\textwidth]{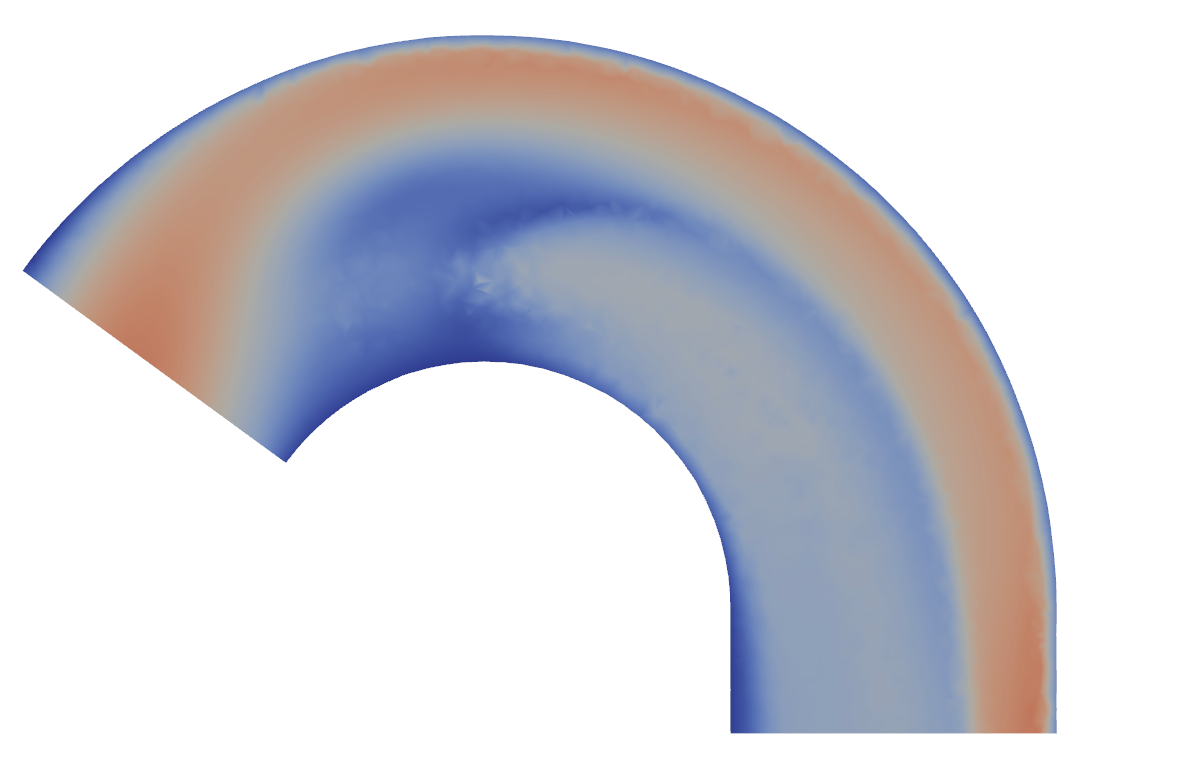} &
     \includegraphics[width=.2\textwidth]{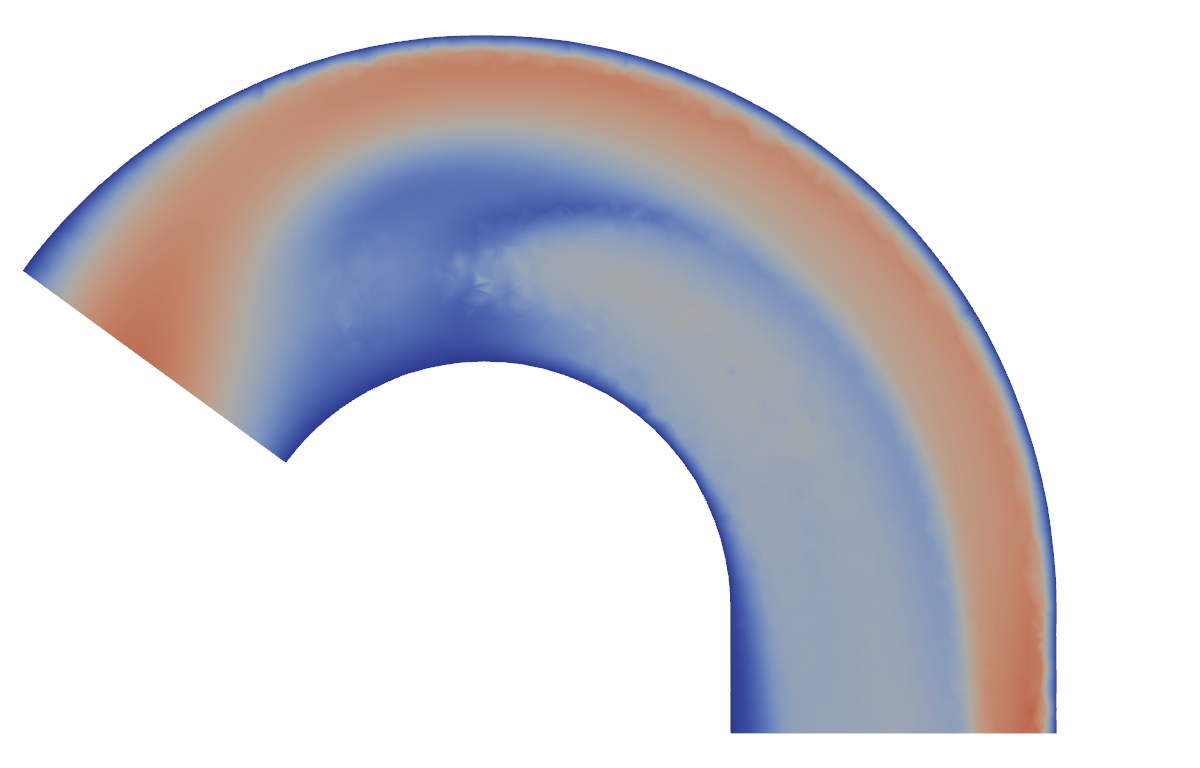} \\
    & $\theta=0.2$ & $\theta=0.5$ & $\theta=0.8$ & $\theta=1.0$ \\ 
     \rotatebox{90}{\begin{tabular}{c} 
     assimilation\\ $V=0.25$
     \end{tabular}} &
     \includegraphics[width=.2\textwidth]{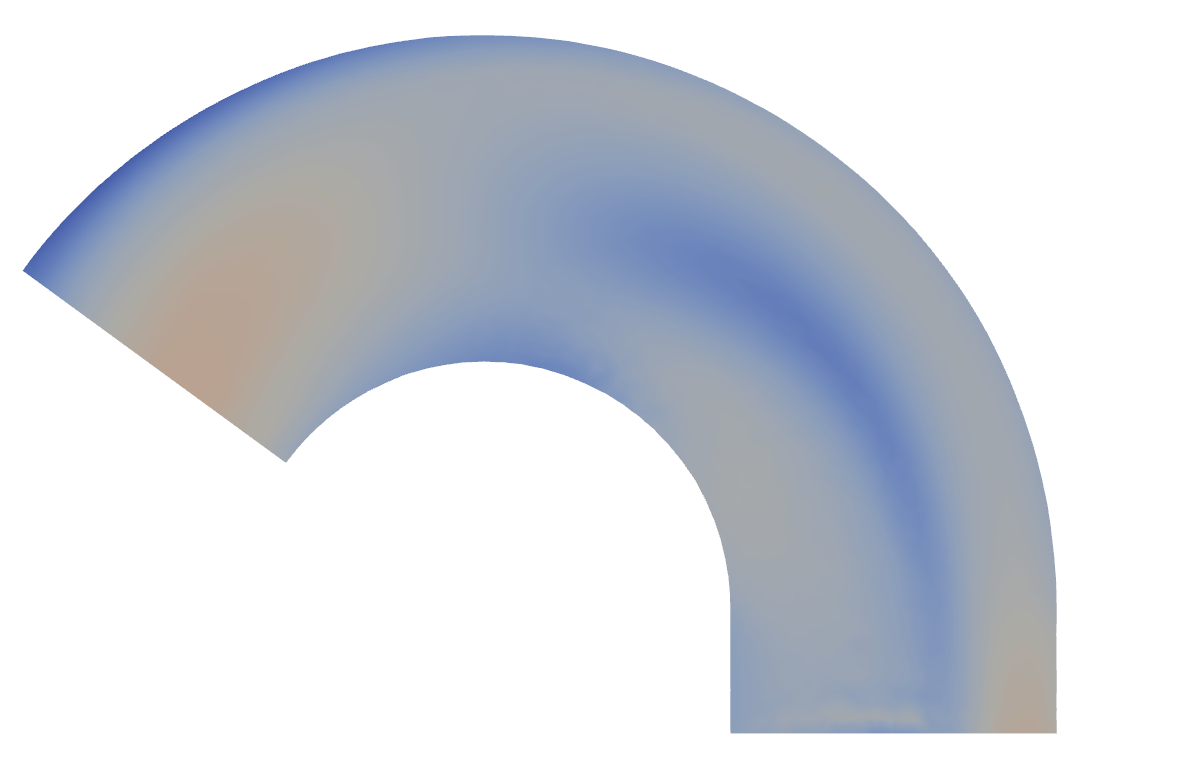} & 
     \includegraphics[width=.2\textwidth]{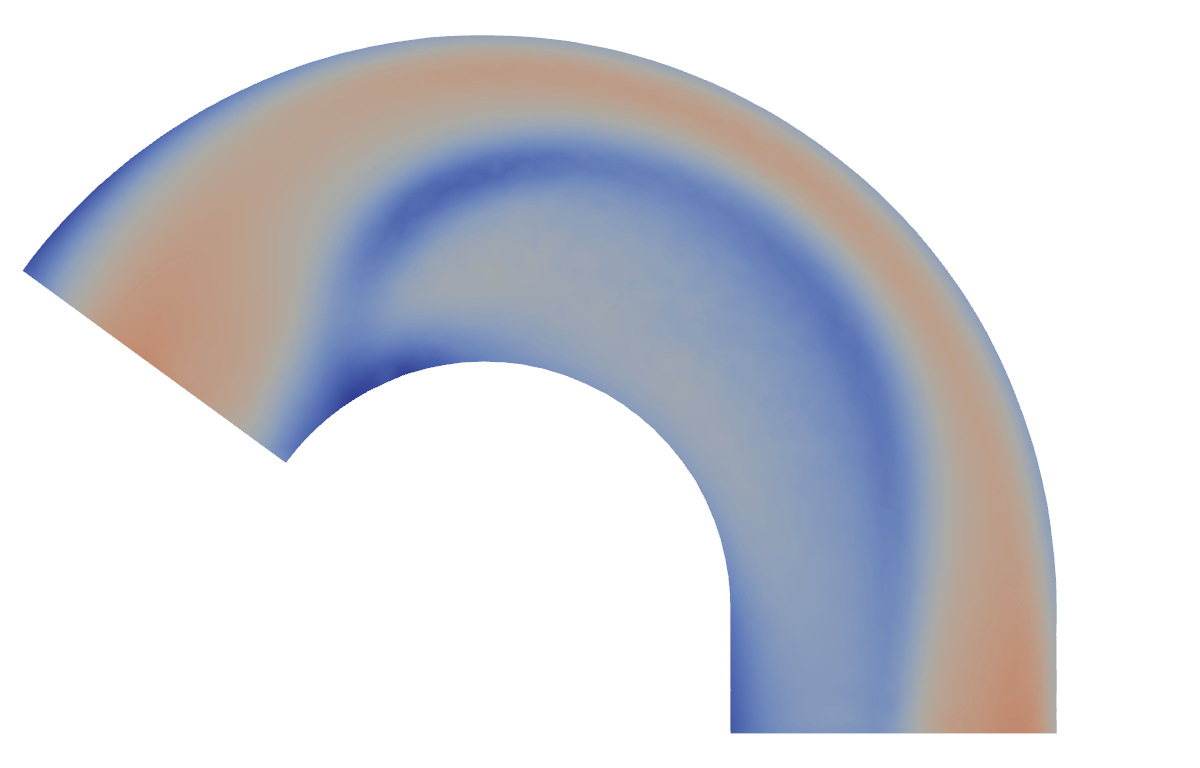} &
     \includegraphics[width=.2\textwidth]{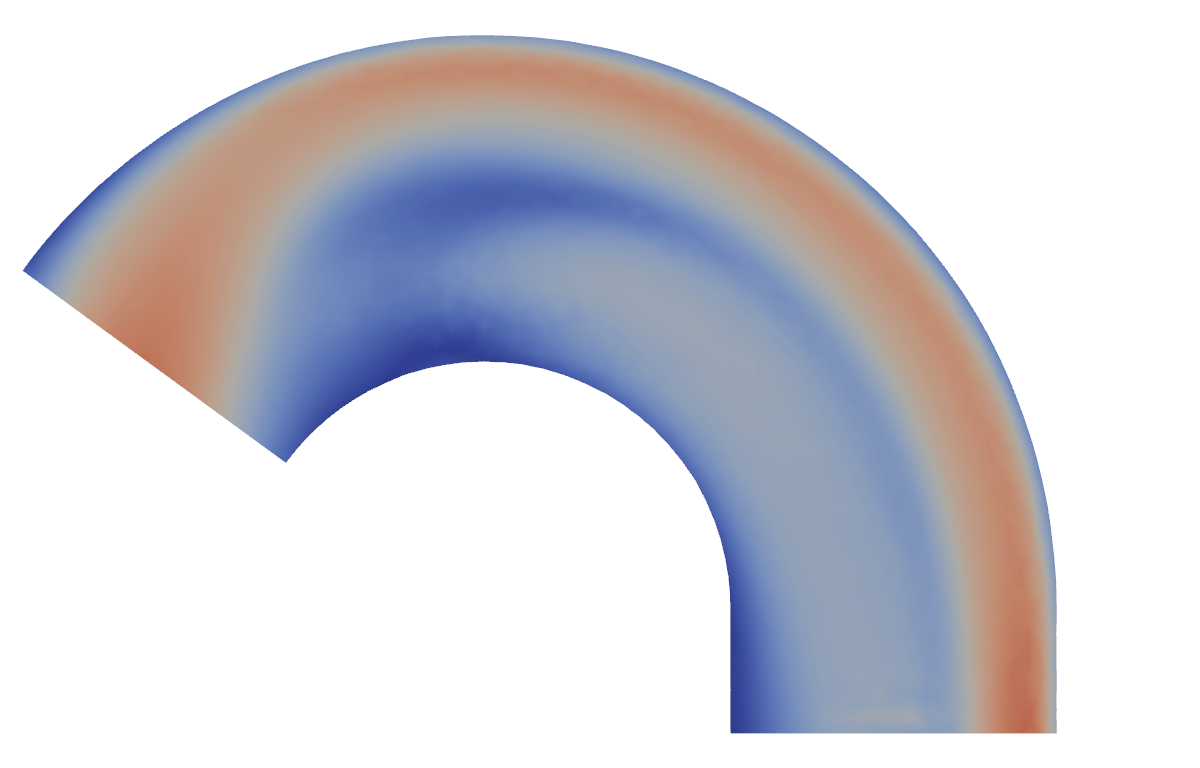} &
     \includegraphics[width=.2\textwidth]{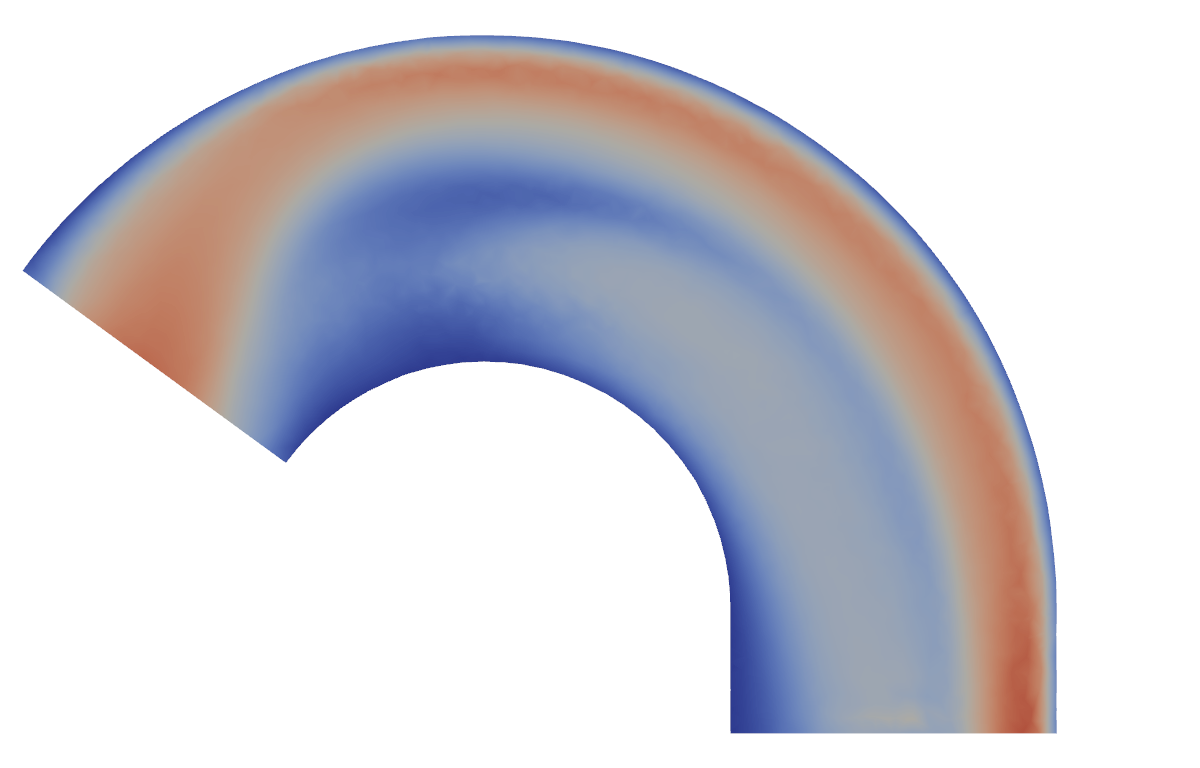} \\
     & $\thet{opt}=0.203$ & $\thet{opt}=0.51$ & $\thet{opt}=0.748$ & $\thet{opt}=0.881$ \\
     & $\mathcal{J}=\rnum{0.001381}$ & $\mathcal{J}=\rnum{0.006826}$ & $\mathcal{J}=\rnum{0.00941}$ & $\mathcal{J}=\rnum{0.01422}$ \\
     & $\mathcal{R}=\rnum{0.0009814}$ & $\mathcal{R}=\rnum{0.002753}$ & $\mathcal{R}=\rnum{0.003096}$ & $\mathcal{R}=\rnum{0.003395}$ \\
     & iterations: 56 & iterations: 83 & iterations: 64 & iterations: 66 \\
     & \multicolumn{4}{c}{\includegraphics[width=0.5\textwidth]{scale0.625.png}}\\
\end{tabular}
\caption{Comparison of data without noise with assimilation velocity results for $V=0.25$ using the stabilized $P_1/P_1$ element with $\alpha_v=\alpha_p=0.01$ on the arch geometry for multiple values of $\theta$.}
\label{fig:velocities_arch}
\end{figure}

Since the motivation of this work is to apply the method to magnetic resonance images, it was tested for multiple velocities typical for blood flow in the descending aorta.
Using the parameters from Table \ref{tab:parameters}, we can compute the Reynolds number as
$$Re = \frac{L\,\rho\,V}{\mu},$$
where $L$ is the characteristic length of the geometry and $V$ is the characteristic flow speed. 
Table \ref{tab:reynolds} lists Reynolds numbers corresponding to different points in the cardiac pulse cycle.
\begin{table}
    \centering
    \begin{tabular}{c c c c}
        \hline\noalign{\smallskip}
        & diastole & average & peak systole \\
        \noalign{\smallskip}\hline\noalign{\smallskip}
        $V$ & 0.1$\frac{\text{m}}{\text{s}}$ & 0.25$\frac{\text{m}}{\text{s}}$ & 0.8$\frac{\text{m}}{\text{s}}$\\
        Re & 270 & 674 & 2156\\
        \noalign{\smallskip}\hline
    \end{tabular}
    \caption{Reynolds numbers for different inflow velocities}
    \label{tab:reynolds}
\end{table}
The assimilation results for the higher velocities are shown in Figures \ref{fig:velocities_bent} and \ref{fig:velocities_arch}.
When increasing the velocity, the main obstacle is the convergence of the nonlinear solver, especially for more complicated geometries, where more intricate flow patterns might appear.
In such a case, using more suitable finite element stabilization and more robust algorithms for nonlinear problems based on continuation or Picard's iterations might be necessary.
Apart from the convergence issues, we observed similar phenomena as described in the previous sections.

\subsection{Pressure reconstruction}
\begin{figure}
\centering
\begin{tabular}{c c c c c}
     \rotatebox{90}{ground truth} & 
     \includegraphics[width=.2\textwidth]{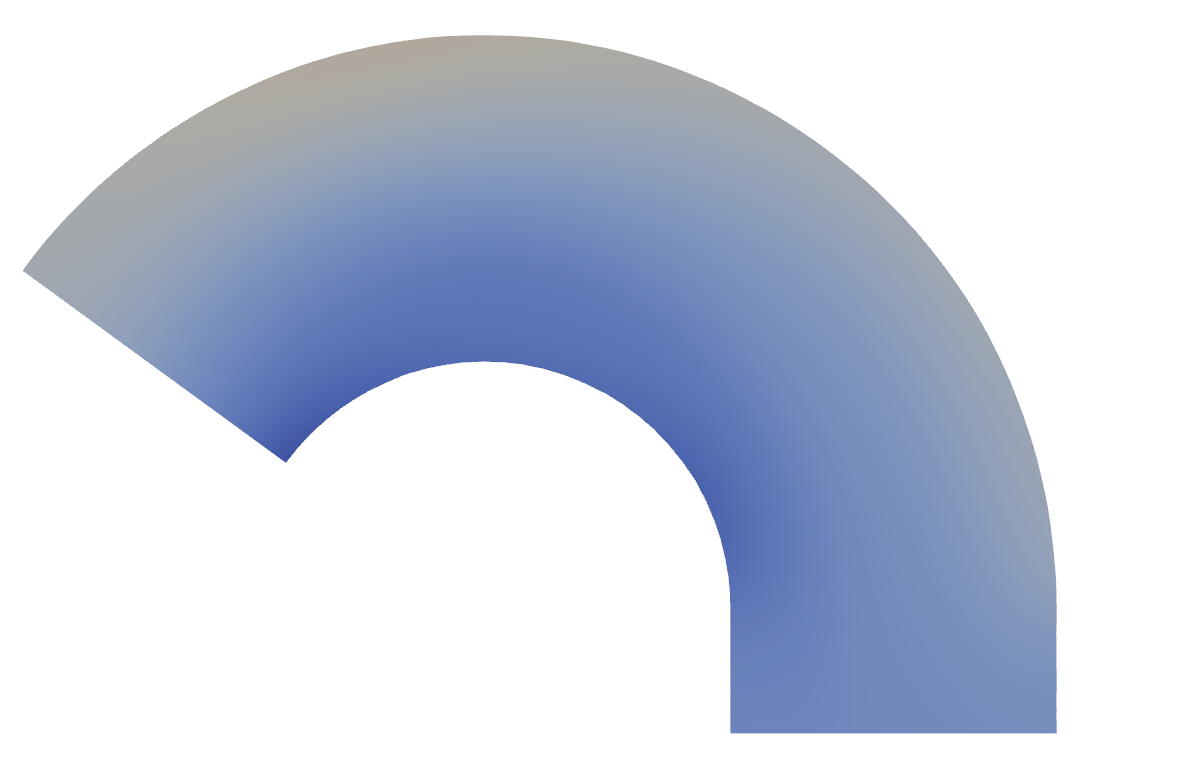} &
     \includegraphics[width=.2\textwidth]{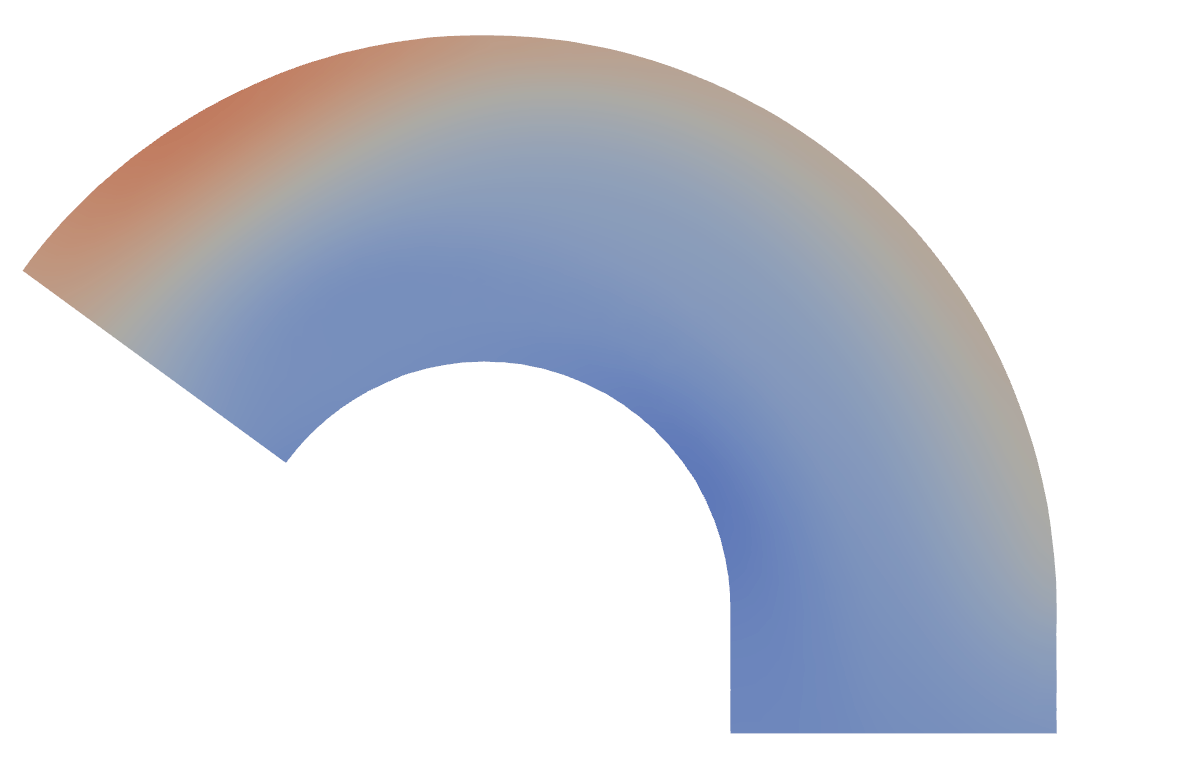} &
     \includegraphics[width=.2\textwidth]{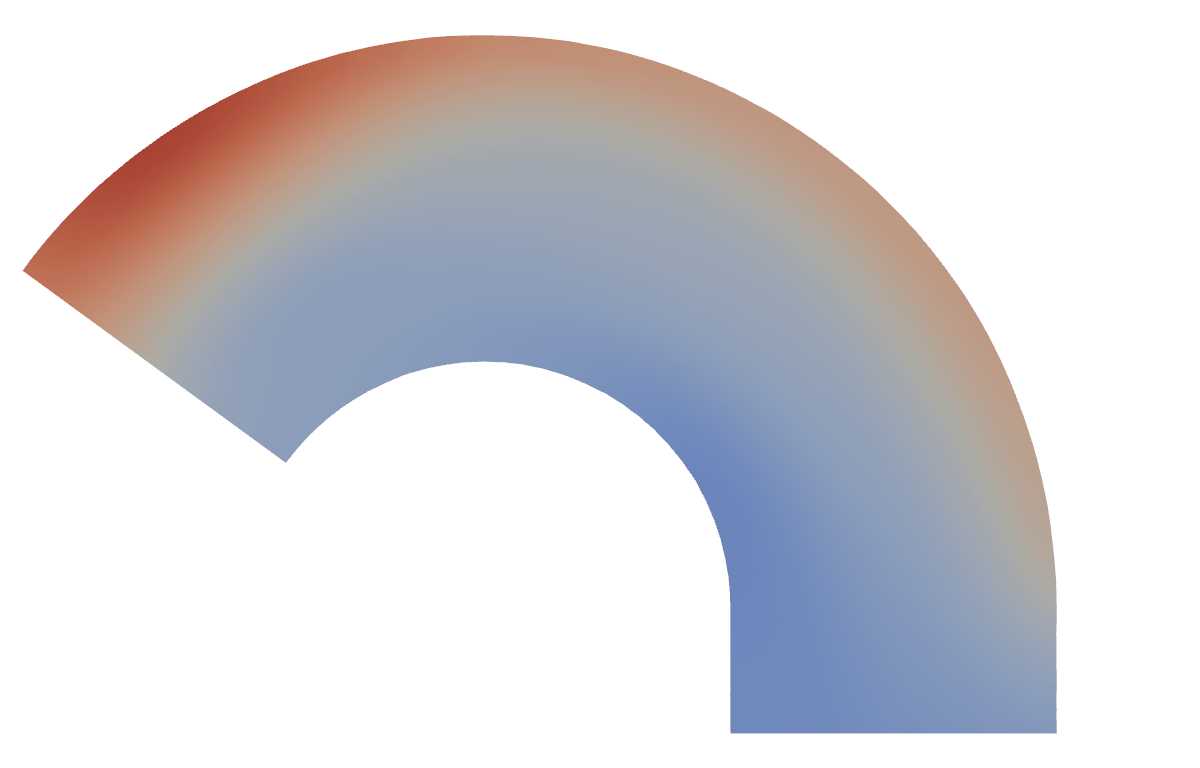} &
     \includegraphics[width=.2\textwidth]{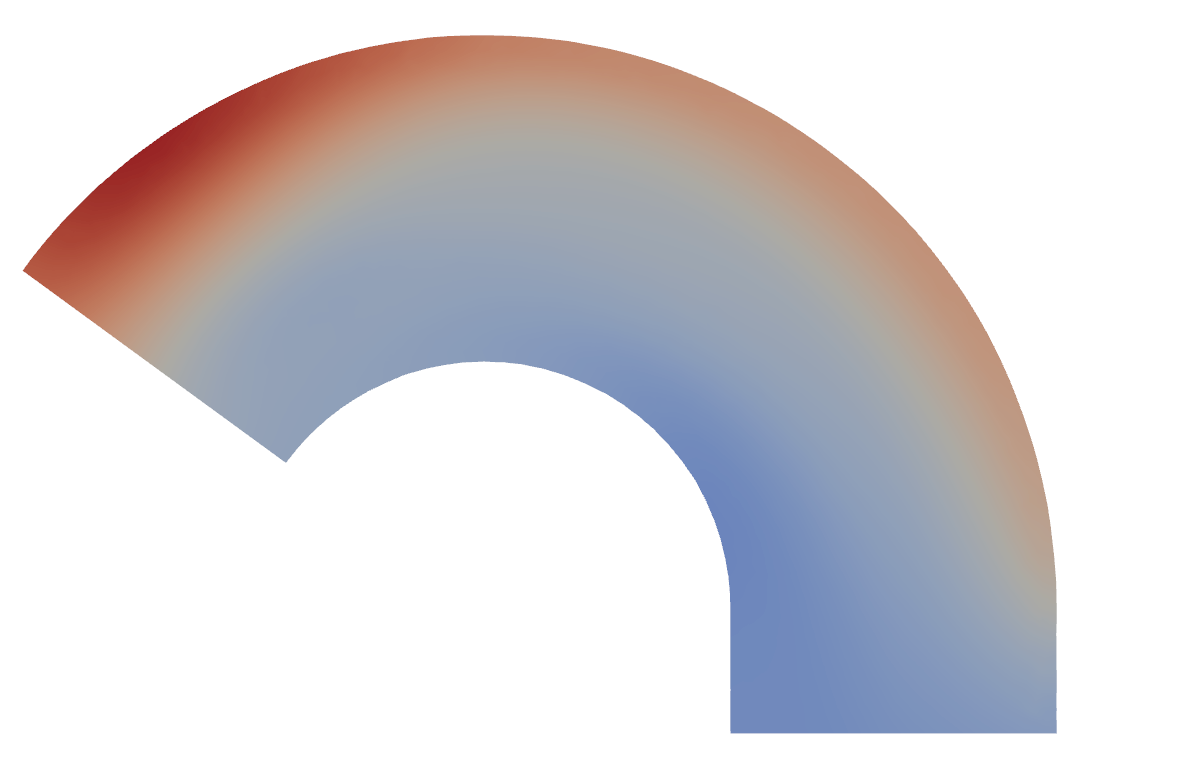} \\
     & $\theta=0.2$ & $\theta=0.5$ & $\theta=0.8$ & $\theta=1.0$ \\ 
     \rotatebox{90}{assimilation} &
     \includegraphics[width=.2\textwidth]{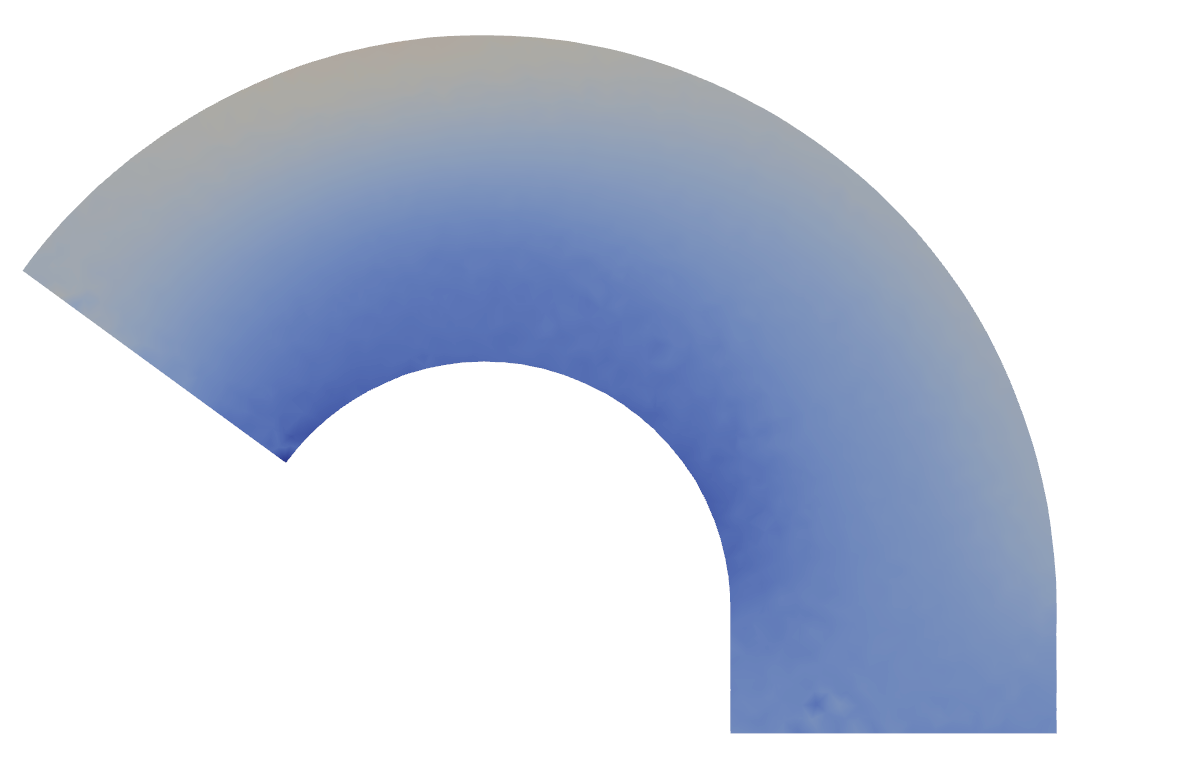} & 
     \includegraphics[width=.2\textwidth]{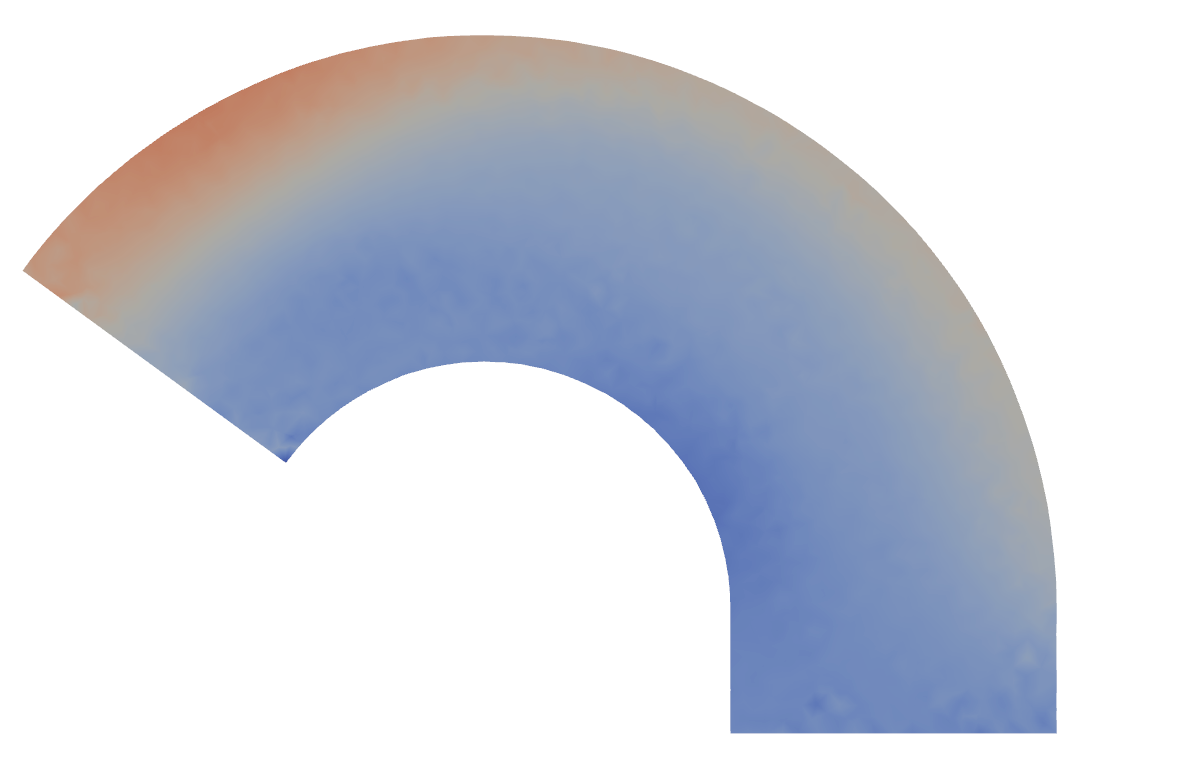} &
     \includegraphics[width=.2\textwidth]{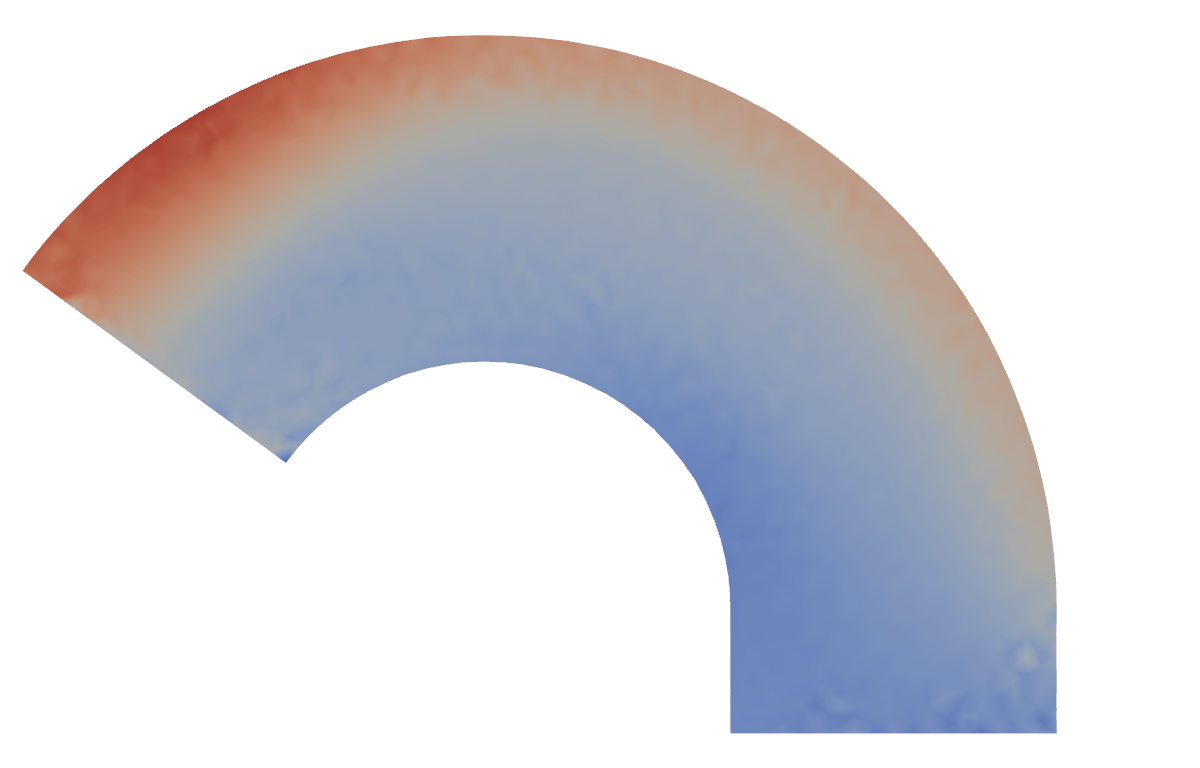} &
     \includegraphics[width=.2\textwidth]{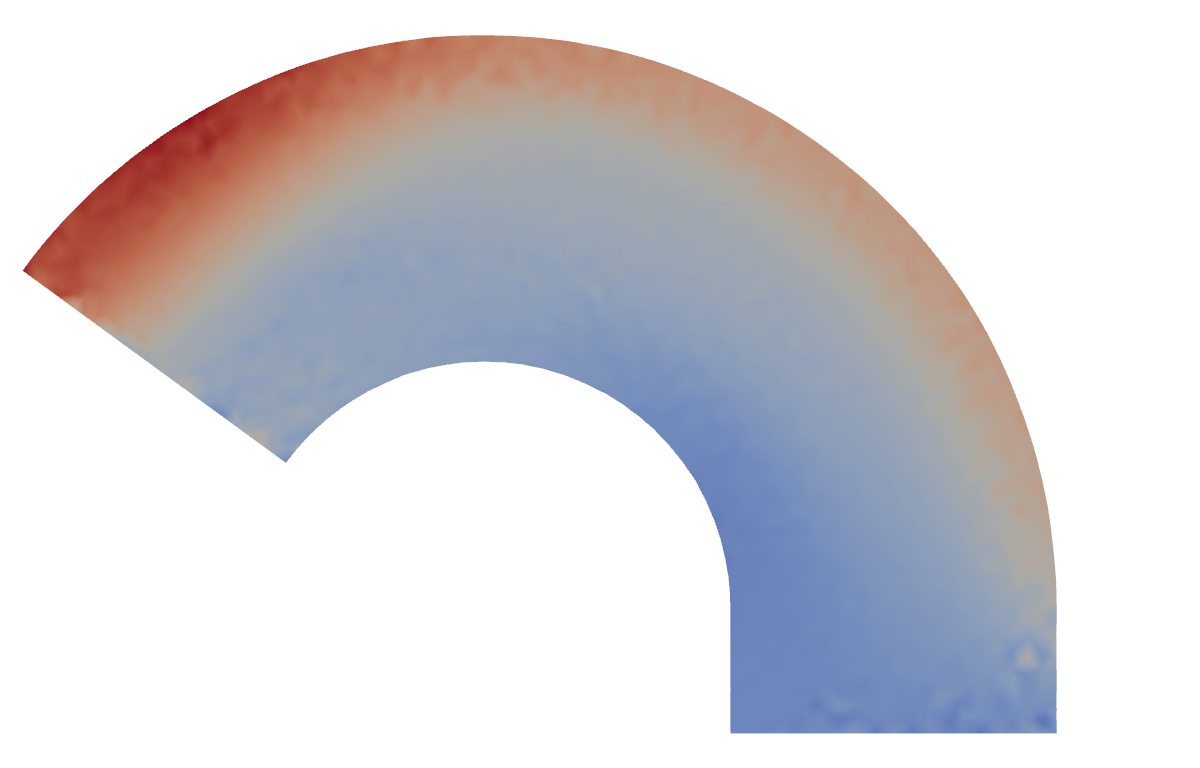} \\
     & $P_{err} = 0.082$ & $P_{err} = 0.11$ & $P_{err} = 0.094$ & $P_{err} = 0.097$ \\
    & \multicolumn{4}{c}{\includegraphics[width=0.5\textwidth]{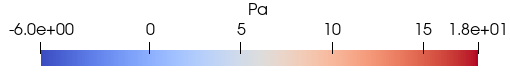}}
\end{tabular}
\caption{Comparison of the reference and reconstructed pressure field using stabilized $P_1/P_1$ element with $\alpha_v=\alpha_p=0.01$ on the arch geometry with edge length $h=1.5\text{ mm}$ for multiple values of $\theta$. The error of pressure reconstruction is measured using a quantity $P_{err}$ defined as $\frac{||p_{opt}-p_{ref}||_{L^2(\Omega)}}{||p_{ref}||_{L^2(\Omega)}}$.}
\label{fig:pressure}
\end{figure}

Since the assimilation uses only the velocity field data in the error functional $\mathcal{J}$, this method can be used to reconstruct the pressure field as a byproduct. 
We compared the reconstructed pressure fields with the ground truth pressure obtained in section \ref{sec:reference velocity fields} and interpolated to the shorter mesh. 
The results are presented in Figure \ref{fig:pressure}, where we can observe that the method was able to reconstruct the pressure field very close to the ground truth. This suggests that this method could be used as an alternative for pressure reconstruction from velocity measurements as in \cite{Svihlov2016}.

\subsection{Analytical tests}
The numerical settings used in this work were also tested using the methodology proposed by Málek and Rajagopal in \cite{MR2}. 
Using a numerical solution of \eqref{weak_form} computed on a $0.1\,\text{m}$ long cylinder with radius $R=0.01\,{\text{m}}$, the volumetric flow rate $Q$ and negative pressure gradient $c$ were computed using the following relations:
\begin{equation}
    Q = \int_{\bndry{out}}\mathbf{v}\cdot\mathbf{n}\,\dx{s}, \quad 
    c = \frac{1}{|\Omega|}\left(\int_{\bndry{out}} p\,\dx{s}-\int_{\bndry{in}} p\,\dx{s}\right).
\end{equation}
As derived in \cite[Sect.~III]{MR2} for Poiseuille flow in a pipe, assuming Navier's slip condition on the impermeable wall, the Navier-Stokes fluid exhibits slip along the wall if the following criterion is fulfilled:
\begin{equation}
    Q > -\frac{c\pi}{8\mu}R^4.
\end{equation}
In addition, see \cite[formula (3.7)]{MR2}, the appropriate slip parameter $\theta$ can be then computed using the relations
\begin{equation}\label{eq: kappa}
    \frac{\theta}{\gamma_*(1-\theta)}=\kappa = \frac{1}{2}\frac{-cR}{\frac{Q}{\pi R^2}+\frac{c}{8\mu}R^2}.
\end{equation}
The comparison of the prescribed slip parameter $\theta$ and its value computed using \eqref{eq: kappa} is presented in Table \ref{tab: thetas}. 
The results show that the selected numerical solutions can capture the relations between the macroscopic quantities well. We also observed that the amount of the finite element stabilization can significantly influence the computed pressure field up to the point where \ref{eq: kappa} no longer holds. 

\begin{table}
    \centering
    \begin{tabular}{l c c c c c}
        \hline\noalign{\smallskip}
         numerical setting & $\theta=0.0$ & $\theta=0.2$ & $\theta=0.5$ & $\theta=0.8$ & $\theta=1.0$ \\
        \noalign{\smallskip}\hline\noalign{\smallskip}
        MINI, $h=1.5\,\text{mm}$ & \rnum{-0.002319993823040269} & \rnum{0.19605029729733556} & \rnum{0.49358901834903907} & \rnum{0.7912218735561806} & \rnum{0.9903478378582109} \\
        MINI, $h=1\,\text{mm}$ & \rnum{-0.001264663864892128} & \rnum{0.19714571244244442} & \rnum{0.49396754708645113} & \rnum{0.7899578131293274} & \rnum{0.9861230385675827} \\
        stab. $P_1/P_1$, $h=1.5\,\text{mm}$ & \rnum{-0.024143199276684988} & \rnum{0.14029829358438736} & \rnum{0.4102090436941758} & \rnum{0.7781413654120942} & \rnum{1.0} \\
        stab. $P_1/P_1$, $h=1\,\text{mm}$ & \rnum{-0.009997807501831385} & \rnum{0.17377579736428153} & \rnum{0.4586249829873764} & \rnum{0.7757886815076586} & \rnum{0.9932946633526345} \\
        stab. $P_1/P_1$, $h=0.8\,\text{mm}$ & \rnum{-0.006544175379058339} & \rnum{0.18272574493638244} & \rnum{0.47067212975609013} & \rnum{0.7738559483524639} & \rnum{0.974602848076632} \\
        \noalign{\smallskip}\hline
    \end{tabular}
    \caption{The comparison of the prescribed slip parameter $\theta$ and its value computed using \eqref{eq: kappa} on a straight tube using different numerical settings.}
    \label{tab: thetas}
\end{table}

\section{Conclusion and future work}\label{sec:conclusion}
We developed and implemented a variational assimilation method to reconstruct the boundary condition parameters $\vin$ and $\theta$ from 3D measured velocity data for Navier--Stokes fluid with Navier's slip boundary condition on the wall. 
The method was tested on three artificial geometries that correspond to real-patient geometries that we intend to study.
Several experiments were performed to study the impact of the choice of finite element, amount of regularization and stabilization, mesh density, amount of noise and velocity magnitude. 
The aim of this study was to assess the robustness of the method with respect to the numerical setting for future use on real patient magnetic resonance data.

We observed that using the interior penalty stabilization resulted in a worse fit of the slip parameter due to the additional diffusion introduced by the stabilization, which affected the cases where $\theta$ is closer to 1 in particular.
On the other hand, stabilized $P_1/P_1$ element proved to be significantly cheaper and more stable computationally and, therefore, more suitable for larger and more complicated problems for which element like MINI or Taylor Hood might be too computationally expensive.
We also observed that the fit of $\theta$ improved with increased mesh density, enabling the stabilized $P_1/P_1$ element to recover more details in the flow pattern.
The experiments also showed that the method is not very sensitive to the amount of Gaussian noise in the data.
The effects of Tikhonov regularization $\mathcal{R}(\mathbf{m})$ were discussed, and a way to choose the regularization weights was proposed.
The method was also tested for higher velocities, which brought complications connected with the higher Reynolds number. 
However, we observed similar effects as in the lower velocity regime.
We also discussed the pressure reconstruction capability of the method. 
The results showed that the pressure was reconstructed well for artificially generated data even with stabilized $P_1/P_1$ elements and a relatively coarse mesh if the amount of stabilization is sufficiently small.

One of the limitations of the method is that the nonlinear solver is not guaranteed to converge, especially in the early stage of the optimization process. 
This can be partially salvaged by providing a suitable initial guess for the control variables or by using continuation and Picard's iteration instead.
Especially for the higher flow velocities it could be improved further by incorporating a more suitable stabilization such as SUPG/PSPG.


\section{Appendix}
The adjoint equations have to be assembled in order to compute the gradient of $\mathcal{J}_R$ with respect to the control variables $\theta$ and $\vin$.
This is done automatically using the dolfin--adjoint library \cite{Mitusch2019}.
However, we derived the adjoint equations by hand as well.
The adjoint equations for \eqref{weak_form} including the stabilization terms \eqref{ip stabilization} are as follows: 
\begin{align}\label{model adjoint equations}
    \begin{split}
        &\text{Find }(\xi, \eta)\in V_h\times P_h \text{ such that}\\
        &\int_\Omega \rho\left((\nabla\mathbf{v})\phi 
        + (\nabla\phi)\mathbf{v}\right)\cdot\xi\,\dx{x} 
        + \int_\Omega\mathbb{T}(\phi, q):\nabla \xi\,\dx{x}
        + \int_\Omega\eta\,\text{div}\,\phi\,\dx{x}\\
        &+\int_{\bndry{wall}}\frac{\theta}{\gamma_*(1-\theta)}\,\phi_\text{t}\cdot\xi_\text{t}\,\dx{s}
        -\int_{\bndry{wall}}(\mathbb{T}(\phi, q)\,\mathbf{n})_\text{n}\cdot\mathbf{\xi}_\text{n}\dx{s}\\
        &+\int_{\bndry{wall}}\phi_{\text{n}}\cdot(\mathbb{T}(\xi, \eta)\,\mathbf{n})_\text{n}\dx{s}+\frac{\beta_*\mu}{h}\int_{\bndry{wall}}\phi_{\text{n}}\cdot\xi_{\text{n}}\dx{s}\\
        &-\int_{\bndry{out}}\rho\,(\mathbf{v}\cdot\mathbf{n})_-(\phi\cdot\mathbf{n})(\xi\cdot\mathbf{n})\,\dx{s}\\
        &-\int_{\bndry{out}}(\mathbb{T}(\phi, q)\,\mathbf{n})_\text{t}\cdot\xi_\text{t}\dx{s}
        +\int_{\bndry{out}}\phi_{\text{t}}\cdot(\mathbb{T}(\xi, \eta)\,\mathbf{n})_\text{t}\dx{s}\\
        &+\frac{\beta_*\mu}{h}\int_{\bndry{out}}\mathbf{\phi}_{\text{t}}\cdot\mathbf{\xi}_{\text{t}}\dx{s}
        + \alpha_v\sum_{K\in\mathcal{F}}\int_K \rho h^2 [\nabla\phi]\cdot[\nabla\xi]\dx{s} \\
        &+ \alpha_p\sum_{K\in\mathcal{F}}\int_K \frac{h^2}{\rho} [\nabla q]\cdot[\nabla \eta]\dx{s}
        =\frac{1}{\ell^3}\int_\Omega(\mathbf{v}-\mathbf{d}_\text{MRI})\cdot\phi\dx{x}\\
        &\text{for all }(\phi, q)\in V_h\times P_h,
    \end{split}
\end{align}
where $V_h\subset\{ \mathbf{v}\in H^1(\Omega_h), \mathbf{v} = 0 \text{ on }\bndry{in} \}$ and $P_h\subset L^2(\Omega_h)$. 
\printbibliography

\end{document}